\documentclass[longauth,useAMS,usenatbib]{aa}  

\usepackage{graphicx}
\usepackage{placeins}
\usepackage{txfonts}
\usepackage{hyperref}
\usepackage{longtable}
\usepackage{appendix}
\newcommand{\kms}{km\,s$^{-1}$}
\newcommand{\ms}{m\,s$^{-1}$}
\newcommand{\cms}{cm\,s$^{-1}$}
\newcommand{\mearth}{$M_\oplus$}
\newcommand{\searth}{$S_\oplus$}
\newcommand{\msun}{$M_\odot$}
\newcommand{\rsun}{$R_\odot$}
\newcommand{\lsun}{$L_\odot$}

\begin{document} 

\title{A sub-Earth-mass planet orbiting Barnard's star \thanks{Based [in part] on Guaranteed
Time Observations collected at the European Southern Observatory under ESO programmes 
1102.C-0744, 1104.C-0350, 106.21M2.001, 106.21M2.004, 106.21M2.006, 108.22GM.001, 108.2254.001, 108.2254.003, 108.2254.004, 108.2254.006, 110.24CD.001, 110.24CD.003 by the ESPRESSO Consortium.} 
\thanks{Tables~\ref{tab:espdataset} and~\ref{tab:alldataset} are only available in electronic form at the CDS via anonymous ftp to cdsarc.u-strasbg.fr (130.79.128.5) or via \url{http://cdsweb.u-strasbg.fr/cgi-bin/qcat?J/A+A/}}
}

\author{J.~I.~Gonz\'alez~Hern\'andez \inst{\ref{iac},\ref{ull}}
    \and
    A. Su\'arez Mascare\~no \inst{\ref{iac},\ref{ull}}
    \and
    A. M. Silva \inst{\ref{IAporto},\ref{uniporto}}
    \and
    A. K. Stefanov \inst{\ref{iac},\ref{ull}}
    \and
    J. P. Faria \inst{\ref{unige},\ref{IAporto},\ref{uniporto}}
    \and
    H. M. Tabernero \inst{\ref{ucm}}
    \and
    A. Sozzetti \inst{\ref{inaf-torino}}
    \and
    R. Rebolo \inst{\ref{iac},\ref{ull},\ref{csic}}
    \and
    F. Pepe \inst{\ref{unige}}
    \and
    N. C. Santos \inst{\ref{IAporto},\ref{uniporto}}
    \and
    S. Cristiani \inst{\ref{inaf-trieste},\ref{unitrieste}}
    \and
    C. Lovis \inst{\ref{unige}}
    \and
    X. Dumusque \inst{\ref{unige}}
    \and
    P. Figueira \inst{\ref{unige},\ref{IAporto}}
    \and
    J. Lillo-Box \inst{\ref{cab}}
    \and
    N. Nari \inst{\ref{iac},\ref{lbridges},\ref{ull}}
    \and
    S. Benatti \inst{\ref{inaf-palermo}}
    \and
    M. J. Hobson \inst{\ref{unige}}
    \and
    A. Castro-Gonz\'alez \inst{\ref{cab}}
    \and
    R. Allart\inst{\ref{montreal},\ref{unige}}
    \and
    V. M. Passegger \inst{\ref{subaru},\ref{iac},\ref{ull},\ref{sw}}
    \and
    M.-R. Zapatero Osorio \inst{\ref{cab}}
    \and
    V. Adibekyan \inst{\ref{IAporto},\ref{uniporto}}
    \and
    Y. Alibert \inst{\ref{bern},\ref{bern2}}
    \and
    C. Allende Prieto \inst{\ref{iac},\ref{ull}}
    \and
    F. Bouchy \inst{\ref{unige}}
    \and
    M. Damasso \inst{\ref{inaf-torino}}
    \and
    V. D'Odorico \inst{\ref{inaf-trieste},\ref{unitrieste}}
    \and
    P. Di Marcantonio \inst{\ref{inaf-trieste}}
    \and
    D. Ehrenreich \inst{\ref{unige}}
    \and
    G. Lo Curto \inst{\ref{eso}}
    \and
    R. G\'enova Santos \inst{\ref{iac},\ref{ull}}
    \and
    C. J. A. P. Martins \inst{\ref{IAporto},\ref{caup}}
    \and
    A. Mehner \inst{\ref{eso}}
    \and
    G. Micela \inst{\ref{inaf-palermo}}
    \and
    P. Molaro \inst{\ref{inaf-trieste}}
    \and
    N. Nunes \inst{\ref{lisbon}}
    \and
    E. Palle \inst{\ref{iac},\ref{ull}}
    \and
    S. G. Sousa \inst{\ref{IAporto},\ref{uniporto}}
    \and
    S. Udry \inst{\ref{unige}}
           }

\institute{Instituto de Astrof{\'i}sica de Canarias, E-38205 La Laguna, Tenerife, Spain \label{iac}\\
           \email{jonay.gonzalez@iac.es}
           \and
           Departamento de Astrof{\'i}sica, Universidad de La Laguna, E-38206 La Laguna, Tenerife, Spain \label{ull}
           \and
           Consejo Superior de Investigaciones Cient\'{i}ficas, Spain \label{csic}
           \and
           Observatoire de Gen\`eve, D\'epartement d'Astronomie, Universit\'e de Genève, Chemin Pegasi 51b, 1290 Versoix, Switzerland \label{unige}
           \and
           Instituto de Astrof\'{i}sica e Ci\^encias do Espa\~co, CAUP, Universidade do Porto, Rua das Estrelas, 4150-762, Porto, Portugal \label{IAporto}
           \and
           Departamento de F\'{i}sica e Astronomia, Faculdade de Ci\^encias, Universidade do Porto, Rua do Campo Alegre, 4169-007, Porto, Portugal \label{uniporto}
           \and
           Centro de Astrof\'{\i}sica da Universidade do Porto, Rua das Estrelas, 4150-762 Porto, Portugal \label{caup}
           \and
           Centro de Astrobiolog\'{i}a (CAB), CSIC-INTA, ESAC campus, Camino Bajo del Castillo s/n, E-28692, Villanueva de la Ca\~nada (Madrid), Spain \label{cab}
           \and
           INAF - Osservatorio Astronomico di Trieste, via G. B. Tiepolo 11, I-34143, Trieste, Italy \label{inaf-trieste}
           \and
           IFPU--Institute for Fundamental Physics of the Universe, via Beirut 2, I-34151 Trieste, Italy \label{unitrieste}
           \and
           INAF - Osservatorio Astrofisico di Torino, Strada Osservatorio 20, I-10025, Pino Torinese (TO), Italy \label{inaf-torino}
           \and
           ESO - European Southern Observatory, Av. Alonso de Cordova 3107, Vitacura, Santiago, Chile \label{eso}
           \and
           D\'epartement de Physique, Institut Trottier de Recherche sur les Exoplan\`etes, Universit\'e de Montr\'eal, Montr\'eal, Qu\'ebec, H3T 1J4, Canada \label{montreal}
           \and
           INAF - Osservatorio Astronomico di Palermo, Piazza del Parlamento 1, I-90134 Palermo, Italy \label{inaf-palermo}
           \and
           Instituto de Astrof\'{i}sica e Ci\^{e}ncias do Espa\c{C}o, Faculdade de Ci\^{e}ncias da Universidade de Lisboa, Campo Grande, 1749-016, Lisboa, Portugal \label{lisbon}
           \and
           Light Bridges S. L., Avda. Alcalde Ram{\'i}rez Bethencourt, 17, E-35004, Las Palmas de Gran Canaria, Canarias, Spain \label{lbridges}
           \and
           Departamento de F{\'i}sica de la Tierra y Astrof{\'i}sica \& IPARCOS-UCM (Instituto de F\'{i}sica de Part\'{i}culas y del Cosmos de la UCM), Facultad de Ciencias F{\'i}sicas, Universidad Complutense de Madrid, E-28040 Madrid, Spain \label{ucm}
           \and
           Center for Space and Habitability, University of Bern, Gesellschaftsstrasse 6, 3012 Bern, Switzerland \label{bern}
           \and
           Weltraumforschung und Planetologie, Physikalisches Institut, University of Bern, Gesellschaftsstrasse 6, 3012 Bern, Switzerland \label{bern2}
           \and
           Subaru Telescope, National Astronomical Observatory of Japan, 650 N Aohoku Place, Hilo, HI 96720, USA \label{subaru}
           \and
           Hamburger Sternwarte, Gojenbergsweg 112, 21029 Hamburg, Germany \label{sw}
           }

\date{Written from March to June 2024}


\abstract
{ESPRESSO guaranteed time observations (GTOs) at the 8.2m VLT telescope were performed to look for Earth-like exoplanets in the habitable zone of nearby stars. Barnard's star is a primary target within the ESPRESSO GTO as it is the second closest neighbour to our Sun after the $\alpha$ Centauri stellar system.
}
{We present here a large set of 156 ESPRESSO observations of Barnard's star carried out over four years with the goal of exploring periods of shorter than 50 days, thus including the habitable zone (HZ).
}
{Our analysis of ESPRESSO data using Gaussian process (GP) to model stellar activity suggests a long-term activity cycle at 3200~d and confirms stellar activity due to rotation at 140~d as the dominant source of radial velocity (RV) variations. These results are in agreement with findings based on publicly available HARPS, HARPS-N, and CARMENES data. ESPRESSO RVs do not support the existence of the previously reported candidate planet at 233~d.
}
{After subtracting the GP model, ESPRESSO RVs reveal several short-period candidate planet signals at periods of 3.15~d, 4.12~d, 2.34~d, and 6.74~d. We confirm the 3.15~d signal as a sub-Earth mass planet, with a semi-amplitude of $55 \pm 7$ \cms, leading to a planet minimum mass $m_p \sin i$ of $0.37 \pm 0.05$ \mearth, which is about three times the mass of Mars. ESPRESSO RVs suggest the possible existence of a candidate system with four sub-Earth mass planets in circular orbits with semi-amplitudes from 20 to 47~\cms, thus corresponding to minimum masses in the range of 0.17--0.32~\mearth.
}
{The sub-Earth mass planet at $3.1533 \pm 0.0006$~d is in a close-to circular orbit with a semi-major axis of $0.0229 \pm 0.0003$~AU, thus located inwards from the HZ of Barnard's star, with an equilibrium temperature of 400~K. Additional ESPRESSO
observations would be required to confirm that the other three candidate signals originate from a compact short-period planet system orbiting Barnard's star inwards from its HZ.
}


\maketitle
%

\section{Introduction\label{sec:intro}}

The field of exoplanet science has been evolving very quickly in recent years towards the detection and characterisation of Earth-like exoplanets  thanks to the combined effort of space missions such as Kepler~\citep{bor09}, TESS~\citep{ric15}, and CHEOPS~\citep{ben21}, and ground-based high-resolution ultrastable spectrographs, such as HARPS~\citep{may03}, HARPS-N~\citep{cos12}, CARMENES~\citep{qui16}, and ESPRESSO~\citep{pep21}. In particular, the exoplanet community is already finding potentially habitable Earth-like planets~\citep[e.g.][]{gil17Natur,dit17Natur,zec19,lil20,sua23,cad24}, paving the path towards the detection of an Earth twin, the ultimate goal of the ESPRESSO project in the long term, and other projects such as the Terra Hunting Experiment with the upcoming HARPS3 spectrograph~\citep{tho16}. The discovery of the temperate Earth-mass planet Proxima Centauri b~\citep{ang16Natur} orbiting the closest star to our Sun has propelled exoplanet studies to focus the search on Earth-like planets in the habitable zone around stars of the solar neighbourhood. These temperate Earth-like planets will be the main targets of future facilities in the next decade, such as ANDES~\citep{mar22} at the Extremely Large Telescope (ELT) in Cerro Armazones within the European Southern Observatory (ESO), with the goal of studying their atmospheres to search for biomarkers using both transmission and reflected light spectroscopy~\citep{pal23}.

During the last decade, the blind radial velocity (RV) search for these Earth-mass exoplanets quickly shifted to the continuous monitoring of M dwarfs, with the development of new instruments in the near-infrared, such as CARMENES~\citep{rib23}, SPIRou~\citep{don20}, and NIRPS~\citep{bou17}. M dwarfs are the most common stellar type in the Galaxy, representing about 80\% of the stars in the solar neighbourhood~\citep{rey21}. M dwarfs are cooler, intrinsically less luminous, and less massive than Sun-like stars, and have habitable zones closer to their host star, making them ideal targets for blind RV searches of Earth-like planets. The ESPRESSO spectrograph at the Very Large Telescope (VLT, ESO) has had a significant impact in exoplanet science since it began regular operations at Paranal Observatory in October 2018~\citep{pep21,gon18}. ESPRESSO has demonstrated unprecedented capabilities, aiming at 10~\cms\ RV precision. ESPRESSO has confirmed, for instance, the Earth-mass planet Proxima b and discovered the sub-Earth Proxima d, a 0.26~\mearth\ planet (approximately twice the  mass of Mars) orbiting Proxima Centauri, from the measurement of a small RV semi-amplitude of $39\pm 7$~\cms~\citep{sua20,far22}. ESPRESSO is opening a new frontier at sub-m/s precision, making it possible to discover and characterise Earth- and sub-Earth-mass and sub-Earth size exoplanets in the solar neighbourhood. ESPRESSO has detected, for instance, the 0.4~\mearth\ planet L98-59~b (half of the mass of Venus) orbiting an M3V star~\citep{dem21}, and one super-Earth and two super-Mercuries HD 23472~d,e,f with masses of 0.54-0.76~\mearth\ orbiting a K4V star~\citep{bar22}.

Barnard's star (GJ 699) is the second closest stellar system to our Sun, after the $\alpha$ Centauri stellar system, and has been investigated in great detail since its discovery~\citep{barnard1916}. It is the nearest single star to our Sun, the closest M dwarf after Proxima Cen, at a distance of about 1.8 parsecs, and is the star with the highest proper motion~\citep{gaia21}, causing significant Doppler shifts due to secular acceleration~\citep{kur03}. \citet{fra20} measured the X-ray flux of GJ699 with the Chandra satellite in the energy range 0.3--10keV at $F_X \sim 4.8\times10^{-14}$ erg~cm$^{-2}$~s$^{-1}$ ($\log_{10}(L_X[{\rm erg\,s^{-1}}]) = 25.3$; $L_X/L_{\rm bol} = 1.6\times10^{-6}$). The X-ray luminosity is within a factor of two of previous ROSAT data ($\log_{10}(L_X[{\rm erg\,s^{-1}}]) = 25.6$). This low X-ray luminosity with $\log_{10} (L_X/L_{\rm bol}) \sim -5.8$~\citep{fra20} indicates a low level of current magnetic activity~\citep{ste13mnras}. Previous spectroscopic works revealed a low level of chromospheric activity with $\log_{10} (R'_{\rm HK}) \sim -5.8$~\citep{sua15,ast17rot,tol19}, suggesting a slow rotation period of $P_{\rm ROT} \sim 140$~d, which was estimated using equation 1 in ~\citet{sua18rot}. This value is in agreement with the photometric value of $P_{\rm ROT} \sim 130$~d derived from HST photometry~\citep{ben98}. \citet{tol19} reported a rotation period of $P_{\rm ROT}=145\pm15$~d from the time-series analysis of spectroscopic activity indexes, and also found evidence of a long-term activity cycle of Barnard's star from a time series of the CaHK index and ASAS-SN $m_V$ photometry with a periodicity of $P_{\rm CYC} \sim 3225-3850$~d. \citet{rei22} estimated a relatively low surface average magnetic field strength at $\langle B \rangle \sim 0.43\pm0.08 $~kG from spectral line fitting of the Zeeman broadening covering a wide range of different Land\'e-$g$ values, consistent with the star's low magnetic activity level, whereas \citet{cri23} found a lower value of $\langle B \rangle \sim 0.21\pm0.08$~kG. \citet{don23} measured longitudinal magnetic fields using SPIRou data and investigated its temporal variations to infer a  rotation period of $P_{\rm ROT}=136\pm16$~d in GJ~699, in agreement with previous estimates.

\citet{rib18} reported the discovery of a 3.3~\mearth\ super-Earth-like planet candidate orbiting Barnard's star with an orbital period of 233~d. This result was challenged by \citet{lub21}, who argues that the signal is transitory in nature and is connected to stellar activity, as the planet candidate period is close to 1 yr alias of the rotation period~\citep{tol19}. More recently, \citet{art22lbl} used SPIRou data to show that a model including a 233d planetary signal with a RV semi-amplitude of $K_p=1.2$~\ms\ is disfavoured when compared to a flat model.

Here we present the ESPRESSO observations of Barnard's star (GJ 699), showing a sub-m/s precision that reveals the presence of a short-period sub-Earth planet and three additional sub-Earth planet candidates. ESPRESSO data allow us to also evaluate the presence of the super-Earth planet candidate reported in \citet{rib18}.

\begin{figure*}
\includegraphics[width=18cm]{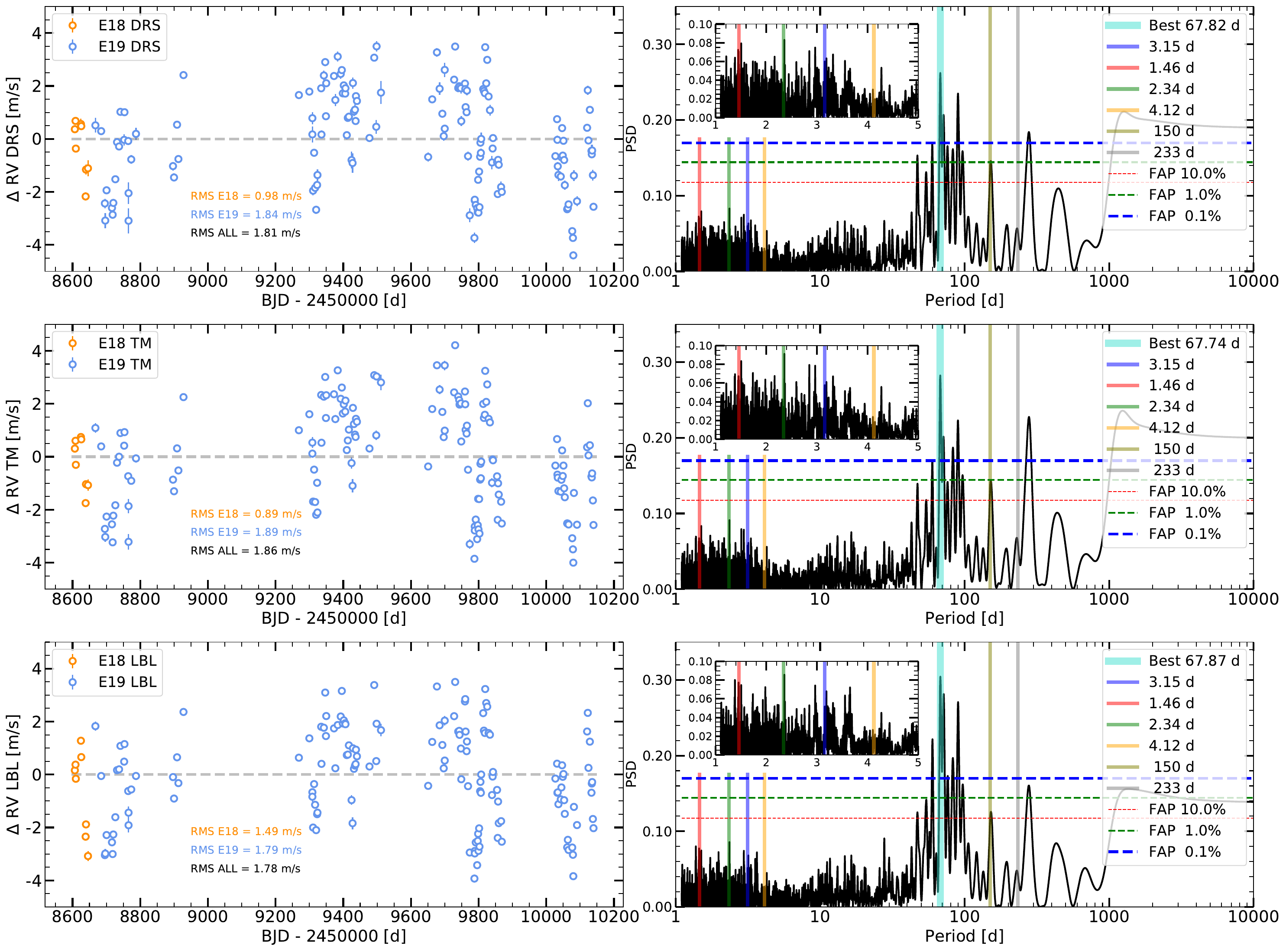}
\caption{
ESPRESSO RV measurements ({\it right}) and GLS periodograms ({\it left}) of GJ~699 after subtracting the median of each dataset before (E18) and after (E19) the intervention in June 2019. Also shown are RV measurements from the ESPRESSO Data Reduction Software (DRS; \textit{top}), from the \texttt{S-BART} template matching (TM) code~(\textit{middle}), and from the line-by-line \texttt{LBL} code (\textit{bottom}).
}
\label{gj699_rv}
\end{figure*}

\begin{figure*}
\includegraphics[width=18cm]{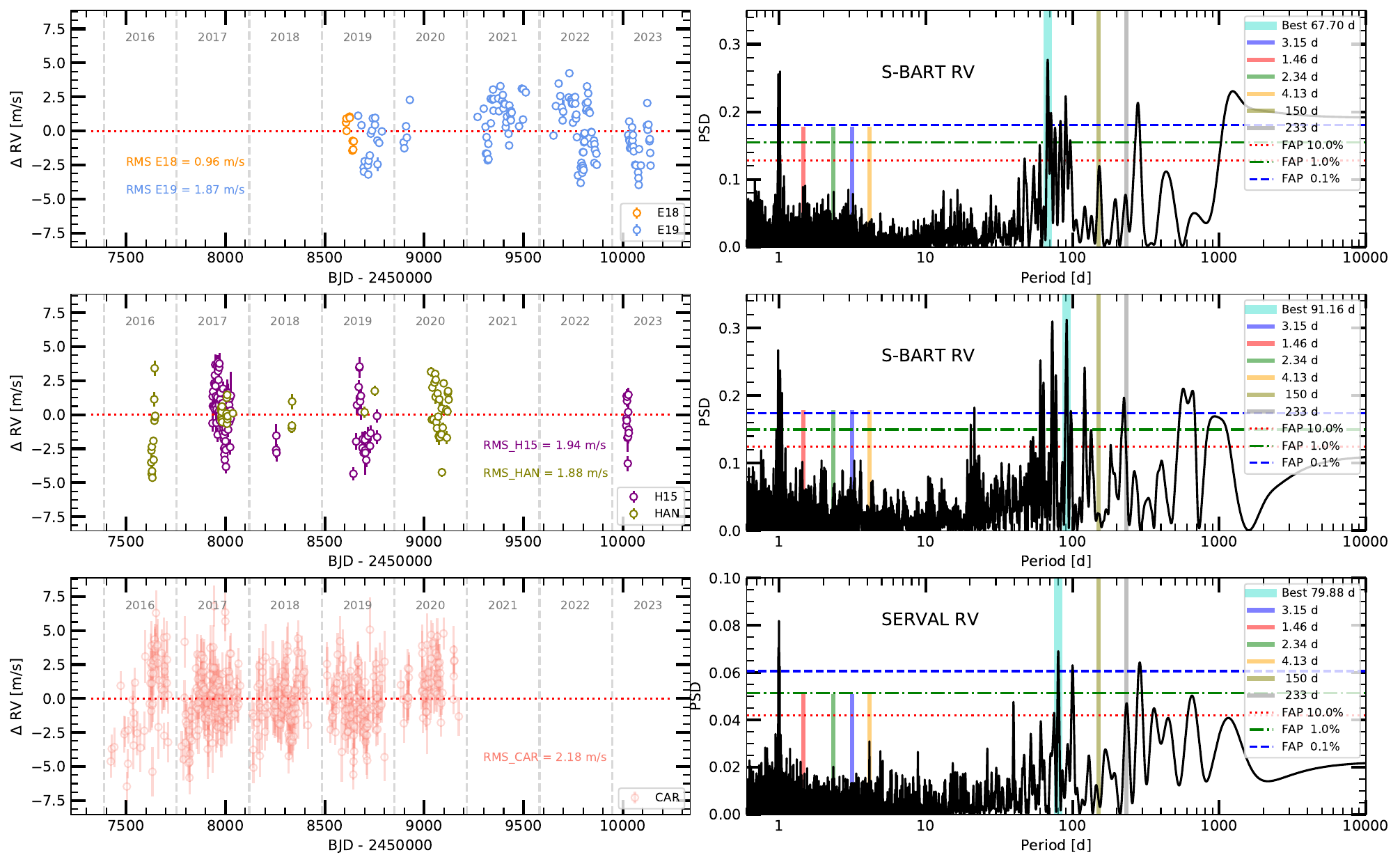}
\caption{
Radial-velocity measurements ({\it right}) and GLS periodograms ({\it left}) of GJ~699 of ESPRESSO ({\it top}), HARPS and HARPS-N ({\it middle}), and CARMENES ({\it bottom}).
}
\label{gj699_rv_all}
\end{figure*}

\section{Observations\label{sec:obs}}


The ESPRESSO
    consortium is the collaborative effort of Switzerland, Italy, Portugal, and Spain, with ESO as an associate partner, to develop, build, and scientifically exploit the ESPRESSO\footnote{\url{https://www.eso.org/sci/facilities/paranal/instruments/espresso.html}} instrument~\citep{pep21}. The ESPRESSO project has mainly been dedicated to the search for and characterisation of exoplanets~\citep[e.g.][]{lil21,far22, sua23,lav23,cas23,sua24} and exoplanet atmospheres~\citep[e.g.][]{ehr20,bor21,aze22}, and the measurement of fundamental constants of the Universe~\citep[e.g.][]{martins22,mur22}.

Barnard's star (GJ 699) is a (main) target of the guaranteed time observations (GTOs) of the ESPRESSO instrument. It was monitored over four years from May 2019 to July 2023. The main goal of the ESPRESSO GTO has been to search for rocky planets in the habitable zone (HZ) of nearby stars. Barnard's star is considered a primary target due to its proximity to our Sun, its relatively low magnetic activity, and the possibility to search for Earth-like planets within its HZ.

This work has also made use of public HARPS, HARPS-N, and CARMENES data, with some of the HARPS and HARPS-N spectra taken by consortium members as part of the follow-up of Barnard's star, as we describe below.

\subsection{ESPRESSO \label{sec:obs_esp}}

The Echelle SPectrograph for Rocky Exoplanets and Stable Spectroscopic Observations~\citep[ESPRESSO;][]{pep21} is a fibre-fed, cross-dispersed, high-resolution \'echelle spectrograph located in the Combined Coud\'e Laboratory (CCL) at the incoherent focus, where the Front-End unit can combine the light from up to four Unit Telescopes (UT) of the Very Large Telescope (VLT) at Paranal Observatory (ESO, Chile). The so-called Coud\'e train optical system feeds the light of each UT to the spectrograph. The Front-End corrects the light beam for atmospheric dispersion with the atmospheric dispersion corrector (ADC) and ensures the light is centered in the fibre with two independent pupil and field stabilisation units. The light from the target and the sky enter the instrument simultaneously through two separate fibres. ESPRESSO, unlike any other ESO instrument, is able to operate simultaneously with either one UT or several of the four 8.2-m UTs. The light of one or several UTs is fed through the Front-End unit into optical fibres that scramble the light within the Fiber-Link unit and provide excellent illumination stability to the spectrograph, using octogonal (1-UT) or square (4-UT) fibres. The instrument, aiming at a long-term 10~cm s$^{-1}$ RV stability, is temperature-controlled and pressure-stabilised within a vacuum vessel (VV). The reference fibre fed simultaneously with stabilised Fabry-P\'erot unit allows the tracking of instrument drifts down to the \cms\ level. In the most used {\it singleHR} (1-UT) mode, a fibre of 140~$\mu$m core, equivalent to 1\arcsec\ on the sky, provides a resolving power of $R\sim 138,000$ in the wavelength range 378--789~nm, sampling properly the resolution element with 4.5 pixels in two different detector binning setups HR11 and HR21.

We obtained 157 ESPRESSO observations of Barnard's star from May 2019 to July 2023. Nine of them were taken before the intervention done at the end of June 2019 to upgrade the fibre link, which increased the photon-detection efficiency reaching more than 10\% at seeing better than 0.75\arcsec~\citep{pep21}. This intervention introduced an RV offset, leading us to consider two separate E18 and E19 datasets at about BJD[d]~$=$~$2458660$. In March 2020, operations at Paranal were interrupted due to the COVID-19 pandemic and ESPRESSO was taken out of operations from March 2020 to December 2020, leading to a large gap after the first year of the ESPRESSO observing campaign. However, following our analysis of the ESPRESSO data of different RV standard stars, we conclude that the change in one of the calibration lamps after the ramp-up of the instrument at the end of 2020 does not justify any additional RV offset (see e.g. Figueira et al., in prep). With the ESPRESSO pipeline\footnote{\url{https://www.eso.org/sci/software/pipelines/espresso/}} version 3.0.0, the wavelength calibration and chromatic drift account for the change of lamp that originally created the need to separate the E19 and E21~\citep[see e.g.][]{far22}. Thus, now there is no offset or appreciable difference, and we do not need to separate these datasets. There has been another significant intervention in the instrument in May 2022 at about BJD[d]~$=$~$2459720$ to repair the blue cryostat. In fact, both the blue and the red cryostat were changed, but the analysis of RV standards does not justify the need to split the E19 dataset into two (see Section~\ref{sec:temp}).

\begin{table}
\begin{center}
\caption{Statistics of difference datasets \label{tab:dataset}}
\begin{tabular}[centre]{lrrrrr}
\hline \hline
Parameter              & E18  & E19  & H15  & HAN  & CAR  \\
\hline
Npoints                & 9    & 140  & 92   & 56   & 479  \\
RMS (RV)               & 0.96 & 1.87 & 1.97 & 1.90 & 2.18 \\
Mean ($\delta$RV)      & 0.10 & 0.11 & 0.55 & 0.36 & 1.41 \\
Median ($\delta$RV)    & 0.07 & 0.10 & 0.53 & 0.32 & 1.34 \\
Maximum ($\delta$RV)   & 0.21 & 0.29 & 0.84 & 0.65 & 2.46 \\
RMS (FWHM)             & 2.46 & 3.34 & 5.80 & 3.40 & 10.9 \\
Mean ($\delta$FWHM)    & 0.32 & 0.32 & 1.40 & 0.81 & 23.4 \\
Median ($\delta$FWHM)  & 0.24 & 0.29 & 1.34 & 0.83 & 23.6 \\
Maximum ($\delta$FWHM) & 0.61 & 0.86 & 2.24 & 1.58 & 25.9 \\
\hline
\end{tabular}
\end{center}
\textbf{Notes:} Statistics of the different datasets with the TM RVs computed with \texttt{S-BART} code for ESPRESSO, HARPS and HARPS-N data, and with \texttt{SERVAL} code for CARMENES data. CCF FWHMs are computed from DRS CCF profiles for ESPRESSO, HARPS and HARPS-N data, and with the \texttt{RACCOON} code code for CARMENES data. CARMENES data is provided in the public DR1 in \citet{rib23}.
\end{table}

The wavelength calibration done by the DRS uses Th-Ar lamp combined with Fabry-P\'erot (FP) etalon exposures. Due to the observed brightness of Barnard's star ($m_V= 9.5$, see Table~\ref{tab:barnard}), the ESPRESSO observations were carried out with the FP as simultaneous calibration in the reference fibre B using the HR11 binning ($R\sim 138,000$) and a typical exposure time of 900~s, with four spectra taken with 1200~s (two in E18 and two in E19) and two spectra taken with 550 and 426~s. The ESPRESSO data covers a time baseline of 1532.7~d (4.2 yr) from BJD[d]~$= 2458606.79918$ (May 2019) to 2460139.51282 (July 2023).

\begin{table}
\begin{center}
\caption{Stellar properties of GJ~699 \label{tab:barnard}}
\begin{tabular}[centre]{l l r}
\hline \hline
Parameter & GJ~699 & Ref. \\ \hline
{\it Gaia} DR3 source id & 4472832130942575872 & 1 \\
$\alpha$ (J2016) [deg] &  269.44850252544 & 1 \\
$\delta$ (J2016) [deg] & +04.73942005111  & 1 \\
error $\alpha$ (J2016) [mas] & 0.0262 & 1 \\
error $\delta$ (J2016) [mas] & 0.0290  & 1 \\
$\alpha$ (J2000) &  17:57:48.4985 & 1 \\
$\delta$ (J2000) & +04:41:36.1139 & 1 \\
$\mu_{\alpha}$ cos $\delta$ [mas yr$^{-1}$] & --801.551 $\pm$ 0.032 & 1 \\
$\mu_{\delta}$ [mas yr$^{-1}$] & 10362.394 $\pm$ 0.036 & 1 \\
$\varpi$ [mas] &  546.9759 $\pm$ 0.0401 & 1\\
d [pc] & 1.82823 $\pm$ 0.00013 & 1 \\
$a_{\rm sec}$ [m s$^{-1}$ d$^{-1}$] & 0.012 & 0 \\
$m_G$ [mag] & 8.1940 $\pm$ 0.0028 & 1 \\
$m_V$ [mag] & 9.511 $\pm$ 0.010 & 2 \\
$m_J$ [mag] & 5.244 $\pm$ 0.020 & 3 \\
$\overline{RV_{\rm CCF}}$[\ms] & --110245.15 $\pm$ 0.15 & 0$^{*}$ \\
\hline
Spectral Type  &  M3.5V-M4V & 4\\
$L_{\star}$ [$10^{-3 }$~\lsun] & 3.558 $\pm$ 0.072 & 5 \\
$T_{\rm eff}$ [K] & 3195 $\pm$ 28 & 0 \\
$\log_{10} g$ [cgs] & 4.90 $\pm$ 0.09 & 0 \\
$[\rm Fe/H]$ [dex] & -0.56 $\pm$ 0.07 & 0 \\
$v_{\rm br}$ [\kms] & 3.17 $\pm$ 0.14 & 0 \\
$R_{\star}$ [\rsun] & 0.185  $\pm$ 0.006 & 5 \\
$M_{\star}$ [\msun] & 0.162  $\pm$ 0.007 & 5 \\
$\Theta_{\rm LD}$ [mas] & 0.952  $\pm$ 0.005 & 6 \\
$R_{\rm interf}$ [\rsun] & 0.187  $\pm$ 0.001 & 5 \\
\hline
$\log_{10} (L_{X}/L_{\rm bol}$) [erg~s$^{-1}$]  & $-5.8$  & 7 \\
$\log_{10} R_{HK}^{'}$ & --5.8 $\pm$ 0.1 & 8  \\
$P_{\rm ROT}$ [d] & 142 $\pm$ 9 & 0$^{**}$ \\
$P_{\rm CYC}$ [d] & 3210 $\pm$ 530 & 0$^{**}$ \\
$HZ_{a}$ [AU] & 0.049--0.129 & 9 \\
$HZ_{p}$ [d] & 9.9--42.0 & 9 \\
\hline
\end{tabular}
\end{center}
\textbf{References:} 0 - This work; 1 -  \citet{gaia21}; 2 - \citet{koe10}; 3 - \citet{cut03}; 4 - \citet{alo15,kir12}; 5 - \citet{sch19}; 6 -  \citet{boy12}; 7 - \citet{fra20}; 8 - \citet{tol19}; 9 -  \citet{kop14}; $^{*}$ Mean DRS RV and uncertainty; $^{**}$Rotation period and long-term activity cycle obtained from the global analysis of the FWHM and RV measurements using the ESPRESSO, HARPS and HARPS-N data.
\end{table}

Fig.~\ref{gj699_rv} shows three different RV computations of the ESPRESSO data reduced using the ESPRESSO pipeline version 3.0.0. In the top panel, we see the RVs provided by the ESPRESSO Data Reduction Software~\citep[DRS;][]{dim18}, which uses the cross-correlation technique; in the middle panel, the RVs computed using the \texttt{S-BART}~\citep{sil22sbart} code\footnote{\url{https://github.com/iastro-pt/sBART}}, a semi-Bayesian radial velocity computation through template matching (TM); and in the bottom panel, the RVs extracted using the line-by-line (LBL) technique applied to these ESPRESSO observations~\citep{art22lbl}. DRS, TM and LBL RVs are not significantly different from each other. We also depict the generalized Lomb-Scargle~\citep[GLS;][]{zec09} periodograms of the three sets of RVs where we only see slightly different power of some peaks, mostly related to stellar activity. The root-mean-square (RMS) of the RVs are very similar, 1.81, 1.86 and 1.78~\ms\ for DRS, TM and LBL, respectively, whereas the mean/median uncertainty of the RVs are 16.5/14.4, 11.0/9.6 and 10.3/9.2~\cms\ for DRS, TM and LBL, respectively. Only eight E18 RVs (instead of nine) are shown since the LBL code crashed for one spectrum with BJD[d]~$= 2458642.75641$. We also remove one spectrum with BJD[d]~$= 2459867.56166$ that has bad quality RV, FWHM measurements, clearly off by about 10~$\sigma$ from the median values.
    
    We note that the ESPRESSO pipeline version 3.0.0 does not include yet the telluric correction from \citet{all22}. Thus DRS RVs have not been computed after telluric correction, which may explain the larger DRS uncertainties compared to TM and LBL techniques. \texttt{S-BART} masks the tellurics at a 1\% threshold, which is a quite conservative mask, thus discarding a considerable amount of RV content. \texttt{S-BART} first constructs a synthetic spectra with the \texttt{TelFit} code~\citep{gul14}, using the weather conditions of the observing block with the highest relative humidity. It determines the continuum level through a median filter and finds where the spectra is more than 1\% away from the continuum, thus masking only deeper features. Then it flags those regions as tellurics accounting for barycentric Earth radial velocity (BERV) changes before continuing with the RV computation~\citep{sil22sbart}. The LBL code also discards features affected by telluric contamination. Given the small differences between the TM and LBL uncertainties, we decided to adopt the \texttt{S-BART} TM RVs as our preferred RV measurements for the rest of the paper. This amounts to nine E18 and 147 E19 data points. Since only a few points are taken the same night, we decided to bin RV and FWHM time-series are subsequently with a 1 d step, and after, two E19 points are discarded with RV uncertainties larger than 50~\cms, the final sample of ESPRESSO data includes nine E18 and 140 E19 points (see Fig.~\ref{gj699_rv_all}). Full width at half maximum (FWHM) measurements from cross-correlation functions (CCFs) were automatically provided by the ESPRESSO DRS (see e.g. Fig.~\ref{gj699_gp}). The RMS of E18 and E19 RVs computed with \texttt{S-BART} are 0.96 and 1.87~\ms. The RMS of the FWHM of E18 and E19 are 2.46 and 3.34~\ms. The statistics of the uncertainties of RV and FWHM measurements are summarised in Table~\ref{tab:dataset}.
    
    We also applied the telluric correction using the code from \citet{all22} only to the 156 useful ESPRESSO spectra. We recomputed a new set of RVs (labelled as TMtc RVs) using the template matching \texttt{S-BART} code, but this time masking out those regions at a 60\% threshold, to avoid including deeper features affected by tellurics that may have not been properly modelled. We use this set of TMtc RVs later in this work (see Section~\ref{sec:planet315}) but we keep TM RVs as our main ESPRESSO RV dataset, which we compare with ESPRESSO DRS and LBL RVs, and that we also use together with the HARPS, HARPS-N and CARMENES datasets described in Sections~\ref{sec:obs_har_han} and~\ref{sec:obs_car}. The TMtc RVs are very similar to the TM RVs, with a minor improvement, showing an RMS of 0.97 and 1.83~\ms\ in TMtc RVs compared with an RMS of 0.96 and 1.87~\ms\ in TM RVs. The mean/median  uncertainties of TMtc RVs are 0.10/0.08 and 0.10/0.09~\ms, thus very similar to those of TM RVs (see Table~\ref{tab:dataset}). This final ESPRESSO dataset of CCF FWHM and TM RV measurements is available at the CDS portal, together with the FWHM and RV HARPS, HARPS-N and CARMENES datasets described in Sections~\ref{sec:obs_har_han} and~\ref{sec:obs_car}.

\subsection{HARPS and HARPS-N \label{sec:obs_har_han}}

The High Accuracy Radial Velocity Planet Searcher~\citep[HARPS;][]{may03} is a fibre-fed \'echelle high-resolution ($R\sim 115,000$) spectrograph installed in 2003 at the 3.6m telescope in La Silla Observatory (ESO, Chile). It covers the wavelength range 378--691 nm, and it is contained in a vacuum vessel to minimise the temperature and pressure variations that may cause spectral drifts. HARPS spectra used in these work  can be downloaded from the ESO archive\footnote{\url{https://archive.eso.org/scienceportal/home/}} from different ESO programs\footnote{HARPS ESO programs: 099.C-0880, 0101.D-0494, 1102.C-0339, 110.242T.001}, and cover a time baseline of 2095.2 d (5.7 yr) from BJD[d]~$= 2457934.65911$ (June 2017) to $2460029.83365$ (March 2023). All the HARPS data used in this work were taken after the fibre-link upgrade in 2015, and thus we label these data as H15. The exposure time varies from 600~s to 1800~s with a typical exposure of 900~s. Wavelengths are calibrated using a Th-Ar lamp combined with FP etalon exposures~\citep{wil10}. Spectra taken in 2017 were taken without any reference calibration in fibre B, and from 2018 with FP simultaneous reference in fibre B. We use a total of 114 useful HARPS spectra, which after 1d binning are turned into 105 HARPS data points.

The High Accuracy Radial velocity Planet Searcher for the Northern hemisphere~\citep[HARPS-N;][]{cos12}  is a fibre-fed \'echelle high-resolution ($R\sim 115,000$) spectrograph installed in 2012 at the 3.6m Telescopio Nazionale {\it Galileo } (TNG) in the Observatorio del Roque de los Muchachos (ORM, La Palma, Spain). Having very similar instrument specifications as HARPS, it can also reach an RV stability better than 1~\ms~\citep{pep14Natur}, and covers the wavelength range 383--693 nm. HARPS-N spectra used in this work cover a time baseline of 1496.9~d (4.1 yr) from BJD[d]~$= 2457626.41888$ (August 2017) to $2459123.34337$ (September 2020). HARPS-N spectra can be accessed at the TNG archive\footnote{\url{http://archives.ia2.inaf.it/tng/}} from different Spanish CAT programs\footnote{HARPS-N programs: CAT16A$\_$109, CAT17A$\_$38, CAT18A$\_$115, CAT20A$\_$121}. As for HARPS data, the wavelength calibration is done using a Th-Ar lamp combined with FP etalon exposures, with the science spectra taken with FP simultaneous reference in fibre B. We label these data as HAN. The total number of HARPS-N spectra is 133 and after binning using 1 d step we finally have 58 HARPS-N data points. The exposure time was 900~s before 2020 and the data taking during the COVID-19 pandemic was taken with three spectra per night of 1200~s each, except for one night with three spectra of 1800~s. The HARPS-N data taken during the COVID-19 pandemic intended to cover the gap of ESPRESSO observations during the ESO Paranal Observatory shutdown in 2020.

Both H15 and HAN spectra were reduced with the standard DRS pipelines at both instruments and the RVs were extracted using the \texttt{S-BART} code (see Fig.~\ref{gj699_rv_all}). The CCF FWHM measurements of HARPS and HARPS-N spectra were computed by adding a colour-correction factor order by order, following \citet{sua23}. After discarding RV and FWHM measurements with uncertainties larger than 0.85~\ms\ and 2.5~\ms, respectively, we are left with 92 data points in H15 and 56 data points in HAN. The RMS of H15 and HAN RVs computed with \texttt{S-BART} are 1.97 and 1.90~\ms. The RMS of H15 and HAN FWHM measurements are 5.8 and 3.4~\ms\ (see Table~\ref{tab:dataset}).

\subsection{CARMENES \label{sec:obs_car}}

The Calar Alto high-Resolution search for M-dwarfs with Exoearths with Near-infrared and optical \'Echelle Spectrographs
\citep[CARMENES;][]{qui16} are visual (VIS) and near-infrared (NIR) vacuum-stabilised spectrographs covering 520--960~nm and 960--1710 nm with a spectral resolution of 94,600 and 80,400, respectively. They are located at the 3.5m telescope of the Centro Astron\'omico Hispano en Andaluc\'{i}a (CAHA) at Observatorio de Calar Alto (Almer\'{i}a, Spain). The wavelength calibration is performed by combining hollow cathode (U-Ar, U-Ne, and Th-Ne) and Fabry-P\'erot etalon exposures. The instrument drift during the nights is tracked with the FP in the simultaneous calibration fibre. We downloaded the CARMENES public data~\citep{rib23} of Barnard's star, which is the CARMENES VIS data of the RV survey within the GTO programme (CARMENES Data Release 1)\footnote{\url{http://carmenes.cab.inta-csic.es/gto/jsp/dr1Public.jsp}}. CARMENES RV measurements were obtained using the template-matching \texttt{SERVAL} algorithm~\citep{zec18}. We use the RVs corrected for nightly zero point (NZP) offsets. The CARMENES spectrograph is usually wavelength calibrated each afternoon and nightly instrumental drifts are measured with the FP etalon, but stellar RVs from the same night often share common systematic effects, producing NZP offsets generally of a few~\ms\ with a median error bar of 0.9~\ms~\citep{rib23}. The useful 520 CARMENES spectra corrected for NZP used in this work cover a time baseline of 1751.5 d (4.8 yr) from BJD[d]~$= 2457422.74662$ (February 2016) to $2459174.25596$ (November 2020). After binning using 1 d step we end up with 501 CARMENES data points. The FWHM measurements of CARMENES CCF profiles were obtained by \citet{laf20} using the \texttt{RACCOON} code, and provided in \citet{rib23}. We discarded those RV and FWHM measurements with uncertainties larger than 2.5~\ms\ and 26~\ms, respectively, to remove only a few points and to slightly clean the dataset. The final number of CARMENES data points, labelled as CAR, is 479. The RMS of the RV measurements is 2.18~\ms (see Fig.~\ref{gj699_rv_all}). The RMS of FWHM measurements is 10.9~\ms (see Table~\ref{tab:dataset}).
    
\section{Stellar properties\label{sec:stepar}} 

Barnard's star (GJ~699) is a bright ($m_V=9.5$) very nearby M3.5-M4 dwarf star located at 1.8~parsecs from the Sun~\citep{gaia21}.
The main stellar properties are provided in Table~\ref{tab:barnard}. We adopted the weighted mean mass estimated ($M_\star = 0.162  \pm 0.007$~\msun) from the three mass determinations (based on the mass-radius relation,  the spectroscopic $\log g$ and 2MASS $K_s$ photometry) in \citet{sch19}. We checked that the updated parallax in~\citet{gaia21} does not change the values given in~\citet{sch19}. We used the 156 ESPRESSO spectra of Barnard's star to create a master mean spectrum~(see Fig.~\ref{gj699_spec}). We used this master ESPRESSO spectrum to derive the stellar parameters ($T_{\rm eff}$, $\log g$ and [Fe/H]) and the total line broadening velocity, $v_{\rm br}$, using the {\sc SteParSyn} code\footnote{\url{https://github.com/hmtabernero/SteParSyn/}} described in~\citet{tab22}.
The derived stellar parameters, given in Table~\ref{tab:barnard}, are compatible with those used in \citet{sch19} which are the spectroscopic parameters ($T_{\rm eff}= 3273\pm51$~K, $\log g = 5.11\pm0.07$,  [Fe/H]~$=-0.15\pm0.16$) determined using CARMENES VIS data~\citep{pas18}. These are also compatible with those derived in~\citet{mar21} using both CARMENES VIS and NIR data ($T_{\rm eff}= 3254\pm32$~K, $\log g = 5.13\pm0.12$, [Fe/H]~$=-0.57\pm0.10$) and \citet{jah23} using SPIRou data ($T_{\rm eff}=3231\pm21$~K, $\log g=5.08\pm0.15$, [Fe/H]~$=-0.39\pm0.03$). 

The luminosity ($L_{\star}$~[$10^{-3 }$~\lsun]~$= 3.558 \pm 0.072$) of GJ~699 from \citet{sch19} and the spectroscopic effective temperature ($T_{\rm eff}$~[K]~$=3195\pm28$) derived from the ESPRESSO master spectrum of GJ~699 was used to estimate the habitable zone (HZ)\footnote{\url{https://github.com/Eelt/HabitableZoneCalculator}}. We find an inner boundary 0.049~AU (recent Venus) and an outer boundary 0.129~AU (early Mars)~\citep{kop14} of the HZ, corresponding to orbital periods of 9.84 and 41.88~d, respectively. The inner edge of the HZ for worlds with very little water content (with 1\% relative humidity and albedo $A=0.2$) could extend inwards to 0.036 AU~\citep{zso13}, or 0.026 AU in the case of high albedo ($A=0.8$), which corresponds to orbital periods of 6.25 and 3.85 days, respectively.
 
\section{Stellar activity}

Stellar activity is possibly the main source of false positive planetary detections from RV time series. Activity signals and their aliases, although not necessarily persistent in time, caused by the presence of long-lived large spots (or spot groups) on the stellar surface can often create periodic signals that can easily mimic planetary signals~\citep{que01,rob14,sua15,sua17rv}. However, in many cases, it is possible to track and study the behaviour of stellar activity with time series of activity indexes and the changing shapes of computed cross-correlation functions, simultaneous to the RV time series.

\begin{figure*}
\includegraphics[width=18cm]{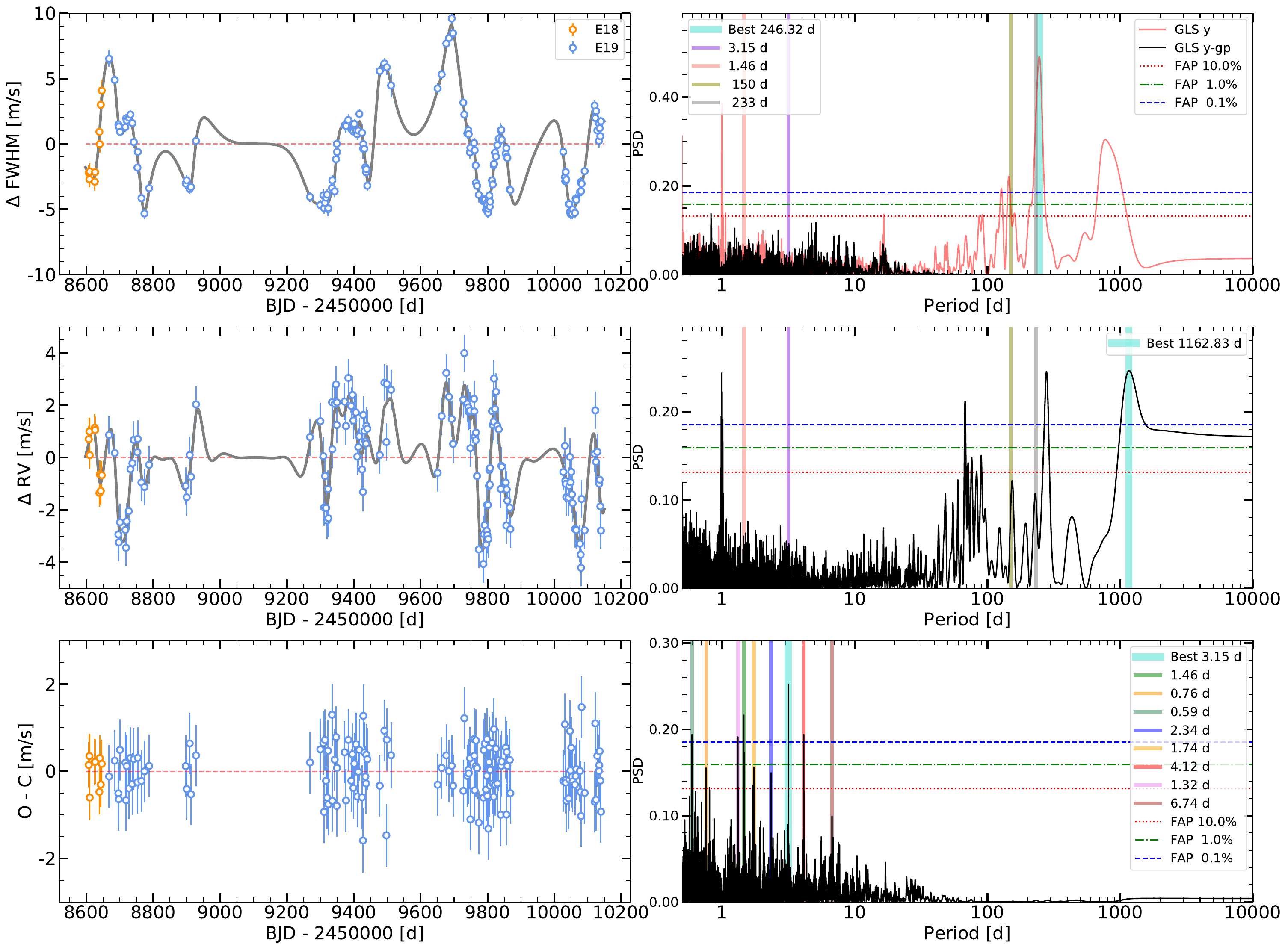}
\caption{
ESPRESSO FWHM measurements ({\it top}), RV measurements ({\it middle}), and RV residuals ({\it bottom}) from the SHO ($P_{\rm ROT}$ and $P_{\rm ROT}/2$) GP model, and GLS periodograms ({\it left}) of GJ~699. The uncertainties include the jitter term coming from the global model $A$ in Table~\ref{tab:logz}.
}
\label{gj699_gp}
\end{figure*}

\subsection{GP model\label{sec:gp}}

To evaluate the behaviour of stellar activity, we model the time series of each activity indicator using the Gaussian process framework~\citep[GP; e.g.][]{ram06}. The GP framework is commonly used in the analysis of stellar activity in RV times series~\citep[e.g.][]{hay14,far16}. The stellar noise is described with a covariance function dependent on a set of parameters, some of them typically associated with a physical quantity. GP models can be used for instance to model the activity signal without requiring a detailed knowledge of the distribution, temperature contrast and lifetime of active regions on the stellar surface. GP models are flexible to model quasi-periodic signals, accounting for changes in the amplitude, phase, or even small period changes. This flexibility can however easily over-fit the data, sometimes suppressing possible planetary signals.
Recently, there have been efforts to better constrain the GP models using the variability of stellar activity indicators, such as training with photometric data or activity indicators~\citep[e.g.][]{hay14} or simultaneous modeling of activity proxies and radial velocity measurements with shared hyper parameters~\citep[e.g.][]{sua20,far22}. A more sophisticated approach is the use of multi-dimensional GPs, which join the fit of all time series under a single covariance matrix~\citep[][]{raj15,barragan22,del22spleaf}. This implementation assumes that there is an underlying function governing the behaviour of stellar activity, $G(t)$, which manifests itself in each time series as a linear combination of itself and its gradient, $G'(t)$, and their amplitudes for each time series $j$ ($\Delta ~TS_{j}$), with $j=0,...,N$ for $N$ times series, following the $FF'$ formalism~\citep{aig12}, as described in equation~\ref{eq_gp_grad}).
\begin{equation} \label{eq_gp_grad}
{\Delta ~TS_{j} = A_{j} \cdot G(t) + B_{j} \cdot G'(t)},
\end{equation}
\noindent \citet{sua20} showed a very good correlation between FWHM of the CCF and the activity-induced RV in the analysis of Proxima using ESPRESSO data. This offers a compelling new approach to study stellar activity signals and separate them from planetary signals in M-dwarfs.

Following \citet{sua23}, we use the \texttt{S+LEAF} code \citep{del22spleaf}, which extends the formalism of semi-separable matrices introduced with \texttt{Celerite}~\citep{fm17celerite2} to allow fast evaluation of GP models even in the case of large datasets. The \texttt{S+LEAF} code allows the simultaneous fit of a GP to several time series, based on a linear combination of the GP and its derivative, with different amplitudes for each time series (see equation~\ref{eq_gp_grad}). The \texttt{S+LEAF} code supports a wide variety of GP kernels with fairly different properties. After testing several kernel functions, based on the shape of posterior sample distributions, we chose a combination of two simple harmonic oscillators (SHO) at the first and second harmonics of the rotation period, $P_{\rm ROT}$ and $P_{\rm ROT}/2$. The selected kernel is defined as:
\begin{equation} \label{act_model}
\begin{split}
\fontsize{8}{11}\selectfont
 k(\tau) = k_{\rm SHO,1}(\tau, P_{1}, S_{1}, Q_{1}) + k_{\rm SHO,2}(\tau, P_{2}, S_{2}, Q_{2})\, ,
 \end{split}
\end{equation}
\noindent with $\tau = t_n - t_{n-1}$, representing the time-lag between measurements. 

Following equation~\ref{eq_gp_grad}, the activity induced signal in every specific time series $j$ is:
\begin{equation} \label{full_gp_model}
\begin{split}
\Delta ~TS_j = A_{11,j} \cdot G_{\rm SHO,1} + A_{12,j} \cdot G'_{\rm SHO,1} \\
+ A_{21,j} \cdot G_{\rm SHO,2} + A_{22,j} \cdot G'_{\rm SHO,2}\, ,
\end{split}
\end{equation}
\noindent where $G_{\rm SHO,i}$ and $G'_{\rm SHO,i}$ is the realisation of a GP with kernel $k_{\rm SHO,i}$ and its first derivative.
Following \citet{fm17celerite2}, the $k_{\rm SHO,i}$ kernel is defined as:
\begin{equation} \label{eq_sho}
\fontsize{8}{11}\selectfont
 k_{i}(\tau) = {C_{i}^{2}} e^{-{{\tau}\over{L}}}{\left\{\begin{array}{rr}\cosh(\eta {{2 \pi \tau}\over{P_{i}}})+{{P_{i}}\over{2 \pi \eta L }}\sinh(\eta {{2 \pi \tau}\over{P_{i}}}) ;~\rm if ~P_{i} > 2 \pi L\\
     2 (1 + {{2 \pi \tau}\over{P_{i}}})\, ; ~\rm if ~P_{i} = 2\pi L\\
     \cos(\eta {{2 \pi \tau}\over{P_{i}}}) + {{P_{i}}\over{2 \pi \eta L}} \sin(\eta {{2 \pi \tau}\over{P_{i}}})\, ; ~\rm if ~P_{i} < 2 \pi L
\end{array}\right\}},
\end{equation}
\noindent with $\eta = (1 - (2L/P_{i})^{-2})^{1/2}$, controlling the damping of the oscillator.
This kernel has a power spectrum density:
\begin{equation} \label{psd_kernel}
S(\omega) = \sqrt {{2} \over {\pi}} {{S_{i} ~\omega_{i}^{4}} \over {(\omega^{2} - \omega_{i}^{2})^2 + \omega_{i}^{2}~\omega^{2} / Q^{2}}}\, ,
\end{equation}
\noindent where $\omega$ is the angular frequency, $\omega_{i}$ is the undamped angular frequency for each component ($\omega_{i}$ = 2 $\pi$ / $P_{i}$), $S_{i}$ is the power at $\omega$ = $\omega_{i}$, and $Q_{i}$ is the quality factor. The parameters $S_{i}$, $P_{i}$ and $Q_{i}$ are sampled in the covariance matrix, related to the amplitude ($C_{i}$), rotation period ($P = P_{\rm ROT}$) and timescale of evolution ($L = T_{\rm ROT}$) as shown in eq.~\ref{eq_params}.
\begin{equation} \label{eq_params}
\begin{split}
P_{1} = P~,~ S_{1} = {{C_{1}} \over {2 \cdot L}} \left( {{P_{1}} \over {\pi}} \right)^{2} ~,~  Q_{1} = {\pi {L}\over{P_{1}}}\, , \\
P_{2} = {{P}\over{2}} ~,~ S_{2} = {{C_{2}} \over {2 \cdot L}} \left( {{P_{2}} \over {\pi}} \right)^{2} ~,~  Q_{2} = {\pi {L}\over{P_{2}}}
\end{split}
.\end{equation}
\noindent The covariance matrix also includes a term of uncorrelated noise ($\sigma$), independent for every instrument. This term is added in quadrature to the diagonal of the covariance matrix to account for all unmodelled noise components, such as uncorrected activity or instrumental instabilities.

The amplitudes $C_{i}$ in equations~\ref{eq_sho} and \ref{eq_params} are related to the amplitude of the underlying function, not to any of the specific time series. We chose to adopt S$_{i}=1$, thus fixing their power at $\omega$ = 0. Thus, the amplitudes of every component will be governed by the parameters $A_{ih}$ shown in equation~\ref{full_gp_model}.

We model the data using Bayesian inference via nested sampling~\citep{ski04}, which in turn allows efficient exploration of large parameter spaces as well as obtaining Bayesian evidence from the model (i.e. marginal likelihood, $\ln \mathcal{Z}$). We used the \texttt{Dynesty} code~\citep{spe20dynesty}, which employs multi-ellipsoidal decomposition \citep{fer09} to more efficiently sample large prior volumes. We use the default configuration, which uses a random walk or random cut sampling strategy~\citep{han15a,han15b} depending on the number of free parameters. We set the number of live points equal to $100  \cdot  N_{\rm par}$, and the number of slices equal to $2  \cdot  N_{\rm par}$, with $N_{\rm par}$, the number of free parameters of the global model, including GP parameters.

\subsection{Activity indicators\label{sec:act_ind}}

Following \citet{lov11,sua15,sua18rot,tol19}, we measure from the ESPRESSO spectra the following activity indexes: $S$-index or $S_{\rm MW}$, defined similar to the original Mount Wilson index, measured from the line core fluxes of Ca~II H\&K stellar lines relative to the continuum fluxes, and the $H_\alpha$ and Na~I-indexes, from the stellar lines $H_\alpha$ and Na~I doublet, all sensitive to chromospheric activity.
As stellar line shape varies with magnetic activity, we also built time series of quantities extracted from the shape of the cross-correlation function: the full width at half maximum, the bisector span (BIS) and the CCF contrast. All these measurements are simultaneous to the RV measurements as they are extracted from the same spectra.

\begin{figure*}
\includegraphics[width=18cm]{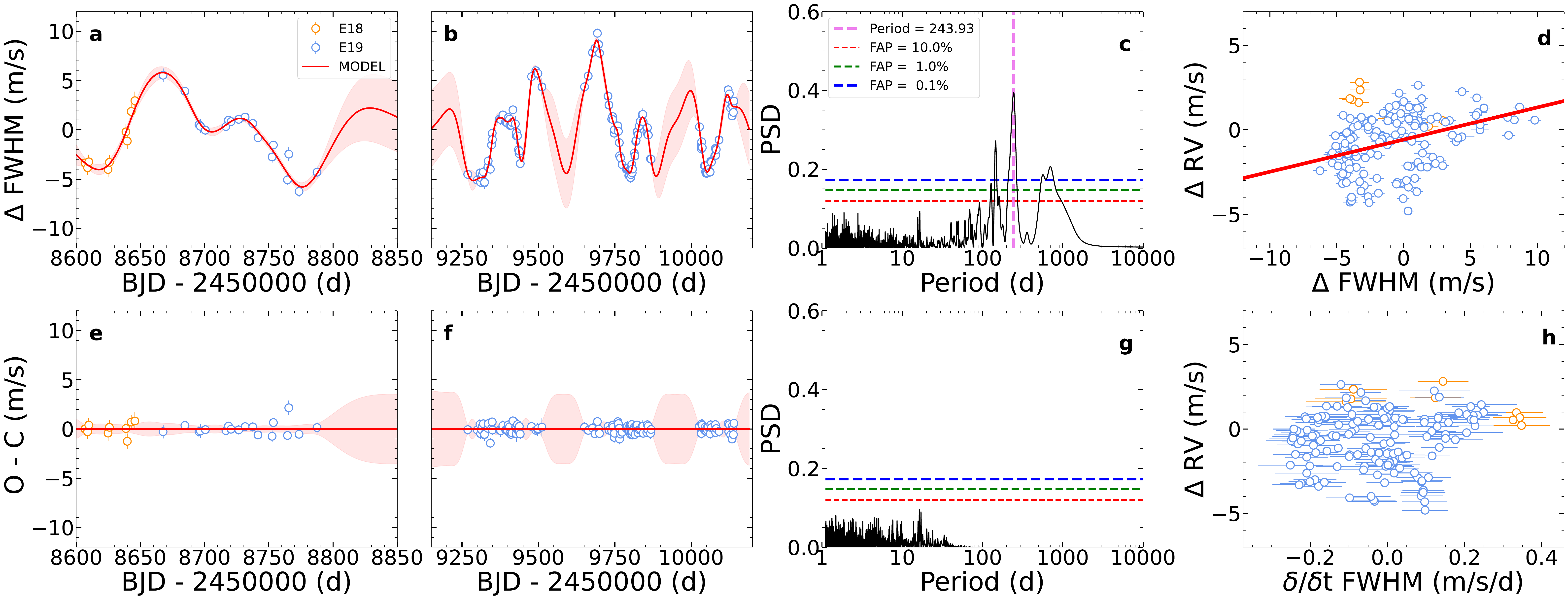}
\caption{
{\bf Analysis of the FWHM of the ESPRESSO CCF.}
\textbf{(Panels a and b)}: FWHM time-series with the best-model fit. The data is split into two panels because of a large period with no observations between the two campaigns. The shaded area shows the variance of the GP model. {\bf (Panel c):} GLS periodogram of the CCF FWHM data. The red vertical dashed line shows the most significant period. \textbf{(Panel d):} Relationship between the CCF RV and CCF FWHM data. The best fit is shown when the slope is $\ge$3$\sigma$ different from zero. \textbf{(Panels e and f):} Residuals of the CCF FWHM after subtracting the best model fit. \textbf{(Panel g):} GLS periodogram of the residuals. \textbf{(Panel h)}: Comparison of the CCF RV and gradient of the CCF FWHM model.}
\label{gj699_act0}
\end{figure*}

We model each of the ESPRESSO time series individually with the adopted GP formalism and the results are shown in Fig.~\ref{gj699_act0} for the FWHM of the ESPRESSO CCF, and Figs.~\ref{gj699_act1} and~\ref{gj699_act2} for the other activity indicators. The timescale of evolution of activity signals typically spans between one and two rotations, sometimes longer for M-dwarfs~\citep{gil17}. We leave the rotation period, $P_{\rm ROT}$, and the timescale, $T_{\rm ROT}$, free in a wide range, with the priors $\mathcal{LU}$(2,1000)~d and $\mathcal{LU}$(4,4000)~d, respectively, using a log-scale to allow for a long tail towards long timescales in the persistence of signals. We use log-normal priors for the amplitudes and jitter terms, centered on $\ln ({\rm RMS})$ of the data and with a sigma of $\ln ({\rm RMS})$ of the data. When using a GP with a completely free amplitude and jitter parameters, on data that includes multiple signals, there is a large risk that the GP absorbs all variations present in the data. Constraining the parameters in this way ensures a smooth GP model, preventing it from over-fitting variations at short timescales without fully excluding any region of the parameter space.

The GP analysis on the time series of the ESPRESSO FWHM measurements provides a $P_{\rm ROT} = 159^{+19}_{-16}$~d and a timescale $T_{\rm ROT} ={101}^{+37}_{-19}$~d. A similar result was found for the bisector span ($P_{\rm ROT} = 174^{+23}_{-52}$~d and $T_{\rm ROT} ={138}^{+43}_{-60}$~d) and the $H_\alpha$-index ($P_{\rm ROT} = 138^{+31}_{-56}$~d and $T_{\rm ROT} ={121}^{+178}_{-52}$~d). These values are consistent with previous $P_{\rm ROT}=145\pm 15$~d~\citep{tol19}, mostly based on the time series of $H_\alpha$-index measurements with a 15-yr baseline derived from seven different high-resolution spectrographs. In all cases the timescale is shorter but consistent with the rotation period. The CCF contrast ($P_{\rm ROT} = 206^{+77}_{-52}$~d and $T_{\rm ROT} ={186}^{+228}_{-76}$~d), $S_{\rm MW}$  ($P_{\rm ROT} = 229^{+35}_{-38}$~d and $T_{\rm ROT} ={180}^{+191}_{-63}$~d) and the  NaI-index ($P_{\rm ROT} = 275^{+193}_{-48}$~d and $T_{\rm ROT} ={249}^{+231}_{-95}$~d), show longer periods and larger uncertainties, thus marginally consistent.  The period measured in the GP analysis of the different activity indicators is close to the period shown in the GLS of the different time series in Figs.~\ref{gj699_act1} and~\ref{gj699_act2}. The structure of peaks in the GLS shows some complexity possibly related to the differential rotation, and with half a rotation, and sometimes the 1-yr alias at 240~d of the rotation period at 145~d. The highest peak of the GLS of BIS and CCF contrast falls at about the rotation period, whereas the H$\alpha$-index peaks at about half the rotation period. $S_{\rm MW}$, Na~I-index and FWHM show the highest peak at about 240~d.

We see clear correlations of the RV measurements with negative slope for $H_\alpha$ index, and positive slopes for FHWM and BIS measurements, in Figs.~\ref{gj699_act0},~\ref{gj699_act1} and~\ref{gj699_act2}. All fits performed to measure the slopes have both horizontal and vertical uncertainties taken into account. Clearly, the cleanest model with minimum residuals is provided by the FWHM, which shows a positive correlation with the CCF RVs. We therefore choose the FWHM in the joint analysis together with RVs to try to search for planetary signals in the RV time series while modeling simultaneously the stellar activity using both the FWHM and RV times series (see e.g. Fig.~\ref{gj699_gp}).
The difference between the correlation RV vs FWHM seen in this work and that of \cite{sua20} could be related to the specific nature of the active regions.
    The signature in RV of spot-induced variations causes a correlation
    between the $\delta/\delta t$ FWHM (or $\delta/\delta t$ Flux) and
    the RV. Variations caused by plages, however, cause a correlation
    between the FWHM (or Flux) and RV, formulas 11 and 12 in \citet{aig12}
    and Figure~3 in \citet{dum14}.
    In this second case, the dominant bulk of the change in RV is due to
    inhibition of convective blueshift.
    In the case of Proxima, the rotation signal is very clear in photometric time-series, and not so much in chromospheric indicators, indicating spot-dominated variations.
    In the case of Barnard's star, the rotation signal is easy to detect in chromospheric indicators (such as H alpha), but not in photometry~\citep{tol19}. This, combined with the correlation between FWHM and RV, hints at inhibition of convective blueshift being the cause of the RV variations.

\subsection{Photometry~\label{sec:phot}}

We used the ASAS-SN Sky Patrol online tool\footnote{\url{https://asas-sn.osu.edu/}} to inspect what photometry was available taken with the same time baseline that the ESPRESSO and other spectroscopic measurements shown in Fig.~\ref{gj699_rv_all}. We were able to acquire 3163 data points that cover the period January 2015 to January 2024 (approx. 2457000-2460300 HJD) and that contain measurements of two different pass-bands: $V$ ($N=722$) and g ($N=2441$). 

\begin{figure}
\includegraphics[width=9cm]{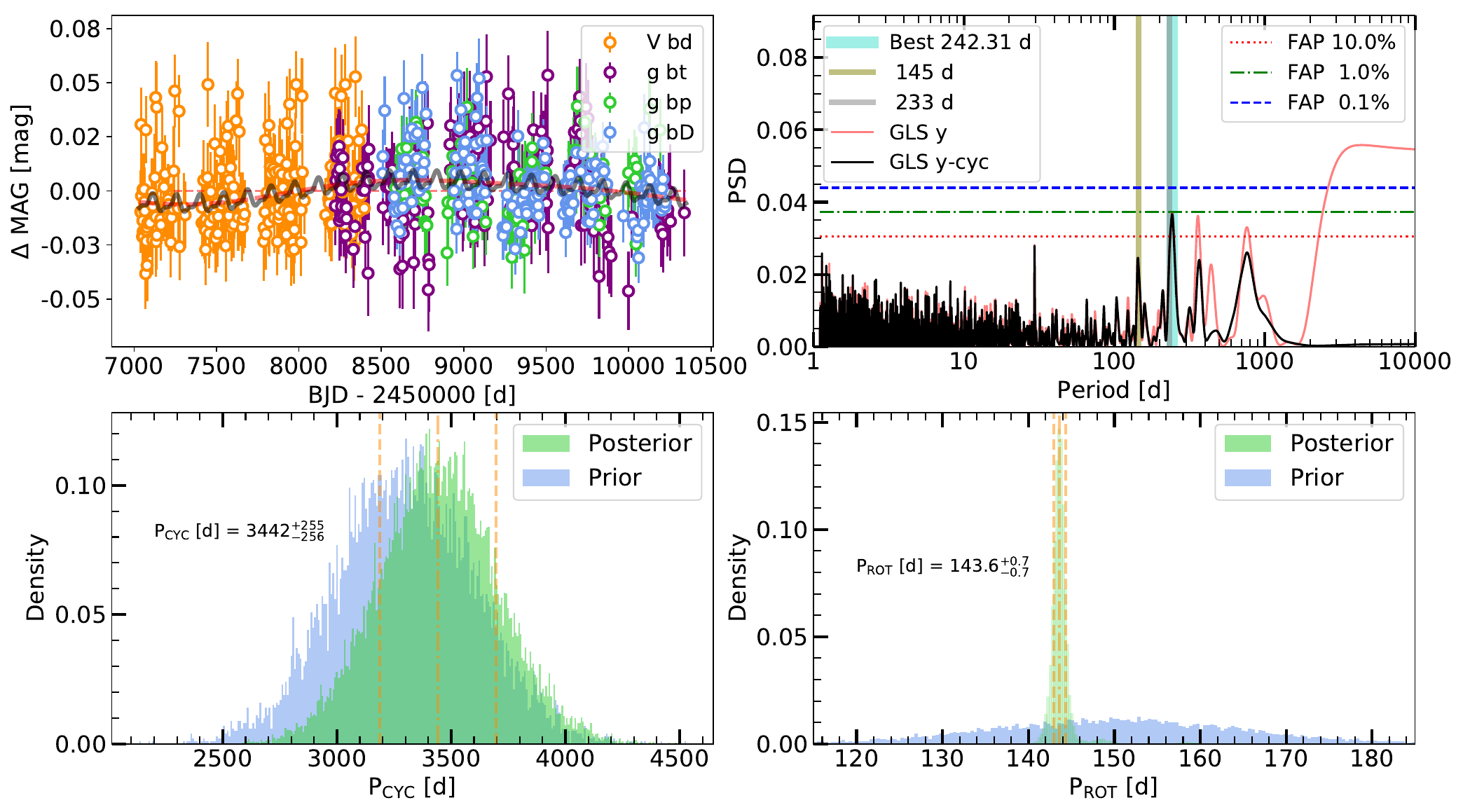}
\caption{
ASAS-SN photometry in the same temporal baseline as the RV data together with the double sinusoidal models of the long-term cycle and rotation signals ({\it top left}) and the GLS periodogram before and after subtracting the long-term cycle ({\it top right}). The posterior and prior distributions of these models including the long-term cycle ({\it bottom left}) and rotation periods ({\it bottom right}).
}
\label{gj699_dasasn}
\end{figure}

The proper motion of GJ~699 exceeds 10 arcsec/yr, while the ASAS-SN detector resolution is 8 arcsec/pixel. This means that the centroid of the star will drift by at least one pixel each year. We account for this in the following manner. Firstly, we took the interval 2457000-2460300 HJD which approximately reflects the ASAS-SN coverage of GJ~699. Then, we did split it in timestamps at every 100 d, including the interval endpoints. For each timestamp, we: (i) compute the expected astrometric position of GJ~699; (ii) use this position to produce the light curve through the ASAS-SN Aperture Photometry Pipeline; (iii) clip this light curve to the region defined by a range of $\pm$~50 d within the timestamp. This query resulted in 34 individual light curves. Then, we concatenated all individual light curves into a common one, and computed the BJD from HJD values using the online tool\footnote{\url{https://astroutils.astronomy.osu.edu/time/hjd2bjd.html}} provided by \citet{eas10}.
The $g$ magnitude measurements were collected using several different cameras (see Fig.~\ref{gj699_dasasn}). We discarded the group of those $g$ magnitude measurements higher than 10.5 which were far from the median, leaving a final set of $N=2215$. We binned the data using a 1 d step. We also discarded the $g$ magnitudes from $bl$ ($N=161$) and $bH$ ($N=97$) cameras which show a slope versus BJD not following the rest of the $g$ measurements, leaving a final set of 1 d binned $V$ ($N=268$) from the $bd$ camera (we discarded the only two $V_{bh}$ points) and $g_{bt}$ ($N=201$), $g_{bp}$ ($N=104$) and $g_{bD}$ ($N=205$) magnitudes (see Fig.~\ref{gj699_dasasn}).

We use the ASAS-SN photometry to verify the stellar activity behaviour we see in the RV and FWHM measurements (see Section~\ref{sec:activity}). We model the photometry with two double sinusoidal models as in equation~\ref{eq_cyc_dsine}: 
\begin{equation} \label{eq_cyc_dsine}
y(t) = A_1 \sin(\omega_1+\phi_1) + A_2 \sin(\omega_2+\phi_2)
,\end{equation} 
\noindent one to account for the long-term cycle and another to model the rotation modulation, where $\omega_2 = 2 \omega_1 = 2 \pi f/P$, with $f = 1/ t$ and $T_{0,i=1,2} =t_{\rm mid}+P  \cdot  \phi_i $, with $t_{\rm mid}$, the mid-time of the observation baseline and $\phi_i$, the phase of the sinusoidal function. We left $A_1$, $A_2$, $\phi_1$, $\phi_2$ and the period $P$ (equal to either the cycle period, $P_{\rm CYC}$, or the rotation period, $P_{\rm ROT}$) as free parameters, together with offsets and jitter terms to each of the magnitudes in the likelihood function. In Fig.~\ref{gj699_dasasn}, we display the ASAS-SN photometry versus BJD with fitted model. We also show GLS periodograms before and after subtracting the model, with the long-term signal and the rotation signal and its 1-yr alias detected. The posterior distributions point to a long-term cycle of $P_{\rm CYC} \sim3440$~d and a rotation signal of $P_{\rm ROT} \sim 144$~d, when assuming Gaussian priors centered on the expected cycle and rotation periods based on the results obtained in the activity analysis presented in Section~\ref{sec:activity} and supported by previous independent results~\citep{tol19}. The adopted time baseline of the photometry analysed here does not allow to get a better result with wide priors on the long-term cycle. We refer to \citet{tol19} for a deeper analysis of photometric data of Barnard's star with a longer baseline of 15 years.

\begin{figure}
\includegraphics[width=9cm]{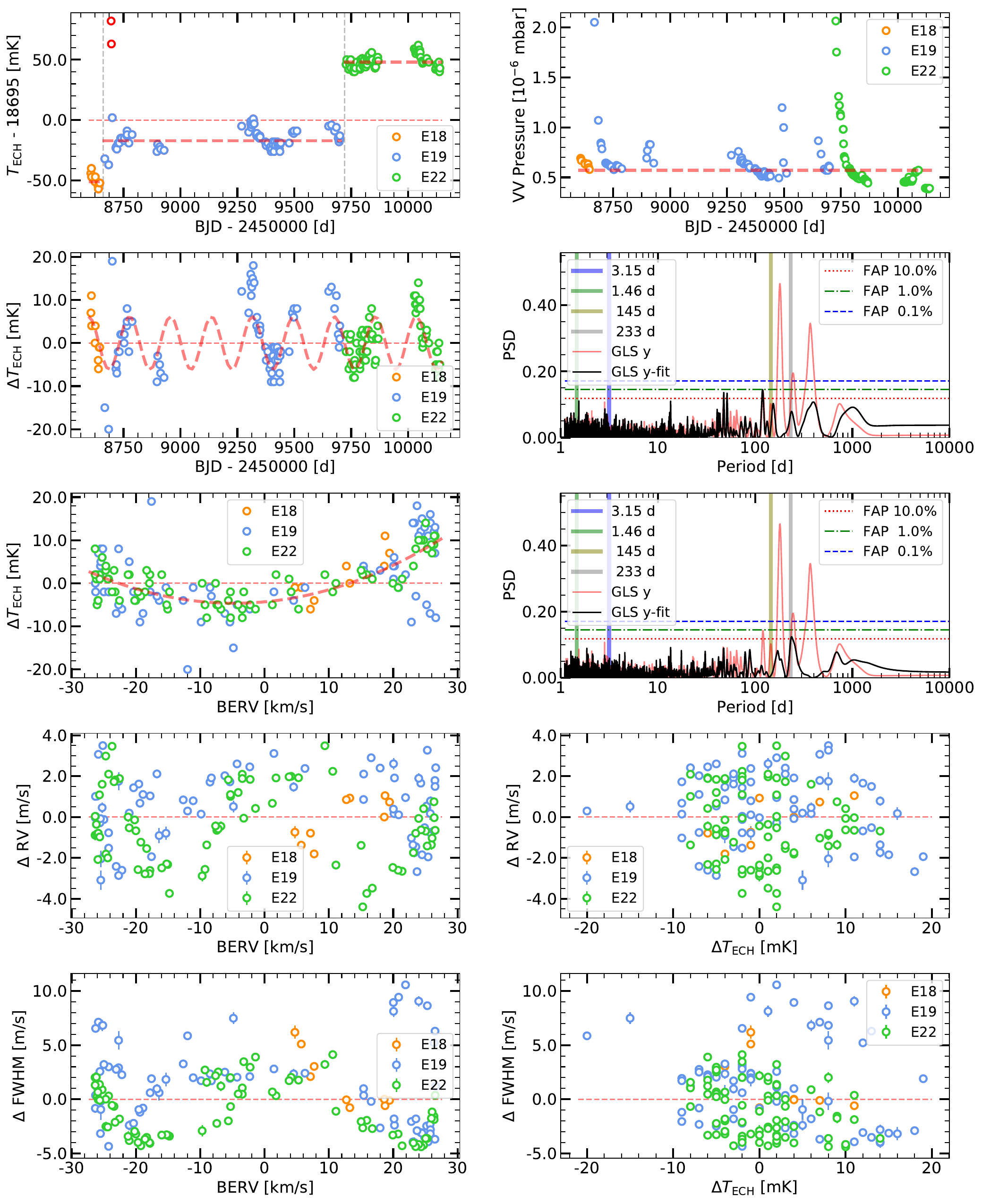}
\caption{
ESPRESSO RV and FWMH measurements, and \'echelle temperature sensor and vacuum vessel pressure sensor measurements versus BJD and barycentric RV correction. The GLS periodograms (left panels) show the result before (red line) and after (black line) subtracting the fitted function (dashed red line) in the left panels.
}
\label{gj699_temp}
\end{figure}

\section{Telemetry data~\label{sec:temp}}

We evaluate the temperature and pressure within the vacuum vessel (VV) to track the stability of the instrument during the ESPRESSO observations of Barnard's star. In Fig.~\ref{gj699_temp}, we display the é\'echelle grating temperature, $T_{\rm ech}$, and the vacuum vessel temperature as a function of BJD and barycentric radial velocity correction. The \'echelle temperature at a median temperature of 18,695 K, shows jumps of $-51$, $-16$ and $+48$~mK for the E18, E19 and E22 datasets. Here, this E22 refers to the jump in é\'echelle temperature happening at about BJD[d]~$=$~$2459720$ in May 2022. We check that this behaviour is shared by other temperature sensors within the VV, where, after the intervention in May 2022, we see jumps in temperature in the \'echelle, in the red cross disperser (RCD) and in the blue cross disperser (BCD) of approximately $+50$~mK, $+80$~mK and $+120$~mK, respectively. This corresponds to another significant intervention in the instrument where both the blue and the red cryostat were changed. However, the analysis of RV standards does not justify the need to split the E19 dataset into two datasets, but we separate with E22 in Fig.~\ref{gj699_temp} for the sake of clarity. We show in the upper-right panel the VV pressure versus BJD keeping reasonably constant at a level of about $0.6 \times 10^{-6}$~mbar and specially sensitive to the interventions coming down to stability after these interventions, in particular just after May 2022. We also display the median values of $T_{\rm ech}$ in the upper-left panel as horizontal dashed lines and subtract them to display the $\Delta T_{\rm ech}$ vs. BJD in the next bottom panel. We perform a sinusoidal fit using the highest peak at 180~d (half a year) of the red GLS periodogram in the right panel to illustrate the seasonal or year dependence. The black GLS periodogram shows how this yearly dependence would disappear from $\Delta T_{\rm ech}$ vs. BJD. In the next panel down, we show again more clearly this yearly dependence versus BERV, and again we perform a second-order polynomial function. We show the corresponding GLS periodogram of $\Delta T_{\rm ech}$ vs BJD in the right panel again  with the result before (red line) and after (black line) subtracting this fit. We do not see any significant peak at periods shorter than 10 days.

However, this dependence of the $\Delta T_{\rm ech}$, which varies smoothly with BERV within the range of [$-10$,$+15$]~mK, with a 4.9~mK RMS around this fit, does not apparently affect the RV and the FHWM as seen in the bottom four panels of Fig.~\ref{gj699_temp}. We do not see any trend of the RV and FWHM versus BERV or $\Delta T_{\rm ech}$. As discussed later in Section~\ref{sec:activity}, the variations found in FHWM and RV measurements are closely related to stellar activity, and the seasonal effect that we see in $\Delta T_{\rm ech}$ versus BERV, is either too small or removed with the drift corrections provided by the simultaneous FP calibration in fibre B during each observing exposure with the science target in fibre A.
The drift corrections are very stable over the whole set of 4.2 yr of observations from May 2019 to July 2023, with a mean blue detector drift of $-0.30$~\ms\ with a RMS of $0.84$~\ms, and a mean red detector drift of $-0.08$~\ms\ with a RMS of $0.33$~\ms. The behaviour of both detectors seems better after the intervention in May 2022, with a mean drift and RMS of $-0.26$~\ms\ ($0.22$~\ms) and $-0.08$~\ms\ ($0.22$~\ms), for the blue and red detector, respectively.

\begin{figure}
\includegraphics[width=9cm]{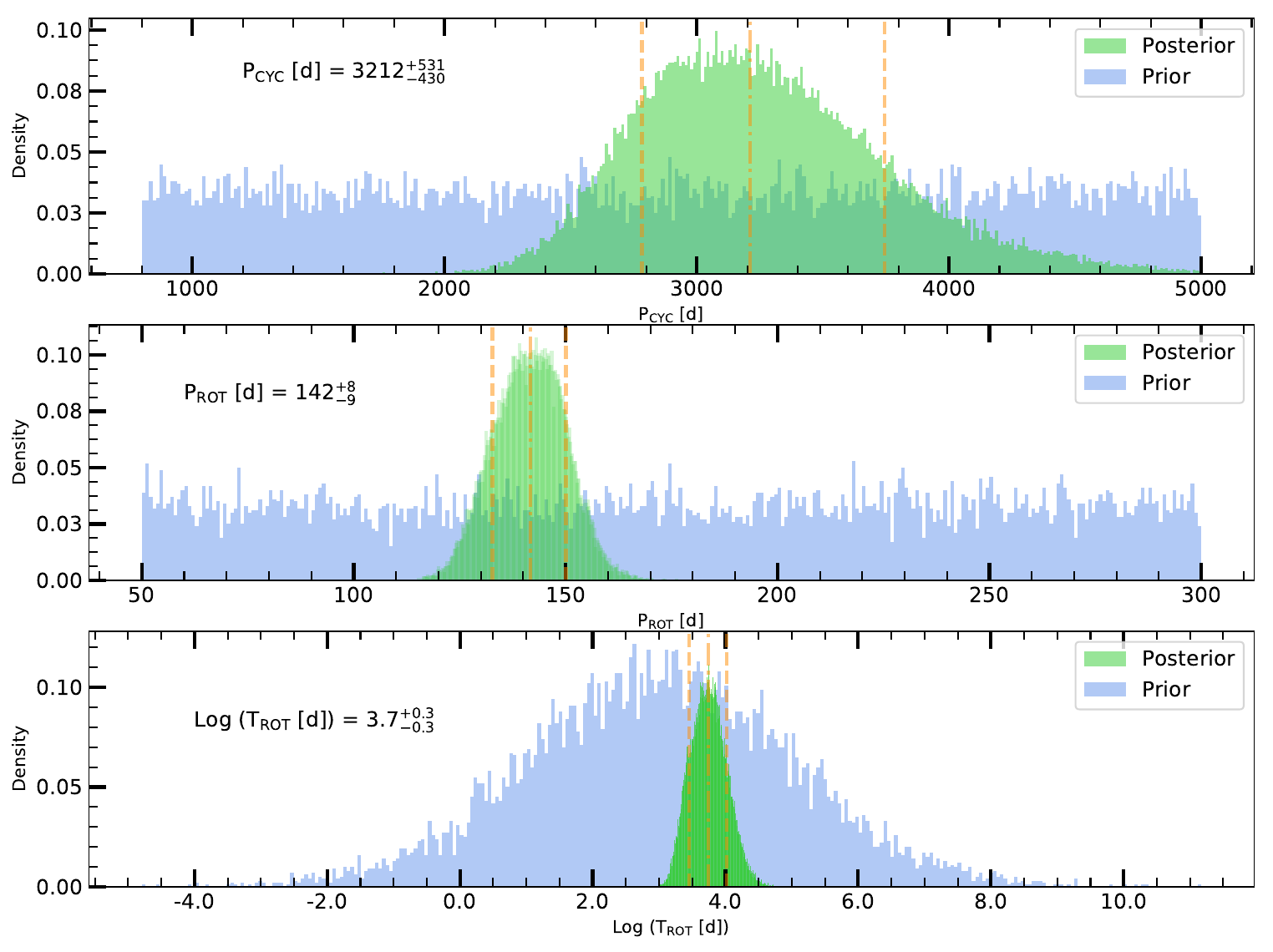}
\caption{
Prior and posterior distributions of the long-term cycle, the GP rotation period, and the timescale from the global analysis of FWHM and RV measurements of ESPRESSO, HARPS, and HARPS-N (model $J1$ in Table~\ref{tab:logz}).}
\label{gj699_cyc_rot}
\end{figure}

\section{Analysis} 

Using the ESPRESSO RV and FWHM data, we build a procedure to search and confirm candidate planets orbiting Barnard's star.

\subsection{Stellar activity\label{sec:activity}}

The magnetic activity of the star is typically the dominant source of RV variations seen in most M dwarf stars, even the most apparently quiet stars such as Barnard's star. The RMS of the ESPRESSO TM RV is 1.86~\ms, and the peak-to-peak RV variation goes from -4 to +4~\ms\ (see Fig.~\ref{gj699_rv}). Following \citet{sua20, far22, sua23} we run a simultaneous model using the ESPRESSO FWHM and RV data only including the GP described in Section~\ref{sec:gp}, together with offsets and jitter parameters for the two ESPRESSO datasets E18 and E19.
We note that the priors adopted for the RV jitter parameters were chosen to be higher than the expected solution to avoid over-fitting in the different model runs, thus forcing the runs if required to converge to lower jitters in the posteriors. In any case, the posterior distributions were significantly narrower than the prior distributions and typically within the prior range.
Fig.~\ref{gj699_gp} depicts the result of this first activity-only model (model $A$ in Table~\ref{tab:logz}). The total number of parameters includes four offsets, four jitters, and ten GP parameters including rotation, timescale and eight GP amplitudes. This includes both FWHM and RV parameters. The GP uses the two simple harmonic oscillators (SHO) at the first and second harmonics of the rotation period, $P_{\rm ROT}$ and $P_{\rm ROT}/2$, and the realisation of the kernel $G$ and its first derivative $G'$ as described in Section~\ref{sec:gp}. We choose this implementation using P and P/2 as we see significant power in the GLS of the RVs at half of the rotation period, not only in ESPRESSO data but also in HARPS, HARPS-N and CARMENES data (see Fig.~\ref{gj699_rv_all}). We use wide priors for both the rotation period, $P_{\rm ROT}$, and the timescale, $T_{\rm ROT}$, with values $\mathcal{U}$(50,300) and $\mathcal{LN}$(3,2) days, respectively. These two hyper-parameters are shared for the RV and FWHM simultaneous analysis. The posterior distributions have a narrow Gaussian shape centered on $P_{\rm ROT} = 152^{+17}_{-14}$~d and $T_{\rm ROT} ={28}^{+10}_{-7}$~d. The timescale is quite short in comparison with the well defined rotation period. It is also shorter than those values found when analyzing individually each activity indicator separately where we found rotation periods in the range $P_{\rm ROT} = [138,275]$~d, and timescales in the range $T_{\rm ROT} =[101,249]$~d. This may indicate that the activity in RV and in the FWHM have different timescales but share the same rotation period. In the right panels of Fig.~\ref{gj699_gp} one can see that the GLS periodograms of the FWHM and RV measurements in this GP-only model, after subtracting the offsets and adding the corresponding jitter terms to both datasets E18 and E19, concentrate the peaks in longer and shorter periods than 100 days for the FWHM and RV respectively. RV data exhibit half a rotation periodicity whereas FWHM data demonstrate the rotation, as shown before in the analysis of the FWHM only with a solution of $P_{\rm ROT} \sim 159$~d and $T_{\rm ROT} \sim 101$~d.

We also tested other GP implementations to model the activity using both the FWHM and RV measurements. We run the one-dimensional GP with a quasi-periodic (QP) kernel implemented within the \texttt{George} package~\citep{fm14george}. We also run the one-dimensional GP with a quasi-periodic and cosine (QPC) kernel, which integrates the period $P$ in the quasi-period function and the period $P/2$ in the cosine function~\citep{per21}. However, for these two implementations, although the result of the median models was similar to the multi-dimensional GP with the SHO $P$ and $P/2$ kernel, the posterior distributions were wider and required a normal Gaussian prior to provide a similar solution. We also tested other kernels within the multi-dimensional implementation, such as ESP and MEP kernels, and similarly to QP and QPC kernels, they all provide worse Bayesian evidence, with $\Delta \ln \mathcal{Z} < -4$ in all cases.

\begin{table*}
\begin{center}
\caption{TM RV and CCF FHWM measurements of ESPRESSO datasets. \label{tab:espdataset}}
\begin{tabular}[centre]{lrrrrr}
\hline \hline
BJD-2450000  & RV (m/s)  & $\delta$RV (m/s)  & FWHM (m/s)  & $\delta$FWHM (m/s)  & SPEC  \\
\hline
8606.799179   &   0.613002  &  0.072661  &  -0.107112  &   0.242575  &  E18 \\
8608.821253   &   0.909167  &  0.071961  &  -0.599065  &   0.240801  &  E18 \\
8609.749449   &   0.000000  &  0.073077  &   0.000000  &   0.236595  &  E18 \\
...  &  ...  &  ...  &  ...  &  ...  &   ... \\
10135.524214  &  -0.598229  &  0.150586  &   2.058090  &   0.436766  &  E19 \\
10137.709347  &  -1.619432  &  0.132375  &   1.448422  &   0.403703  &  E19 \\
10139.512816  &  -2.547420  &  0.078461  &   2.590661  &   0.226382  &  E19 \\
\hline
\end{tabular}
\end{center}
\textbf{Notes:} Portion of the complete table at CDS of ESPRESSO datasets used in Fig.~\ref{gj699_cyc_gp} (model $E1e$ in Table~\ref{tab:logz} with prior and posterior parameters given in Table~\ref{tab:modGPcyc1pe}). The TM RVs are computed with \texttt{S-BART} code and CCF FWHMs are computed from DRS CCF profiles. Labels corresponding to the different spectra are given in column SPEC as E18 and E19 for ESPRESSO before and after the intervention in June 2019, respectively.
\end{table*}

\begin{figure}
\includegraphics[width=9cm]{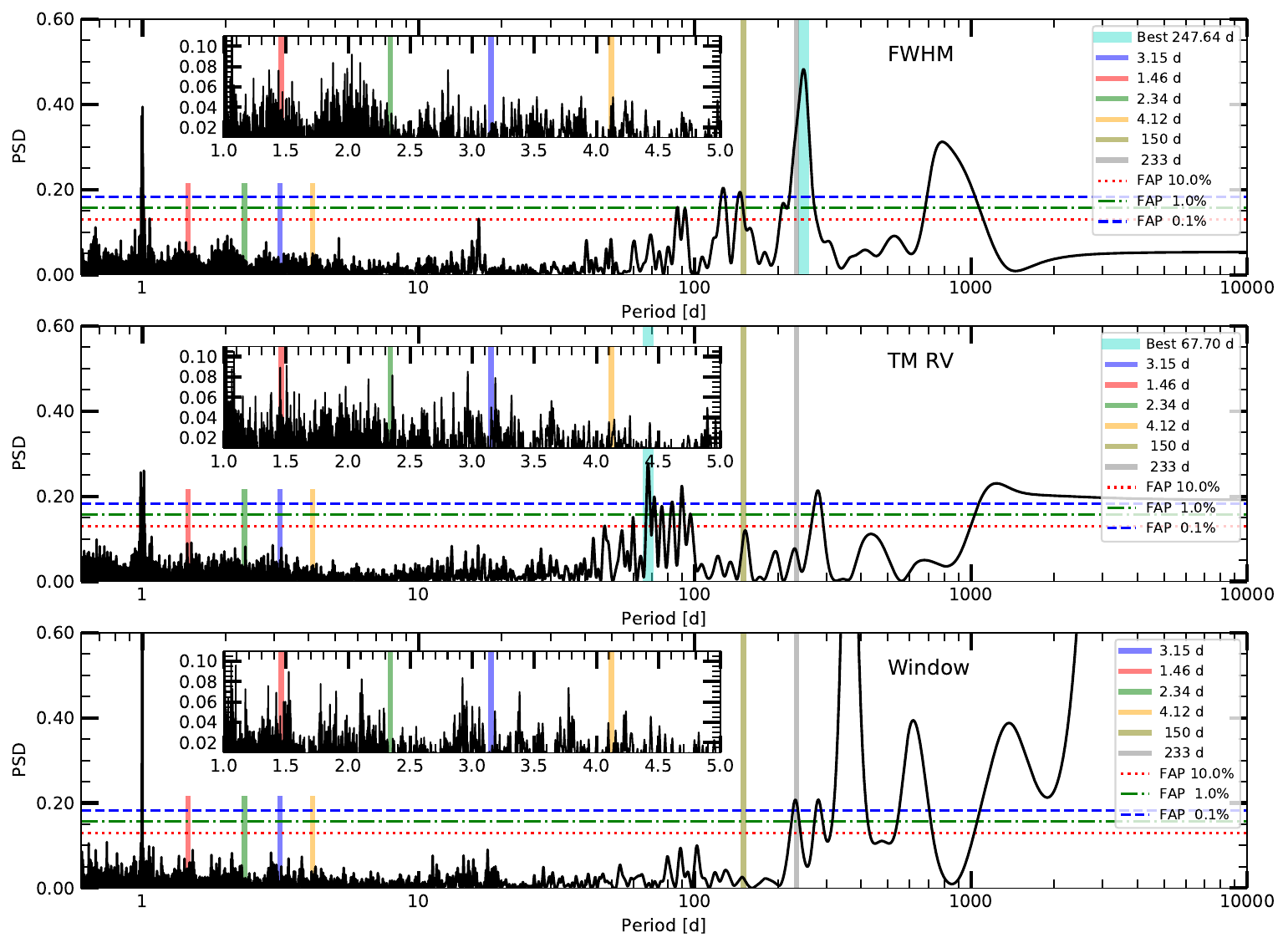}
\caption{
GLS periodograms of ESPRESSO CCF FWHM measurements ({\it top}), TM RV measurements ({\it middle}), and the window function ({\it bottom}) of GJ~699 after subtracting the median of each dataset before (E18) and after (E19) the intervention in June 2019. 
}
\label{gj699_gls}
\end{figure}

The ESPRESSO RV measurements reveal a long-term variation clearly seen in the E19 dataset in Figs.~\ref{gj699_rv} and~\ref{gj699_rv_all}. These variations, difficult to see by eye in other datasets such as HARPS, HARPS-N, and CARMENES due to instrument limitations, may indicate the presence of a long-term activity signal that we also see in the GLS periodograms in Figs.~\ref{gj699_rv},~\ref{gj699_rv_all}, and~\ref{gj699_gp}. We use a longer dataset composed of ESPRESSO, HARPS, and HARPS-N data to verify this possibility. Thus, we run a simultaneous model of FWHM and RV measurements including the GP model previously described and adding a double sinusoidal model as used in the analysis of the ASAS-SN photometric data given in equation~\ref{eq_cyc_dsine}. This global model is displayed in Fig.~\ref{gj699_esp_har}, corresponding to model $J1$ in Table~\ref{tab:logz}, including the double sinusoidal long-term cycle model, the GP model and a Keplerian model discussed in Section~\ref{sec:planet}. The wide prior distributions, with values $P_{\rm CYC}$ $\mathcal{U}(800,5000)$, $P_{\rm ROT}$ $\mathcal{U}(50,300)$, and  $T_{\rm ROT}$ $\mathcal{LN}(3,2)$, and relatively narrow posterior distributions of the activity parameters are displayed in Fig.~\ref{gj699_cyc_rot}. This run gives the median values of the long-term cycle, $P_{\rm CYC} = 3212^{+531}_{-430}$~d, the rotation, $P_{\rm ROT} = 142^{+8}_{-9}$~d, and the timescale, $T_{\rm ROT} ={40}^{+14}_{-10}$~d. These spectroscopically derived values perfectly match previous determination of the long-term cycle period and the stellar rotation period~\citep{tol19}. We note a slightly longer timescale in this model although still at about one third of the rotation period. The global analysis including the CARMENES data without including any Keplerian model (model $K1$ in Table~\ref{tab:logz}) provides almost the same result, with median values of the long-term cycle, $P_{\rm CYC} = 3179^{+508}_{-364}$~d, the rotation, $P_{\rm ROT} = 144^{+11}_{-9}$~d, and the timescale, $T_{\rm ROT} ={28}^{+11}_{-8}$~d.

\begin{figure}
\includegraphics[width=9cm]{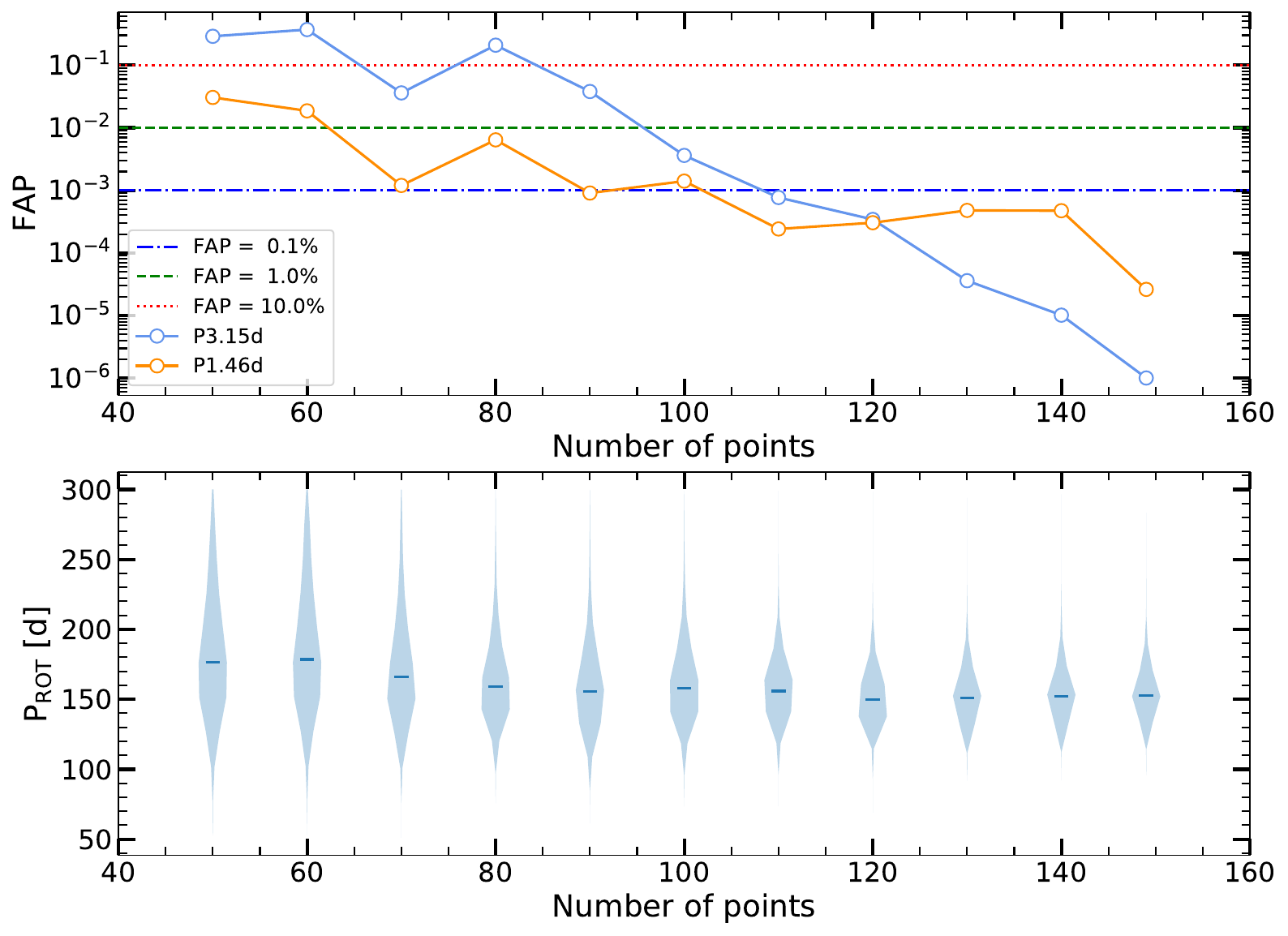}
\caption{
Evolution of the false-alarm probability of the 3.15~d and 1.46~d (1 day alias) signals ({\it top}) and the GP rotation period ({\it bottom}) with the number of observations. The 0.1\%, 1\%, and 10\% FAP lines are computed using the full 149 ESPRESSO dataset.
}
\label{gj699_pall}
\end{figure}

\begin{table}[!h]
\begin{center}
\caption{Bayesian evidence of different models\label{tab:logz}}
\begin{tabular}[centre]{llrrrr}
\hline \hline
Name & Model  & $N_{\rm pl}$ & $N_{\rm par}$ & $\ln \mathcal{Z}$ & $\Delta \ln \mathcal{Z}$ \\
\hline
\multicolumn{6}{c}{E18,E19 ($N_{\rm point}=149 \times 2$)} \\
\hline
A            & GP          & 0 & 18 &  $-449.3$ & $-2.1$  \\
AD           & GP          & 0 & 18 &  $-466.5$ & $-19.3$ \\
AL           & GP          & 0 & 18 &  $-452.3$ & $-5.1$  \\
AT           & GP          & 0 & 18 &  $-450.5$ & $-3.3$  \\
C1c          & A+1pcLU50   & 1 & 21 &  $-443.7$ & $+3.5$  \\
C1e          & A+1peLU50   & 1 & 23 &  $-445.5$ & $+1.7$  \\
{\bf D}      & {\bf A+cycN} & {\bf 0} & {\bf 26} & {\bf --447.2} & {\bf 0.0} \\
DD           & AD+cycN     & 0 & 26 &  $-463.8$ & $-16.6$ \\
DL           & AL+cycN     & 0 & 26 &  $-453.6$ & $-6.4$ \\
DT           & AT+cycN     & 0 & 26 &  $-448.2$ & $-1.0$ \\
E1c          & D+1pcLU50   & 1 & 29 &  $-439.5$ & $+7.7$ \\
E1cT         & DT+1pcLU50  & 1 & 29 &  $-440.7$ & $+6.5$ \\
E1e          & D+1peLU50   & 1 & 31 &  $-438.6$ & $+8.6$ \\
E1eD         & DD+1peLU50  & 1 & 31 &  $-460.1$ & $-12.9$ \\
E1eL         & DL+1peLU50  & 1 & 31 &  $-446.7$ & $+0.5$ \\
E1eT         & DT+1peLU50  & 1 & 31 &  $-442.2$ & $+5.0$ \\
E12c         & D+1pcLU20   & 1 & 29 &  $-440.5$ & $+6.7$ \\
E12e         & D+1peLU20   & 1 & 31 &  $-440.2$ & $+7.0$ \\
E22c         & D+2pcLU20   & 2 & 32 &  $-443.8$ & $+3.4$ \\
E22e         & D+2peLU20   & 2 & 36 &  $-443.6$ & $+3.6$ \\
\hline
\multicolumn{6}{c}{E18, E19, H15, HAN ($N_{\rm point}=298 \times 2$)} \\
\hline
{\bf I1}    & {\bf GP+cycU} & {\bf 0} & {\bf 35} & {\bf --1074.1} & {\bf 0.0}  \\
I2          & GP+cycN     & 0 & 35 & $-1073.1$ & $+1$    \\
J1          & I1+1peLU50  & 1 & 40 & $-1061.7$ & $+12.4$ \\
J2          & I2+2peLU50  & 2 & 45 & $-1063.9$ & $+10.2$ \\
\hline
\multicolumn{6}{c}{E18, E19, H15, HAN, CAR ($N_{\rm point}=792 \times 2$)} \\
\hline
{\bf K1}  & {\bf GP+cycU} & {\bf 0} & {\bf 39} & {\bf --4123.0} & {\bf 0.0}  \\
K2          & GP+cycN     & 0 & 39 & $-4122.0$ & $+1.0$  \\
L1          & K1+1peLU50  & 1 & 44 & $-4110.7$ & $+12.3$ \\
L2          & K2+1peLU50  & 1 & 44 & $-4109.9$ & $+13.1$ \\
M2          & K2+2peLU20  & 2 & 49 & $-4108.8$ & $+14.2$ \\
\hline
\end{tabular}
\end{center}
\textbf{Notes:} Model selection based on Bayesian evidence of the analysis of CCF FWHM and TM RV measurements. $A$, $AD,$ and $AL$ are models for ESPRESSO TM, DRS, and LBL RVs, respectively. Models with $T$ are for TMtc RVs computed using telluric-corrected ESPRESSO spectra (see Section~\ref{sec:obs_esp}). Different models: {\it cyc} indicates cycle, and $N$, $U,$ and $LU$ indicate normal, uniform, and log-uniform priors. $LUx$ indicates priors $\mathcal{LU}(0.5,x)$ with $x = 20$ and $50$~d. $npe$ and $npc$ indicate $n$ Keplerian and circular orbits. We highlight in bold font the reference activity-only model in each group of datasets.
\end{table}

\begin{table}[!h]
\begin{center}
\caption{Bayesian evidence of models evaluating the 233~d candidate\label{tab:logz233}}
\begin{tabular}[centre]{llrrrr}
\hline \hline
Name & Model  & $N_{\rm pl}$ & $N_{\rm par}$ & $\ln \mathcal{Z}$ & $\Delta \ln \mathcal{Z}$ \\
\hline
\multicolumn{6}{c}{P233: E18, E19 ($N_{\rm point}=149 \times 2$)} \\
\hline
{\bf D}      & {\bf A+cycN} & {\bf 0} & {\bf 26} & {\bf --447.2} & {\bf 0.0} \\
N1          & D+1peN0.5   & 1 & 31 & $-449.4$ & $-2.2$  \\
N2          & D+1peU50    & 1 & 31 & $-449.6$ & $-2.4$  \\
S1          & D+1peU100   & 1 & 31 & -- & -- \\
S2          & D+1peU100   & 1 & 31 & -- & -- \\
\hline
\multicolumn{6}{c}{P233: E18, E19, H15, HAN, CAR ($N_{\rm point}=792 \times 2$)} \\
\hline
{\bf K1}  & {\bf GP+cycU} & {\bf 0} & {\bf 39} & {\bf --4123.0} & {\bf 0.0}  \\
O1          & K1+1peU50   & 1 & 44 & $-4123.5$ & $-0.5$ \\
\hline
\end{tabular}
\end{center}
\textbf{Notes:} Model selection based on Bayesian evidence evaluating the 233~d candidate planet. Different models as in Table~\ref{tab:logz}. Model S1 and S2 are simulations with injected planets. $N$ and $U$ indicate normal and uniform priors, $N0.5$ for $\mathcal{N}(P_{\rm orb},\sigma_P)$ with $P_{\rm orb}=233$~d and $\sigma_P=0.5$~d. $U50$ and $U100$ for $\mathcal{U}(200,x)$ with $x=250$ and 300~d. Here, $npe$ indicates $n$ Keplerian orbits.
\end{table}

\subsection{Candidate planetary signals~\label{sec:pl}}

In the RV residuals of the simplest activity-only model shown in Fig.~\ref{gj699_gp} (model $A$ in Table~\ref{tab:logz}), described in Section~\ref{sec:activity}, we find several signals in the GLS periodogram with a false-alarm probability (FAP) of less than 1\%, which we cannot attribute to any activity process, and we tentatively associated them with candidate planetary signals. We identify three of them as main signals at 3.15~d (with 1.46~d and 0.76~d as 1 d aliases, and 0.59~d as 1 d alias of the 1.46~d signal), 4.12~d (with 1.32~d as 1 d alias), and 2.34~d (with 1.74~d as 1 d alias). Figure~\ref{gj699_gls} shows the GLS periodogram of the ESPRESSO FWHM and RV measurements, and the window function computed as the periodogram of all values equal to 1 at the BJD values of the ESPRESSO FWHM and RV measurements. We note strong peaks at 1 d and 1 yr in the window function and we do not see any significant peak at the position of the main candidate planetary signals. The GLS of the FWHM and the RV measurements are dominated by the activity signals at about 247~d and 67~d, respectively, but there are no significant peaks at the position of the candidate planetary signals. In Fig.~\ref{gj699_gp}, we display the GLS after adding the jitter of the corresponding activity-only model which explain the slight differences with those in Fig.~\ref{gj699_gls}. The GLS after subtracting the GP model in the FWHM does not show either any significant peak in the whole period range from 1 to 10,000~d (see Fig.~\ref{gj699_gp}).

\subsection{Evaluating the 3.15 d signal~\label{sec:planet}}

We evaluate the strongest signal at 3.15~d in the RV residuals after subtracting the GP model (see Fig.~\ref{gj699_gp}). In Fig.~\ref{gj699_pall} we show the evolution of both the 3.15~d signal and the 1 d alias at 1.46~d as a function of number of RV points. To do that, we run the activity-only model for both the FWHM and RV measurements from 50 RV points and adding ten points in each run until we end up with all 149 points. We clearly see a steady increase in confidence of both signals. In particular, the 3.15~d signal, reaches in the last run a FAP of about $10^{-6}$, being already very significant (FAP~$< 0.1\%$) after 110 ESPRESSO observations. In all these runs, we adopt wide priors on $P_{\rm ROT}$ $\mathcal{U}(50,300)$~d and on $T_{\rm ROT}$ $\mathcal{LN}(3,2)$. We see how the estimated rotation period is converging towards the final value of $P_{\rm ROT} = 152^{+17}_{-14}$~d of this model $A$ in Table~\ref{tab:logz}. The 3.15~d signal is certainly slightly diminished due to the GP modeling since we are not including a Keplerian model targeting the 3.15~d signal in this run.

We first include a circular planet model (model $C1c$ in Table~\ref{tab:logz}) in a new run to search for that particular signal but again with a wide prior $\mathcal{LU}(0.5,50)$~d on orbital period, $P_{\rm orb}$. We use the time of conjunction given by the phase, $\phi$, with a prior $\mathcal{U}$(-0.5,0.5), centered around the maximum time, $t_{\rm max}$, of the observation baseline as in eq.~\ref{eq_tc}:
\begin{equation} \label{eq_tc}
T_{0} = t_{\rm max}+P_{\rm orb}   \cdot  \phi,
\end{equation} 
\noindent and the semi-amplitude velocity, $k_p$, with a uniform prior $\mathcal{U}$(0,5)~\ms. 
    We detect unequivocally the 3.15~d signal as a planetary signal with a Bayesian evidence $\ln \mathcal{Z}=-443.7$ (model $C1c$ in Table~\ref{tab:logz}), and $\Delta \ln \mathcal{Z} = +5.6$, with respect to the activity-only model $A$, which corresponds to a $1/e^{+5.6} = 0.37$\% false alarm probability for the activity+planet model. We run this model five times to check any possible $\ln \mathcal{Z}$ variance and found $\ln \mathcal{Z}$ in the range [-442.7,-445.3], with a mean $\ln \mathcal{Z}$ of -444.2 ($\sigma =1.0$). This run converges to $P_{\rm orb} = 3.1533 \pm 0.0005$~d and $k_p = 54^{+8}_{-9}$~\cms. We note that in Table~\ref{tab:logz} all the differences $\Delta \ln \mathcal{Z}$ are always referred for simplicity to the activity-only model $D$, although in some cases, as in this particular case, it is another model (model $A$) the reference activity-only model.

We also run a Keplerian planet model (model $C1e$ in Table~\ref{tab:logz}) given in eq.~\ref{eq_kepler}:
\begin{equation} \label{eq_kepler}
  y(t)= k_p \left(\cos(\eta+\omega) + e \ \cos(\omega)\right),
\end{equation} 
\noindent where the true anomaly $\eta$, which is the angle between periastron and the planet, measured from the barycenter of the system~\citep[e.g.][]{eas13}, is related to the solution of the Kepler's equation, and depends on the orbital period of the planet $P_{\rm orb}$, the orbital phase $\phi$, and the eccentricity of the orbit $e$.
We use the \texttt{RadVel} toolkit to implement the Keplerian model~\citep{ful18radvel}.
We associate the orbital phase $\phi$ with the time at inferior conjunction, $T_c$, which depends on the maximum time, $t_{\rm max}$, of the observation baseline as in eq.~\ref{eq_tc}. Then, we convert this $T_c$ time into time of periastron, $T_p$, which depends on the argument of periastron $\omega$ and the eccentricity $e$. Following \citet{eas13}, we parameterise the eccentricity as in eq.~\ref{eq_ecc}:
\begin{equation} \label{eq_ecc}
\begin{split}
e = (\sqrt{e} ~\cos(\omega))^{2} + (\sqrt{e} ~\sin(\omega))^{2} \, , \\
\omega = \arctan2(\sqrt{e} ~sin(\omega),\sqrt{e} ~\cos(\omega)).
\end{split}
\end{equation}
Initially we tried an eccentricity with uniform prior $\mathcal{U}$(0,1), but the solution converged to a very low value consistent with zero. Thus we decided to sample $\sqrt{e} ~cos(\omega)$ and $\sqrt{e} ~\sin(\omega)$ with normal priors $\mathcal{N}$(0,0.3), still allowing the possibility of high eccentricity values but favoring low eccentricity values, expected for the short periods of the candidate planets. Eccentricity is typically overestimated in noisy data and datasets with unmodelled sources of variability \citep{har19}. We measure a $\ln \mathcal{Z}$ of --445.5 (model $C1e$ in Table~\ref{tab:logz}), which corresponds to a $\Delta \ln \mathcal{Z}$ of $+3.8$ (equivalent to 2.2\% FAP), with respect to the reference activity-only model (model $A$ in Table~\ref{tab:logz}).

\begin{figure}
\includegraphics[width=9cm]{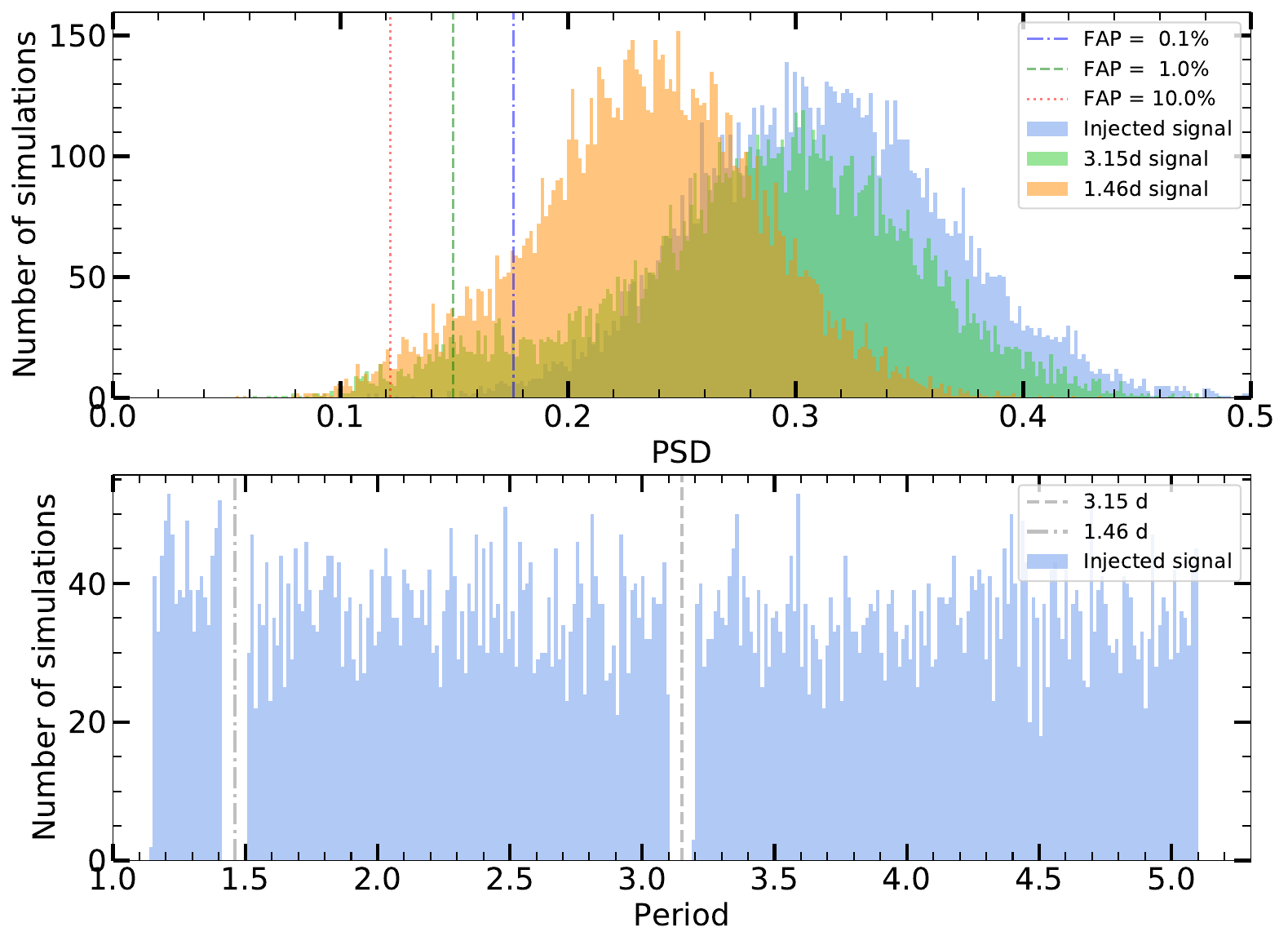}
\caption{
    GLS power spectral density of the observed signals 3.15d and 1.46d after removing the GP compared to a simulated 54~\cms\ injected sinusoidal signal at different orbital periods.
}
\label{gj699_k50_315}
\end{figure}

\begin{figure*}
\includegraphics[width=18cm]{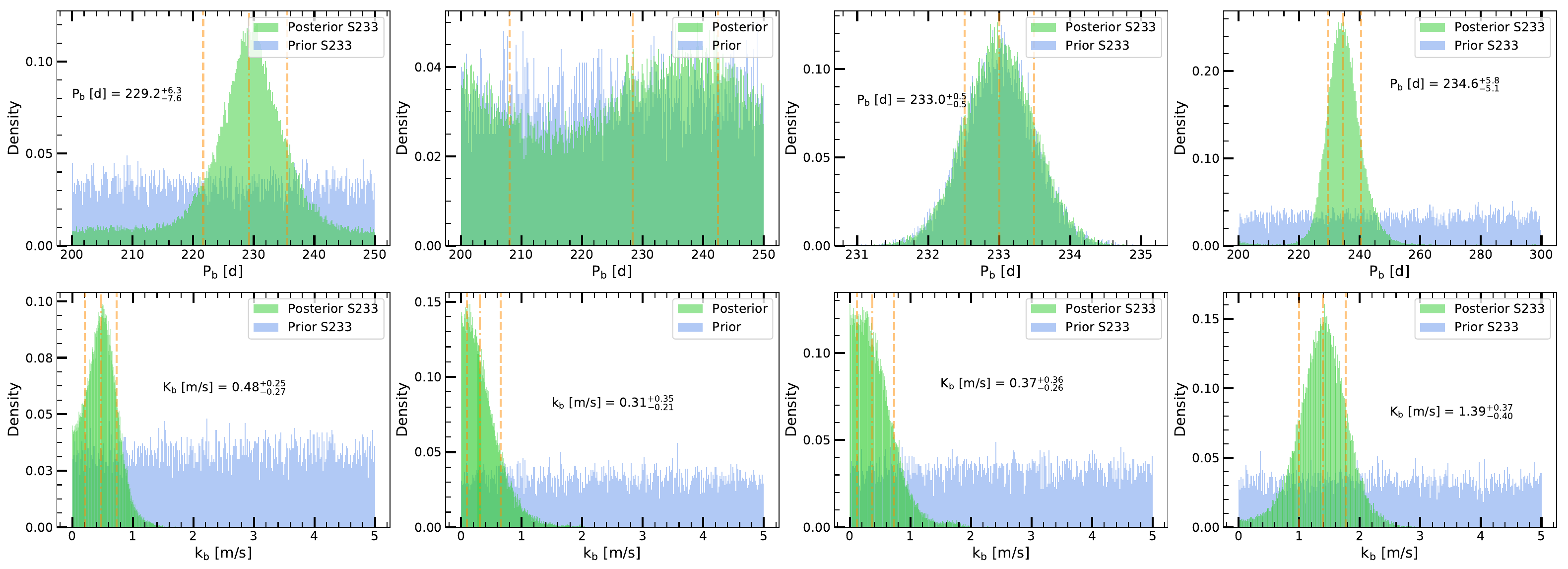}
\caption{
    Prior and posterior distributions of the period and semi-amplitude from the global analysis of FWHM and RV measurements of the blind search of the 233d candidate planet using ESPRESSO, HARPS, HARPS-N and CARMENES data ({\it left}, model $O1$ in Table~\ref{tab:logz233}), a blind search ({\it centre-left}, model $N2$ in Table~\ref{tab:logz233}) and a guided search ({\it centre-right}, model $N1$ in Table~\ref{tab:logz233}) of the 233d candidate planet using ESPRESSO only, and a blind search of the 233d simulated signal ({\it right}, model $S1$ in Table~\ref{tab:logz233}).
    }
\label{gj699_p233}
\end{figure*}

\subsection{Statistical tests to validate the 3.15 d signal~\label{sec:test315}}

We ran several tests to validate the 3.15~d signal, using the model $C1e$ in Table~\ref{tab:logz} as a reference model. Firstly, following \citet{raj16}, we tested the injection and recovery of different sinusoidal signals with $k_p = 54$~\cms, with random phases $\mathcal{U}$(-0.5,0.5), and periods uniformly distributed within $3.15\pm2.0$~d, excluding 0.1~d around 3.15~d and its 1 d alias 1.46~d. We performed 10,000 simulations and the result is displayed in Fig.~\ref{gj699_k50_315}. We randomly selected 10,000 different samples among the resulting 124524 samples of the run of model $C1e$. From each sample, we built the GP model using the parameters of that sample, subtracted it from the observed RVs and injected one simulated sinusoidal signal with period $P_x$, random phase, and a semi-amplitude of $k_p = 54$~\cms. We then computed the GLS periodogram and measured the power spectral density (PSD) of the signals at the 3.15~d, 1.46~d, and $P_x$. The distribution of PSD of the simulated planet $x$ appears to perfectly match that of signal 3.15~d, and both show larger values of PSD distributions than that of the 1.46~d signal. This indicates that indeed the 1.46~d signal is just the 1 d alias of the main signal at 3.15~d. We note that the PSD distribution of the 3.15~d signal has  a long tail below PSD~$=0.2$ due to the fact that about 10\% of the samples converge to a detection of 4.12~d signal or the 1.46~d signal as the main signal in the GLS periodogram, thus decreasing the PSD of the 3.15~d signal. This, and the fact that we adopted a Keplerian model in this run, may explain why the PSD of the injected planet $x$ is larger than that of the 3.15~d signal in about 60\% of the simulations.

Secondly, we ran the search of a Keplerian solution with the same priors as in model $C1e$ in Table~\ref{tab:logz}, but we randomly removed 15 points (10\% of data points) from the datasets of FWHM and RV measurements. We repeated this 25 times and recovered the 3.15~d signal in 18 cases (72\% with the 1 d alias 1.46~d in 3 cases) and the 4.12~d signal in 7 cases (28\% with the 1 d alias 1.32~d in 1 case).

Thirdly, we ran another simulation using again the 10,000 samples of the run model $C1e$ in Table~\ref{tab:logz}, where we built the GP using each sample parameter. For each iteration of these 10,000 samples, we computed the GLS of the RVs, corrected for offsets, and, with the jitter added to the RV uncertainties and after subtracting the GP, measured the PSD at 3.15 and 1.46~d. The distributions of these PSD values are labelled as `observed' in Fig.~\ref{gj699_s315}. We see that the observed PSD distributions of the RVs  measured at 3.15~d and 1.46~d have some structure at lower PSD values, which is related to a significant fraction of the samples where the solution peaks at 4.12~d (10.85\%) and at 2.34~d (0.08\%) instead of 3.15~d (89\%). This is consistent with the aforementioned test. As expected, a blind search for the planetary signal recovers in most cases the 3.15~d signal and thus the median of the PSD distribution of the observed 3.15~d signal is larger than that of the PSD distribution of the observed 1.46~d signal. We refer to Appendix~\ref{sec:ap:315} for further discussion of Fig.~\ref{gj699_s315}.

\begin{figure*}
\includegraphics[width=18cm]{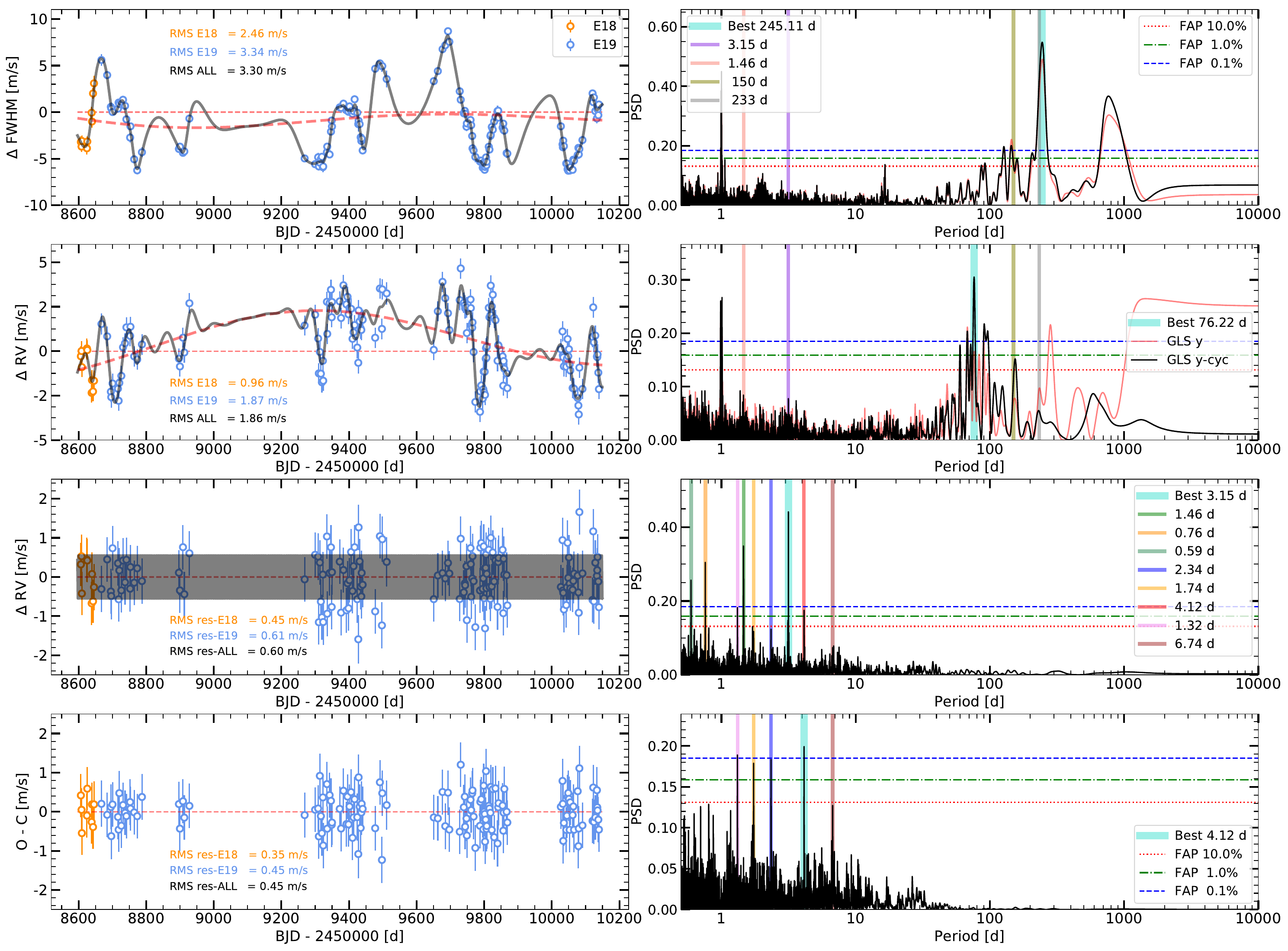}
\caption{
    ESPRESSO FWHM measurements ({\it top}), RV measurements ({\it middle}), RV residuals ({\it next bottom}) from long-term cycle (red thick dashed line) and SHO ($P_{\rm ROT}$ and $P_{\rm ROT}/2$) GP model (grey solid line), and RV residuals ({\it bottom}) from the Keplerian model (grey shaded area), and GLS periodograms ({\it left}) of GJ~699. The uncertainties include the jitter term coming from the global model $E1e$ in Table~\ref{tab:logz}, with prior and posterior parameters given in Table~\ref{tab:modGPcyc1pe}.
}
\label{gj699_cyc_gp}
\end{figure*}

Fourthly, we run another test where we model the activity of the star using only the RVs with a multi-sinusoidal function composed of six sinusoidals with periods in the range [60,255]~d (see Fig.~\ref{gj699_s315sin}). We found that six sinusoidal functions were sufficient to be able to reproduce all possible RV variations due to stellar activity. The fitting procedure was done following a prewhitening methodology, which is based on a single sinusoidal fit every time at the period of highest peak (lowest FAP) in the GLS periodogram followed by subtraction of this fit and repeating this sequence. The fitted periods were 79.1~d, 254.9~d, 68.2~d, 89.8~d, 59.5~d, and 244.9~d. This activity-only model resembles the GP-only model displayed in Fig.~\ref{gj699_gp} (model $A$ in Table~\ref{tab:logz}). The GLS periodogram of the RV residuals after subtracting the multi-sinusoidal fit also exhibits the detection of the 3.15~d signal above the 0.1\% FAP line.

We subsequently performed an additional test, where we modelled the RVs and all activity indicators using a moving average model with an exponential decay using a 5.25~d window. The aim of this test was to run a simplified model with less freedom than GP models. To choose this 5.25~d window of the moving average, we used the FWHM, trying to avoid over-fitting. This 5.25~d exponential moving average model was then applied to the RVs and all activity indicators. The result of this test is depicted in Fig.~\ref{gj699_s315ewma_real}, where we are able to recover again the 3.15~d signal in the GLS periodogram of the RV residuals after subtracting this activity-only model using the exponential moving average. This 3.15~d signal does not appear in the residuals of any of the activity indicators after fitting the same activity-only model using the 5.25~d exponential moving average. All GLS periodograms of all activity indicators have all GLS peaks below 10\% FAP line.

Finally, we performed a last test where we build several simulated RV time series: (i) a first flat model to evaluate the sampling, with flat result in the GLS periodogram; (ii) a second one using the activity-only model with injected white noise at 30~\cms level, resembling the moving average model, again resulting in a flat GLS periodogram with all peaks below the 10\% FAP line; and (iii) a third one by injecting a planetary signal at 3.15~d (see Fig.~\ref{gj699_s315ewma_sim}), thus in this case recovering the 3.15~d signal. We note the difference between the GLS periodograms of the residuals of real and simulated RVs, pointing that the real data shows additional peaks such as the 4.12~d signal although at the 10\% FAP level.

All these tests allow us to confidently confirm the 3.15~d signal as a planet signal.

\subsection{Evaluating the 233~d candidate super-Earth~\label{sec:test233}}

\citet{rib18} reported the detection of a candidate super-Earth-like planet orbiting in an slightly eccentric orbit with a period of $232.8\pm0.4$~d, with a semi-amplitude velocity of $k_p = 1.2\pm0.1$~\ms. This result has been challenged by \citet{lub21,art22lbl} which argue that the signal is connected to stellar activity as the planet candidate period is close to 1 yr alias of the rotation signal~\citep{tol19}. We run several models with the RV and FWMH data to search for the 233~d signal. Firstly, we use the whole dataset of ESPRESSO, HARPS, HARPS-N, and CARMENES to run a blind search for the 233~d signal (model $O1$ in Table~\ref{tab:logz233}), with a uniform prior $\mathcal{U}$(200,250) on the orbital period. The posterior distribution of the orbital period and semi-amplitude velocity are depicted in Fig.~\ref{gj699_p233}. This model is not satisfactory since the semi-amplitude, $k_p=0.48^{+0.25}_{-0.27}$~\ms, is consistent with zero at 1.8~$\sigma$, and significantly lower than that of the reported candidate planet. The model also exhibit a Bayesian evidence $\Delta \ln \mathcal{Z} = -0.5$ with respect to activity-only model, but the orbital period, $P_{\rm orb}=229^{+6}_{-8}$~d, is shorter than that of the candidate planet although still consistent within the error bars.

\begin{figure}
\includegraphics[width=9cm]{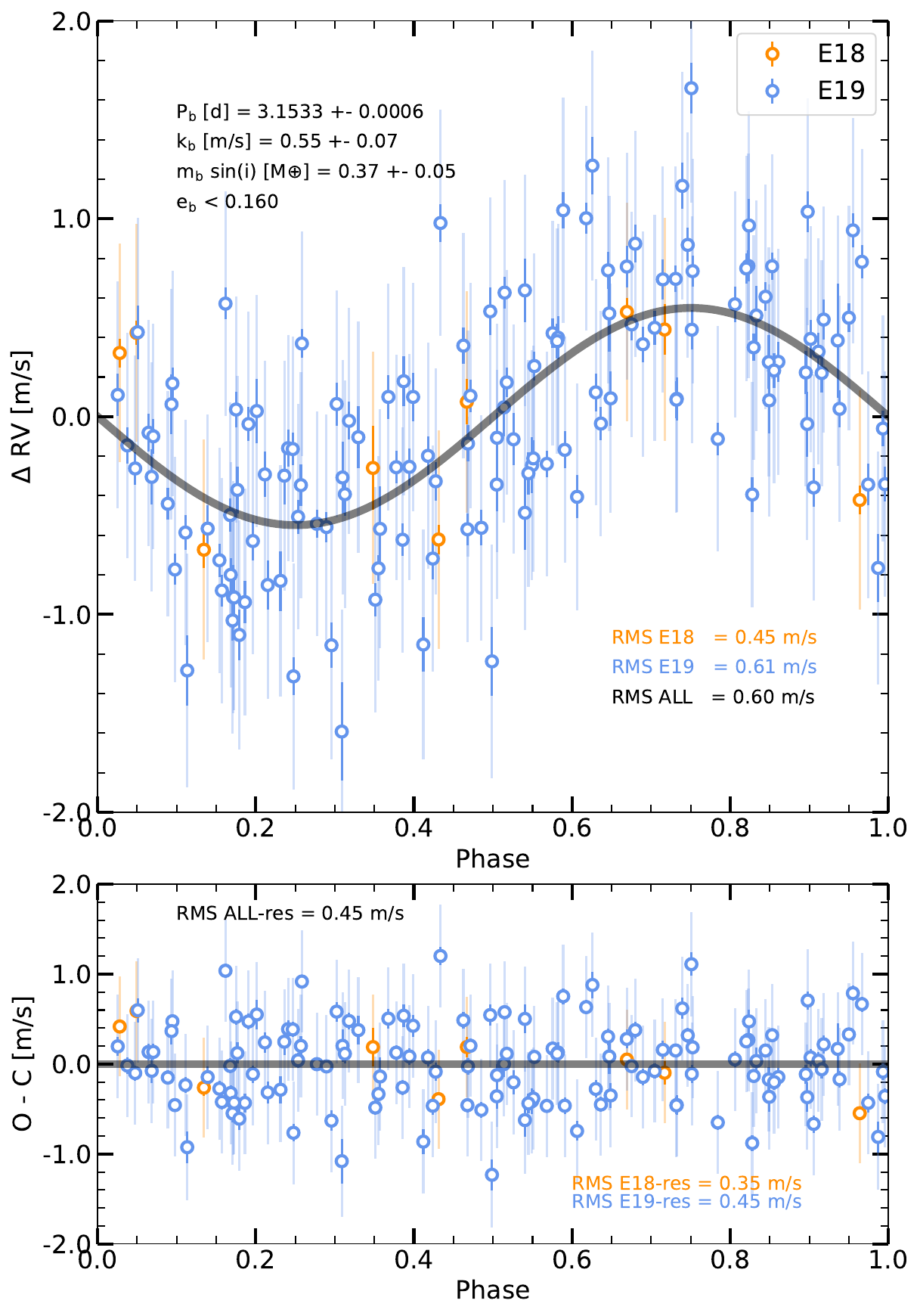}
\caption{
    RV curve of the sub-Earth-mass planet of GJ~699 with a 3.15~d orbital period together with ESPRESSO RVs, with uncertainties with (light colour) and without the jitter term (dark colour) coming from the global model $E1e$, with prior and posterior parameters given in Table~\ref{tab:modGPcyc1pe}.
}
\label{gj699_1pe_phase}
\end{figure}

Secondly, we run a blind search only with ESPRESSO data, with the period free (model $N2$ in Table~\ref{tab:logz233}) in the same range $\mathcal{U}$(200,250), and in this case the posterior of the orbital period is flat as the prior, and the semi-amplitude is even lower and consistent with zero (see central left panels in Fig.~\ref{gj699_p233}). Thirdly, we run a guided search (model $N1$ in Table~\ref{tab:logz233}) with ESPRESSO only with the period prior at $\mathcal{N}$(233,0.5) and the result is similar to the blind search, the planet is not detected with a worse Bayesian evidence in both cases with $\Delta \ln \mathcal{Z} = -2.5$. One could argue that the number of data points and baseline is not enough to detect such a long orbital period, but the semi-amplitude is really significant for ESPRESSO to have missed the candidate planet.

Finally, we run a simulation by adding a planetary signal in the activity model with injected white noise using the uncertainties including the jitter terms (with a mean value of 0.7~\ms) of the previous run (model $N1$ in Table~\ref{tab:logz233}). We injected a Keplerian with the same eccentricity (very close to zero) of the Keplerian solution of model $N1$, a period of $P_{\rm orb} = 233$~d, with a semi-amplitude velocity of $k_p = 1.2$~\ms. We run a blind search with priors of $P_{\rm orb}$ $\mathcal{U}$(200,300)~d and $k_p$ $\mathcal{U}$(0,5)~\ms. The posteriors are $P_{\rm orb} = 235\pm 6$~d and $k_p = 1.4\pm0.4$~\ms (with the $k_p$ only detected at 3.5$\sigma$) which are consistent with the injected Keplerian signal but not exactly peaking at right values (see right panels in Fig.~\ref{gj699_p233}). This result (model $S1$ in Table~\ref{tab:logz233}) is probably related to the baseline and number of RV epochs, as well as to the search for a signal at longer period than the activity-induced signal by stellar rotation. We explore if using a conservative white noise of the input model had a significant impact, so we run one last simulation (model $S2$ in Table~\ref{tab:logz233}) similar to the previous one but with the injected white noise at 0.3~\ms, and we get a slightly improved result whose posteriors are very similar with median values $P_{\rm orb} = 235\pm 5$~d and  $k_p = 1.3\pm0.3$~\ms (at 4.3$\sigma$). Therefore, the quality of the ESPRESSO RVs would have allowed us to clearly detect such a relatively strong signal as that of the candidate planet reported in \citet{rib18}. Thus, we conclude that ESPRESSO data do not support the existence of the candidate planetary signal.

\begin{figure}
\includegraphics[width=9cm]{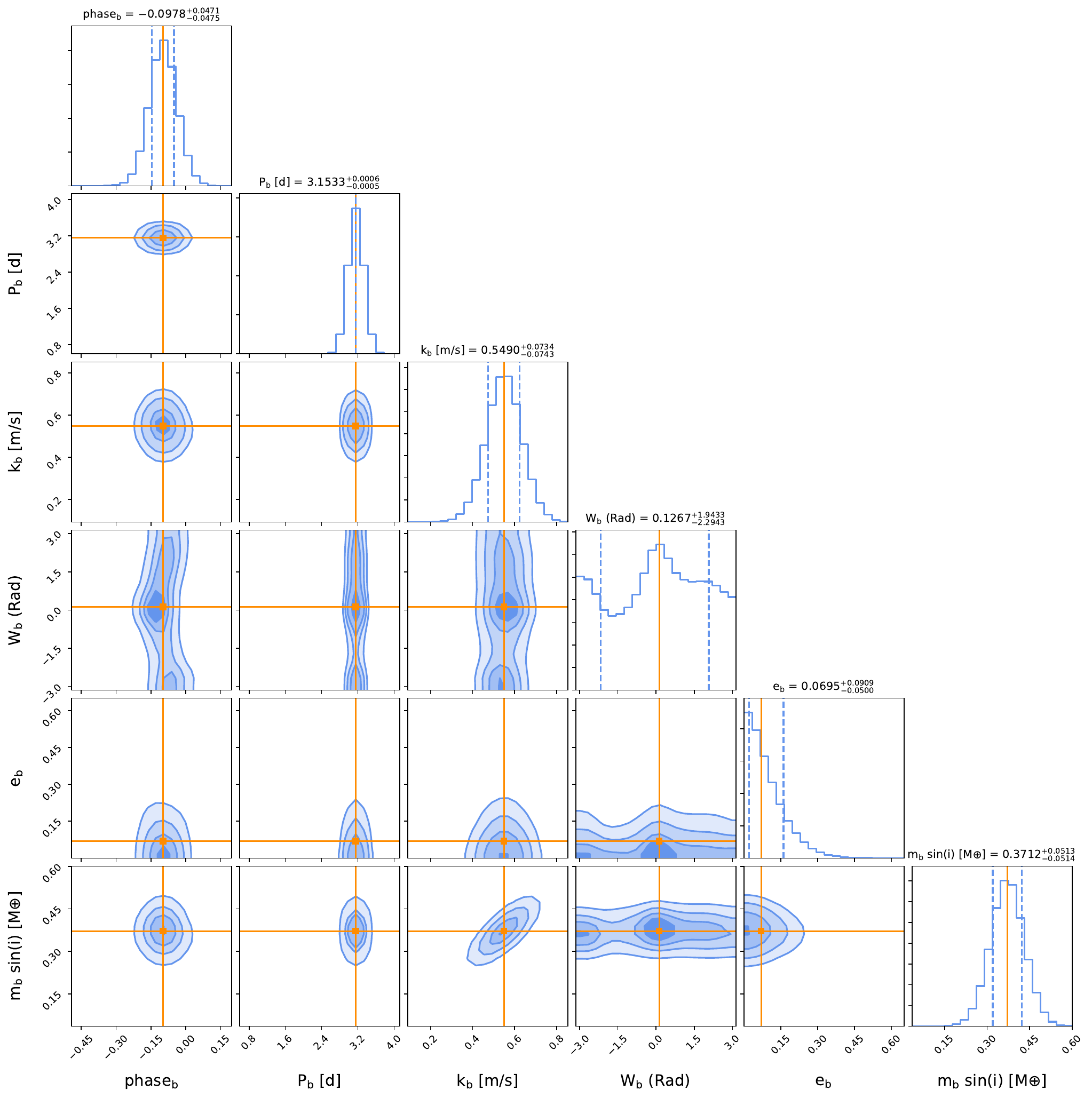}
\caption{
    Corner plot with the posterior distributions of the orbital parameters of the sub-Earth-mass planet of GJ~699 with a 3.15~d orbital period, with prior and posterior parameters given in Table~\ref{tab:modGPcyc1pe}.
}
\label{gj699_1pe_corner}
\end{figure}

\begin{table}
\begin{center}
\caption{Parameters of planet Barnard b. \label{tab:planet}}
\begin{tabular}[centre]{l c}
\hline \hline
Parameter  & GJ~699 b  \\
\hline
$T_{0}$ -- 2460139 [d]        & 0.204 $\pm$ 0.149     \\
$P_{\rm orb}$ [d]             & 3.1533 $\pm$ 0.0006   \\
$k_p$ [\ms]                   & 0.55 $\pm$ 0.07       \\
$m_p \sin i$ [\mearth]        & 0.37 $\pm$ 0.05       \\
$a_p$ [AU]                    & 0.02294 $\pm$ 0.00033 \\
$e_p$                         & \textless~0.16        \\
$S_p$ [\searth]               & 6.76 $\pm$ 0.05       \\
$T_{\rm eq,}$$_{A = 0.3}$ [K] & 400 $\pm$ 7           \\
$\mathcal{P}_{\rm transit}$   &  3.7\%                \\
\hline
\end{tabular}
\end{center}
\end{table}

\begin{table}
\begin{center}
\caption{Prior and posteriors of the activity+planet ESPRESSO model \label{tab:modGPcyc1pe}}
\begin{tabular}[centre]{lrr}
\hline \hline
Parameter  & Prior & Posterior   \\
\hline
\multicolumn{3}{c}{Offsets and Jitters} \\
\hline
V0 FWHM$_{E18}$         [\ms] & $\mathcal{N}  (0,10) $     & $  3.10^{+2.68}_{-2.32} $ \\
V0 FWHM$_{E19}$         [\ms] & $\mathcal{N}  (0,10) $     & $  1.77^{+1.41}_{-1.10} $ \\
V0 RV$_{E18}$           [\ms] & $\mathcal{N}  (0,3)  $     & $  0.90^{+0.92}_{-0.90} $ \\
V0 RV$_{E19}$           [\ms] & $\mathcal{N}  (0,3)  $     & $ -0.41^{+0.42}_{-0.53} $ \\
$\ln$ Jit. FWHM$_{E18}$ [\ms] & $\mathcal{N} (1.5,3) $     & $ -0.64^{+0.44}_{-0.50} $ \\
$\ln$ Jit. FWHM$_{E19}$ [\ms] & $\mathcal{N} (1.5,3) $     & $ -1.30^{+0.23}_{-0.47} $ \\
$\ln$ Jit. RV$_{E18}$   [\ms] & $\mathcal{N} (0.5,1) $     & $ -0.60^{+0.40}_{-0.34} $ \\
$\ln$ Jit. RV$_{E19}$   [\ms] & $\mathcal{N} (0.5,1) $     & $ -0.57^{+0.09}_{-0.09} $ \\
\hline
\multicolumn{3}{c}{Long-term cycle} \\
\hline
$\ln$ ACYC1 FWHM        [\ms] & $\mathcal{N} (1.5,1.5)  $  & $  0.33^{+0.72}_{-1.04} $ \\
$\ln$ ACYC1 RV          [\ms] & $\mathcal{N} (0.5,0.5)  $  & $  0.17^{+0.42}_{-0.35} $ \\
$\ln$ ACYC2 FWHM        [\ms] & $\mathcal{N} (1.5,1.5)  $  & $  0.11^{+0.68}_{-0.93} $ \\
$\ln$ ACYC2 RV          [\ms] & $\mathcal{N} (0.5,0.5)  $  & $  0.16^{+0.26}_{-0.35} $ \\

$P_{\rm CYC}$             [d] & $\mathcal{N} (3250,300) $  & $ 3325^{+276}_{-226} $ \\

PH1 CYC FW                    & $\mathcal{U} (-0.5,0.5) $  & $ -0.22^{+0.12}_{-0.14} $ \\
PH1 CYC RV                    & $\mathcal{U} (-0.1,0.6) $  & $  0.28^{+0.07}_{-0.09} $ \\
PH2 CYC FW                    & $\mathcal{U} ( 0.0,0.4) $  & $  0.20^{+0.08}_{-0.09} $ \\
PH2 CYC RV                    & $\mathcal{U} ( 0.0,0.3) $  & $  0.09^{+0.06}_{-0.03} $ \\
\hline
\multicolumn{3}{c}{SHO (P and P/2) GP} \\
\hline
$\ln$ A11 GP FWHM       [\ms] & $\mathcal{N} (1.5,3) $    & $  1.36^{+0.13}_{-0.13} $ \\
$\ln$ A12 GP FWHM       [\ms] & $\mathcal{N} (1.5,3) $    & $  1.16^{+1.49}_{-2.56} $ \\
$\ln$ A21 GP FWHM       [\ms] & $\mathcal{N} (1.5,3) $    & $ -2.36^{+1.04}_{-1.57} $ \\
$\ln$ A22 GP FWHM       [\ms] & $\mathcal{N} (1.5,3) $    & $  0.23^{+1.03}_{-1.98} $ \\
$\ln$ A11 GP RV         [\ms] & $\mathcal{N} (0.5,1) $    & $ -0.49^{+0.29}_{-0.38} $ \\
$\ln$ A12 GP RV         [\ms] & $\mathcal{N} (0.5,2) $    & $ -0.29^{+1.15}_{-1.63} $ \\
$\ln$ A21 GP RV         [\ms] & $\mathcal{N} (0.5,1) $    & $  0.32^{+0.14}_{-0.14} $ \\
$\ln$ A22 GP RV         [\ms] & $\mathcal{N} (0.5,2) $    & $  0.29^{+1.09}_{-1.75} $ \\

$P_{\rm ROT}$             [d] & $\mathcal{U} (50,300) $    & $ 136.2^{+10.5}_{-9.4} $ \\
$\ln T_{\rm ROT}$         [d] & $\mathcal{N} (3,2)    $    & $ 3.60^{+0.32}_{-0.35} $ \\
\hline
\multicolumn{3}{c}{Keplerian orbit} \\
\hline
$\phi_b$                          & $\mathcal{U}  (-0.5,1) $ & $ -0.098^{+0.047 }_{-0.047 } $ \\
$\ln P_b$                     [d] & $\mathcal{LU} (0.5,50) $ & $ 1.14844^{+0.00017}_{-0.00016} $ \\
$k_b$                       [\ms] & $\mathcal{U}  (0,5)    $ & $ 0.549   ^{+0.073   }_{-0.074} $ \\
$\sqrt{e_b} \cos(\omega_b)$       & $\mathcal{N}  (0,0.3)  $ & $ 0.02   ^{+0.24   }_{-0.23} $ \\
$\sqrt{e_b} \sin(\omega_b)$       & $\mathcal{N}  (0,0.3)  $ & $ 0.02   ^{+0.21   }_{-0.19} $ \\
\hline
\multicolumn{3}{c}{Model statistics} \\
\hline
$\ln \mathcal{Z}$              & & $-438.6$ \\
$\Delta \ln \mathcal{Z}$       & & $+8.6$   \\
$N_{\rm par}$                  & & $32$   \\
$N_{\rm samples}$              & & $186985$   \\
RMS FWHM                [\ms]  & &  3.32    \\
RMS RV                  [\ms]  & &  1.83    \\
RMS (O--C) FWHM         [\ms]  & &  0.35    \\
RMS (O--C) RV           [\ms]  & &  0.45    \\
\hline
\end{tabular}
\end{center}
\textbf{Notes:} Parameters with prior and posterior values of model $E1e$ in Table~\ref{tab:logz} of the ESPRESSO RV and FWHM datasets, including a GP, cycle and Keplerian model.
\end{table}

\subsection{The 3.15~d sub-Earth mass planet~\label{sec:planet315}}

ESPRESSO RVs show a long-term variation, in particular, in the E19 dataset, which may be interpreted as a long-term activity cycle. We then run a model for the ESPRESSO dataset (model $D$ in Table~\ref{tab:logz}) with the GP and the cycle as done in Section~\ref{sec:activity} but with priors on $P_{\rm CYC}$ $\mathcal{N}(3250,300)$ (normal prior centered on $P_{\rm CYC}=3250$~d with a $\sigma = 300$~d), on $P_{\rm ROT}$ $\mathcal{U}(50,300)$ (uniform prior on $P_{\rm ROT}$ in the range [50,300]~d), and on $T_{\rm ROT}$ $\mathcal{LN}(3,2)$ (log-normal prior centered on $\log T_{\rm ROT}[{\rm d}]=3$ with a $\sigma (\log T_{\rm ROT}[{\rm d}])=2$). This is supported by models $I1$ and $J1$ in Table~\ref{tab:logz}, which use a longer baseline of ESPRESSO, HARPS and HARPS-N (see Fig.~\ref{gj699_cyc_rot}). Model $D$ results in a Bayesian evidence of $\ln \mathcal{Z} = -447.2$ which we adopt as our reference model for the ESPRESSO dataset. The posteriors of this model give $P_{\rm CYC} = 3362^{+284}_{-283}$~d, the rotation, $P_{\rm ROT} = 139^{+12}_{-10}$~d, and the timescale, $T_{\rm ROT} ={43}^{+18}_{-12}$~d. We run three times this activity-only model to check any possible variance on $\ln \mathcal{Z}$ values and we found for these three models equivalent $\ln \mathcal{Z}$ values at $-447.1\pm0.1$.   We also explore other possibilities as testing the GP plus first, second and third order polynomials (models $B1$, $B2$ and $B3$ in Table~\ref{tab:logzP}), but they provide worse Bayesian evidence with $\Delta \ln \mathcal{Z} = -6.2$, $-6.7$ and $-15.5$, respectively. The same behaviour is seen with ESPRESSO, HARPS and HARPS-N dataset with the model $G3$ compared to $I1$ and $I2$ in Table~\ref{tab:logzP}.

We run a blind search of a planet using either a circular or a Keplerian model ($E1c$ and $E1e$ in Table~\ref{tab:logz}) with a prior $P_{\rm orb}$ $\mathcal{LU}(0.5,50)$~d. The cycle+GP+Keplerian model, $E1e$, displayed in Fig.~\ref{gj699_cyc_gp}, uses the $\sqrt{e} ~cos(\omega)$ and $\sqrt{e} ~\sin(\omega)$ with normal priors $\mathcal{N}$(0,0.3). We note that even if this eccentricity prior favors low eccentricity values, it still leaves the possibility to choose high eccentricity values if needed. In Table~\ref{tab:espdataset} we provide RV and FWHM measurements of ESPRESSO datasets used in Fig.~\ref{gj699_cyc_gp}.
We detect unequivocally the 3.15~d signal as a planetary signal with a Bayesian evidence $\ln \mathcal{Z} = -438.6$, and $\Delta \ln \mathcal{Z} = +8.6$ with respect to the activity-only model (model $D$ in Table~\ref{tab:logz}), which corresponds to a 0.018\% FAP for the activity+planet model.
We repeated 10 times the run for model $E1e$ and found some variance in the $\ln \mathcal{Z}$ values in the range [-438.2,-444.7], with a mean of -442.2 ($\sigma=2.1$). In all these runs, we clearly detect the 3.15d planet but the planet posteriors are slightly different. The runs with lower $\ln \mathcal{Z}$ values produce slightly asymmetric posterior distribution for the planet semi-amplitude $k_p$ and also few per cent of the planet period samples (typically less than 5--10\%) corresponds to either aliases of 3.15d or other signals (typically 4.12d and its alias). We believe this is the main reason why we find some variance in the $\ln \mathcal{Z}$ values. We adopted as our final solution for model $E1e$ one of the runs with higher $\ln \mathcal{Z}$ value at -438.6 which provides a fairly symmetric posterior for the planet semi-amplitude $k_p$ and all the posterior period samples at 3.15d (see Fig.~\ref{gj699_1pe_corner}).

Fig.~\ref{gj699_cyc_gp} depicts the FWHM and RV ESPRESSO measurements together with the activity model and the GLS periodograms before and after subtracting the long-term cycle. We note that the SHO (P and P/2) GP (model $E1e$, with priors and posteriors parameters given in Table~\ref{tab:modGPcyc1pe}) is efficiently absorbing all GLS peaks at periods longer than 50~d, even with the $P_{\rm ROT}=136 \pm 10$~d, for both the FWHM and RV ESPRESSO datasets. In the lower left panels of Fig.~\ref{gj699_cyc_gp} we show the RVs after subtracting the activity model and below that panel the RV residuals after subtracting the Keplerian model. In the lower right panels, the GLS periodograms show clearly the 3.15~d signal with PSD greater than 0.4 well above the 0.1\% FAP line. In the GLS periodogram of the RV residuals after subtracting the Keplerian model (lower right panel of of Fig.~\ref{gj699_cyc_gp}), the 3.15~d signal together with the 1 d alias at 1.46~d and 0.76~d and 0.59~d (1 d alias of 1.46~d) have disappeared. Other signals, already mentioned before, appear at 4.12~d (with 1 d alias at 1.32~d), 2.34~d (with 1 d alias at 1.74~d), still above the 0.1\% FAP line. At periods below 1~d, a few new peaks that have grown are all 1 d alias of the signals at 4.12~d and 2.34~d and their 1 d alias, but still below the 10\% FAP line. The RMS of the RVs goes from the original value at 1.86~\ms\ to 0.60~\ms\ after subtracting the activity model, ending at 0.45~\ms\ in the RV residuals after subtracting the Keplerian model. The median jitter of the E18 and E19 datasets in model $E1e$ are 0.56~\ms, compared with the median \texttt{S-BART} RV uncertainties of 0.10~\ms. It is remarkable that the RMS of the RV residuals is 0.45~\ms, greater than the RMS of the FWHM residuals of 0.35~\ms, thus requiring an additional 0.56~\ms\ RV jitter while a 0.30~\ms\ FWHM jitter in the $E19$ dataset (see Table~\ref{tab:modGPcyc1pe}). This suggest that this remaining RMS in RV may be possibly related to additional candidate planetary signals present in the data and not included in this model $E1e$.
    
Figure~\ref{gj699_1pe_phase} displays the RV measurements versus orbital phase. The planetary signal at 3.15~d is not eccentric. The $E1e$ model converges to $P_{\rm orb} = 3.1533 \pm 0.0006$~d and $k_p = 55 \pm 7$~\cms\ ($\sim 7.9 \sigma$ detection) and an eccentricity less than 0.16, consistent with zero. The planet parameters, hereinafter referred as {\it Barnard b}, are provided in Table~\ref{tab:planet}. Fig.~\ref{gj699_1pe_corner} shows the corner plot with the minimum mass of the planet equal to $m_p \sin i = 0.37 \pm 0.05$~\mearth, i.e. a sub-Earth mass planet with about half of the mass of Venus and about three times the mass of Mars.
    
We verified that we were able to recover the planet using the DRS RVs and the \texttt{LBL} RVs, with semi-amplitude velocity $k_p = 52^{+9}_{-7}$ and $55^{+7}_{-9}$~\cms for DRS and LBL runs (models $E1eD$ and $E1eD$ in Table~\ref{tab:logz}), respectively, thus all compatible with TM run within the error bars (see also Section~\ref{sec:ap:315} and Table~\ref{tab:logzP}). We run several additional models with the TMtc RVs derived using the telluric corrected ESPRESSO spectra (see Section~\ref{sec:obs_esp}). We first run an activity-only model with only a GP (model $AT$ in Table~\ref{tab:logz}). We note that the Bayesian evidence cannot be directly compared between models $A$, $AD$, $AL$ and $AT$, but we provide $\Delta \ln \mathcal{Z}$ always with respect to model $D$ in Table~\ref{tab:logz} for simplicity. Activity-only model $DT$ shows better Bayesian evidence than model $AT$. We also recover the 3.15~d planet using the TMtc RVs in the runs $E1cT$ and $E1eT$ with $\Delta \ln \mathcal{Z} = 6.0$ and 7.5 with respect to the activity-only model $DT$, corresponding to 0.25 and 0.05\% FAP. In these cases, the semi-amplitude velocity is $55^{+8}_{-8}$ and $55^{+8}_{-10}$~\cms for the circular and Keplerian orbits respectively. In the following we keep using the TM RVs as reference ESPRESSO RVs, since we do not see significant improvement, and we do consistently the RV computation with HARPS and HARPS-N, without using telluric corrected spectra, but masking out the regions possibly affected by telluric features taking into account BERV changes.

We also perform several runs with the \texttt{S-BART} code \citep{sil22sbart}, to explore any possible chromatic effects on the 3.15~d signal. We compute RVs using either only the blue-detector or the red-detector spectra, or by dividing the whole ESPRESSO spectral coverage into three parts with approximately identical RV content, being the blue, green and red RVs. We run models using the red-detector RVs, with priors on orbital period $P_{\rm orb}$ $\mathcal{LU}(0.5,50)$~d and $\mathcal{LU}(0.5,20)$~d, providing clear detections of the 3.15~d signal, with posteriors of the planet and activity parameters with Gaussian shapes peaking at the values very similar to reference model $E1e$ in Table~\ref{tab:logz}. The blue-detector RVs, and the blue, green, and red RVs, with the prior $P_{\rm orb}$ $\mathcal{LU}(0.5,20)$~d, were also providing a detection of the 3.15~d signal. In all these models the activity parameters give median values in the ranges for the cycle $P_{\rm CYC} = [3254,3498]$~d, and the rotation $P_{\rm ROT} = [136,144]$~d, whereas for the semi-amplitude velocity these models give $k_p=[0.40,0.54]$~\ms. We consider these results reasonable given the low semi-amplitude of the signal, which makes it difficult to be detected. The red detector seems to contain most of the RV content, according to the shape of the posterior distributions on semi-amplitude. These two runs provide a result very similar to the reference model, which seems reasonable given the late spectral type of the star that shifts the stellar flux towards the red part of the spectrum.

The analysis of the ESPRESSO, HARPS and HARPS-N data together reinforces the confirmation of the 3.15~d signal as a sub-Earth mass planet. Fig.~\ref{gj699_esp_har} shows the result of a model consisting of long-term cycle, a GP and a Keplerian model. This run (model $J1$ in Table~\ref{tab:logz}) uses wide priors on $P_{\rm CYC}$ $\mathcal{U}(800,5000)$~d, $P_{\rm ROT}$ $\mathcal{U}(50,300)$~d, $P_{\rm orb}$ $\mathcal{LU}(0.5,50)$~d. This run $J1$ gives a $\Delta \ln \mathcal{Z} = +12.4$ with respect to the activity-only model $I1$, equivalent to 0.0004\% FAP. This model leaves in the GLS RV residuals (bottom right panel in Fig.~\ref{gj699_esp_har}) the tentative signal at 4.12~d above 1\% FAP line. We believe that this is the result of the balance between including more RV data with a longer baseline but, on the other hand, poorer quality data according to the lower RV precision of the HARPS and HARPS-N spectrographs. Figure~\ref{gj699_esp_har_1pe} displays the RV data versus orbital phase, where the RMS of the original ESPRESSO, HARPS and HARPS-N RVs of 1.93~\ms goes down to 0.97~\ms, after subtracting the activity-only model, and down to 0.91~\ms after subtracting the 3.15~d Keplerian RV curve. Figure~\ref{gj699_esp_har_1pe_corner} shows the corner plot of the posteriors of the planet parameters of this run, very similar to the ESPRESSO-only run $E1e$. The posteriors give again a practically null eccentricity, and the values $P_{\rm orb} = 3.1537 \pm 0.0004$~d and $k_p = 48\pm8$~\cms\ ($6 \sigma$ detection), with $m_p \sin i = 0.33 \pm 0.06$~\mearth, consistent with ESPRESSO-only model $E1e$ within the uncertainties. Figure~\ref{gj699_esp_har_1pe_corner_all} depicts the prior and posteriors distributions of the 40 parameters of model $J1$.

Finally, we run the model of FWHM and RV measurements including the CARMENES data (see Fig.~\ref{gj699_esp_har_car}) that also confirms the detection of a planetary signal at 3.15~d. This model $L1$ in Table~\ref{tab:logz} shows again a difference on the Bayesian evidence of $\Delta \ln \mathcal{Z} = +12.3$ with respect to the activity-only model $K1$. We also verify that using uniform or normal priors for the cycle provides very similar results. Now, again, the 4.12~d and 2.34~d signals appear in the GLS of the RV residuals above the 0.1\% FAP line. The RMS of the RV measurements in this run $K1$ involving ESPRESSO, HARPS, HARPS-N and CARMENES data goes from the original values at 2.08~\ms down to 1.49~\ms, after subtracting the activity-only model, and down to 1.47~\ms, after subtracting the Keplerian 3.15~d signal.
The 3.15~d signal is too weak to be independently confirmed using CARMENES data only, using H15+HAN data only, or using both datasets together. A search with a very narrow prior at 3.15~d~$\pm 0.1$~d provides a tentative detection but weaker than 10\% FAP. Other candidate signals at 4.12~d and 2.34~d are also present, but still less significant than 10\% FAP.

\subsection{Candidate multi-planetary system~\label{sec:2planet}}

We investigate the possibility to detect and confirm in a blind search these additional signals seen in the GLS periodogram of the ESPRESSO RV residuals. We note that the high required jitter of 56~\cms\ invites one to believe that this is the consequence of including only one Keplerian signal in the model. Thus we first run several models including two Keplerian signals with priors on orbital period $P_{\rm orb}$ $\mathcal{LU}(0.5,20)$~d, where we only put the restriction that planet b should have shorter period, and therefore shorter orbital distance, than planet c. In addition, we only permit those orbits where the eccentricity of planet b is such that excludes the orbital collision~\citep[e.g.][]{las17}, given by the eq.~\ref{eq_orb_col}:
\begin{equation} \label{eq_orb_col}
{P_b < P_c\,\,,\,\,a_b   \cdot  (1+e_b) < a_c   \cdot   (1-e_c)},
\end{equation}
\noindent where $P_i$, $a_i$ and $e_i$ are the orbital period, distance and eccentricity of planet $b$ and $c$. We note that the prior on eccentricity $e$ with $\sqrt{e} ~\cos(\omega)$ and $\sqrt{e} ~\sin(\omega)$ $\mathcal{N}(0,0.3)$, though it favors low eccentricity orbits, it leaves some freedom to reach high eccentricity values. Models $E22c$ and $E22e$ in Table~\ref{tab:logz} shows a satisfactory result in terms of the shape of the posteriors of the activity and planetary parameters of planet $b$, but not for planet $c$, where the semi-amplitude gets lower median value than expected and consistent with zero in model $E22e$. The result is better for the circular model $E22c$ with the posteriors with better shape and median values of $P_b = 3.1533\pm 0.0006$~d and $P_c = 4.1249\pm 0.0014$~d, and $k_b =  0.47^{+0.08}_{-0.12}$~\ms\ and $k_c =  0.40^{+0.07}_{-0.15}$~\ms.
However, the Bayesian evidence of these models is worse than that of the models $E12c$ and $E12e$.
We note here that we computed, as we did for model $E1e$, five times the model $E12c$ and $E12e$ and we see there is some (slightly lower) variance in the $\ln \mathcal{Z}$ in the range [-438.8,-442.5] for $E12c$, with a mean of -441.1 ($\sigma=1.5$), and in the range [-439.8,-442.3] for $E12e$, with a mean of -440.9 ($\sigma=1.1$).
Therefore, from a blind search of two planetary signal using ESPRESSO only FWHM and RV data we cannot confirm the presence of the planet $c$ signal at 4.12~d. We also tried additional two-planet model runs by adding the datasets of HARPS and HARPS-N (model $J2$) and adding the CARMENES dataset (model $M2$). The Bayesian evidence of these models were a bit worse (model $J2$) or a bit better (model $M2$) than the single-planet model in each case, but not significant. It is remarkable that for the $M2$ model the periods and the semi-amplitudes, with median values of $P_b = 3.1543\pm 0.0003$~d and $P_c = 4.1245\pm 0.0005$~d, and $k_b =  0.42^{+0.05}_{-0.05}$~\ms\ and $k_c =  0.39^{+0.06}_{-0.11}$~\ms, are marginally compatible with the ESPRESSO-only run $E22c$, within the uncertainties.

In Appendix~\ref{sec:4planet} we describe the exercise to run a guided search of the four candidate signals at 3.15, 4.12, 2.34 and 6.74~d, corresponding to models $F1$, $F2$, $F3$, and $F4$ in Table~\ref{tab:logzN}.

\begin{figure}
\includegraphics[width=9cm]{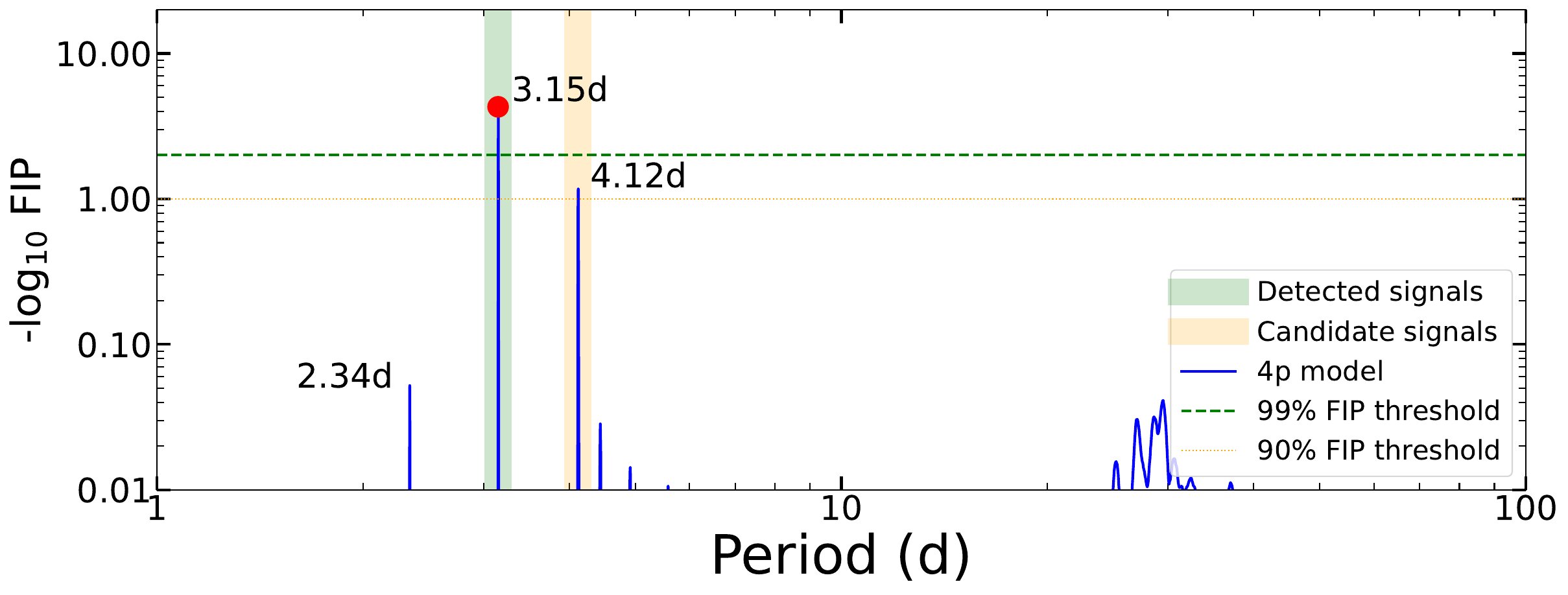}
\caption{FIP periodogram of the 4-planet model of the ESPRESSO data of GJ 699. The periods of the detected signals are indicated in red solid circles. The $-\log_{10}FIP$ as a function of the center of the period is represented as a blue line. The green dashed and dotted orange lines  show the 1\% and 10\% FIP thresholds, respectively.
}
\label{gj699_esp_fip}
\end{figure}

\begin{figure}
\includegraphics[width=9cm]{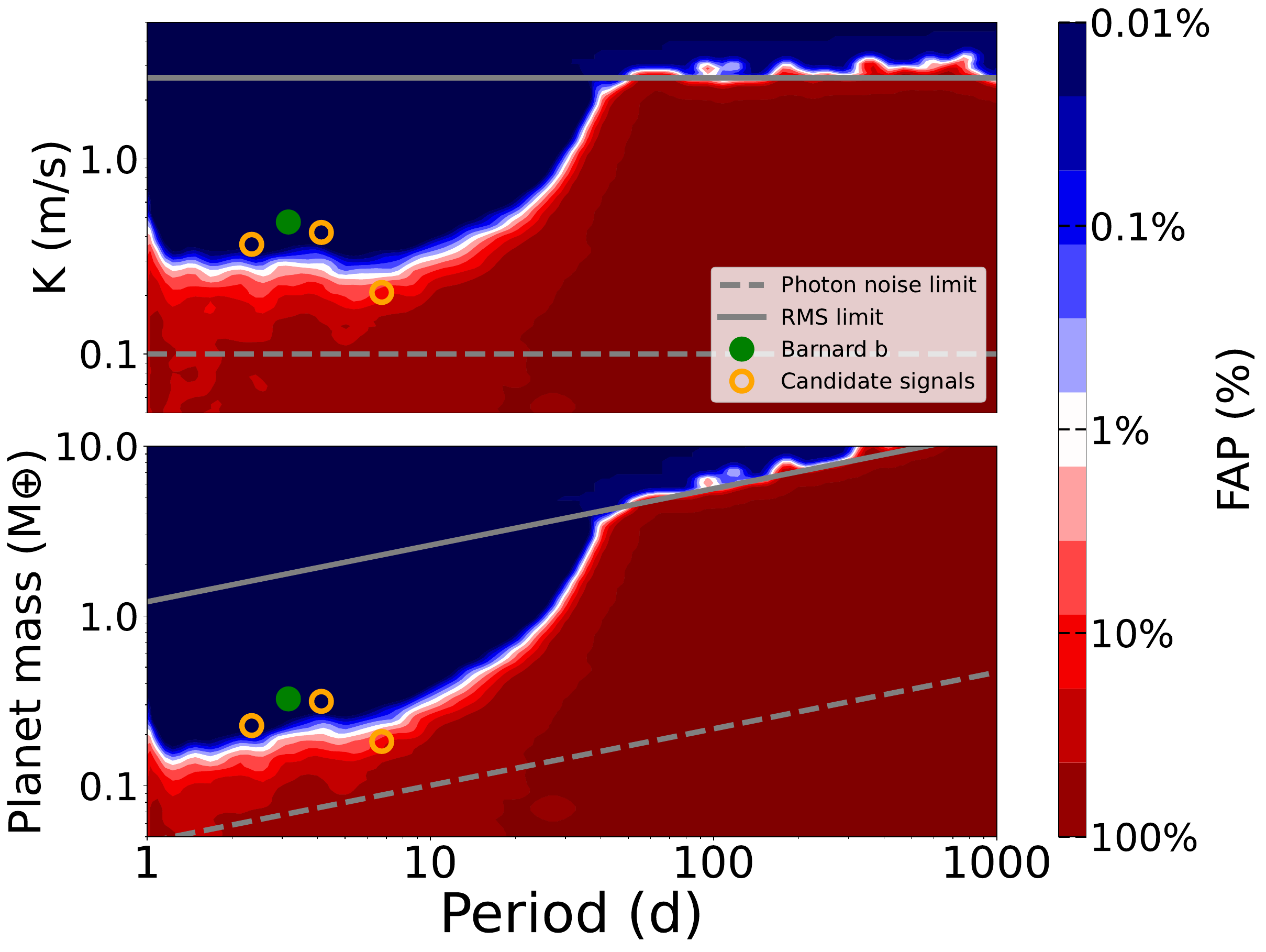}
\caption{Detection limits. {\it Top panel}: semi-amplitude $k_p$ versus period $P_{\rm orb}$ for our detection limits based on injection and recovery tests. The white region shows the signals that we would detect in a periodogram with 1\% FAP. The blue region shows signals we would detect with a FAP lower than 1\% FAP. With the current dataset and activity model, we would miss the red region signals. The top grey dashed line shows the maximum semi-amplitudes that would still be compatible with the data. The bottom grey dashed line shows the minimum amplitude that we expect to detect with ESPRESSO data and the current exposure times. The green solid circle shows the confirmed planet and the orange circles show the position of the candidate planets. {\it Bottom panel}: same as top panel but in mass versus period.
}
\label{gj699_esp_det_lim}
\end{figure}

\section{Discussion}

We take advantage of the result of run $F4$ in Table~\ref{tab:logzN} (described in Section~\ref{sec:4planet}) to test the presence of planets, and assess their significance with the False Inclusion Probability~\citep[FIP;][]{har22}. This framework uses the posterior distribution of the nested sampling run, and computes the probability of having a planet within a certain orbital frequency interval based on the fraction of samples within that frequency interval. It can test the presence of an arbitrary number of planetary signals. To produce the posterior distribution, we run a ESPRESSO-only model composed of the same activity model as described before, and four circular planetary signals. For all these signals we use the same priors, with wide log-uniform priors for periods $P_{\rm orb}$ $\mathcal{LU}(2,40)$~d days, and uniform prior on semi-amplitudes $k_p$ $\mathcal{U}(0,5)$~\ms. Fig.~\ref{gj699_esp_fip} shows the results of the $-\log_{10} FIP$ as a function of period. The results support the detection of a planet at a period of a 3.15~d, and a solid hint of the presence of a second planet at 4.12~d. This result is expected from the previous discussion in Sections~\ref{sec:test315},~\ref{sec:planet315} and~\ref{sec:4planet}.
 
We also estimate our detection limits for additional companions in the ESPRESSO data by performing a simple injection-recovery test. We construct activity-only RVs by generating random samples from the GP model, and adding white noise with the same standard deviation of the residuals of the 4-planet model (see Section~\ref{sec:4planet}). We then inject 100,000 random sinusoidal signals with periods between 1 and 1100~d, semi-amplitudes between 1~\cms\ and 11~\ms, and random phases. We reject those combinations that would create an RMS of the data significantly higher than the real RMS of our data. Then we subtract the stellar activity using the same GP hyper-parameters as measured for our four-planet model, but recomputing the activity model for each specific dataset. Then we generate the periodogram of the residuals and check the false alarm probability of the highest peak at periods within 10\% of the injected period.

In Fig.~\ref{gj699_esp_det_lim} we display the result of this exercise.  With this dataset and activity model, we are sensitive to signals of $\sim 0.3$~\ms\ for periods from 1.2~d up to $\sim 20$~d. Our sensitivity drops to $\sim 1$~\ms at 35~d and then quickly plummets to 2.5~\ms\ at periods longer than 40~d. These values correspond to minimum masses of $\sim 0.5$~\mearth\ at 15~d, $\sim 1.5$~\mearth\ at 35~d  and $> 4$~\mearth\ at periods longer than 40~d. This is a typical effect of GP-only models, which try to account for all variations at low frequencies up to the characteristic value specified by the kernel. To extract possible low-amplitude signals at longer periods we would need either a very significant amount of new data or a different strategy to mitigate stellar activity. This also shows that finding the signal of {\it Barnard b} (3.15~d) is within expectations. Finding the signals at 2.34~d and 4.12~d is also within expectations, with the 4.12~d signal being easier to find. Finding the signal at 6.74~d, however, is not within expectations. Exploring the region of periods longer than 30--40~d likely requires simultaneous modeling of the GP and activity signals using a Nested Sampling, or MCMC, algorithm. We discuss in Section~\ref{sec:4planet} the candidate multi-planetary system.

\begin{figure}
\includegraphics[width=9cm]{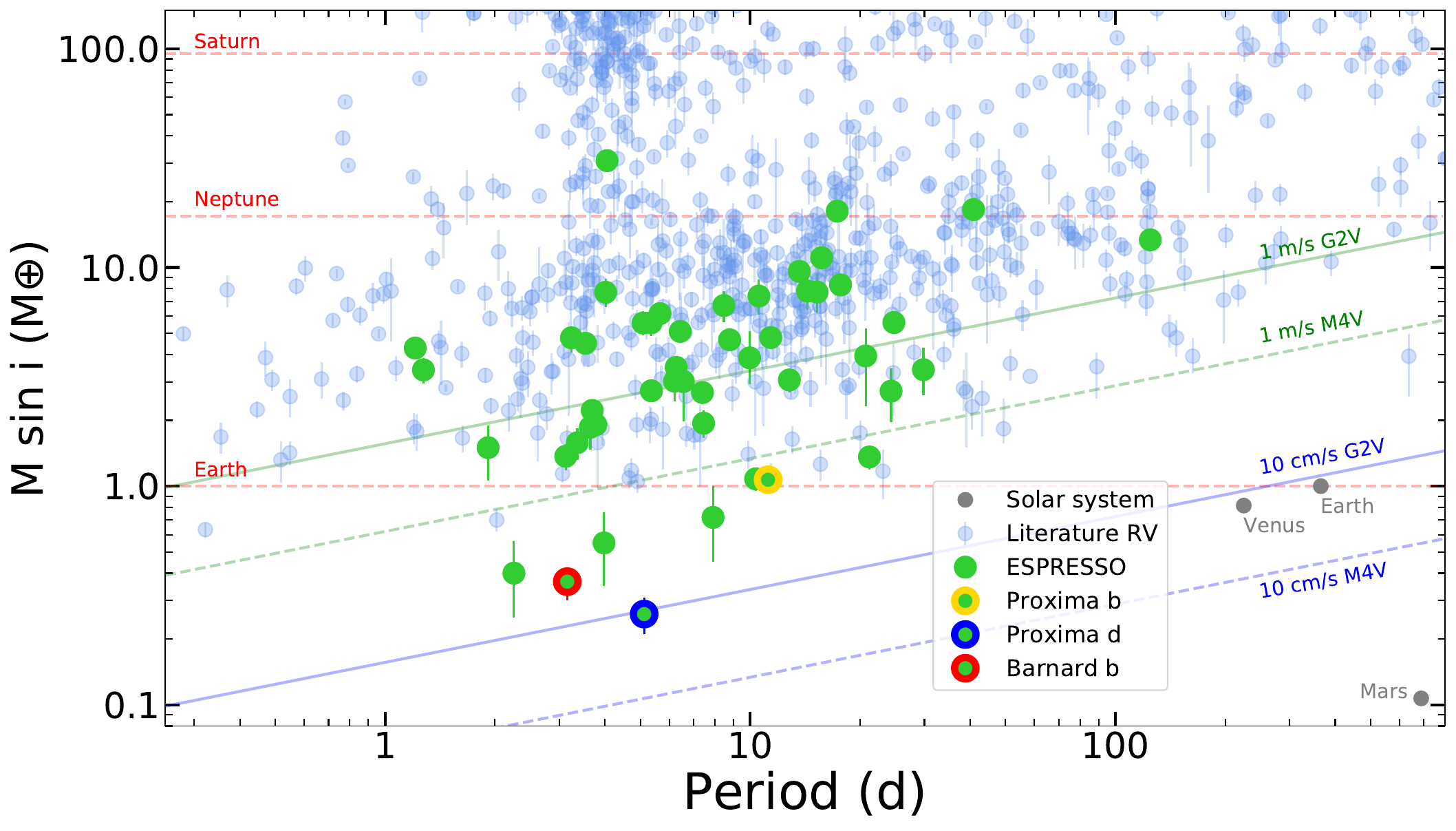}
\caption{
Minimum mass vs orbital period diagram for known planets from the NASA Exoplanet archive as of December 2023 orbiting solar-type stars, together with those discovered and confirmed using ESPRESSO (green circles). Confirmed planet Barnard b (red), and planet candidates (purple) from the 4-planet candidate system orbiting Barnard's star, together with the two planets orbiting Proxima Cen, Proxima b (yellow) and d (blue) are highlighted. Planets of the solar system (grey circles) are also labelled. Inclined solid and dashed lines show the RV semi- amplitude of planets orbiting a late M dwarf star with 0.25~\msun\ and a G dwarf star with 1~\msun\ star assuming a RV semi-amplitude of 1~\cms\ (green line) and 10~\cms\ (blue line), respectively, and null eccentricity.
}
\label{gj699_planet_mass}
\end{figure}

The sub-Earth mass planet orbiting at period 3.15~d our second closest neighbour Barnard's star has a minimum mass of $0.37\pm0.05$~\mearth. Fig.~\ref{gj699_planet_mass} compares the confirmed planet Barnard b with those other planets in the exoplanet database. We highlight the ESPRESSO planet discoveries and confirmations with particular emphasis on our closest neighbour star, hosting the two planets Proxima d and b~\citep{sua20,far22}. ESPRESSO has gone beyond the Earth mass frontier, thus starting to discover sub-Earth mass planets orbiting nearby stars to our Sun such as Proxima Cen and Barnard's star, although still at short orbital periods. These discoveries have required significant investment with more than 100 measurements in an 8-m class telescope. The endeavor of filling the region in Fig.~\ref{gj699_planet_mass} below the 1~\ms\ precision limit at periods larger than 10~d towards finding long-period Earth-like planets would require many years of continuous, well planned, sufficiently sampled observations and a deep understanding of magnetic activity of every nearby star. This would open the opportunity to sample the habitable zone of nearby stars with the aim of further investigating the atmosphere of Earth-like planets in the future with ground-based facilities such as ANDES at ELT~\citep{mar22} and with space missions such as LIFE~\citep{qua22}.

\begin{figure}
\includegraphics[width=9cm]{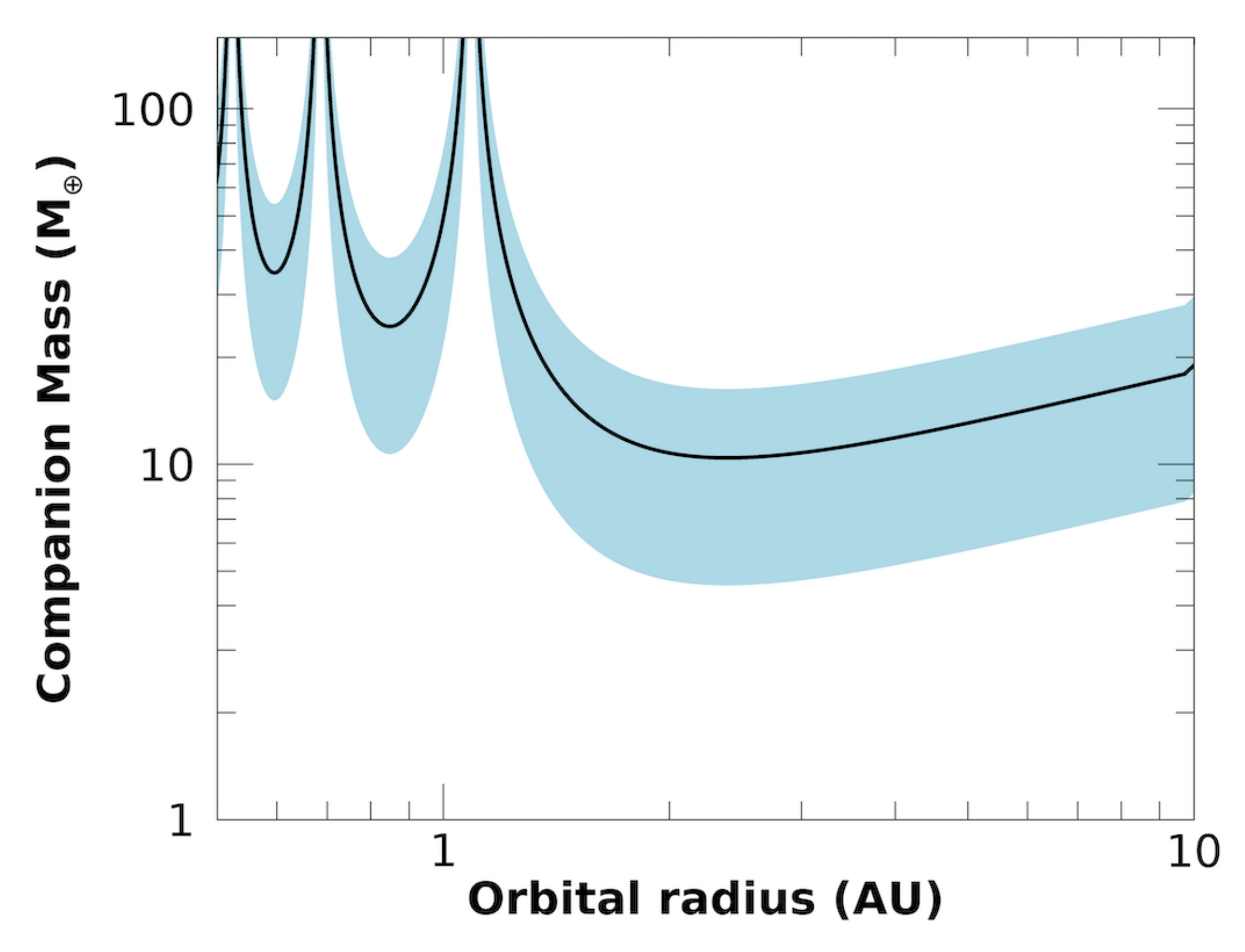}
\caption{
    Hipparcos-{\it Gaia} PMA sensitivity to companions of a given mass (in \mearth) as a function of the orbital semi-major axis (in AU) orbiting Barnard's star. The solid black line identifies the combinations of mass and separation explaining the observed proper-motion anomaly (PMA) at the mean epoch of {\it Gaia} DR3~\citep[Eq.~15 in][]{ker19}. The shaded light-blue region corresponds to the 1$\sigma$ uncertainty region.
}
\label{GJ699_PMA}
\end{figure}

Stringent upper limits on the presence of planetary mass companions at larger separations can be placed using the catalogues of Hipparcos-{\it Gaia} astrometric accelerations produced by \citet{bra21} and \citet{ker22}. For example, using the analytical formulation of \citet{ker19}, Fig.~\ref{GJ699_PMA} shows the sensitivity curve for Barnard's star given the measured PMA in the Hipparcos-{\it Gaia} DR3 catalogue of \citet{ker22}. The peaks, corresponding to regimes of orbital separation with reduced sensitivity, are a direct consequence of the observing window smearing effect of the orbital motion of any companion. This effect is intrinsic to the proper-motion difference technique, where the catalogue proper-motion values for Hipparcos and {\it Gaia} are averaged over the time-span of the observations, thus with smearing orbital effects for shorter periods. The last peak in sensitivity loss appears, in fact, at the exact time-span of the observations of the astrometric catalogue we consider. For {\it Gaia} DR3, this means 2.83 years, i.e. a separation of 1.08 AU for a star with the mass of Barnard's star. Therefore, there are also changes to the precise location of the regions of sensitivity loss due to the fact that the time-span from DR2 to DR3 over which the {\it Gaia} proper-motion values are computed varies. In the region of uniform sensitivity, the gain in mass limits is a factor of several, almost one order of magnitude in comparison with figure A.3 \citet{ker19}. Thus Fig.~\ref{GJ699_PMA} shows that in the regime of approximately uniform sensitivity $2-10$~AU, the presence of super-Earths with true masses around 10 M$_\oplus$ can be ruled out by the lack of statistically significant PMA. The combination of ESPRESSO RVs and Hipparcos-{\it Gaia} absolute astrometry thus allows to infer that super-Earths or larger planets are not present in the GJ~699 system out of $\sim10$~AU.

\section{Conclusions}

We analysed 156 ESPRESSO spectra of Barnard's star, the second closest stellar system to the Sun after the $\alpha$ Centauri stellar system, at a distance of about 1.8 parsecs. These spectra have been taken with the ESPRESSO spectrograph at the VLT as part of the GTOs over the last five years. The ESPRESSO RV data show a 1.8~\ms RMS with about 8~\ms peak-to-peak RV measurements, which mainly reflect the magnetic activity of this old middle-M-type dwarf star. We modelled these RV measurements together with CCF FWHM measurements, with a 15~\ms peak to peak and 3.3~\ms RMS, using a long-term cycle described with a double sinusoidal model, and a rotation-induced activity model, which is described using a Gaussian process approach. The activity cycle is better constrained using additional HARPS, HARPS-N, and CARMENES data, which help to extend the baseline of the observations up to eight years. We obtained a well-described activity model with a cycle period of $P_{\rm CYC} = 3210^{+530}_{-430}$~d and a rotation period of $P_{\rm ROT} = 142^{+8}_{-9}$~d, which are consistent with previous results in the literature~\citep[e.g.][]{tol19}.

The high quality of the ESPRESSO data means that we can evaluate the presence of the candidate super-Earth-like planet at 233~d reported in \citet{rib18}. We are unable to recover the 233~d signal in either a blind search or a guided search with a narrow prior on the planet period $\mathcal{N}$(233,0.5)~d. We do not recover a clean signal above 0.50~\ms , with all the models giving solutions with semi-amplitudes consistent with zero. We also tested simulated data at the ESPRESSO precision and conclude that ESPRESSO data do not support the existence of the 233~d candidate planet.

The RV residuals of the activity model reveal several signals at periods of shorter than 10~d; in particular four signals at periods of 3.15~d, 4.12~d, 2.34~d, and 6.74~d, sorted here according to signal strength. We assessed the nature of the main signal at 3.15~d, and confirm it as a planetary signal. We modelled the 3.15~d signal with a Keplerian, resulting in an almost circular orbit with a semi-amplitude of $55\pm 7$~\cms, thus uncovering a sub-Earth-mass planet of $0.37 \pm 0.05$ \mearth, which is about half of the mass of Venus or three times that of Mars. This sub-Earth-mass planet is located inwards from the HZ of the star, with an equilibrium temperature of 400~K ---assuming an albedo of 0.3.

We are unable to confirm the other signals in a blind search. However, we ran a guided search of these signals modelled as Keplerian with narrow priors on each of the four signals at $P_{\rm orb} = 3.15$, 4.12, 2.34, and 6.74~d, assuming very low eccentricities, recovering a candidate four-planet system with semi-amplitudes of $k_p = 47$, 41, 35, and 20~\cms, which would correspond to a system of four sub-Earth-mass planets with $m_p\sin i = 0.32$, 0.31, 0.22, and 0.17~\mearth. All candidate planetary orbits would be located inwards from the HZ of the star, with orbital semi-major axes of between 0.019~AU and 0.038~AU. Thus, all the candidate planets would be irradiated to a greater extent than the Earth, with incident fluxes of between 2.4~\searth\ and 10.1~\searth, and their equilibrium temperatures, assuming an albedo of 0.3, would be in between 440~K for the inner planet and 310~K for the outer planet.

Confirming the presence of a compact four-planet system orbiting Barnard's star, similar to other planetary systems orbiting nearby stars, would require many more ESPRESSO observations. These observations would need to be carried out with sufficient cadence to sample the planet periods, and with a sufficiently long baseline to be able to properly model the activity of the star; in particular, those activity signals associated with stellar rotation. This result further stimulates the search for Earth- and sub-Earth-mass planets in the nearest stars of the solar neighbourhood, and encourages new detailed studies with current and future facilities, such as ANDES at ELT~\citep{mar22,pal23}.

\begin{acknowledgements}
JIGH, ASM, VMP, NN, AKS, CAP and RR acknowledge financial support from the Spanish Ministry of Science, Innovation and Universities (MICIU) project PID2020-117493GB-I00.
JIGH, ASM, VMP and RR acknowledge financial support from the Government of the Canary Islands project ProID2020010129.
AKS acknowledges financial support from La Caixa Foundation (ID 100010434) under the grant LCF/BQ/DI23/11990071.
NN acknowledges financial support by Light Bridges S.L, Las Palmas de Gran Canaria.

FPE and CLO would like to acknowledge the Swiss National Science Foundation (SNSF) for supporting research with ESPRESSO through the SNSF grants nr. 140649, 152721, 166227, 184618 and 215190. The ESPRESSO Instrument Project was partially funded through SNSF's FLARE Programme for large infrastructures.

Co-funded by the European Union (ERC, FIERCE, 101052347). Views and opinions expressed are however those of the author(s) only and do not necessarily reflect those of the European Union or the European Research Council. Neither the European Union nor the granting authority can be held responsible for them. This work was supported by FCT - Funda\c c\~ao para a Ci\^encia e a Tecnologia through national funds and by FEDER through COMPETE2020 - Programa Operacional Competitividade e Internacionaliza\c c\~ao by these grants: UIDB/04434/2020; UIDP/04434/2020.

HMT acknowledges support from the "Tecnologías avanzadas para la exploración de universo y sus componentes" (PR47/21 TAU) project funded by Comunidad de Madrid, by the Recovery, Transformation and Resilience Plan from the Spanish State, and by NextGenerationEU from the European Union through the Recovery and Resilience Facility.
HMT acknowledges financial support from the Agencia Estatal de Investigaci\'on (AEI/10.13039/501100011033) of the Ministerio de Ciencia e Innovaci\'on  through projects  PID2022-137241NB-C41 y PID2022-137241NB-C44.

This work was financed by Portuguese funds through FCT (Funda\c c\~ao para a Ci\^encia e a Tecnologia) in the framework of the project 2022.04048.PTDC (Phi in the Sky, DOI 10.54499/2022.04048.PTDC). CJM also acknowledges FCT and POCH/FSE (EC) support through Investigador FCT Contract 2021.01214.CEECIND/CP1658/CT0001 (DOI 10.54499/2021.01214.CEECIND/CP1658/CT0001).

We thank the TNG Director E.~Poretti and the TNG staff for the observational effort made to guarantee the continuity of the survey by using HARPS-N during the closure of the Paranal site due to pandemic.

This work has been carried out within the framework of the National Centre of Competence in Research PlanetS supported by the Swiss National Science Foundation under grants 51NF40\_182901 and 51NF40\_205606.

The INAF authors acknowledge financial support of the Italian Ministry of Education, University, and Research with PRIN 201278X4FL and the "Progetti Premiali" funding scheme.

This work was supported by FCT - Funda\c c\~ao para a Ci\^encia e a Tecnologia through national funds by these grants: UIDB/04434/2020 (DOI: 10.54499/UIDB/04434/2020); UIDP/04434/2020 (DOI: 10.54499/UIDP/04434/2020). Funded/Co-funded by the European Union (ERC, FIERCE, 101052347). Views and opinions expressed are however those of the author(s) only and do not necessarily reflect those of the European Union or the European Research Council. Neither the European Union nor the granting authority can be held responsible for them. A.M.S acknowledges support from the Fundação para a Ciência e a Tecnologia (FCT) through the Fellowship 2020.05387.BD (DOI: 10.54499/2020.05387.BD)

A.C.-G. is funded by the Spanish Ministry of Science through MCIN/AEI/10.13039/501100011033 grant PID2019-107061GB-C61.

This work has made use of data from the European Space Agency (ESA) mission {\it Gaia} (\url{https://www.cosmos.esa.int/gaia}), processed by the {\it Gaia} Data Processing and Analysis Consortium (DPAC,
    \url{https://www.cosmos.esa.int/web/gaia/dpac/consortium}). Funding for the DPAC has been provided by national institutions, in particular the institutions participating in the {\it Gaia} Multilateral Agreement.

Main analysis performed in \texttt{Python3} \citep{vanros09python3} running on \texttt{Ubuntu} \citep{sob15ubuntu} systems and \texttt{Mac OSX} systems.
Radial velocity computation with \texttt{S-BART} \citep{sil22sbart}.
Second RV computation with \texttt{LBL} \citep{art22lbl}.
Extensive use of the DACE platform \footnote{\url{https://dace.unige.ch/}}
Extensive usage of \texttt{Numpy} \citep{har20numpy}.
Extensive usage of \texttt{Scipy} \citep{vir20scipy}.
All figures built with \texttt{Matplotlib} \citep{hun07matplotlib}.
Periodograms and phase folded curves built using \texttt{Pyastronomy} \citep{cze19pya}.
Gaussian processes modelled with \texttt{S+LEAF} \citep{del22spleaf}, \texttt{Celerite2} \citep{fm17celerite2} and \texttt{George} \citep{fm14george}.
Nested sampling with \texttt{Dynesty} \citep{spe20dynesty}.  

\end{acknowledgements}

%
%
\bibliographystyle{aa}
\bibliography{barnard_esp_jonay}

\begin{thebibliography}{122}
\expandafter\ifx\csname natexlab\endcsname\relax\def\natexlab#1{#1}\fi

\bibitem[{{Agol} {et~al.}(2021){Agol}, {Dorn}, {Grimm}, {Turbet}, {Ducrot},
  {Delrez}, {Gillon}, {Demory}, {Burdanov}, {Barkaoui}, {Benkhaldoun},
  {Bolmont}, {Burgasser}, {Carey}, {de Wit}, {Fabrycky}, {Foreman-Mackey},
  {Haldemann}, {Hernandez}, {Ingalls}, {Jehin}, {Langford}, {Leconte},
  {Lederer}, {Luger}, {Malhotra}, {Meadows}, {Morris}, {Pozuelos}, {Queloz},
  {Raymond}, {Selsis}, {Sestovic}, {Triaud}, \& {Van Grootel}}]{ago21}
{Agol}, E., {Dorn}, C., {Grimm}, S.~L., {et~al.} 2021, Planet. Sci. J., 2, 1

\bibitem[{{Aigrain} {et~al.}(2012){Aigrain}, {Pont}, \& {Zucker}}]{aig12}
{Aigrain}, S., {Pont}, F., \& {Zucker}, S. 2012, \mnras, 419, 3147

\bibitem[{{Allart} {et~al.}(2022){Allart}, {Lovis}, {Faria}, {Dumusque},
  {Sosnowska}, {Figueira}, {Silva}, {Mehner}, {Pepe}, {Cristiani}, {Rebolo},
  {Santos}, {Adibekyan}, {Cupani}, {Di Marcantonio}, {D'Odorico}, {Gonz{\'a}lez
  Hern{\'a}ndez}, {Martins}, {Milakovi{\'c}}, {Nunes}, {Sozzetti}, {Su{\'a}rez
  Mascare{\~n}o}, {Tabernero}, \& {Zapatero Osorio}}]{all22}
{Allart}, R., {Lovis}, C., {Faria}, J., {et~al.} 2022, \aap, 666, A196

\bibitem[{{Alonso-Floriano} {et~al.}(2015){Alonso-Floriano}, {Morales},
  {Caballero}, {Montes}, {Klutsch}, {Mundt}, {Cort{\'e}s-Contreras}, {Ribas},
  {Reiners}, {Amado}, {Quirrenbach}, \& {Jeffers}}]{alo15}
{Alonso-Floriano}, F.~J., {Morales}, J.~C., {Caballero}, J.~A., {et~al.} 2015,
  \aap, 577, A128

\bibitem[{{Anglada-Escud{\'e}} {et~al.}(2016){Anglada-Escud{\'e}}, {Amado},
  {Barnes}, {Berdi{\~n}as}, {Butler}, {Coleman}, {de La Cueva}, {Dreizler},
  {Endl}, {Giesers}, {Jeffers}, {Jenkins}, {Jones}, {Kiraga}, {K{\"u}rster},
  {L{\'o}pez-Gonz{\'a}lez}, {Marvin}, {Morales}, {Morin}, {Nelson}, {Ortiz},
  {Ofir}, {Paardekooper}, {Reiners}, {Rodr{\'{\i}}guez},
  {Rodr{\'{\i}}guez-L{\'o}pez}, {Sarmiento}, {Strachan}, {Tsapras}, {Tuomi}, \&
  {Zechmeister}}]{ang16Natur}
{Anglada-Escud{\'e}}, G., {Amado}, P.~J., {Barnes}, J., {et~al.} 2016, \nat,
  536, 437

\bibitem[{{Artigau} {et~al.}(2022){Artigau}, {Cadieux}, {Cook}, {Doyon},
  {Vandal}, {Donati}, {Moutou}, {Delfosse}, {Fouqu{\'e}}, {Martioli}, {Bouchy},
  {Parsons}, {Carmona}, {Dumusque}, {Astudillo-Defru}, {Bonfils}, \&
  {Mignon}}]{art22lbl}
{Artigau}, {\'E}., {Cadieux}, C., {Cook}, N.~J., {et~al.} 2022, \aj, 164, 84

\bibitem[{{Astudillo-Defru} {et~al.}(2017{\natexlab{a}}){Astudillo-Defru},
  {Delfosse}, {Bonfils}, {Forveille}, {Lovis}, \& {Rameau}}]{ast17rot}
{Astudillo-Defru}, N., {Delfosse}, X., {Bonfils}, X., {et~al.}
  2017{\natexlab{a}}, \aap, 600, A13

\bibitem[{{Astudillo-Defru} {et~al.}(2017{\natexlab{b}}){Astudillo-Defru},
  {Forveille}, {Bonfils}, {S{\'e}gransan}, {Bouchy}, {Delfosse}, {Lovis},
  {Mayor}, {Murgas}, {Pepe}, {Santos}, {Udry}, \& {W{\"u}nsche}}]{ast17}
{Astudillo-Defru}, N., {Forveille}, T., {Bonfils}, X., {et~al.}
  2017{\natexlab{b}}, \aap, 602, A88

\bibitem[{{Azevedo Silva} {et~al.}(2022){Azevedo Silva}, {Demangeon}, {Santos},
  {Allart}, {Borsa}, {Cristo}, {Esparza-Borges}, {Seidel}, {Palle}, {Sousa},
  {Tabernero}, {Zapatero Osorio}, {Cristiani}, {Pepe}, {Rebolo}, {Adibekyan},
  {Alibert}, {Barros}, {Bouchy}, {Bourrier}, {Lo Curto}, {Di Marcantonio},
  {D'Odorico}, {Ehrenreich}, {Figueira}, {Gonz{\'a}lez Hern{\'a}ndez}, {Lovis},
  {Martins}, {Mehner}, {Micela}, {Molaro}, {Mounzer}, {Nunes}, {Sozzetti},
  {Su{\'a}rez Mascare{\~n}o}, \& {Udry}}]{aze22}
{Azevedo Silva}, T., {Demangeon}, O.~D.~S., {Santos}, N.~C., {et~al.} 2022,
  \aap, 666, L10

\bibitem[{{Barnard}(1916)}]{barnard1916}
{Barnard}, E.~E. 1916, \aj, 29, 181

\bibitem[{{Barrag{\'a}n} {et~al.}(2022){Barrag{\'a}n}, {Aigrain}, {Rajpaul}, \&
  {Zicher}}]{barragan22}
{Barrag{\'a}n}, O., {Aigrain}, S., {Rajpaul}, V.~M., \& {Zicher}, N. 2022,
  \mnras, 509, 866

\bibitem[{{Barros} {et~al.}(2022){Barros}, {Demangeon}, {Alibert}, {Leleu},
  {Adibekyan}, {Lovis}, {Bossini}, {Sousa}, {Hara}, {Bouchy}, {Lavie},
  {Rodrigues}, {Gomes da Silva}, {Lillo-Box}, {Pepe}, {Tabernero}, {Zapatero
  Osorio}, {Sozzetti}, {Su{\'a}rez Mascare{\~n}o}, {Micela}, {Allende Prieto},
  {Cristiani}, {Damasso}, {Di Marcantonio}, {Ehrenreich}, {Faria}, {Figueira},
  {Gonz{\'a}lez Hern{\'a}ndez}, {Jenkins}, {Lo Curto}, {Martins}, {Micela},
  {Nunes}, {Pall{\'e}}, {Santos}, {Rebolo}, {Seager}, {Twicken}, {Udry},
  {Vanderspek}, \& {Winn}}]{bar22}
{Barros}, S.~C.~C., {Demangeon}, O.~D.~S., {Alibert}, Y., {et~al.} 2022, \aap,
  665, A154

\bibitem[{{Benedict} {et~al.}(1998){Benedict}, {McArthur}, {Nelan}, {Story},
  {Whipple}, {Shelus}, {Jefferys}, {Hemenway}, {Franz}, {Wasserman},
  {Duncombe}, {van Altena}, \& {Fredrick}}]{ben98}
{Benedict}, G.~F., {McArthur}, B., {Nelan}, E., {et~al.} 1998, \aj, 116, 429

\bibitem[{{Benz} {et~al.}(2021){Benz}, {Broeg}, {Fortier}, {Rando}, {Beck},
  {Beck}, {Queloz}, {Ehrenreich}, {Maxted}, {Isaak}, {Billot}, {Alibert},
  {Alonso}, {Ant{\'o}nio}, {Asquier}, {Bandy}, {B{\'a}rczy}, {Barrado},
  {Barros}, {Baumjohann}, {Bekkelien}, {Bergomi}, {Biondi}, {Bonfils},
  {Borsato}, {Brandeker}, {Busch}, {Cabrera}, {Cessa}, {Charnoz}, {Chazelas},
  {Collier Cameron}, {Corral Van Damme}, {Cortes}, {Davies}, {Deleuil},
  {Deline}, {Delrez}, {Demangeon}, {Demory}, {Erikson}, {Farinato}, {Fossati},
  {Fridlund}, {Futyan}, {Gandolfi}, {Garcia Munoz}, {Gillon}, {Guterman},
  {Gutierrez}, {Hasiba}, {Heng}, {Hernandez}, {Hoyer}, {Kiss}, {Kovacs},
  {Kuntzer}, {Laskar}, {Lecavelier des Etangs}, {Lendl}, {L{\'o}pez}, {Lora},
  {Lovis}, {L{\"u}ftinger}, {Magrin}, {Malvasio}, {Marafatto}, {Michaelis}, {de
  Miguel}, {Modrego}, {Munari}, {Nascimbeni}, {Olofsson}, {Ottacher},
  {Ottensamer}, {Pagano}, {Palacios}, {Pall{\'e}}, {Peter}, {Piazza}, {Piotto},
  {Pizarro}, {Pollaco}, {Ragazzoni}, {Ratti}, {Rauer}, {Ribas}, {Rieder},
  {Rohlfs}, {Safa}, {Salatti}, {Santos}, {Scandariato}, {S{\'e}gransan},
  {Simon}, {Smith}, {Sordet}, {Sousa}, {Steller}, {Szab{\'o}}, {Szoke},
  {Thomas}, {Tschentscher}, {Udry}, {Van Grootel}, {Viotto}, {Walter},
  {Walton}, {Wildi}, \& {Wolter}}]{ben21}
{Benz}, W., {Broeg}, C., {Fortier}, A., {et~al.} 2021, Experimental Astronomy,
  51, 109

\bibitem[{{Borsa} {et~al.}(2021){Borsa}, {Allart}, {Casasayas-Barris},
  {Tabernero}, {Zapatero Osorio}, {Cristiani}, {Pepe}, {Rebolo}, {Santos},
  {Adibekyan}, {Bourrier}, {Demangeon}, {Ehrenreich}, {Pall{\'e}}, {Sousa},
  {Lillo-Box}, {Lovis}, {Micela}, {Oshagh}, {Poretti}, {Sozzetti}, {Allende
  Prieto}, {Alibert}, {Amate}, {Benz}, {Bouchy}, {Cabral}, {Dekker},
  {D'Odorico}, {Di Marcantonio}, {Figueira}, {Genova Santos}, {Gonz{\'a}lez
  Hern{\'a}ndez}, {Lo Curto}, {Manescau}, {Martins}, {M{\'e}gevand}, {Mehner},
  {Molaro}, {Nunes}, {Riva}, {Su{\'a}rez Mascare{\~n}o}, {Udry}, \&
  {Zerbi}}]{bor21}
{Borsa}, F., {Allart}, R., {Casasayas-Barris}, N., {et~al.} 2021, \aap, 645,
  A24

\bibitem[{{Borucki} {et~al.}(2009){Borucki}, {Koch}, {Batalha}, {Caldwell},
  {Christensen-Dalsgaard}, {Cochran}, {Dunham}, {Gautier}, {Geary},
  {Gilliland}, {Jenkins}, {Kjeldsen}, {Lissauer}, \& {Rowe}}]{bor09}
{Borucki}, W., {Koch}, D., {Batalha}, N., {et~al.} 2009, in IAU Symposium, Vol.
  253, Transiting Planets, ed. F.~{Pont}, D.~{Sasselov}, \& M.~J. {Holman},
  289--299

\bibitem[{{Bouchy} {et~al.}(2017){Bouchy}, {Doyon}, {Artigau}, {Melo},
  {Hernandez}, {Wildi}, {Delfosse}, {Lovis}, {Figueira}, {Canto Martins},
  {Gonz{\'a}lez Hern{\'a}ndez}, {Thibault}, {Reshetov}, {Pepe}, {Santos}, {de
  Medeiros}, {Rebolo}, {Abreu}, {Adibekyan}, {Bandy}, {Benz}, {Blind},
  {Bohlender}, {Boisse}, {Bovay}, {Broeg}, {Brousseau}, {Cabral}, {Chazelas},
  {Cloutier}, {Coelho}, {Conod}, {Cumming}, {Delabre}, {Genolet}, {Hagelberg},
  {Jayawardhana}, {K{\"a}ufl}, {Lafreni{\`e}re}, {de Castro Le{\~a}o}, {Malo},
  {de Medeiros Martins}, {Matthews}, {Metchev}, {Oshagh}, {Ouellet}, {Parro},
  {Rasilla Pi{\~n}eiro}, {Santos}, {Sarajlic}, {Segovia}, {Sordet}, {Udry},
  {Valencia}, {Vall{\'e}e}, {Venn}, {Wade}, \& {Saddlemyer}}]{bou17}
{Bouchy}, F., {Doyon}, R., {Artigau}, {\'E}., {et~al.} 2017, The Messenger,
  169, 21

\bibitem[{{Boyajian} {et~al.}(2012){Boyajian}, {von Braun}, {van Belle},
  {McAlister}, {ten Brummelaar}, {Kane}, {Muirhead}, {Jones}, {White},
  {Schaefer}, {Ciardi}, {Henry}, {L{\'o}pez-Morales}, {Ridgway}, {Gies}, {Jao},
  {Rojas-Ayala}, {Parks}, {Sturmann}, {Sturmann}, {Turner}, {Farrington},
  {Goldfinger}, \& {Berger}}]{boy12}
{Boyajian}, T.~S., {von Braun}, K., {van Belle}, G., {et~al.} 2012, \apj, 757,
  112

\bibitem[{{Brandt}(2021)}]{bra21}
{Brandt}, T.~D. 2021, \apjs, 254, 42

\bibitem[{{Cadieux} {et~al.}(2024){Cadieux}, {Plotnykov}, {Doyon}, {Valencia},
  {Jahandar}, {Dang}, {Turbet}, {Fauchez}, {Cloutier}, {Cherubim}, {Artigau},
  {Cook}, {Edwards}, {Hallatt}, {Charnay}, {Bouchy}, {Allart}, {Mignon},
  {Baron}, {Barros}, {Benneke}, {Canto Martins}, {Cowan}, {De Medeiros},
  {Delfosse}, {Delgado-Mena}, {Dumusque}, {Ehrenreich}, {Frensch},
  {Gonz{\'a}lez Hern{\'a}ndez}, {Hara}, {Lafreni{\`e}re}, {Lo Curto}, {Malo},
  {Melo}, {Mounzer}, {Passeger}, {Pepe}, {Poulin-Girard}, {Santos},
  {Sosnowska}, {Su{\'a}rez Mascare{\~n}o}, {Thibault}, {Vaulato}, {Wade}, \&
  {Wildi}}]{cad24}
{Cadieux}, C., {Plotnykov}, M., {Doyon}, R., {et~al.} 2024, \apjl, 960, L3

\bibitem[{{Castro-Gonz{\'a}lez} {et~al.}(2023){Castro-Gonz{\'a}lez},
  {Demangeon}, {Lillo-Box}, {Lovis}, {Lavie}, {Adibekyan}, {Acu{\~n}a},
  {Deleuil}, {Aguichine}, {Zapatero Osorio}, {Tabernero}, {Davoult}, {Alibert},
  {Santos}, {Sousa}, {Antoniadis-Karnavas}, {Borsa}, {Winn}, {Allende Prieto},
  {Figueira}, {Jenkins}, {Sozzetti}, {Damasso}, {Silva}, {Astudillo-Defru},
  {Barros}, {Bonfils}, {Cristiani}, {Di Marcantonio}, {Gonz{\'a}lez
  Hern{\'a}ndez}, {Curto}, {Martins}, {Nunes}, {Palle}, {Pepe}, {Seager}, \&
  {Su{\'a}rez Mascare{\~n}o}}]{cas23}
{Castro-Gonz{\'a}lez}, A., {Demangeon}, O.~D.~S., {Lillo-Box}, J., {et~al.}
  2023, \aap, 675, A52

\bibitem[{{Chambers} {et~al.}(1996){Chambers}, {Wetherill}, \& {Boss}}]{cha96}
{Chambers}, J.~E., {Wetherill}, G.~W., \& {Boss}, A.~P. 1996, \icarus, 119, 261

\bibitem[{{Cosentino} {et~al.}(2012){Cosentino}, {Lovis}, {Pepe}, {Collier
  Cameron}, {Latham}, {Molinari}, {Udry}, {Bezawada}, {Black}, {Born},
  {Buchschacher}, {Charbonneau}, {Figueira}, {Fleury}, {Galli}, {Gallie},
  {Gao}, {Ghedina}, {Gonzalez}, {Gonzalez}, {Guerra}, {Henry}, {Horne},
  {Hughes}, {Kelly}, {Lodi}, {Lunney}, {Maire}, {Mayor}, {Micela}, {Ordway},
  {Peacock}, {Phillips}, {Piotto}, {Pollacco}, {Queloz}, {Rice}, {Riverol},
  {Riverol}, {San Juan}, {Sasselov}, {Segransan}, {Sozzetti}, {Sosnowska},
  {Stobie}, {Szentgyorgyi}, {Vick}, \& {Weber}}]{cos12}
{Cosentino}, R., {Lovis}, C., {Pepe}, F., {et~al.} 2012, in \procspie, Vol.
  8446, Ground-based and Airborne Instrumentation for Astronomy IV, 84461V

\bibitem[{{Cristofari} {et~al.}(2023){Cristofari}, {Donati}, {Moutou},
  {Lehmann}, {Charpentier}, {Fouqu{\'e}}, {Folsom}, {Masseron}, {Carmona},
  {Delfosse}, {Petit}, {Artigau}, {Cook}, \& {SLS Consortium}}]{cri23}
{Cristofari}, P.~I., {Donati}, J.~F., {Moutou}, C., {et~al.} 2023, \mnras, 526,
  5648

\bibitem[{{Cutri} {et~al.}(2003){Cutri}, {Skrutskie}, {van Dyk}, {Beichman},
  {Carpenter}, {Chester}, {Cambresy}, {Evans}, {Fowler}, {Gizis}, {Howard},
  {Huchra}, {Jarrett}, {Kopan}, {Kirkpatrick}, {Light}, {Marsh}, {McCallon},
  {Schneider}, {Stiening}, {Sykes}, {Weinberg}, {Wheaton}, {Wheelock}, \&
  {Zacarias}}]{cut03}
{Cutri}, R.~M., {Skrutskie}, M.~F., {van Dyk}, S., {et~al.} 2003, {2MASS All
  Sky Catalog of point sources.}

\bibitem[{{Czesla} {et~al.}(2019){Czesla}, {Schr{\"o}ter}, {Schneider},
  {Huber}, {Pfeifer}, {Andreasen}, \& {Zechmeister}}]{cze19pya}
{Czesla}, S., {Schr{\"o}ter}, S., {Schneider}, C.~P., {et~al.} 2019, {PyA:
  Python astronomy-related packages}

\bibitem[{{Delisle} {et~al.}(2022){Delisle}, {Unger}, {Hara}, \&
  {S{\'e}gransan}}]{del22spleaf}
{Delisle}, J.~B., {Unger}, N., {Hara}, N.~C., \& {S{\'e}gransan}, D. 2022,
  \aap, 659, A182

\bibitem[{{Demangeon} {et~al.}(2021){Demangeon}, {Zapatero Osorio}, {Alibert},
  {Barros}, {Adibekyan}, {Tabernero}, {Antoniadis-Karnavas}, {Camacho},
  {Su{\'a}rez Mascare{\~n}o}, {Oshagh}, {Micela}, {Sousa}, {Lovis}, {Pepe},
  {Rebolo}, {Cristiani}, {Santos}, {Allart}, {Allende Prieto}, {Bossini},
  {Bouchy}, {Cabral}, {Damasso}, {Di Marcantonio}, {D'Odorico}, {Ehrenreich},
  {Faria}, {Figueira}, {G{\'e}nova Santos}, {Haldemann}, {Hara}, {Gonz{\'a}lez
  Hern{\'a}ndez}, {Lavie}, {Lillo-Box}, {Lo Curto}, {Martins}, {M{\'e}gevand},
  {Mehner}, {Molaro}, {Nunes}, {Pall{\'e}}, {Pasquini}, {Poretti}, {Sozzetti},
  \& {Udry}}]{dem21}
{Demangeon}, O.~D.~S., {Zapatero Osorio}, M.~R., {Alibert}, Y., {et~al.} 2021,
  \aap, 653, A41

\bibitem[{{Di Marcantonio} {et~al.}(2018){Di Marcantonio}, {Sosnowska},
  {Cupani}, {D'Odorico}, {Lovis}, {Segovia}, {Sousa}, {Gonz{\'a}lez
  Hern{\'a}ndez}, {Calderone}, {Cirami}, {Modigliani}, {Lo Curto}, {Cristiani},
  {Molaro}, {Pepe}, \& {M{\'e}gevand}}]{dim18}
{Di Marcantonio}, P., {Sosnowska}, D., {Cupani}, G., {et~al.} 2018, in Society
  of Photo-Optical Instrumentation Engineers (SPIE) Conference Series, Vol.
  10704, Observatory Operations: Strategies, Processes, and Systems VII,
  107040F

\bibitem[{{Dittmann} {et~al.}(2017){Dittmann}, {Irwin}, {Charbonneau},
  {Bonfils}, {Astudillo-Defru}, {Haywood}, {Berta-Thompson}, {Newton},
  {Rodriguez}, {Winters}, {Tan}, {Almenara}, {Bouchy}, {Delfosse}, {Forveille},
  {Lovis}, {Murgas}, {Pepe}, {Santos}, {Udry}, {W{\"u}nsche}, {Esquerdo},
  {Latham}, \& {Dressing}}]{dit17Natur}
{Dittmann}, J.~A., {Irwin}, J.~M., {Charbonneau}, D., {et~al.} 2017, \nat, 544,
  333

\bibitem[{{Donati} {et~al.}(2020){Donati}, {Kouach}, {Moutou}, {Doyon},
  {Delfosse}, {Artigau}, {Baratchart}, {Lacombe}, {Barrick}, {H{\'e}brard},
  {Bouchy}, {Saddlemyer}, {Par{\`e}s}, {Rabou}, {Micheau}, {Dolon}, {Reshetov},
  {Challita}, {Carmona}, {Striebig}, {Thibault}, {Martioli}, {Cook},
  {Fouqu{\'e}}, {Vermeulen}, {Wang}, {Arnold}, {Pepe}, {Boisse}, {Figueira},
  {Bouvier}, {Ray}, {Feugeade}, {Morin}, {Alencar}, {Hobson}, {Castilho},
  {Udry}, {Santos}, {Hernandez}, {Benedict}, {Vall{\'e}e}, {Gallou}, {Dupieux},
  {Larrieu}, {Perruchot}, {Sottile}, {Moreau}, {Usher}, {Baril}, {Wildi},
  {Chazelas}, {Malo}, {Bonfils}, {Loop}, {Kerley}, {Wevers}, {Dunn}, {Pazder},
  {Macdonald}, {Dubois}, {Carri{\'e}}, {Valentin}, {Henault}, {Yan}, \&
  {Steinmetz}}]{don20}
{Donati}, J.~F., {Kouach}, D., {Moutou}, C., {et~al.} 2020, \mnras, 498, 5684

\bibitem[{{Donati} {et~al.}(2023){Donati}, {Lehmann}, {Cristofari},
  {Fouqu{\'e}}, {Moutou}, {Charpentier}, {Ould-Elhkim}, {Carmona}, {Delfosse},
  {Artigau}, {Alencar}, {Cadieux}, {Arnold}, {Petit}, {Morin}, {Forveille},
  {Cloutier}, {Doyon}, {H{\'e}brard}, \& {SLS Collaboration}}]{don23}
{Donati}, J.~F., {Lehmann}, L.~T., {Cristofari}, P.~I., {et~al.} 2023, \mnras,
  525, 2015

\bibitem[{{Dreizler} {et~al.}(2024){Dreizler}, {Luque}, {Ribas}, {Koseleva},
  {Ruh}, {Nagel}, {Pozuelos}, {Zechmeister}, {Reiners}, {Caballero}, {Amado},
  {B{\'e}jar}, {Bean}, {Brady}, {Cifuentes}, {Gillon}, {Hatzes}, {Henning},
  {Kasper}, {Montes}, {Morales}, {Murray}, {Pall{\'e}}, {Quirrenbach},
  {Seifahrt}, {Schweitzer}, {St{\"u}rmer}, {Stef{\'a}nsson}, \&
  {Linares}}]{dre24}
{Dreizler}, S., {Luque}, R., {Ribas}, I., {et~al.} 2024, \aap, 684, A117

\bibitem[{{Dumusque} {et~al.}(2014){Dumusque}, {Boisse}, \& {Santos}}]{dum14}
{Dumusque}, X., {Boisse}, I., \& {Santos}, N.~C. 2014, \apj, 796, 132

\bibitem[{{Eastman} {et~al.}(2013){Eastman}, {Gaudi}, \& {Agol}}]{eas13}
{Eastman}, J., {Gaudi}, B.~S., \& {Agol}, E. 2013, \pasp, 125, 83

\bibitem[{{Eastman} {et~al.}(2010){Eastman}, {Siverd}, \& {Gaudi}}]{eas10}
{Eastman}, J., {Siverd}, R., \& {Gaudi}, B.~S. 2010, \pasp, 122, 935

\bibitem[{{Ehrenreich} {et~al.}(2020){Ehrenreich}, {Lovis}, {Allart}, {Zapatero
  Osorio}, {Pepe}, {Cristiani}, {Rebolo}, {Santos}, {Borsa}, {Demangeon},
  {Dumusque}, {Gonz{\'a}lez Hern{\'a}ndez}, {Casasayas-Barris},
  {S{\'e}gransan}, {Sousa}, {Abreu}, {Adibekyan}, {Affolter}, {Allende Prieto},
  {Alibert}, {Aliverti}, {Alves}, {Amate}, {Avila}, {Baldini}, {Bandy}, {Benz},
  {Bianco}, {Bolmont}, {Bouchy}, {Bourrier}, {Broeg}, {Cabral}, {Calderone},
  {Pall{\'e}}, {Cegla}, {Cirami}, {Coelho}, {Conconi}, {Coretti}, {Cumani},
  {Cupani}, {Dekker}, {Delabre}, {Deiries}, {D'Odorico}, {Di Marcantonio},
  {Figueira}, {Fragoso}, {Genolet}, {Genoni}, {G{\'e}nova Santos}, {Hara},
  {Hughes}, {Iwert}, {Kerber}, {Knudstrup}, {Landoni}, {Lavie}, {Lizon},
  {Lendl}, {Lo Curto}, {Maire}, {Manescau}, {Martins}, {M{\'e}gevand},
  {Mehner}, {Micela}, {Modigliani}, {Molaro}, {Monteiro}, {Monteiro},
  {Moschetti}, {M{\"u}ller}, {Nunes}, {Oggioni}, {Oliveira}, {Pariani},
  {Pasquini}, {Poretti}, {Rasilla}, {Redaelli}, {Riva}, {Santana Tschudi},
  {Santin}, {Santos}, {Segovia Milla}, {Seidel}, {Sosnowska}, {Sozzetti},
  {Span{\`o}}, {Su{\'a}rez Mascare{\~n}o}, {Tabernero}, {Tenegi}, {Udry},
  {Zanutta}, \& {Zerbi}}]{ehr20}
{Ehrenreich}, D., {Lovis}, C., {Allart}, R., {et~al.} 2020, \nat, 580, 597

\bibitem[{{Faria} {et~al.}(2016){Faria}, {Haywood}, {Brewer}, {Figueira},
  {Oshagh}, {Santerne}, \& {Santos}}]{far16}
{Faria}, J.~P., {Haywood}, R.~D., {Brewer}, B.~J., {et~al.} 2016, \aap, 588,
  A31

\bibitem[{{Faria} {et~al.}(2022){Faria}, {Su{\'a}rez Mascare{\~n}o},
  {Figueira}, {Silva}, {Damasso}, {Demangeon}, {Pepe}, {Santos}, {Rebolo},
  {Cristiani}, {Adibekyan}, {Alibert}, {Allart}, {Barros}, {Cabral},
  {D'Odorico}, {Di Marcantonio}, {Dumusque}, {Ehrenreich}, {Gonz{\'a}lez
  Hern{\'a}ndez}, {Hara}, {Lillo-Box}, {Lo Curto}, {Lovis}, {Martins},
  {M{\'e}gevand}, {Mehner}, {Micela}, {Molaro}, {Nunes}, {Pall{\'e}},
  {Poretti}, {Sousa}, {Sozzetti}, {Tabernero}, {Udry}, \& {Zapatero
  Osorio}}]{far22}
{Faria}, J.~P., {Su{\'a}rez Mascare{\~n}o}, A., {Figueira}, P., {et~al.} 2022,
  \aap, 658, A115

\bibitem[{{Feroz} {et~al.}(2009){Feroz}, {Hobson}, \& {Bridges}}]{fer09}
{Feroz}, F., {Hobson}, M.~P., \& {Bridges}, M. 2009, \mnras, 398, 1601

\bibitem[{{Foreman-Mackey} {et~al.}(2017){Foreman-Mackey}, {Agol},
  {Ambikasaran}, \& {Angus}}]{fm17celerite2}
{Foreman-Mackey}, D., {Agol}, E., {Ambikasaran}, S., \& {Angus}, R. 2017, \aj,
  154, 220

\bibitem[{{Foreman-Mackey} {et~al.}(2014){Foreman-Mackey}, {Hoyer}, {Bernhard},
  \& {Angus}}]{fm14george}
{Foreman-Mackey}, D., {Hoyer}, S., {Bernhard}, J., \& {Angus}, R. 2014,
  {george: George (v0.2.0)}

\bibitem[{{France} {et~al.}(2020){France}, {Duvvuri}, {Egan}, {Koskinen},
  {Wilson}, {Youngblood}, {Froning}, {Brown}, {Alvarado-G{\'o}mez},
  {Berta-Thompson}, {Drake}, {Garraffo}, {Kaltenegger}, {Kowalski}, {Linsky},
  {Loyd}, {Mauas}, {Miguel}, {Pineda}, {Rugheimer}, {Schneider}, {Tian}, \&
  {Vieytes}}]{fra20}
{France}, K., {Duvvuri}, G., {Egan}, H., {et~al.} 2020, \aj, 160, 237

\bibitem[{{Fulton} {et~al.}(2018){Fulton}, {Petigura}, {Blunt}, \&
  {Sinukoff}}]{ful18radvel}
{Fulton}, B.~J., {Petigura}, E.~A., {Blunt}, S., \& {Sinukoff}, E. 2018, \pasp,
  130, 044504

\bibitem[{{Gaia Collaboration} {et~al.}(2021){Gaia Collaboration}, {Brown},
  {Vallenari}, {Prusti}, {de Bruijne}, {Babusiaux}, {Biermann}, {Creevey},
  {Evans}, {Eyer}, {Hutton}, {Jansen}, {Jordi}, {Klioner}, {Lammers},
  {Lindegren}, {Luri}, {Mignard}, {Panem}, {Pourbaix}, {Randich}, {Sartoretti},
  {Soubiran}, {Walton}, {Arenou}, {Bailer-Jones}, {Bastian}, {Cropper},
  {Drimmel}, {Katz}, {Lattanzi}, {van Leeuwen}, {Bakker}, {Cacciari},
  {Casta{\~n}eda}, {De Angeli}, {Ducourant}, {Fabricius}, {Fouesneau},
  {Fr{\'e}mat}, {Guerra}, {Guerrier}, {Guiraud}, {Jean-Antoine Piccolo},
  {Masana}, {Messineo}, {Mowlavi}, {Nicolas}, {Nienartowicz}, {Pailler},
  {Panuzzo}, {Riclet}, {Roux}, {Seabroke}, {Sordo}, {Tanga}, {Th{\'e}venin},
  {Gracia-Abril}, {Portell}, {Teyssier}, {Altmann}, {Andrae}, {Bellas-Velidis},
  {Benson}, {Berthier}, {Blomme}, {Brugaletta}, {Burgess}, {Busso}, {Carry},
  {Cellino}, {Cheek}, {Clementini}, {Damerdji}, {Davidson}, {Delchambre},
  {Dell'Oro}, {Fern{\'a}ndez-Hern{\'a}ndez}, {Galluccio}, {Garc{\'\i}a-Lario},
  {Garcia-Reinaldos}, {Gonz{\'a}lez-N{\'u}{\~n}ez}, {Gosset}, {Haigron},
  {Halbwachs}, {Hambly}, {Harrison}, {Hatzidimitriou}, {Heiter},
  {Hern{\'a}ndez}, {Hestroffer}, {Hodgkin}, {Holl}, {Jan{\ss}en}, {Jevardat de
  Fombelle}, {Jordan}, {Krone-Martins}, {Lanzafame}, {L{\"o}ffler}, {Lorca},
  {Manteiga}, {Marchal}, {Marrese}, {Moitinho}, {Mora}, {Muinonen}, {Osborne},
  {Pancino}, {Pauwels}, {Petit}, {Recio-Blanco}, {Richards}, {Riello},
  {Rimoldini}, {Robin}, {Roegiers}, {Rybizki}, {Sarro}, {Siopis}, {Smith},
  {Sozzetti}, {Ulla}, {Utrilla}, {van Leeuwen}, {van Reeven}, {Abbas}, {Abreu
  Aramburu}, {Accart}, {Aerts}, {Aguado}, {Ajaj}, {Altavilla}, {{\'A}lvarez},
  {{\'A}lvarez Cid-Fuentes}, {Alves}, {Anderson}, {Anglada Varela}, {Antoja},
  {Audard}, {Baines}, {Baker}, {Balaguer-N{\'u}{\~n}ez}, {Balbinot}, {Balog},
  {Barache}, {Barbato}, {Barros}, {Barstow}, {Bartolom{\'e}}, {Bassilana},
  {Bauchet}, {Baudesson-Stella}, {Becciani}, {Bellazzini}, {Bernet}, {Bertone},
  {Bianchi}, {Blanco-Cuaresma}, {Boch}, {Bombrun}, {Bossini}, {Bouquillon},
  {Bragaglia}, {Bramante}, {Breedt}, {Bressan}, {Brouillet}, {Bucciarelli},
  {Burlacu}, {Busonero}, {Butkevich}, {Buzzi}, {Caffau}, {Cancelliere},
  {C{\'a}novas}, {Cantat-Gaudin}, {Carballo}, {Carlucci}, {Carnerero},
  {Carrasco}, {Casamiquela}, {Castellani}, {Castro-Ginard}, {Castro Sampol},
  {Chaoul}, {Charlot}, {Chemin}, {Chiavassa}, {Cioni}, {Comoretto}, {Cooper},
  {Cornez}, {Cowell}, {Crifo}, {Crosta}, {Crowley}, {Dafonte}, {Dapergolas},
  {David}, {David}, {de Laverny}, {De Luise}, {De March}, {De Ridder}, {de
  Souza}, {de Teodoro}, {de Torres}, {del Peloso}, {del Pozo}, {Delbo},
  {Delgado}, {Delgado}, {Delisle}, {Di Matteo}, {Diakite}, {Diener},
  {Distefano}, {Dolding}, {Eappachen}, {Edvardsson}, {Enke}, {Esquej}, {Fabre},
  {Fabrizio}, {Faigler}, {Fedorets}, {Fernique}, {Fienga}, {Figueras},
  {Fouron}, {Fragkoudi}, {Fraile}, {Franke}, {Gai}, {Garabato},
  {Garcia-Gutierrez}, {Garc{\'\i}a-Torres}, {Garofalo}, {Gavras}, {Gerlach},
  {Geyer}, {Giacobbe}, {Gilmore}, {Girona}, {Giuffrida}, {Gomel}, {Gomez},
  {Gonzalez-Santamaria}, {Gonz{\'a}lez-Vidal}, {Granvik},
  {Guti{\'e}rrez-S{\'a}nchez}, {Guy}, {Hauser}, {Haywood}, {Helmi}, {Hidalgo},
  {Hilger}, {H{\l}adczuk}, {Hobbs}, {Holland}, {Huckle}, {Jasniewicz},
  {Jonker}, {Juaristi Campillo}, {Julbe}, {Karbevska}, {Kervella}, {Khanna},
  {Kochoska}, {Kontizas}, {Kordopatis}, {Korn}, {Kostrzewa-Rutkowska},
  {Kruszy{\'n}ska}, {Lambert}, {Lanza}, {Lasne}, {Le Campion}, {Le Fustec},
  {Lebreton}, {Lebzelter}, {Leccia}, {Leclerc}, {Lecoeur-Taibi}, {Liao},
  {Licata}, {Lindstr{\o}m}, {Lister}, {Livanou}, {Lobel}, {Madrero Pardo},
  {Managau}, {Mann}, {Marchant}, {Marconi}, {Marcos Santos}, {Marinoni},
  {Marocco}, {Marshall}, {Martin Polo}, {Mart{\'\i}n-Fleitas}, {Masip},
  {Massari}, {Mastrobuono-Battisti}, {Mazeh}, {McMillan}, {Messina},
  {Michalik}, {Millar}, {Mints}, {Molina}, {Molinaro}, {Moln{\'a}r},
  {Montegriffo}, {Mor}, {Morbidelli}, {Morel}, {Morris}, {Mulone}, {Munoz},
  {Muraveva}, {Murphy}, {Musella}, {Noval}, {Ord{\'e}novic}, {Orr{\`u}},
  {Osinde}, {Pagani}, {Pagano}, {Palaversa}, {Palicio}, {Panahi}, {Pawlak},
  {Pe{\~n}alosa Esteller}, {Penttil{\"a}}, {Piersimoni}, {Pineau}, {Plachy},
  {Plum}, {Poggio}, {Poretti}, {Poujoulet}, {Pr{\v{s}}a}, {Pulone}, {Racero},
  {Ragaini}, {Rainer}, {Raiteri}, {Rambaux}, {Ramos}, {Ramos-Lerate}, {Re
  Fiorentin}, {Regibo}, {Reyl{\'e}}, {Ripepi}, {Riva}, {Rixon}, {Robichon},
  {Robin}, {Roelens}, {Rohrbasser}, {Romero-G{\'o}mez}, {Rowell}, {Royer},
  {Rybicki}, {Sadowski}, {Sagrist{\`a} Sell{\'e}s}, {Sahlmann}, {Salgado},
  {Salguero}, {Samaras}, {Sanchez Gimenez}, {Sanna}, {Santove{\~n}a},
  {Sarasso}, {Schultheis}, {Sciacca}, {Segol}, {Segovia}, {S{\'e}gransan},
  {Semeux}, {Shahaf}, {Siddiqui}, {Siebert}, {Siltala}, {Slezak}, {Smart},
  {Solano}, {Solitro}, {Souami}, {Souchay}, {Spagna}, {Spoto}, {Steele},
  {Steidelm{\"u}ller}, {Stephenson}, {S{\"u}veges}, {Szabados}, {Szegedi-Elek},
  {Taris}, {Tauran}, {Taylor}, {Teixeira}, {Thuillot}, {Tonello}, {Torra},
  {Torra}, {Turon}, {Unger}, {Vaillant}, {van Dillen}, {Vanel}, {Vecchiato},
  {Viala}, {Vicente}, {Voutsinas}, {Weiler}, {Wevers}, {Wyrzykowski}, {Yoldas},
  {Yvard}, {Zhao}, {Zorec}, {Zucker}, {Zurbach}, \& {Zwitter}}]{gaia21}
{Gaia Collaboration}, {Brown}, A.~G.~A., {Vallenari}, A., {et~al.} 2021, \aap,
  649, A1

\bibitem[{{Giles} {et~al.}(2017){Giles}, {Collier Cameron}, \&
  {Haywood}}]{gil17}
{Giles}, H. A.~C., {Collier Cameron}, A., \& {Haywood}, R.~D. 2017, \mnras,
  472, 1618

\bibitem[{{Gillon} {et~al.}(2017){Gillon}, {Triaud}, {Demory}, {Jehin}, {Agol},
  {Deck}, {Lederer}, {de Wit}, {Burdanov}, {Ingalls}, {Bolmont}, {Leconte},
  {Raymond}, {Selsis}, {Turbet}, {Barkaoui}, {Burgasser}, {Burleigh}, {Carey},
  {Chaushev}, {Copperwheat}, {Delrez}, {Fernandes}, {Holdsworth}, {Kotze}, {Van
  Grootel}, {Almleaky}, {Benkhaldoun}, {Magain}, \& {Queloz}}]{gil17Natur}
{Gillon}, M., {Triaud}, A.~H.~M.~J., {Demory}, B.-O., {et~al.} 2017, \nat, 542,
  456

\bibitem[{{Gladman}(1993)}]{gla93}
{Gladman}, B. 1993, \icarus, 106, 247

\bibitem[{{Gonz{\'a}lez Hern{\'a}ndez} {et~al.}(2018){Gonz{\'a}lez
  Hern{\'a}ndez}, {Pepe}, {Molaro}, \& {Santos}}]{gon18}
{Gonz{\'a}lez Hern{\'a}ndez}, J.~I., {Pepe}, F., {Molaro}, P., \& {Santos},
  N.~C. 2018, in Handbook of Exoplanets, ed. H.~J. {Deeg} \& J.~A. {Belmonte},
  157

\bibitem[{{Gratia} \& {Lissauer}(2021)}]{gra21}
{Gratia}, P. \& {Lissauer}, J.~J. 2021, \icarus, 358, 114038

\bibitem[{{Gullikson} {et~al.}(2014){Gullikson}, {Dodson-Robinson}, \&
  {Kraus}}]{gul14}
{Gullikson}, K., {Dodson-Robinson}, S., \& {Kraus}, A. 2014, \aj, 148, 53

\bibitem[{{Handley} {et~al.}(2015{\natexlab{a}}){Handley}, {Hobson}, \&
  {Lasenby}}]{han15a}
{Handley}, W.~J., {Hobson}, M.~P., \& {Lasenby}, A.~N. 2015{\natexlab{a}},
  \mnras, 450, L61

\bibitem[{{Handley} {et~al.}(2015{\natexlab{b}}){Handley}, {Hobson}, \&
  {Lasenby}}]{han15b}
{Handley}, W.~J., {Hobson}, M.~P., \& {Lasenby}, A.~N. 2015{\natexlab{b}},
  \mnras, 453, 4384

\bibitem[{{Hara} {et~al.}(2019){Hara}, {Bou{\'e}}, {Laskar}, {Delisle}, \&
  {Unger}}]{har19}
{Hara}, N.~C., {Bou{\'e}}, G., {Laskar}, J., {Delisle}, J.~B., \& {Unger}, N.
  2019, \mnras, 489, 738

\bibitem[{{Hara} {et~al.}(2022){Hara}, {Unger}, {Delisle}, {D{\'\i}az}, \&
  {S{\'e}gransan}}]{har22}
{Hara}, N.~C., {Unger}, N., {Delisle}, J.-B., {D{\'\i}az}, R.~F., \&
  {S{\'e}gransan}, D. 2022, \aap, 663, A14

\bibitem[{{Harris} {et~al.}(2020){Harris}, {Millman}, {van der Walt},
  {Gommers}, {Virtanen}, {Cournapeau}, {Wieser}, {Taylor}, {Berg}, {Smith},
  {Kern}, {Picus}, {Hoyer}, {van Kerkwijk}, {Brett}, {Haldane}, {del R{\'\i}o},
  {Wiebe}, {Peterson}, {G{\'e}rard-Marchant}, {Sheppard}, {Reddy}, {Weckesser},
  {Abbasi}, {Gohlke}, \& {Oliphant}}]{har20numpy}
{Harris}, C.~R., {Millman}, K.~J., {van der Walt}, S.~J., {et~al.} 2020, \nat,
  585, 357

\bibitem[{{Haywood} {et~al.}(2014){Haywood}, {Collier Cameron}, {Queloz},
  {Barros}, {Deleuil}, {Fares}, {Gillon}, {Lanza}, {Lovis}, {Moutou}, {Pepe},
  {Pollacco}, {Santerne}, {S{\'e}gransan}, \& {Unruh}}]{hay14}
{Haywood}, R.~D., {Collier Cameron}, A., {Queloz}, D., {et~al.} 2014, \mnras,
  443, 2517

\bibitem[{Hunter(2007)}]{hun07matplotlib}
Hunter, J.~D. 2007, Computing in Science \& Engineering, 9, 90

\bibitem[{{Hussain} \& {Tamayo}(2020)}]{hus20}
{Hussain}, N. \& {Tamayo}, D. 2020, \mnras, 491, 5258

\bibitem[{{Jahandar} {et~al.}(2023){Jahandar}, {Doyon}, {Artigau}, {Cook},
  {Cadieux}, {Lafreni{\`e}re}, {Forveille}, {Donati}, {Fouqu{\'e}}, {Carmona},
  {Cloutier}, {Cristofari}, {Gaidos}, {Gomes da Silva}, {Malo}, {Martioli}, {do
  Nascimento}, {Pelletier}, {Vandal}, \& {Venn}}]{jah23}
{Jahandar}, F., {Doyon}, R., {Artigau}, {\'E}., {et~al.} 2023, arXiv e-prints,
  arXiv:2310.12125

\bibitem[{{Kervella} {et~al.}(2019){Kervella}, {Arenou}, {Mignard}, \&
  {Th{\'e}venin}}]{ker19}
{Kervella}, P., {Arenou}, F., {Mignard}, F., \& {Th{\'e}venin}, F. 2019, \aap,
  623, A72

\bibitem[{{Kervella} {et~al.}(2022){Kervella}, {Arenou}, \&
  {Th{\'e}venin}}]{ker22}
{Kervella}, P., {Arenou}, F., \& {Th{\'e}venin}, F. 2022, \aap, 657, A7

\bibitem[{{Kirkpatrick} {et~al.}(2012){Kirkpatrick}, {Gelino}, {Cushing},
  {Mace}, {Griffith}, {Skrutskie}, {Marsh}, {Wright}, {Eisenhardt}, {McLean},
  {Mainzer}, {Burgasser}, {Tinney}, {Parker}, \& {Salter}}]{kir12}
{Kirkpatrick}, J.~D., {Gelino}, C.~R., {Cushing}, M.~C., {et~al.} 2012, \apj,
  753, 156

\bibitem[{{Koen} {et~al.}(2010){Koen}, {Kilkenny}, {van Wyk}, \&
  {Marang}}]{koe10}
{Koen}, C., {Kilkenny}, D., {van Wyk}, F., \& {Marang}, F. 2010, \mnras, 403,
  1949

\bibitem[{{Kopparapu} {et~al.}(2014){Kopparapu}, {Ramirez}, {SchottelKotte},
  {Kasting}, {Domagal-Goldman}, \& {Eymet}}]{kop14}
{Kopparapu}, R.~K., {Ramirez}, R.~M., {SchottelKotte}, J., {et~al.} 2014,
  \apjl, 787, L29

\bibitem[{{K{\"u}rster} {et~al.}(2003){K{\"u}rster}, {Endl}, {Rouesnel}, {Els},
  {Kaufer}, {Brillant}, {Hatzes}, {Saar}, \& {Cochran}}]{kur03}
{K{\"u}rster}, M., {Endl}, M., {Rouesnel}, F., {et~al.} 2003, \aap, 403, 1077

\bibitem[{{Lafarga} {et~al.}(2020){Lafarga}, {Ribas}, {Lovis}, {Perger},
  {Zechmeister}, {Bauer}, {K{\"u}rster}, {Cort{\'e}s-Contreras}, {Morales},
  {Herrero}, {Rosich}, {Baroch}, {Reiners}, {Caballero}, {Quirrenbach},
  {Amado}, {Alacid}, {B{\'e}jar}, {Dreizler}, {Hatzes}, {Henning}, {Jeffers},
  {Kaminski}, {Montes}, {Pedraz}, {Rodr{\'\i}guez-L{\'o}pez}, \&
  {Schmitt}}]{laf20}
{Lafarga}, M., {Ribas}, I., {Lovis}, C., {et~al.} 2020, \aap, 636, A36

\bibitem[{{Laskar} \& {Petit}(2017)}]{las17}
{Laskar}, J. \& {Petit}, A.~C. 2017, \aap, 605, A72

\bibitem[{{Lavie} {et~al.}(2023){Lavie}, {Bouchy}, {Lovis}, {Zapatero Osorio},
  {Deline}, {Barros}, {Figueira}, {Sozzetti}, {Gonz{\'a}lez Hern{\'a}ndez},
  {Lillo-Box}, {Rodrigues}, {Mehner}, {Damasso}, {Adibekyan}, {Alibert},
  {Allende Prieto}, {Cristiani}, {D'Odorico}, {Di Marcantonio}, {Ehrenreich},
  {G{\'e}nova Santos}, {Lo Curto}, {Martins}, {Micela}, {Molaro}, {Nunes},
  {Palle}, {Pepe}, {Poretti}, {Rebolo}, {Santos}, {Sousa}, {Su{\'a}rez
  Mascare{\~n}o}, {Tabrenero}, \& {Udry}}]{lav23}
{Lavie}, B., {Bouchy}, F., {Lovis}, C., {et~al.} 2023, \aap, 673, A69

\bibitem[{{Lillo-Box} {et~al.}(2021){Lillo-Box}, {Faria}, {Su{\'a}rez
  Mascare{\~n}o}, {Figueira}, {Sousa}, {Tabernero}, {Lovis}, {Silva},
  {Demangeon}, {Benatti}, {Santos}, {Mehner}, {Pepe}, {Sozzetti}, {Zapatero
  Osorio}, {Gonz{\'a}lez Hern{\'a}ndez}, {Micela}, {Hojjatpanah}, {Rebolo},
  {Cristiani}, {Adibekyan}, {Allart}, {Allende Prieto}, {Cabral}, {Damasso},
  {Di Marcantonio}, {Lo Curto}, {Martins}, {Megevand}, {Molaro}, {Nunes},
  {Pall{\'e}}, {Pasquini}, {Poretti}, \& {Udry}}]{lil21}
{Lillo-Box}, J., {Faria}, J.~P., {Su{\'a}rez Mascare{\~n}o}, A., {et~al.} 2021,
  \aap, 654, A60

\bibitem[{{Lillo-Box} {et~al.}(2020){Lillo-Box}, {Figueira}, {Leleu},
  {Acu{\~n}a}, {Faria}, {Hara}, {Santos}, {Correia}, {Robutel}, {Deleuil},
  {Barrado}, {Sousa}, {Bonfils}, {Mousis}, {Almenara}, {Astudillo-Defru},
  {Marcq}, {Udry}, {Lovis}, \& {Pepe}}]{lil20}
{Lillo-Box}, J., {Figueira}, P., {Leleu}, A., {et~al.} 2020, \aap, 642, A121

\bibitem[{{Lovis} {et~al.}(2011){Lovis}, {S{\'e}gransan}, {Mayor}, {Udry},
  {Benz}, {Bertaux}, {Bouchy}, {Correia}, {Laskar}, {Lo Curto}, {Mordasini},
  {Pepe}, {Queloz}, \& {Santos}}]{lov11}
{Lovis}, C., {S{\'e}gransan}, D., {Mayor}, M., {et~al.} 2011, \aap, 528, A112

\bibitem[{{Lubin} {et~al.}(2021){Lubin}, {Robertson}, {Stefansson}, {Ninan},
  {Mahadevan}, {Endl}, {Ford}, {Wright}, {Beard}, {Bender}, {Cochran},
  {Diddams}, {Fredrick}, {Halverson}, {Kanodia}, {Metcalf}, {Ramsey}, {Roy},
  {Schwab}, \& {Terrien}}]{lub21}
{Lubin}, J., {Robertson}, P., {Stefansson}, G., {et~al.} 2021, \aj, 162, 61

\bibitem[{{Marconi} {et~al.}(2022){Marconi}, {Abreu}, {Adibekyan}, {Alberti},
  {Albrecht}, {Alcaniz}, {Aliverti}, {Allende Prieto}, {Alvarado G{\'o}mez},
  {Amado}, {Amate}, {Andersen}, {Artigau}, {Baker}, {Baldini}, {Balestra},
  {Barnes}, {Baron}, {Barros}, {Bauer}, {Beaulieu}, {Bellido-Tirado},
  {Benneke}, {Bensby}, {Bergin}, {Biazzo}, {Bik}, {Birkby}, {Blind}, {Boisse},
  {Bolmont}, {Bonaglia}, {Bonfils}, {Borsa}, {Brandeker}, {Brandner}, {Broeg},
  {Brogi}, {Brousseau}, {Brucalassi}, {Brynnel}, {Buchhave}, {Buscher},
  {Cabral}, {Calderone}, {Calvo-Ortega}, {Canto Martins}, {Cantalloube},
  {Carbonaro}, {Chauvin}, {Chazelas}, {Cheffot}, {Cheng}, {Chiavassa},
  {Christensen}, {Cirami}, {Cook}, {Cooke}, {Coretti}, {Covino}, {Cowan},
  {Cresci}, {Cristiani}, {Cunha Parro}, {Cupani}, {D'Odorico}, {de Castro
  Le{\~a}o}, {De Cia}, {De Medeiros}, {Debras}, {Debus}, {Demangeon},
  {Dessauges-Zavadsky}, {Di Marcantonio}, {Dionies}, {Doyon}, {Dunn},
  {Ehrenreich}, {Faria}, {Feruglio}, {Fisher}, {Fontana}, {Fumagalli}, {Fusco},
  {Fynbo}, {Gabella}, {Gaessler}, {Gallo}, {Gao}, {Genolet}, {Genoni},
  {Giacobbe}, {Giro}, {Gon{\c{c}}alves}, {Gonzalez}, {Gonz{\'a}lez
  Hern{\'a}ndez}, {Gracia T{\'e}mich}, {Haehnelt}, {Haniff}, {Hatzes},
  {Helled}, {Hoeijmakers}, {Huke}, {J{\"a}rvinen}, {J{\"a}rvinen}, {Kaminski},
  {Korn}, {Kouach}, {Kowzan}, {Kreidberg}, {Landoni}, {Lanotte}, {Lavail},
  {Li}, {Liske}, {Lovis}, {Lucatello}, {Lunney}, {MacIntosh}, {Madhusudhan},
  {Magrini}, {Maiolino}, {Malo}, {Man}, {Marquart}, {Marques}, {Martins},
  {Martins}, {Maslowski}, {Mason}, {Mason}, {McCracken}, {Mergo}, {Micela},
  {Mitchell}, {Molli{\`e}re}, {Monteiro}, {Montgomery}, {Mordasini}, {Morin},
  {Mucciarelli}, {Murphy}, {N'Diaye}, {Neichel}, {Niedzielski}, {Niemczura},
  {Nortmann}, {Noterdaeme}, {Nunes}, {Oggioni}, {Oliva}, {{\"O}nel}, {Origlia},
  {{\"O}stlin}, {Palle}, {Papaderos}, {Pariani}, {Pe{\~n}ate Castro}, {Pepe},
  {Perreault Levasseur}, {Petit}, {Pino}, {Piqueras}, {Pollo}, {Poppenhaeger},
  {Quirrenbach}, {Rauscher}, {Rebolo}, {Redaelli}, {Reffert}, {Reid},
  {Reiners}, {Richter}, {Riva}, {Rivoire}, {Rodr{\'\i}guez-L{\'o}pez},
  {Roederer}, {Romano}, {Rousseau}, {Rowe}, {Salvadori}, {Santos}, {Santos
  Diaz}, {Sanz-Forcada}, {Sarajlic}, {Sauvage}, {Sch{\"a}fer}, {Schiavon},
  {Schmidt}, {Selmi}, {Sivanandam}, {Sordet}, {Sordo}, {Sortino}, {Sosnowska},
  {Sousa}, {Stempels}, {Strassmeier}, {Su{\'a}rez Mascare{\~n}o}, {Sulich},
  {Sun}, {Tanvir}, {Tenegi-Sangin{\'e}s}, {Thibault}, {Thompson}, {Tozzi},
  {Turbet}, {Vall{\'e}e}, {Varas}, {Venn}, {V{\'e}ran}, {Verma}, {Viel},
  {Wade}, {Waring}, {Weber}, {Weder}, {Wehbe}, {Weingrill}, {Woche}, {Xompero},
  {Zackrisson}, {Zanutta}, {Zapatero Osorio}, {Zechmeister}, \&
  {Zimara}}]{mar22}
{Marconi}, A., {Abreu}, M., {Adibekyan}, V., {et~al.} 2022, in Society of
  Photo-Optical Instrumentation Engineers (SPIE) Conference Series, Vol. 12184,
  Ground-based and Airborne Instrumentation for Astronomy IX, ed. C.~J.
  {Evans}, J.~J. {Bryant}, \& K.~{Motohara}, 1218424

\bibitem[{{Marfil} {et~al.}(2021){Marfil}, {Tabernero}, {Montes}, {Caballero},
  {L{\'a}zaro}, {Gonz{\'a}lez Hern{\'a}ndez}, {Nagel}, {Passegger},
  {Schweitzer}, {Ribas}, {Reiners}, {Quirrenbach}, {Amado}, {Cifuentes},
  {Cort{\'e}s-Contreras}, {Dreizler}, {Duque-Arribas},
  {Galad{\'\i}-Enr{\'\i}quez}, {Henning}, {Jeffers}, {Kaminski}, {K{\"u}rster},
  {Lafarga}, {L{\'o}pez-Gallifa}, {Morales}, {Shan}, \& {Zechmeister}}]{mar21}
{Marfil}, E., {Tabernero}, H.~M., {Montes}, D., {et~al.} 2021, \aap, 656, A162

\bibitem[{{Martins} {et~al.}(2022){Martins}, {Cristiani}, {Cupani},
  {D'Odorico}, {G{\'e}nova Santos}, {Leite}, {Marques}, {Milakovi{\'c}},
  {Molaro}, {Murphy}, {Nunes}, {Schmidt}, {Adibekyan}, {Alibert}, {Di
  Marcantonio}, {Gonz{\'a}lez Hern{\'a}ndez}, {M{\'e}gevand}, {Palle}, {Pepe},
  {Santos}, {Sousa}, {Sozzetti}, {Su{\'a}rez Mascare{\~n}o}, \& {Zapatero
  Osorio}}]{martins22}
{Martins}, C.~J.~A.~P., {Cristiani}, S., {Cupani}, G., {et~al.} 2022, \prd,
  105, 123507

\bibitem[{{Mayor} {et~al.}(2003){Mayor}, {Pepe}, {Queloz}, {Bouchy},
  {Rupprecht}, {Lo Curto}, {Avila}, {Benz}, {Bertaux}, {Bonfils}, {Dall},
  {Dekker}, {Delabre}, {Eckert}, {Fleury}, {Gilliotte}, {Gojak}, {Guzman},
  {Kohler}, {Lizon}, {Longinotti}, {Lovis}, {Megevand}, {Pasquini}, {Reyes},
  {Sivan}, {Sosnowska}, {Soto}, {Udry}, {van Kesteren}, {Weber}, \&
  {Weilenmann}}]{may03}
{Mayor}, M., {Pepe}, F., {Queloz}, D., {et~al.} 2003, The Messenger, 114, 20

\bibitem[{{Murphy} {et~al.}(2022){Murphy}, {Molaro}, {Leite}, {Cupani},
  {Cristiani}, {D'Odorico}, {G{\'e}nova Santos}, {Martins}, {Milakovi{\'c}},
  {Nunes}, {Schmidt}, {Pepe}, {Rebolo}, {Santos}, {Sousa}, {Zapatero Osorio},
  {Amate}, {Adibekyan}, {Alibert}, {Allende Prieto}, {Baldini}, {Benz},
  {Bouchy}, {Cabral}, {Dekker}, {Di Marcantonio}, {Ehrenreich}, {Figueira},
  {Gonz{\'a}lez Hern{\'a}ndez}, {Landoni}, {Lovis}, {Lo Curto}, {Manescau},
  {M{\'e}gevand}, {Mehner}, {Micela}, {Pasquini}, {Poretti}, {Riva},
  {Sozzetti}, {Mascare{\~n}o}, {Udry}, \& {Zerbi}}]{mur22}
{Murphy}, M.~T., {Molaro}, P., {Leite}, A. C.~O., {et~al.} 2022, \aap, 658,
  A123

\bibitem[{{Palle} {et~al.}(2023){Palle}, {Biazzo}, {Bolmont}, {Molliere},
  {Poppenhaeger}, {Birkby}, {Brogi}, {Chauvin}, {Chiavassa}, {Hoeijmakers},
  {Lellouch}, {Lovis}, {Maiolino}, {Nortmann}, {Parviainen}, {Pino}, {Turbet},
  {Wender}, {Albrecht}, {Antoniucci}, {Barros}, {Beaudoin}, {Benneke},
  {Boisse}, {Bonomo}, {Borsa}, {Brandeker}, {Brandner}, {Buchhave}, {Cheffot},
  {Deborde}, {Debras}, {Doyon}, {Di Marcantonio}, {Giacobbe}, {Gonzalez
  Hernandez}, {Helled}, {Kreidberg}, {Machado}, {Maldonado}, {Marconi}, {Canto
  Martins}, {Miceli}, {Mordasini}, {N'Diaye}, {Niedzielski}, {Nisini},
  {Origlia}, {Peroux}, {Pietrow}, {Pinna}, {Rauscher}, {Reffert}, {Rousselot},
  {Sanna}, {Simonnin}, {Suarez Mascareno}, {Zanutta}, \& {Zechmeister}}]{pal23}
{Palle}, E., {Biazzo}, K., {Bolmont}, E., {et~al.} 2023, arXiv e-prints,
  arXiv:2311.17075

\bibitem[{{Passegger} {et~al.}(2018){Passegger}, {Reiners}, {Jeffers},
  {Wende-von Berg}, {Sch{\"o}fer}, {Caballero}, {Schweitzer}, {Amado},
  {B{\'e}jar}, {Cort{\'e}s-Contreras}, {Hatzes}, {K{\"u}rster}, {Montes},
  {Pedraz}, {Quirrenbach}, {Ribas}, \& {Seifert}}]{pas18}
{Passegger}, V.~M., {Reiners}, A., {Jeffers}, S.~V., {et~al.} 2018, \aap, 615,
  A6

\bibitem[{{Pepe} {et~al.}(2021){Pepe}, {Cristiani}, {Rebolo}, {Santos},
  {Dekker}, {Cabral}, {Di Marcantonio}, {Figueira}, {Lo Curto}, {Lovis},
  {Mayor}, {M{\'e}gevand}, {Molaro}, {Riva}, {Zapatero Osorio}, {Amate},
  {Manescau}, {Pasquini}, {Zerbi}, {Adibekyan}, {Abreu}, {Affolter}, {Alibert},
  {Aliverti}, {Allart}, {Allende Prieto}, {{\'A}lvarez}, {Alves}, {Avila},
  {Baldini}, {Bandy}, {Barros}, {Benz}, {Bianco}, {Borsa}, {Bourrier},
  {Bouchy}, {Broeg}, {Calderone}, {Cirami}, {Coelho}, {Conconi}, {Coretti},
  {Cumani}, {Cupani}, {D'Odorico}, {Damasso}, {Deiries}, {Delabre},
  {Demangeon}, {Dumusque}, {Ehrenreich}, {Faria}, {Fragoso}, {Genolet},
  {Genoni}, {G{\'e}nova Santos}, {Gonz{\'a}lez Hern{\'a}ndez}, {Hughes},
  {Iwert}, {Kerber}, {Knudstrup}, {Landoni}, {Lavie}, {Lillo-Box}, {Lizon},
  {Maire}, {Martins}, {Mehner}, {Micela}, {Modigliani}, {Monteiro}, {Monteiro},
  {Moschetti}, {Murphy}, {Nunes}, {Oggioni}, {Oliveira}, {Oshagh}, {Pall{\'e}},
  {Pariani}, {Poretti}, {Rasilla}, {Rebord{\~a}o}, {Redaelli}, {Santana
  Tschudi}, {Santin}, {Santos}, {S{\'e}gransan}, {Schmidt}, {Segovia},
  {Sosnowska}, {Sozzetti}, {Sousa}, {Span{\`o}}, {Su{\'a}rez Mascare{\~n}o},
  {Tabernero}, {Tenegi}, {Udry}, \& {Zanutta}}]{pep21}
{Pepe}, F., {Cristiani}, S., {Rebolo}, R., {et~al.} 2021, \aap, 645, A96

\bibitem[{{Pepe} {et~al.}(2014){Pepe}, {Ehrenreich}, \& {Meyer}}]{pep14Natur}
{Pepe}, F., {Ehrenreich}, D., \& {Meyer}, M.~R. 2014, \nat, 513, 358

\bibitem[{{Perger} {et~al.}(2021){Perger}, {Anglada-Escud{\'e}}, {Ribas},
  {Rosich}, {Herrero}, \& {Morales}}]{per21}
{Perger}, M., {Anglada-Escud{\'e}}, G., {Ribas}, I., {et~al.} 2021, \aap, 645,
  A58

\bibitem[{{Quanz} {et~al.}(2022){Quanz}, {Ottiger}, {Fontanet}, {Kammerer},
  {Menti}, {Dannert}, {Gheorghe}, {Absil}, {Airapetian}, {Alei}, {Allart},
  {Angerhausen}, {Blumenthal}, {Buchhave}, {Cabrera},
  {Carri{\'o}n-Gonz{\'a}lez}, {Chauvin}, {Danchi}, {Dandumont}, {Defr{\'e}re},
  {Dorn}, {Ehrenreich}, {Ertel}, {Fridlund}, {Garc{\'\i}a Mu{\~n}oz},
  {Gasc{\'o}n}, {Girard}, {Glauser}, {Grenfell}, {Guidi}, {Hagelberg},
  {Helled}, {Ireland}, {Janson}, {Kopparapu}, {Korth}, {Kozakis}, {Kraus},
  {L{\'e}ger}, {Leedj{\"a}rv}, {Lichtenberg}, {Lillo-Box}, {Linz}, {Liseau},
  {Loicq}, {Mahendra}, {Malbet}, {Mathew}, {Mennesson}, {Meyer}, {Mishra},
  {Molaverdikhani}, {Noack}, {Oza}, {Pall{\'e}}, {Parviainen}, {Quirrenbach},
  {Rauer}, {Ribas}, {Rice}, {Romagnolo}, {Rugheimer}, {Schwieterman},
  {Serabyn}, {Sharma}, {Stassun}, {Szul{\'a}gyi}, {Wang}, {Wunderlich},
  {Wyatt}, \& {LIFE Collaboration}}]{qua22}
{Quanz}, S.~P., {Ottiger}, M., {Fontanet}, E., {et~al.} 2022, \aap, 664, A21

\bibitem[{{Queloz} {et~al.}(2001){Queloz}, {Henry}, {Sivan}, {Baliunas},
  {Beuzit}, {Donahue}, {Mayor}, {Naef}, {Perrier}, \& {Udry}}]{que01}
{Queloz}, D., {Henry}, G.~W., {Sivan}, J.~P., {et~al.} 2001, \aap, 379, 279

\bibitem[{{Quirrenbach} {et~al.}(2016){Quirrenbach}, {Amado}, {Caballero},
  {Mundt}, {Reiners}, {Ribas}, {Seifert}, {Abril}, {Aceituno},
  {Alonso-Floriano}, {Anwand-Heerwart}, {Azzaro}, {Bauer}, {Barrado},
  {Becerril}, {Bejar}, {Benitez}, {Berdinas}, {Brinkm{\"o}ller}, {Cardenas},
  {Casal}, {Claret}, {Colom{\'e}}, {Cortes-Contreras}, {Czesla}, {Doellinger},
  {Dreizler}, {Feiz}, {Fernandez}, {Ferro}, {Fuhrmeister}, {Galadi},
  {Gallardo}, {G{\'a}lvez-Ortiz}, {Garcia-Piquer}, {Garrido}, {Gesa},
  {G{\'o}mez Galera}, {Gonz{\'a}lez Hern{\'a}ndez}, {Gonzalez Peinado},
  {Gr{\"o}zinger}, {Gu{\`a}rdia}, {Guenther}, {de Guindos}, {Hagen}, {Hatzes},
  {Hauschildt}, {Helmling}, {Henning}, {Hermann}, {Hern{\'a}ndez Arabi},
  {Hern{\'a}ndez Casta{\~n}o}, {Hern{\'a}ndez Hernando}, {Herrero}, {Huber},
  {Huber}, {Huke}, {Jeffers}, {de Juan}, {Kaminski}, {Kehr}, {Kim}, {Klein},
  {Kl{\"u}ter}, {K{\"u}rster}, {Lafarga}, {Lara}, {Lamert}, {Laun},
  {Launhardt}, {Lemke}, {Lenzen}, {Llamas}, {Lopez del Fresno},
  {L{\'o}pez-Puertas}, {L{\'o}pez-Santiago}, {Lopez Salas}, {Magan
  Madinabeitia}, {Mall}, {Mandel}, {Mancini}, {Marin Molina}, {Maroto
  Fern{\'a}ndez}, {Mart{\'{\i}}n}, {Mart{\'{\i}}n-Ruiz}, {Marvin}, {Mathar},
  {Mirabet}, {Montes}, {Morales}, {Morales Mu{\~n}oz}, {Nagel}, {Naranjo},
  {Nowak}, {Palle}, {Panduro}, {Passegger}, {Pavlov}, {Pedraz}, {Perez},
  {P{\'e}rez-Medialdea}, {Perger}, {Pluto}, {Ram{\'o}n}, {Rebolo}, {Redondo},
  {Reffert}, {Reinhart}, {Rhode}, {Rix}, {Rodler}, {Rodr{\'{\i}}guez},
  {Rodr{\'{\i}}guez L{\'o}pez}, {Rohloff}, {Rosich}, {Sanchez Carrasco},
  {Sanz-Forcada}, {Sarkis}, {Sarmiento}, {Sch{\"a}fer}, {Schiller}, {Schmidt},
  {Schmitt}, {Sch{\"o}fer}, {Schweitzer}, {Shulyak}, {Solano}, {Stahl},
  {Storz}, {Tabernero}, {Tala}, {Tal-Or}, {Ulbrich}, {Veredas}, {Vico Linares},
  {Vilardell}, {Wagner}, {Winkler}, {Zapatero Osorio}, {Zechmeister},
  {Ammler-von Eiff}, {Anglada-Escud{\'e}}, {del Burgo}, {Garcia-Vargas},
  {Klutsch}, {Lizon}, {Lopez-Morales}, {Ofir}, {P{\'e}rez-Calpena}, {Perryman},
  {S{\'a}nchez-Blanco}, {Strachan}, {St{\"u}rmer}, {Su{\'a}rez}, {Trifonov},
  {Tulloch}, \& {Xu}}]{qui16}
{Quirrenbach}, A., {Amado}, P.~J., {Caballero}, J.~A., {et~al.} 2016, in
  \procspie, Vol. 9908, Ground-based and Airborne Instrumentation for Astronomy
  VI, 990812

\bibitem[{{Rajpaul} {et~al.}(2015){Rajpaul}, {Aigrain}, {Osborne}, {Reece}, \&
  {Roberts}}]{raj15}
{Rajpaul}, V., {Aigrain}, S., {Osborne}, M.~A., {Reece}, S., \& {Roberts}, S.
  2015, \mnras, 452, 2269

\bibitem[{{Rajpaul} {et~al.}(2016){Rajpaul}, {Aigrain}, \& {Roberts}}]{raj16}
{Rajpaul}, V., {Aigrain}, S., \& {Roberts}, S. 2016, \mnras, 456, L6

\bibitem[{{Rasmussen} \& {Williams}(2006)}]{ram06}
{Rasmussen}, C.~E. \& {Williams}, C. K.~I. 2006, {Gaussian Processes for
  Machine Learning}

\bibitem[{{Rein} \& {Tamayo}(2017)}]{rei17}
{Rein}, H. \& {Tamayo}, D. 2017, \mnras, 467, 2377

\bibitem[{{Reiners} {et~al.}(2022){Reiners}, {Shulyak}, {K{\"a}pyl{\"a}},
  {Ribas}, {Nagel}, {Zechmeister}, {Caballero}, {Shan}, {Fuhrmeister},
  {Quirrenbach}, {Amado}, {Montes}, {Jeffers}, {Azzaro}, {B{\'e}jar},
  {Chaturvedi}, {Henning}, {K{\"u}rster}, \& {Pall{\'e}}}]{rei22}
{Reiners}, A., {Shulyak}, D., {K{\"a}pyl{\"a}}, P.~J., {et~al.} 2022, \aap,
  662, A41

\bibitem[{{Reyl{\'e}} {et~al.}(2021){Reyl{\'e}}, {Jardine}, {Fouqu{\'e}},
  {Caballero}, {Smart}, \& {Sozzetti}}]{rey21}
{Reyl{\'e}}, C., {Jardine}, K., {Fouqu{\'e}}, P., {et~al.} 2021, \aap, 650,
  A201

\bibitem[{{Ribas} {et~al.}(2023){Ribas}, {Reiners}, {Zechmeister}, {Caballero},
  {Morales}, {Sabotta}, {Baroch}, {Amado}, {Quirrenbach}, {Abril}, {Aceituno},
  {Anglada-Escud{\'e}}, {Azzaro}, {Barrado}, {B{\'e}jar}, {Ben{\'\i}tez de
  Haro}, {Bergond}, {Bluhm}, {Calvo Ortega}, {Cardona Guill{\'e}n},
  {Chaturvedi}, {Cifuentes}, {Colom{\'e}}, {Cont}, {Cort{\'e}s-Contreras},
  {Czesla}, {D{\'\i}ez-Alonso}, {Dreizler}, {Duque-Arribas}, {Espinoza},
  {Fern{\'a}ndez}, {Fuhrmeister}, {Galad{\'\i}-Enr{\'\i}quez},
  {Garc{\'\i}a-L{\'o}pez}, {Gonz{\'a}lez-{\'A}lvarez}, {Gonz{\'a}lez
  Hern{\'a}ndez}, {Guenther}, {de Guindos}, {Hatzes}, {Henning}, {Herrero},
  {Hintz}, {Huelmo}, {Jeffers}, {Johnson}, {de Juan}, {Kaminski}, {Kemmer},
  {Khaimova}, {Khalafinejad}, {Kossakowski}, {K{\"u}rster}, {Labarga},
  {Lafarga}, {Lalitha}, {Lamp{\'o}n}, {Lillo-Box}, {Lodieu}, {L{\'o}pez
  Gonz{\'a}lez}, {L{\'o}pez-Puertas}, {Luque}, {Mag{\'a}n}, {Mancini},
  {Marfil}, {Mart{\'\i}n}, {Mart{\'\i}n-Ruiz}, {Molaverdikhani}, {Montes},
  {Nagel}, {Nortmann}, {Nowak}, {Pall{\'e}}, {Passegger}, {Pavlov}, {Pedraz},
  {Perdelwitz}, {Perger}, {Ram{\'o}n-Ballesta}, {Reffert}, {Revilla},
  {Rodr{\'\i}guez}, {Rodr{\'\i}guez-L{\'o}pez}, {Sadegi}, {S{\'a}nchez
  Carrasco}, {S{\'a}nchez-L{\'o}pez}, {Sanz-Forcada}, {Sch{\"a}fer},
  {Schlecker}, {Schmitt}, {Sch{\"o}fer}, {Schweitzer}, {Seifert}, {Shan},
  {Skrzypinski}, {Solano}, {Stahl}, {Stangret}, {Stock}, {St{\"u}rmer},
  {Tabernero}, {Tal-Or}, {Trifonov}, {Vanaverbeke}, {Yan}, \& {Zapatero
  Osorio}}]{rib23}
{Ribas}, I., {Reiners}, A., {Zechmeister}, M., {et~al.} 2023, \aap, 670, A139

\bibitem[{{Ribas} {et~al.}(2018){Ribas}, {Tuomi}, {Reiners}, {Butler},
  {Morales}, {Perger}, {Dreizler}, {Rodr{\'\i}guez-L{\'o}pez}, {Gonz{\'a}lez
  Hern{\'a}ndez}, {Rosich}, {Feng}, {Trifonov}, {Vogt}, {Caballero}, {Hatzes},
  {Herrero}, {Jeffers}, {Lafarga}, {Murgas}, {Nelson}, {Rodr{\'\i}guez},
  {Strachan}, {Tal-Or}, {Teske}, {Toledo-Padr{\'o}n}, {Zechmeister},
  {Quirrenbach}, {Amado}, {Azzaro}, {B{\'e}jar}, {Barnes}, {Berdi{\~n}as},
  {Burt}, {Coleman}, {Cort{\'e}s-Contreras}, {Crane}, {Engle}, {Guinan},
  {Haswell}, {Henning}, {Holden}, {Jenkins}, {Jones}, {Kaminski}, {Kiraga},
  {K{\"u}rster}, {Lee}, {L{\'o}pez-Gonz{\'a}lez}, {Montes}, {Morin}, {Ofir},
  {Pall{\'e}}, {Rebolo}, {Reffert}, {Schweitzer}, {Seifert}, {Shectman},
  {Staab}, {Street}, {Su{\'a}rez Mascare{\~n}o}, {Tsapras}, {Wang}, \&
  {Anglada-Escud{\'e}}}]{rib18}
{Ribas}, I., {Tuomi}, M., {Reiners}, A., {et~al.} 2018, \nat, 563, 365

\bibitem[{{Rice} \& {Steffen}(2023)}]{ric23}
{Rice}, D.~R. \& {Steffen}, J.~H. 2023, \mnras, 520, 4057

\bibitem[{{Ricker} {et~al.}(2015){Ricker}, {Winn}, {Vanderspek}, {Latham},
  {Bakos}, {Bean}, {Berta-Thompson}, {Brown}, {Buchhave}, {Butler}, {Butler},
  {Chaplin}, {Charbonneau}, {Christensen-Dalsgaard}, {Clampin}, {Deming},
  {Doty}, {De Lee}, {Dressing}, {Dunham}, {Endl}, {Fressin}, {Ge}, {Henning},
  {Holman}, {Howard}, {Ida}, {Jenkins}, {Jernigan}, {Johnson}, {Kaltenegger},
  {Kawai}, {Kjeldsen}, {Laughlin}, {Levine}, {Lin}, {Lissauer}, {MacQueen},
  {Marcy}, {McCullough}, {Morton}, {Narita}, {Paegert}, {Palle}, {Pepe},
  {Pepper}, {Quirrenbach}, {Rinehart}, {Sasselov}, {Sato}, {Seager},
  {Sozzetti}, {Stassun}, {Sullivan}, {Szentgyorgyi}, {Torres}, {Udry}, \&
  {Villasenor}}]{ric15}
{Ricker}, G.~R., {Winn}, J.~N., {Vanderspek}, R., {et~al.} 2015, Journal of
  Astronomical Telescopes, Instruments, and Systems, 1, 014003

\bibitem[{{Robertson} {et~al.}(2014){Robertson}, {Mahadevan}, {Endl}, \&
  {Roy}}]{rob14}
{Robertson}, P., {Mahadevan}, S., {Endl}, M., \& {Roy}, A. 2014, Science, 345,
  440

\bibitem[{{Schweitzer} {et~al.}(2019){Schweitzer}, {Passegger}, {Cifuentes},
  {B{\'e}jar}, {Cort{\'e}s-Contreras}, {Caballero}, {del Burgo}, {Czesla},
  {K{\"u}rster}, {Montes}, {Zapatero Osorio}, {Ribas}, {Reiners},
  {Quirrenbach}, {Amado}, {Aceituno}, {Anglada-Escud{\'e}}, {Bauer},
  {Dreizler}, {Jeffers}, {Guenther}, {Henning}, {Kaminski}, {Lafarga},
  {Marfil}, {Morales}, {Schmitt}, {Seifert}, {Solano}, {Tabernero}, \&
  {Zechmeister}}]{sch19}
{Schweitzer}, A., {Passegger}, V.~M., {Cifuentes}, C., {et~al.} 2019, \aap,
  625, A68

\bibitem[{{Silva} {et~al.}(2022){Silva}, {Faria}, {Santos}, {Sousa}, {Viana},
  {Martins}, {Figueira}, {Lovis}, {Pepe}, {Cristiani}, {Rebolo}, {Allart},
  {Cabral}, {Mehner}, {Sozzetti}, {Su{\'a}rez Mascare{\~n}o}, {Martins},
  {Ehrenreich}, {M{\'e}gevand}, {Palle}, {Lo Curto}, {Tabernero}, {Lillo-Box},
  {Gonz{\'a}lez Hern{\'a}ndez}, {Zapatero Osorio}, {Hara}, {Nunes}, {Di
  Marcantonio}, {Udry}, {Adibekyan}, \& {Dumusque}}]{sil22sbart}
{Silva}, A.~M., {Faria}, J.~P., {Santos}, N.~C., {et~al.} 2022, \aap, 663, A143

\bibitem[{{Skilling}(2004)}]{ski04}
{Skilling}, J. 2004, in American Institute of Physics Conference Series, Vol.
  735, Bayesian Inference and Maximum Entropy Methods in Science and
  Engineering: 24th International Workshop on Bayesian Inference and Maximum
  Entropy Methods in Science and Engineering, ed. R.~{Fischer}, R.~{Preuss}, \&
  U.~V. {Toussaint} (AIP), 395--405

\bibitem[{Sobell(2015)}]{sob15ubuntu}
Sobell, M.~G. 2015, A practical guide to Ubuntu Linux (Pearson Education)

\bibitem[{{Speagle}(2020)}]{spe20dynesty}
{Speagle}, J.~S. 2020, \mnras, 493, 3132

\bibitem[{{Stelzer} {et~al.}(2013){Stelzer}, {Marino}, {Micela},
  {L{\'o}pez-Santiago}, \& {Liefke}}]{ste13mnras}
{Stelzer}, B., {Marino}, A., {Micela}, G., {L{\'o}pez-Santiago}, J., \&
  {Liefke}, C. 2013, \mnras, 431, 2063

\bibitem[{{Stock} {et~al.}(2020){Stock}, {Kemmer}, {Reffert}, {Trifonov},
  {Kaminski}, {Dreizler}, {Quirrenbach}, {Caballero}, {Reiners}, {Jeffers},
  {Anglada-Escud{\'e}}, {Ribas}, {Amado}, {Barrado}, {Barnes}, {Bauer},
  {Berdi{\~n}as}, {B{\'e}jar}, {Coleman}, {Cort{\'e}s-Contreras},
  {D{\'\i}ez-Alonso}, {Dom{\'\i}nguez-Fern{\'a}ndez}, {Espinoza}, {Haswell},
  {Hatzes}, {Henning}, {Jenkins}, {Jones}, {Kossakowski}, {K{\"u}rster},
  {Lafarga}, {Lee}, {L{\'o}pez Gonz{\'a}lez}, {Montes}, {Morales}, {Morales},
  {Pall{\'e}}, {Pedraz}, {Rodr{\'\i}guez}, {Rodr{\'\i}guez-L{\'o}pez}, \&
  {Zechmeister}}]{sto20}
{Stock}, S., {Kemmer}, J., {Reffert}, S., {et~al.} 2020, \aap, 636, A119

\bibitem[{{Su{\'a}rez Mascare{\~n}o} {et~al.}(2020){Su{\'a}rez Mascare{\~n}o},
  {Faria}, {Figueira}, {Lovis}, {Damasso}, {Gonz{\'a}lez Hern{\'a}ndez},
  {Rebolo}, {Cristiani}, {Pepe}, {Santos}, {Zapatero Osorio}, {Adibekyan},
  {Hojjatpanah}, {Sozzetti}, {Murgas}, {Abreu}, {Affolter}, {Alibert},
  {Aliverti}, {Allart}, {Allende Prieto}, {Alves}, {Amate}, {Avila}, {Baldini},
  {Bandi}, {Barros}, {Bianco}, {Benz}, {Bouchy}, {Broeng}, {Cabral},
  {Calderone}, {Cirami}, {Coelho}, {Conconi}, {Coretti}, {Cumani}, {Cupani},
  {D'Odorico}, {Deiries}, {Delabre}, {Di Marcantonio}, {Dumusque},
  {Ehrenreich}, {Fragoso}, {Genolet}, {Genoni}, {G{\'e}nova Santos}, {Hughes},
  {Iwert}, {Kerber}, {Knusdstrup}, {Landoni}, {Lavie}, {Lillo-Box}, {Lizon},
  {Lo Curto}, {Maire}, {Manescau}, {Martins}, {M{\'e}gevand}, {Mehner},
  {Micela}, {Modigliani}, {Molaro}, {Monteiro}, {Monteiro}, {Moschetti},
  {Mueller}, {Nunes}, {Oggioni}, {Oliveira}, {Pall{\'e}}, {Pariani},
  {Pasquini}, {Poretti}, {Rasilla}, {Redaelli}, {Riva}, {Santana Tschudi},
  {Santin}, {Santos}, {Segovia}, {Sosnowska}, {Sousa}, {Span{\`o}}, {Tenegi},
  {Udry}, {Zanutta}, \& {Zerbi}}]{sua20}
{Su{\'a}rez Mascare{\~n}o}, A., {Faria}, J.~P., {Figueira}, P., {et~al.} 2020,
  \aap, 639, A77

\bibitem[{{Su{\'a}rez Mascare{\~n}o} {et~al.}(2023){Su{\'a}rez Mascare{\~n}o},
  {Gonz{\'a}lez-{\'A}lvarez}, {Zapatero Osorio}, {Lillo-Box}, {Faria},
  {Passegger}, {Gonz{\'a}lez Hern{\'a}ndez}, {Figueira}, {Sozzetti}, {Rebolo},
  {Pepe}, {Santos}, {Cristiani}, {Lovis}, {Silva}, {Ribas}, {Amado},
  {Caballero}, {Quirrenbach}, {Reiners}, {Zechmeister}, {Adibekyan}, {Alibert},
  {B{\'e}jar}, {Benatti}, {D'Odorico}, {Damasso}, {Delisle}, {Di Marcantonio},
  {Dreizler}, {Ehrenreich}, {Hatzes}, {Hara}, {Henning}, {Kaminski},
  {L{\'o}pez-Gonz{\'a}lez}, {Martins}, {Micela}, {Montes}, {Pall{\'e}},
  {Pedraz}, {Rodr{\'\i}guez}, {Rodr{\'\i}guez-L{\'o}pez}, {Tal-Or}, {Sousa}, \&
  {Udry}}]{sua23}
{Su{\'a}rez Mascare{\~n}o}, A., {Gonz{\'a}lez-{\'A}lvarez}, E., {Zapatero
  Osorio}, M.~R., {et~al.} 2023, \aap, 670, A5

\bibitem[{{Su{\'a}rez Mascare{\~n}o} {et~al.}(2024){Su{\'a}rez Mascare{\~n}o},
  {Passegger}, {Gonz{\'a}lez Hern{\'a}ndez}, {Armstrong}, {Nielsen}, {Lovis},
  {Lavie}, {Sousa}, {Silva}, {Allart}, {Rebolo}, {Pepe}, {Santos}, {Cristiani},
  {Sozzetti}, {Zapatero Osorio}, {Tabernero}, {Dumusque}, {Udry}, {Adibekyan},
  {Allende Prieto}, {Alibert}, {Barros}, {Bouchy}, {Castro-Gonz{\'a}lez},
  {Collins}, {Damasso}, {D'Odorico}, {Demangeon}, {Di Marcantonio},
  {Ehrenreich}, {Hadjigeorghiou}, {Hara}, {Hawthorn}, {Jenkins}, {Lillo-Box},
  {Lo Curto}, {Martins}, {Mehner}, {Micela}, {Molaro}, {Nunes}, {Nari},
  {Osborn}, {Pall{\'e}}, {Ricker}, {Rodrigues}, {Rowden}, {Seager}, {Stefanov},
  {Str{\o}m}, {Villase{\~n}or}, {Watkins}, {Winn}, {Wohler}, \&
  {Zambelli}}]{sua24}
{Su{\'a}rez Mascare{\~n}o}, A., {Passegger}, V.~M., {Gonz{\'a}lez
  Hern{\'a}ndez}, J.~I., {et~al.} 2024, \aap, 685, A56

\bibitem[{{Su{\'a}rez Mascare{\~n}o} {et~al.}(2015){Su{\'a}rez Mascare{\~n}o},
  {Rebolo}, {Gonz{\'a}lez Hern{\'a}ndez}, \& {Esposito}}]{sua15}
{Su{\'a}rez Mascare{\~n}o}, A., {Rebolo}, R., {Gonz{\'a}lez Hern{\'a}ndez},
  J.~I., \& {Esposito}, M. 2015, \mnras, 452, 2745

\bibitem[{{Su{\'a}rez Mascare{\~n}o} {et~al.}(2017){Su{\'a}rez Mascare{\~n}o},
  {Rebolo}, {Gonz{\'a}lez Hern{\'a}ndez}, \& {Esposito}}]{sua17rv}
{Su{\'a}rez Mascare{\~n}o}, A., {Rebolo}, R., {Gonz{\'a}lez Hern{\'a}ndez},
  J.~I., \& {Esposito}, M. 2017, \mnras, 468, 4772

\bibitem[{{Su{\'a}rez Mascare{\~n}o} {et~al.}(2018){Su{\'a}rez Mascare{\~n}o},
  {Rebolo}, {Gonz{\'a}lez Hern{\'a}ndez}, {Toledo-Padr{\'o}n}, {Perger},
  {Ribas}, {Affer}, {Micela}, {Damasso}, {Maldonado}, {Gonz{\'a}lez-Alvarez},
  {Leto}, {Pagano}, {Scandariato}, {Sozzetti}, {Lanza}, {Malavolta}, {Claudi},
  {Cosentino}, {Desidera}, {Giacobbe}, {Maggio}, {Rainer}, {Esposito},
  {Benatti}, {Pedani}, {Morales}, {Herrero}, {Lafarga}, {Rosich}, \&
  {Pinamonti}}]{sua18rot}
{Su{\'a}rez Mascare{\~n}o}, A., {Rebolo}, R., {Gonz{\'a}lez Hern{\'a}ndez},
  J.~I., {et~al.} 2018, \aap, 612, A89

\bibitem[{{Tabernero} {et~al.}(2022){Tabernero}, {Marfil}, {Montes}, \&
  {Gonz{\'a}lez Hern{\'a}ndez}}]{tab22}
{Tabernero}, H.~M., {Marfil}, E., {Montes}, D., \& {Gonz{\'a}lez
  Hern{\'a}ndez}, J.~I. 2022, \aap, 657, A66

\bibitem[{{Tamayo} {et~al.}(2020){Tamayo}, {Cranmer}, {Hadden}, {Rein},
  {Battaglia}, {Obertas}, {Armitage}, {Ho}, {Spergel}, {Gilbertson}, {Hussain},
  {Silburt}, {Jontof-Hutter}, \& {Menou}}]{tam20}
{Tamayo}, D., {Cranmer}, M., {Hadden}, S., {et~al.} 2020, Proceedings of the
  National Academy of Science, 117, 18194

\bibitem[{{Thompson} {et~al.}(2016){Thompson}, {Queloz}, {Baraffe}, {Brake},
  {Dolgopolov}, {Fisher}, {Fleury}, {Geelhoed}, {Hall}, {Gonz{\'a}lez
  Hern{\'a}ndez}, {ter Horst}, {Kragt}, {Navarro}, {Naylor}, {Pepe},
  {Piskunov}, {Rebolo}, {Sander}, {S{\'e}gransan}, {Seneta}, {Sing}, {Snellen},
  {Snik}, {Spronck}, {Stempels}, {Sun}, {Santana Tschudi}, \& {Young}}]{tho16}
{Thompson}, S.~J., {Queloz}, D., {Baraffe}, I., {et~al.} 2016, in Society of
  Photo-Optical Instrumentation Engineers (SPIE) Conference Series, Vol. 9908,
  Ground-based and Airborne Instrumentation for Astronomy VI, ed. C.~J.
  {Evans}, L.~{Simard}, \& H.~{Takami}, 99086F

\bibitem[{{Toledo-Padr{\'o}n} {et~al.}(2019){Toledo-Padr{\'o}n}, {Gonz{\'a}lez
  Hern{\'a}ndez}, {Rodr{\'\i}guez-L{\'o}pez}, {Su{\'a}rez Mascare{\~n}o},
  {Rebolo}, {Butler}, {Ribas}, {Anglada-Escud{\'e}}, {Johnson}, {Reiners},
  {Caballero}, {Quirrenbach}, {Amado}, {B{\'e}jar}, {Morales}, {Perger},
  {Jeffers}, {Vogt}, {Teske}, {Shectman}, {Crane}, {D{\'\i}az}, {Arriagada},
  {Holden}, {Burt}, {Rodr{\'\i}guez}, {Herrero}, {Murgas}, {Pall{\'e}},
  {Morales}, {L{\'o}pez-Gonz{\'a}lez}, {D{\'\i}ez Alonso}, {Tuomi}, {Kiraga},
  {Engle}, {Guinan}, {Strachan}, {Aceituno}, {Aceituno}, {Casanova},
  {Mart{\'\i}n-Ruiz}, {Montes}, {Ortiz}, {Sota}, {Briol}, {Barbieri},
  {Cervini}, {Deldem}, {Dubois}, {Hambsch}, {Harris}, {Kotnik}, {Logie},
  {Lopez}, {McNeely}, {Ogmen}, {P{\'e}rez}, {Rau}, {Rodr{\'\i}guez}, {Urquijo},
  \& {Vanaverbeke}}]{tol19}
{Toledo-Padr{\'o}n}, B., {Gonz{\'a}lez Hern{\'a}ndez}, J.~I.,
  {Rodr{\'\i}guez-L{\'o}pez}, C., {et~al.} 2019, \mnras, 488, 5145

\bibitem[{Van~Rossum \& Drake(2009)}]{vanros09python3}
Van~Rossum, G. \& Drake, F.~L. 2009, Python 3 Reference Manual (Scotts Valley,
  CA: CreateSpace)

\bibitem[{Virtanen {et~al.}(2020)Virtanen, Gommers, Oliphant, Haberland, Reddy,
  Cournapeau, Burovski, Peterson, Weckesser, Bright, {van der Walt}, Brett,
  Wilson, Millman, Mayorov, Nelson, Jones, Kern, Larson, Carey, Polat, Feng,
  Moore, {VanderPlas}, Laxalde, Perktold, Cimrman, Henriksen, Quintero, Harris,
  Archibald, Ribeiro, Pedregosa, {van Mulbregt}, \& {SciPy 1.0
  Contributors}}]{vir20scipy}
Virtanen, P., Gommers, R., Oliphant, T.~E., {et~al.} 2020, Nature Methods, 17,
  261

\bibitem[{{Weiss} {et~al.}(2018){Weiss}, {Marcy}, {Petigura}, {Fulton},
  {Howard}, {Winn}, {Isaacson}, {Morton}, {Hirsch}, {Sinukoff}, {Cumming},
  {Hebb}, \& {Cargile}}]{wei18}
{Weiss}, L.~M., {Marcy}, G.~W., {Petigura}, E.~A., {et~al.} 2018, \aj, 155, 48

\bibitem[{{Wildi} {et~al.}(2010){Wildi}, {Pepe}, {Chazelas}, {Lo Curto}, \&
  {Lovis}}]{wil10}
{Wildi}, F., {Pepe}, F., {Chazelas}, B., {Lo Curto}, G., \& {Lovis}, C. 2010,
  in Society of Photo-Optical Instrumentation Engineers (SPIE) Conference
  Series, Vol. 7735, Ground-based and Airborne Instrumentation for Astronomy
  III, ed. I.~S. {McLean}, S.~K. {Ramsay}, \& H.~{Takami}, 77354X

\bibitem[{{Zechmeister} {et~al.}(2019){Zechmeister}, {Dreizler}, {Ribas},
  {Reiners}, {Caballero}, {Bauer}, {B{\'e}jar}, {Gonz{\'a}lez-Cuesta},
  {Herrero}, {Lalitha}, {L{\'o}pez-Gonz{\'a}lez}, {Luque}, {Morales},
  {Pall{\'e}}, {Rodr{\'\i}guez}, {Rodr{\'\i}guez L{\'o}pez}, {Tal-Or},
  {Anglada-Escud{\'e}}, {Quirrenbach}, {Amado}, {Abril}, {Aceituno},
  {Aceituno}, {Alonso-Floriano}, {Ammler-von Eiff}, {Antona Jim{\'e}nez},
  {Anwand-Heerwart}, {Arroyo-Torres}, {Azzaro}, {Baroch}, {Barrado},
  {Becerril}, {Ben{\'\i}tez}, {Berdi{\~n}as}, {Bergond}, {Bluhm},
  {Brinkm{\"o}ller}, {del Burgo}, {Calvo Ortega}, {Cano}, {Cardona
  Guill{\'e}n}, {Carro}, {C{\'a}rdenas V{\'a}zquez}, {Casal},
  {Casasayas-Barris}, {Casanova}, {Chaturvedi}, {Cifuentes}, {Claret},
  {Colom{\'e}}, {Cort{\'e}s-Contreras}, {Czesla}, {D{\'\i}ez-Alonso}, {Dorda},
  {Fern{\'a}ndez}, {Fern{\'a}ndez-Mart{\'\i}n}, {Fuhrmeister}, {Fukui},
  {Galad{\'\i}-Enr{\'\i}quez}, {Gallardo Cava}, {Garcia de la Fuente},
  {Garcia-Piquer}, {Garc{\'\i}a Vargas}, {Gesa}, {G{\'o}ngora Rueda},
  {Gonz{\'a}lez-{\'A}lvarez}, {Gonz{\'a}lez Hern{\'a}ndez},
  {Gonz{\'a}lez-Peinado}, {Gr{\"o}zinger}, {Gu{\`a}rdia}, {Guijarro}, {de
  Guindos}, {Hatzes}, {Hauschildt}, {Hedrosa}, {Helmling}, {Henning},
  {Hermelo}, {Hern{\'a}ndez Arabi}, {Hern{\'a}ndez Casta{\~n}o}, {Hern{\'a}ndez
  Otero}, {Hintz}, {Huke}, {Huber}, {Jeffers}, {Johnson}, {de Juan},
  {Kaminski}, {Kemmer}, {Kim}, {Klahr}, {Klein}, {Kl{\"u}ter}, {Klutsch},
  {Kossakowski}, {K{\"u}rster}, {Labarga}, {Lafarga}, {Llamas}, {Lamp{\'o}n},
  {Lara}, {Launhardt}, {L{\'a}zaro}, {Lodieu}, {L{\'o}pez del Fresno},
  {L{\'o}pez-Puertas}, {L{\'o}pez Salas}, {L{\'o}pez-Santiago}, {Mag{\'a}n
  Madinabeitia}, {Mall}, {Mancini}, {Mandel}, {Marfil}, {Mar{\'\i}n Molina},
  {Maroto Fern{\'a}ndez}, {Mart{\'\i}n}, {Mart{\'\i}n-Fern{\'a}ndez},
  {Mart{\'\i}n-Ruiz}, {Marvin}, {Mirabet}, {Monta{\~n}{\'e}s-Rodr{\'\i}guez},
  {Montes}, {Moreno-Raya}, {Nagel}, {Naranjo}, {Narita}, {Nortmann}, {Nowak},
  {Ofir}, {Oshagh}, {Panduro}, {Parviainen}, {Pascual}, {Passegger}, {Pavlov},
  {Pedraz}, {P{\'e}rez-Calpena}, {P{\'e}rez Medialdea}, {Perger}, {Perryman},
  {Rabaza}, {Ram{\'o}n Ballesta}, {Rebolo}, {Redondo}, {Reffert}, {Reinhardt},
  {Rhode}, {Rix}, {Rodler}, {Rodr{\'\i}guez Trinidad}, {Rosich}, {Sadegi},
  {S{\'a}nchez-Blanco}, {S{\'a}nchez Carrasco}, {S{\'a}nchez-L{\'o}pez},
  {Sanz-Forcada}, {Sarkis}, {Sarmiento}, {Sch{\"a}fer}, {Schmitt},
  {Sch{\"o}fer}, {Schweitzer}, {Seifert}, {Shulyak}, {Solano}, {Sota}, {Stahl},
  {Stock}, {Strachan}, {Stuber}, {St{\"u}rmer}, {Su{\'a}rez}, {Tabernero},
  {Tala Pinto}, {Trifonov}, {Veredas}, {Vico Linares}, {Vilardell}, {Wagner},
  {Wolthoff}, {Xu}, {Yan}, \& {Zapatero Osorio}}]{zec19}
{Zechmeister}, M., {Dreizler}, S., {Ribas}, I., {et~al.} 2019, \aap, 627, A49

\bibitem[{{Zechmeister} \& {K{\"u}rster}(2009)}]{zec09}
{Zechmeister}, M. \& {K{\"u}rster}, M. 2009, \aap, 496, 577

\bibitem[{{Zechmeister} {et~al.}(2018){Zechmeister}, {Reiners}, {Amado},
  {Azzaro}, {Bauer}, {B{\'e}jar}, {Caballero}, {Guenther}, {Hagen}, {Jeffers},
  {Kaminski}, {K{\"u}rster}, {Launhardt}, {Montes}, {Morales}, {Quirrenbach},
  {Reffert}, {Ribas}, {Seifert}, {Tal-Or}, \& {Wolthoff}}]{zec18}
{Zechmeister}, M., {Reiners}, A., {Amado}, P.~J., {et~al.} 2018, \aap, 609, A12

\bibitem[{{Zsom} {et~al.}(2013){Zsom}, {Seager}, {de Wit}, \&
  {Stamenkovi{\'c}}}]{zso13}
{Zsom}, A., {Seager}, S., {de Wit}, J., \& {Stamenkovi{\'c}}, V. 2013, \apj,
  778, 109

\end{thebibliography}

\begin{appendix}

\section{ESPRESSO spectrum of Barnard's star}

\begin{figure}[h!]
\includegraphics[width=8.5cm]{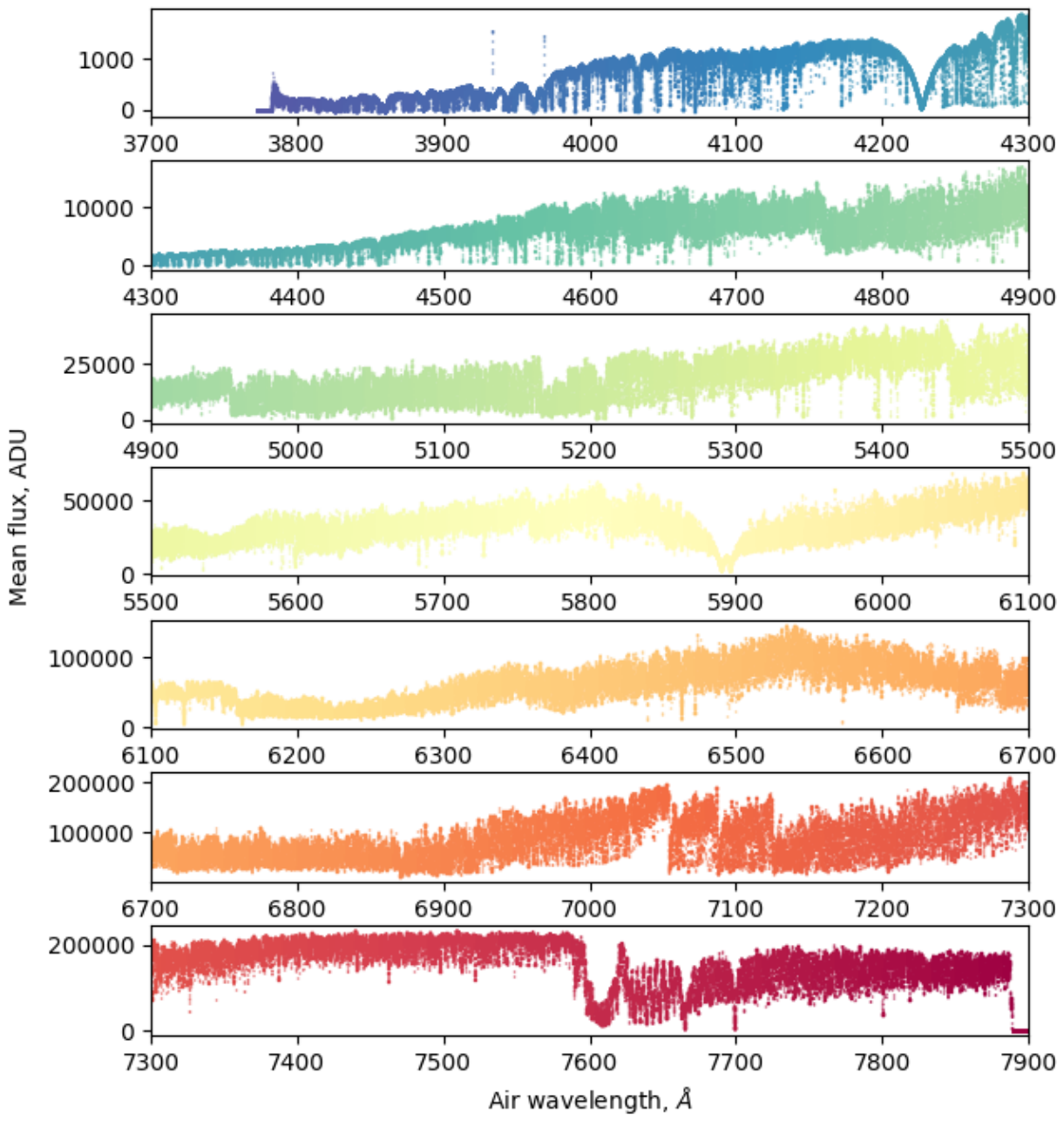}
\caption{
Mean ESPRESSO HR11 spectrum obtained from the 156 individual S1D$\_$A spectra reduced using the DRS pipeline v3.0.0. 
}
\label{gj699_spec}
\end{figure}

The ESPRESSO spectra were reduced using the DRS pipeline v3.0.0. The 156 S1D spectra were combined using the mean values at each pixel after interpolating all spectra to a common wavelength array. The resulting mean spectrum is display in Fig.~\ref{gj699_spec}. The individual ESPRESSO spectra have a signal-to-noise ratio (S/N), from the flux over flux error pixel, of $\sim 10$, 100, 200, and more than 300 at 4000, 5000, 6000 and 7500~$\AA$, respectively. The mean spectrum has a S/N following approximately Poisson statistics, thus, about 12.5 times the S/N of individual spectra.

\FloatBarrier


\section{Evaluating 3.15~d signal\label{sec:ap:315}}

\begin{figure}[!h]
\includegraphics[width=9cm]{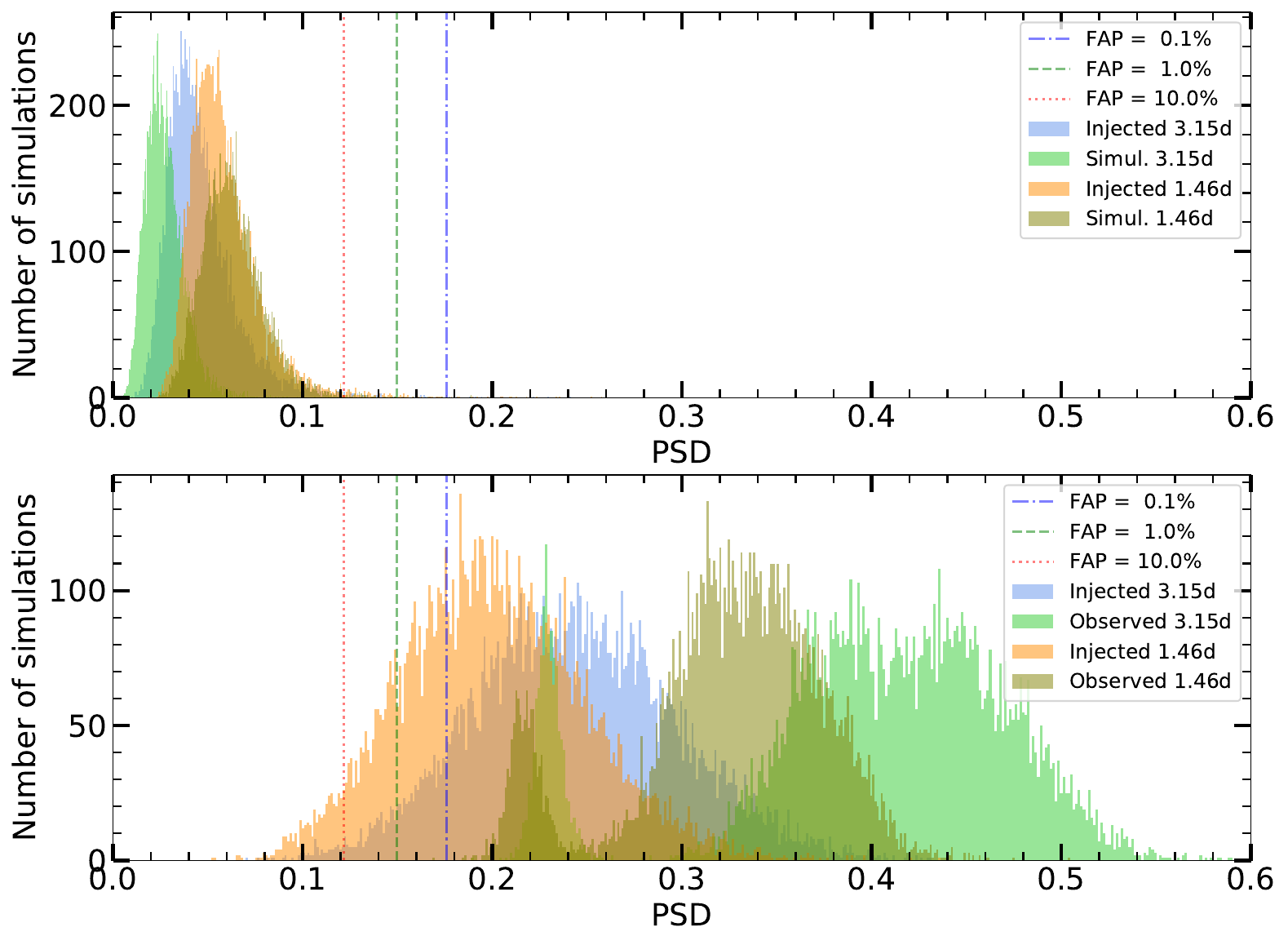}
\caption{
Power spectral density (PSD) distributions of GLS periodograms of 10,000 samples extracted from the run of ESPRESSO FWHM and RV measurements using a model with the GP and a Keplerian orbital model to search for the 3.15~d signal. The PSD distributions are extracted from the GLS periodograms at 3.15~d and the 1 d alias 1.46~d periods from simulated RV time series of the GP values plus white noise before ({\it upper panel}) and after ({\it lower panel}) subtracting the GP and injecting a planet signal in the simulated data. The observed PSD distributions ({\it lower panel}) are measured after subtracted the GP to the original RVs.
}
\label{gj699_s315}
\end{figure}

In order to further investigate the origin of the 3.15~d signal we perform 10,000 simulations from the 124254 samples of analysis of model including activity and planetary signal. We build a simulated RV time series by injecting white noise to the GP model using a random normal distribution with a $\sigma$ equal to the RV uncertainties including the jitter term (with a median value of 0.61~\ms). In top panel of Fig.~\ref{gj699_s315} we show the PSD distributions measured at the 3.15~d and 1.46~d periods of this simulated RV time series (labelled as "simul").  We finally add to this simulated RV time series the three Keplerian signals (with the parameters from model $F3$ in Table~\ref{tab:logzN}) associated with the three main signals (3.15, 4.12 and 2.34~d) we see above the 1\% FAP line in Fig.~\ref{gj699_gp}, measured the PSD distributions (labelled as "injected") at 3.15~d and 1.46~d, before and after subtracting the corresponding GP model of each sample (upper and lower panel in Fig.~\ref{gj699_s315}). Since most of the PSD of GP simulated time series is concentrated in strong activity signals at periods larger than 60~d,  the PSD at these short periods stays below the 10\% FAP line (upper panel). We note that slight PSD increase of the 3.15~d injected signal as compared with the simulated PSD values. We also note that the 1 d alias 1.46~d signal is stronger in the top panel possibly due to the sampling of the time series, as this signal is also stronger in the simulated RV times series that do not contain any planet model. On the other hand, it is also remarkable that the observed signals are stronger that those of the injected signals in the simulated time series, which may be related to maybe too conservative injected white noise (RV uncertainties plus jitter with a median value of 0.61~\ms), the lower semi-amplitude of the 3.15~d signal in the three planet model and the different GP model not adapted for this three planet model.

We provide in Table~\ref{tab:logzP} the blind searches of the 3.15~d planet signal using polynomial functions to model the long-term activity signal. We show in this table different runs computed for TM, DRS and LBL datasets.

\FloatBarrier

\begin{figure}[!h]
\includegraphics[width=9cm]{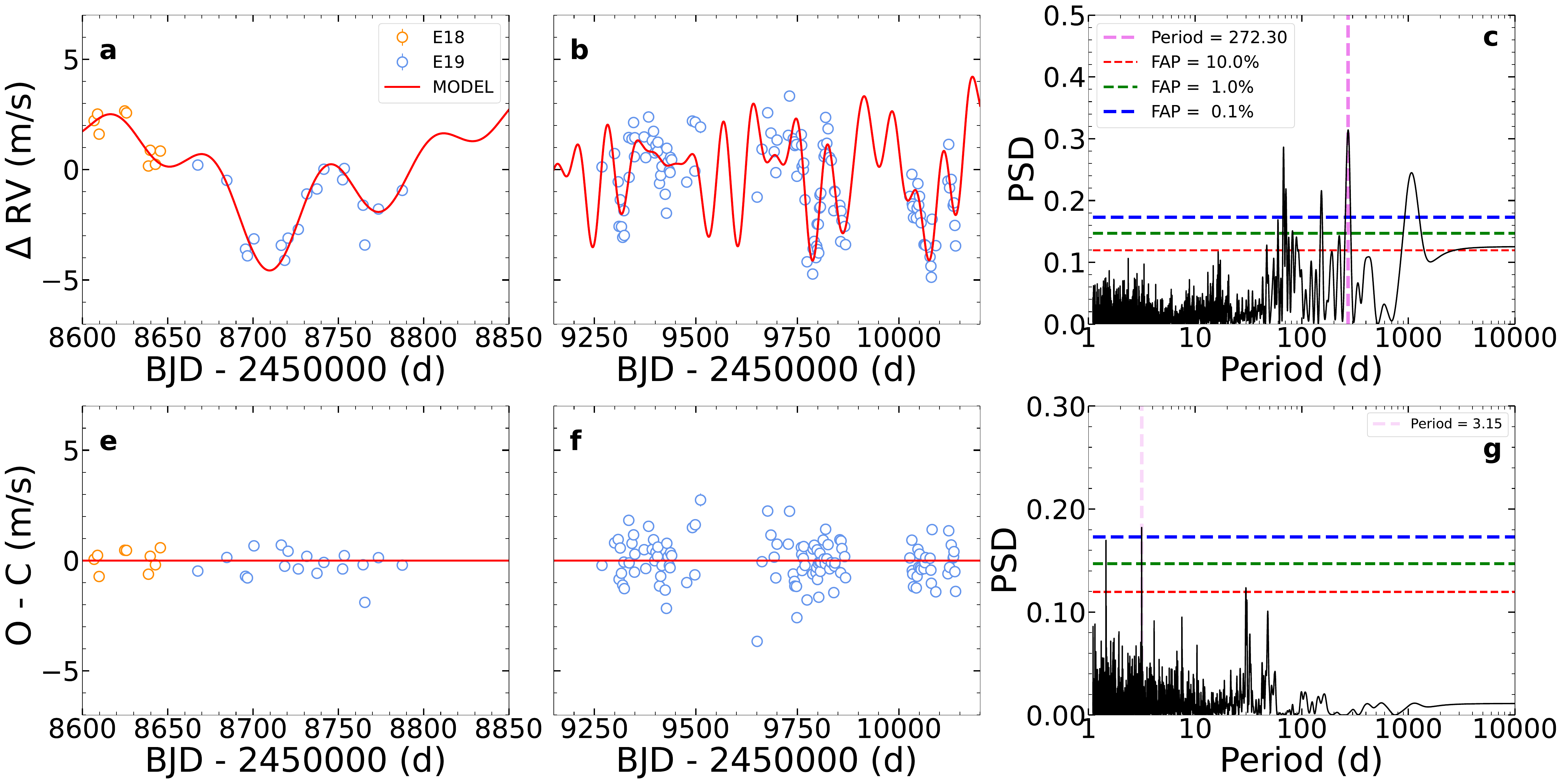}
\caption{
Activity model of the ESPRESSO RV measurements ({\it upper panels}) using a multi-sinusoidal function that uses up to six sinusoids with periods in the range [59,255]~d, and RV residuals after subtracting this activity-only model ({\it lower panels}). The periods fitted are 79.1, 254.9, 68.2, 89.8, 59.5, and 244.9~d. The GLS periodograms of both the RVs and RV residuals are also displayed ({\it left panels}), with the detection of the 3.15~d signal({\it lower right panel}).
}
\label{gj699_s315sin}
\end{figure}

\FloatBarrier

\begin{figure}[!h]
\includegraphics[width=9cm]{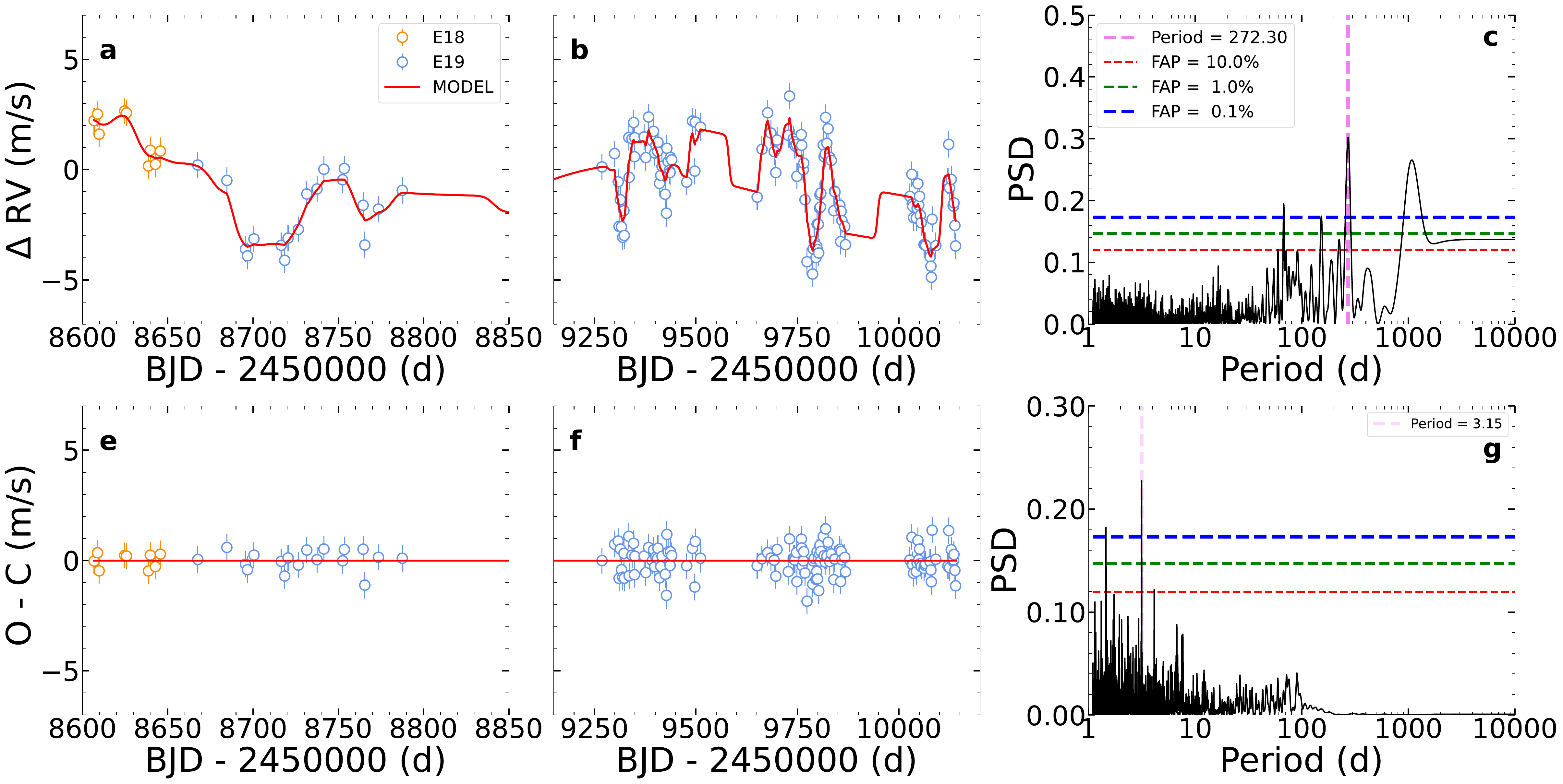}
\caption{
Activity model of the ESPRESSO RV measurements ({\it upper panels}) using moving average with an exponential decay, and RV residuals after subtracting this activity-only model ({\it lower panels}). The GLS periodograms of both the RVs and RV residuals are also displayed ({\it left panels}), with the detection of the 3.15~d signal({\it lower right panel}).
}
\label{gj699_s315ewma_real}
\end{figure}

\FloatBarrier

\begin{figure}[!h]
\includegraphics[width=9cm]{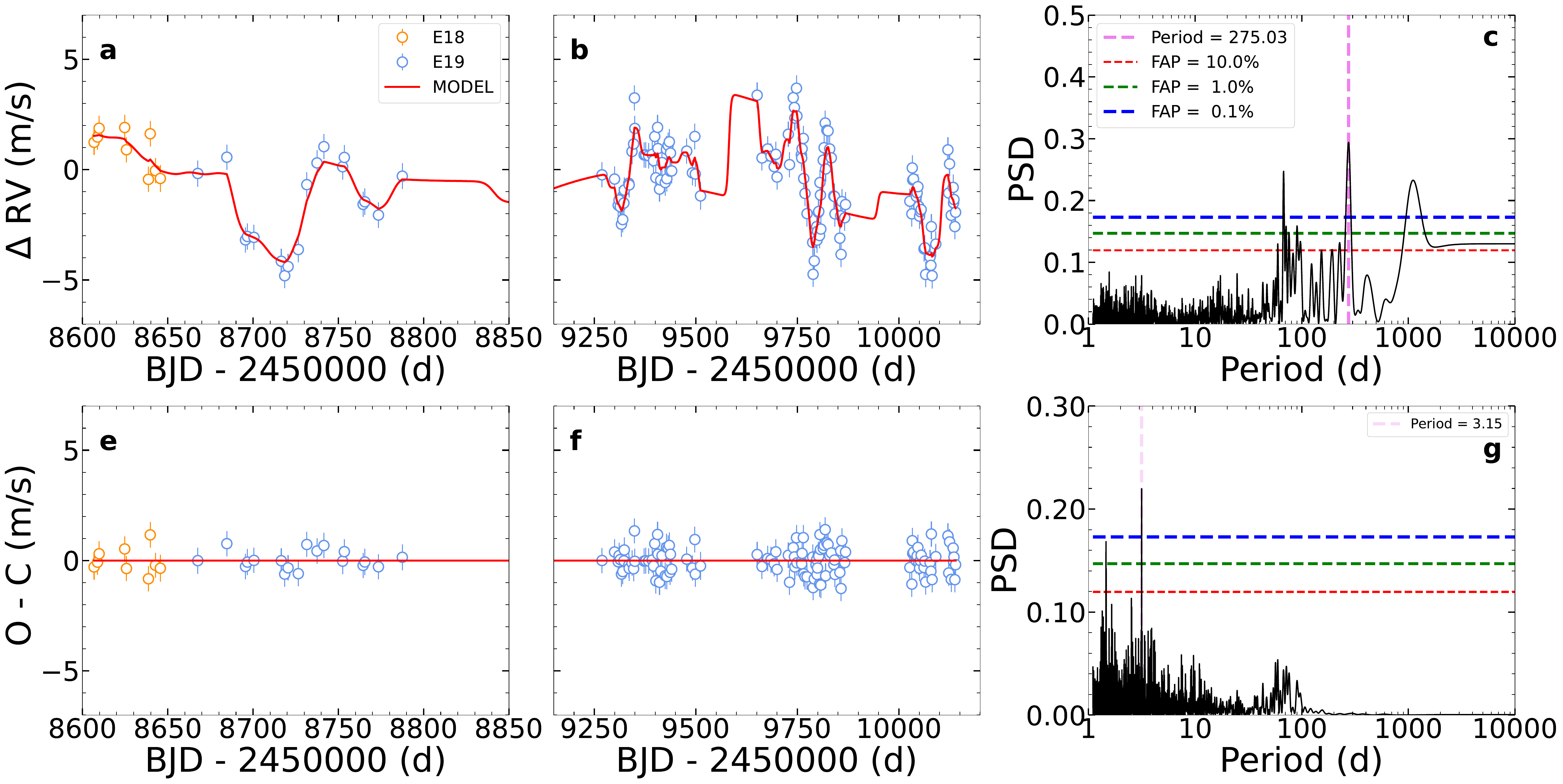}
\caption{
Same as Fig.~\ref{gj699_s315ewma_real}, but using simulated RV data ({\it upper panels}) from the activity-only model with injected white noise and injecting a planet signal at 3.15~d, and RV residuals after subtracting this activity-only model ({\it lower panels}). The GLS periodograms of both the RVs and RV residuals are also displayed ({\it left panels}), with the detection of the 3.15~d signal({\it lower right panel}). 
}
\label{gj699_s315ewma_sim}
\end{figure}

\FloatBarrier

\begin{table}[!h]
\begin{center}
\caption{Bayesian evidence of different models\label{tab:logzP}}
\begin{tabular}[centre]{llrrrr}
\hline \hline
Name & Model  & $N_{\rm pl}$ & $N_{\rm par}$ & $\ln \mathcal{Z}$ & $\Delta \ln \mathcal{Z}$ \\
\hline
\multicolumn{6}{c}{E18,E19 ($N_{\rm point}=149 \times 2$)} \\
\hline
A            & GP          & 0 & 18 &  $-449.3$ & $-2.1$  \\
AD           & GP          & 0 & 18 &  $-466.5$ & $-19.3$ \\
AL           & GP          & 0 & 18 &  $-452.3$ & $-5.1$  \\
B1           & A+poly1     & 0 & 20 &  $-453.4$ & $-6.2$  \\
B2           & A+poly2     & 0 & 22 &  $-453.9$ & $-6.7$  \\
B3           & A+poly3     & 0 & 24 &  $-462.7$ & $-15.5$ \\
B3D          & AD+poly3    & 0 & 24 &  $-479.3$ & $-32.1$ \\
B3L          & AL+poly3    & 0 & 24 &  $-468.7$ & $-21.5$ \\
C2           & B2+1peLU50  & 1 & 29 &  $-450.3$ & $-3.1$  \\
C3c          & B3+1pcLU50  & 1 & 29 &  $-459.3$ & $-12.1$ \\
C3e          & B3+1peLU50  & 1 & 29 &  $-458.4$ & $-11.2$ \\
C3eD         & B3D+1peLU50 & 1 & 29 &  $-474.2$ & $-27.0$ \\
C3eL         & B3L+1peLU50 & 1 & 29 &  $-462.8$ & $-15.6$ \\
C32          & B3+2peLU50  & 2 & 34 &  $-458.5$ & $-11.3$ \\
{\bf D}      & {\bf A+cycN} & {\bf 0} & {\bf 26} & {\bf --447.2} & {\bf 0.0} \\
\hline
\multicolumn{6}{c}{E18, E19, H15, HAN ($N_{\rm point}=298 \times 2$)} \\
\hline
G3          & GP+poly3    & 0 & 35 & $-1096.4$ & $-22.3$ \\
H3          & G3+1peLU50  & 0 & 35 & $-1084.1$ & $-10.0$ \\
{\bf I1}    & {\bf GP+cycU} & {\bf 0} & {\bf 35} & {\bf --1074.1} & {\bf 0.0}  \\
\hline
\end{tabular}
\end{center}
\textbf{Notes:} Model selection based on Bayesian evidence of the analysis of CCF FWHM and TM RV measurements. $A$, $AD$ and $AL$ are models for ESPRESSO TM, DRS and LBL RVs, respectively. The same as in Table~\ref{tab:logz} but for models including polynomial functions. Different models: {\it polyn} indicates $n$th-order polynomials and $N$, $U$ and $LU$ indicate normal, uniform and log-uniform priors. $LU50$ indicate priors $\mathcal{LU}(0.5,50)$~d. $npe$ and $npc$ indicate $n$ Keplerian and circular orbits. We highlight in bold fonts the reference activity-only model in each group of datasets, the same reference models as in Table~\ref{tab:logz}.
\end{table}

\FloatBarrier

\clearpage

\section{Activity indicators\label{sec:ap:act}}

\begin{figure*}
\includegraphics[width=18cm]{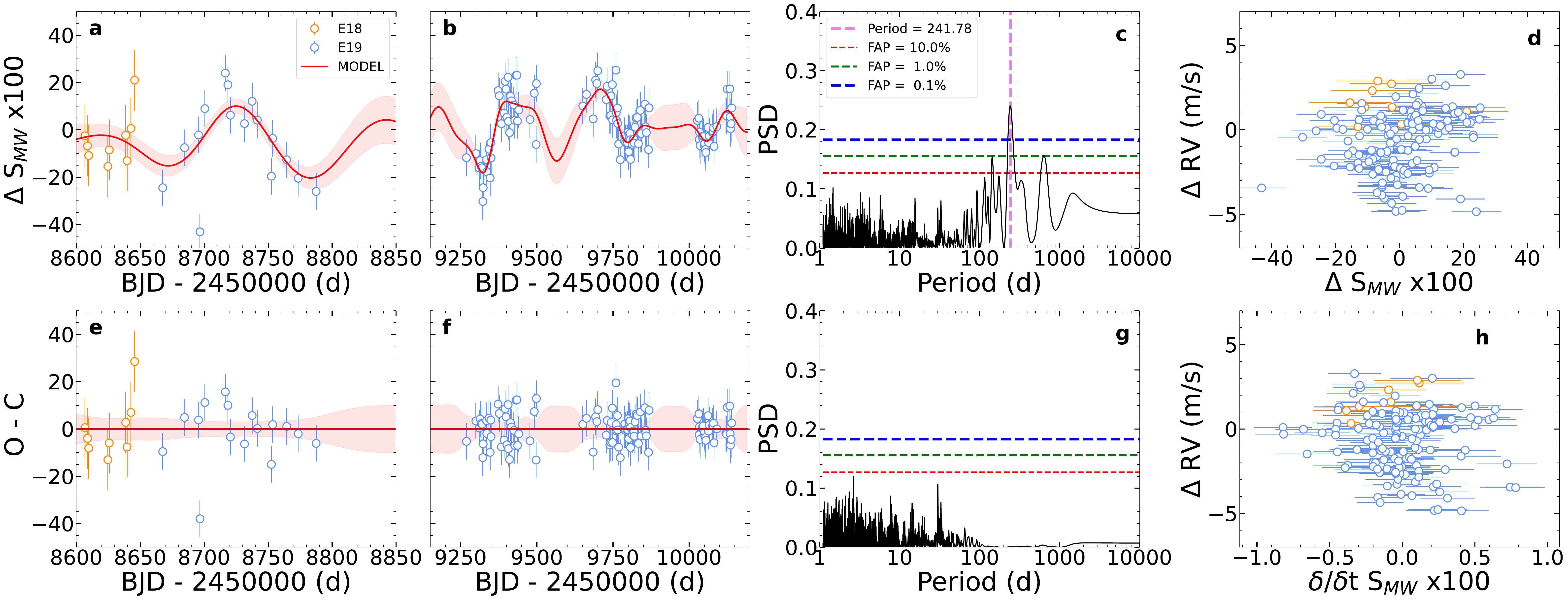}
\includegraphics[width=18cm]{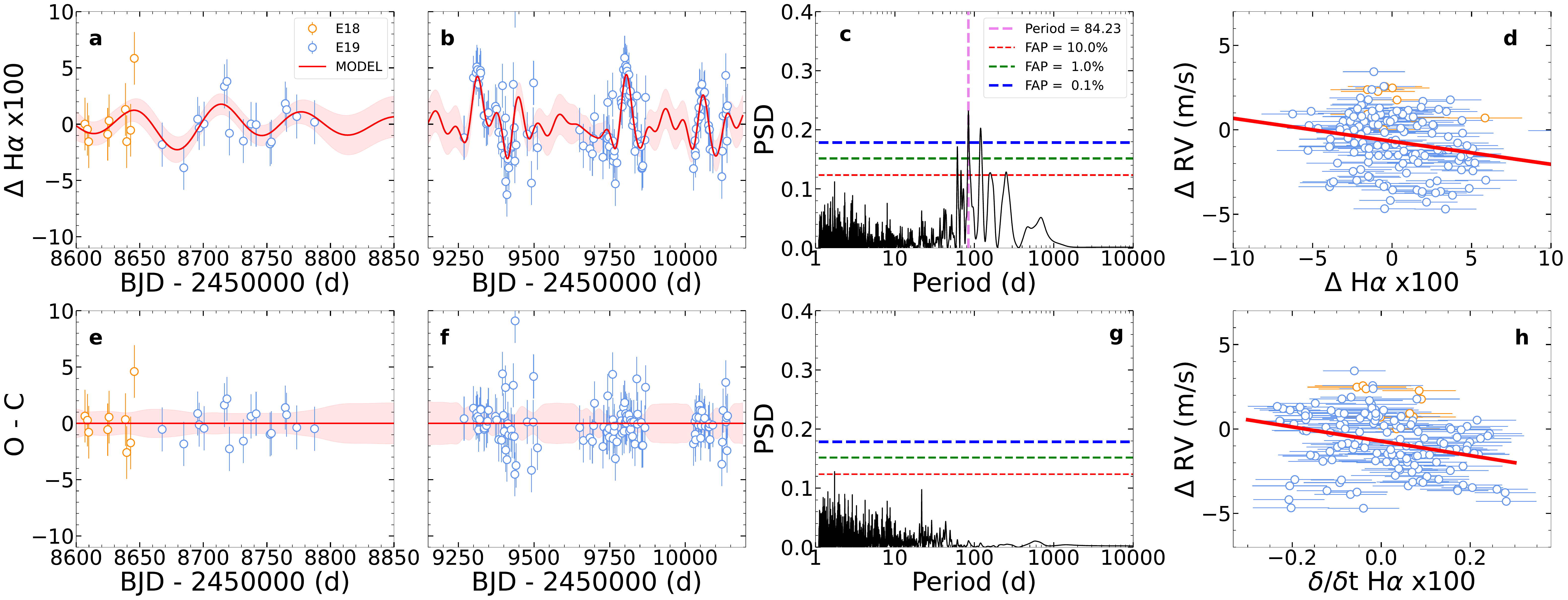}
\includegraphics[width=18cm]{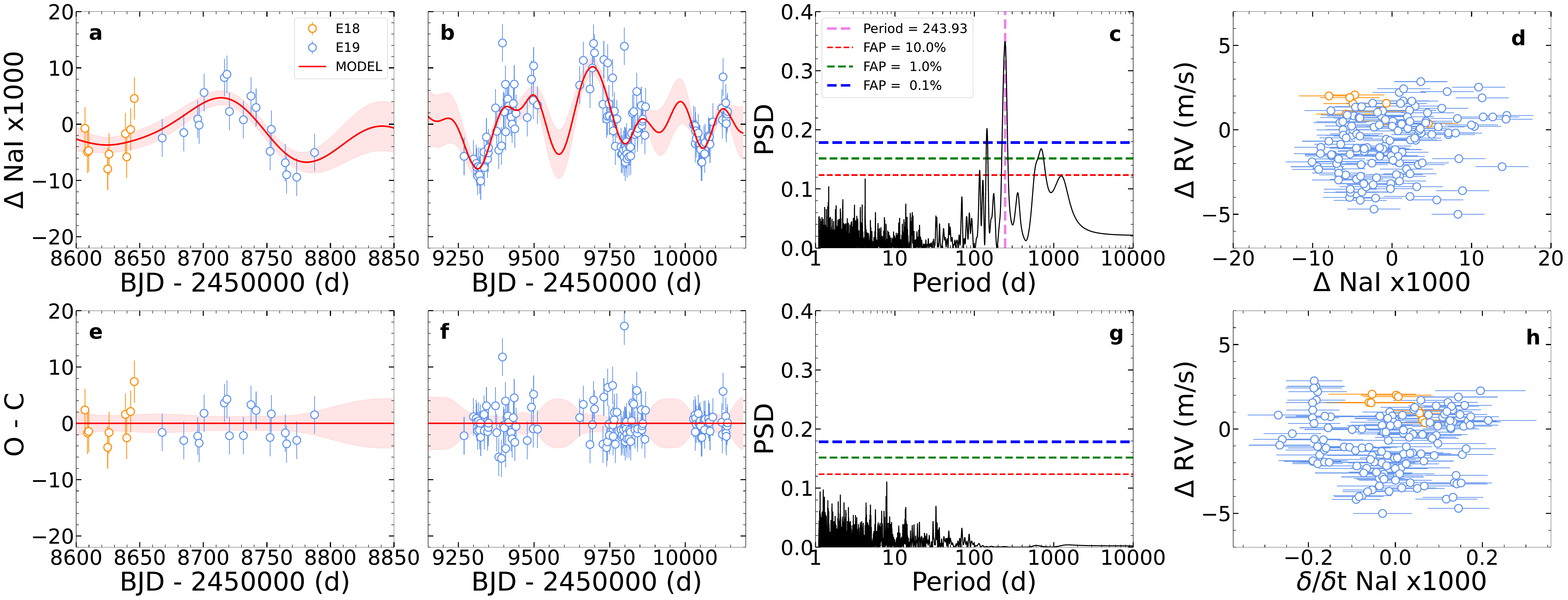}
\caption{{\bf Analysis of the ESPRESSO $S_{\rm MW}$, H$\alpha$ and Na~I spectroscopic indexes.}
\textbf{a, b}:  $S_{\rm MW}$, H$\alpha$ and Na~I-index time-series with the best-model fit. The data is split into two panels because of a large period with no observations between the two campaigns. The shaded area shows the variance of the GP model. {\bf c:} GLS periodogram of the $S_{\rm MW}$, H$\alpha$ and Na~I-index data. The red vertical dashed line shows the most significant period. \textbf{d:} Relationship between the $S_{\rm MW}$, H$\alpha$ and Na~I-index data. The best fit is shown when the slope is $\ge$3$\sigma$ different from zero. \textbf{e, f:} Residuals of the $S_{\rm MW}$, H$\alpha$ and Na~I-index after subtracting the best model fit. \textbf{g:} GLS periodogram of the residuals. \textbf{h}: Comparison of the CCF RV and gradient of the $S_{\rm MW}$, H$\alpha$ and Na~I-index model.
}
\label{gj699_act1}
\end{figure*}

\begin{figure*}
\includegraphics[width=18cm]{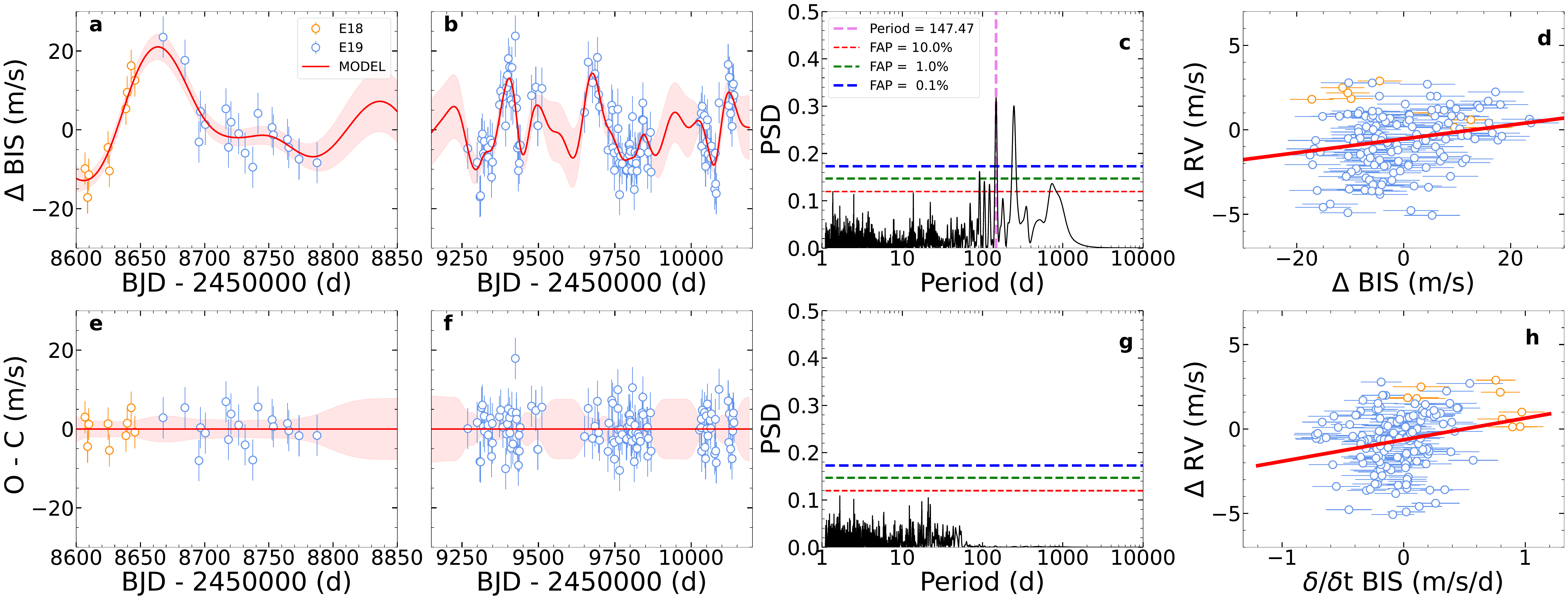}
\includegraphics[width=18cm]{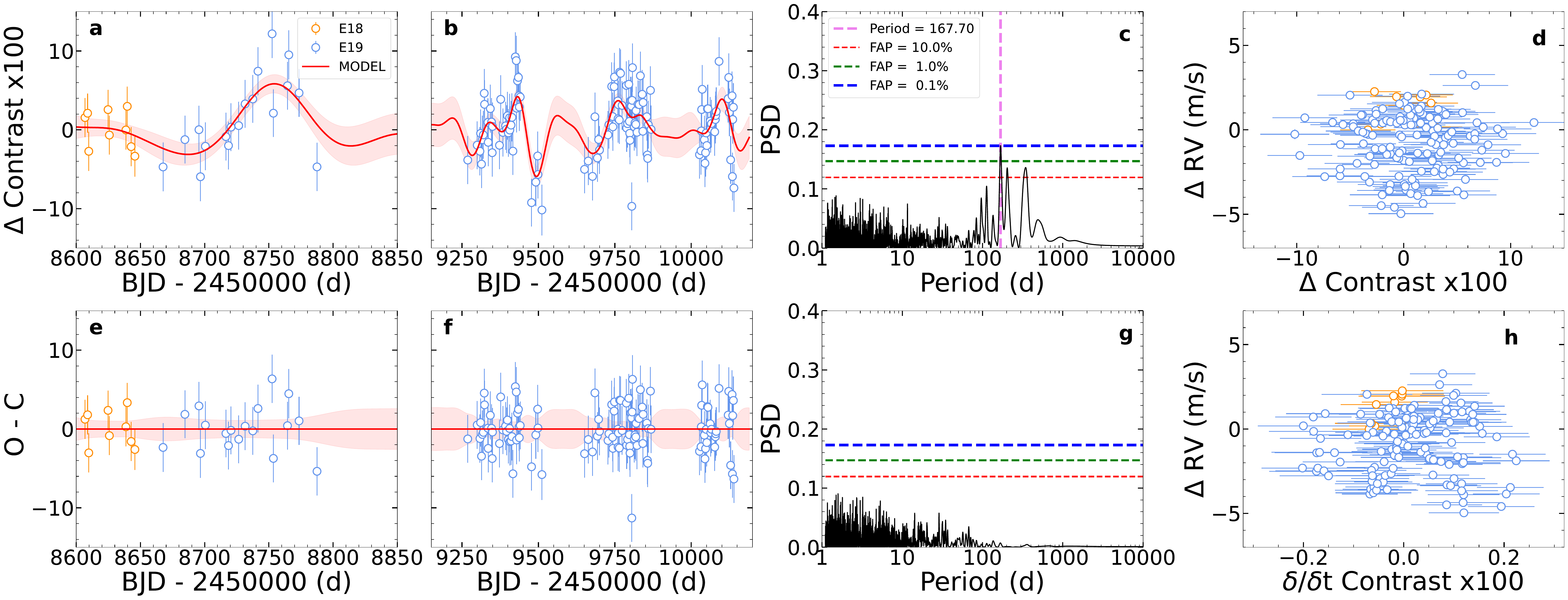}
\caption{
{\bf Analysis of the BIS and contrast of the ESPRESSO CCF.}
\textbf{a, b}: BIS and contrast time-series with the best-model fit. The data is split into two panels because of a large period with no observations between the two campaigns. The shaded area shows the variance of the GP model. {\bf c:} GLS periodogram of the CCF BIS and contrast data. The red vertical dashed line shows the most significant period. \textbf{d:} Relationship between the CCF RV and CCF BIS and contrast data. The best fit is shown when the slope is $\ge$3$\sigma$ different from zero. \textbf{e, f:} Residuals of the CCF BIS and contrast after subtracting the best model fit. \textbf{g:} GLS periodogram of the residuals. \textbf{h}: Comparison of the CCF RV and gradient of the CCF BIS and contrast model.}
\label{gj699_act2}
\end{figure*}

We measured the activity indexes $S_{\rm MW}$, Na~I doublet, and H$\alpha$ from the ESPRESSO spectra, and from the ESPRESSO DRS cross-correlation functions, we measured the full width at half maximum (FWHM), the contrast, and the bisector span (BIS). We modelled the time series of these activity indicators using the GP formalism described in Section~\ref{sec:gp} to evaluate their possible correlations with the time series of RV measurements.
Figs.~\ref{gj699_act1} and~\ref{gj699_act2} show the model of each index for which we display  the index (and its derivative) against the RV. The uncertainties include the jitter. The correlation trend in the index versus RV plots is only drawn in cases where the slope is significant at least at 3~$\sigma$ (slope / error~$ > 3$).

We summarise the analysis in Table~\ref{tab:rv_corr}, which includes Spearman's correlation index, p-value (the lower the value, the better the correlation), and the slope we obtain when doing an adjustment (least squares).


\begin{table}[!h]
\begin{center}
\caption{Spearman's correlation coefficient, and measured slopes,
    between activity indicators, their gradient, and the
    RV measurements.\label{tab:rv_corr}}
\begin{tabular}[center]{l c c c c}
\hline
& Correlation & p-value & Slope   \\ \hline
RV vs FWHM & 0.36 & \textless 0.01 & 16.2 $\pm$ 4.6\\
RV vs ${d}\over{dt}$ FWHM  & 0.15 & 0.07 & 190 $\pm$ 100 \\

RV vs BIS & 0.22 & \textless 0.01  & 3.4 $\pm$ 1.7\\
RV vs ${d}\over{dt}$BIS  & 0.21 & 0.01 & 149 $\pm$ 47 \\

RV vs Cont. $_{\times100}$ & -0.02 & 0.77  & 2.2 $\pm$ 4.4\\
RV vs ${d}\over{dt}$ Cont. $_{\times100}$  & 0.07 & 0.36 & -70 $\pm$ 160 \\

RV vs S-index $_{\times100}$ & 0.22 & 0.01  & 2.5 $\pm$ 1.4\\
RV vs ${d}\over{dt}$ S-index $_{\times100}$  & -0.06 & 0.46 & -57 $\pm$ 55 \\

RV vs H$\alpha$ $_{\times100}$ & -0.24 & 0.04  & -14.5 $\pm$ 5.2\\
RV vs ${d}\over{dt}$ H$\alpha$ $_{\times100}$  & -0.30 & \textless 0.01 & -430 $\pm$ 110 \\

RV vs Na I $_{\times1000}$ & 0.19 & 0.02  & 4.2 $\pm$ 2.8\\
RV vs ${d}\over{dt}$ Na I $_{\times1000}$  & 0.08 & 0.35 & 60 $\pm$ 140 \\
\hline
\end{tabular}
\end{center}
\end{table}

\FloatBarrier

\FloatBarrier

\FloatBarrier


\section{Analysis of all datasets}

\begin{figure}[!h]
\includegraphics[width=9cm]{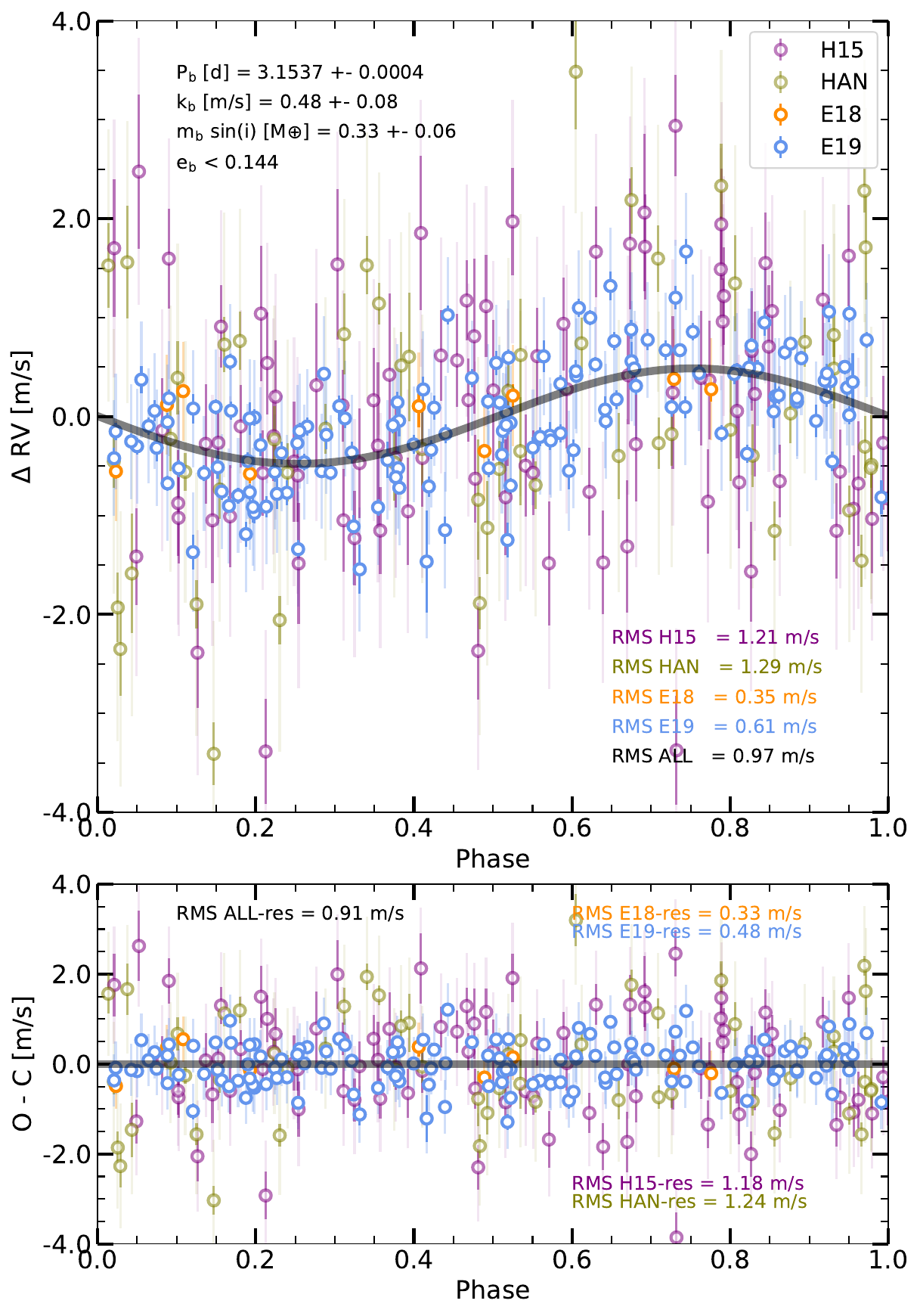}
\caption{
    RV curve of the sub-Earth-mass planet of GJ~699 with a 3.15~d orbital period together with ESPRESSO, HARPS and HARPS-N RVs with uncertainties with (light colour) and without including the jitter term (dark colour) coming from the global model.
}
\label{gj699_esp_har_1pe}
\end{figure}

\begin{figure}[!h]
\includegraphics[width=9cm]{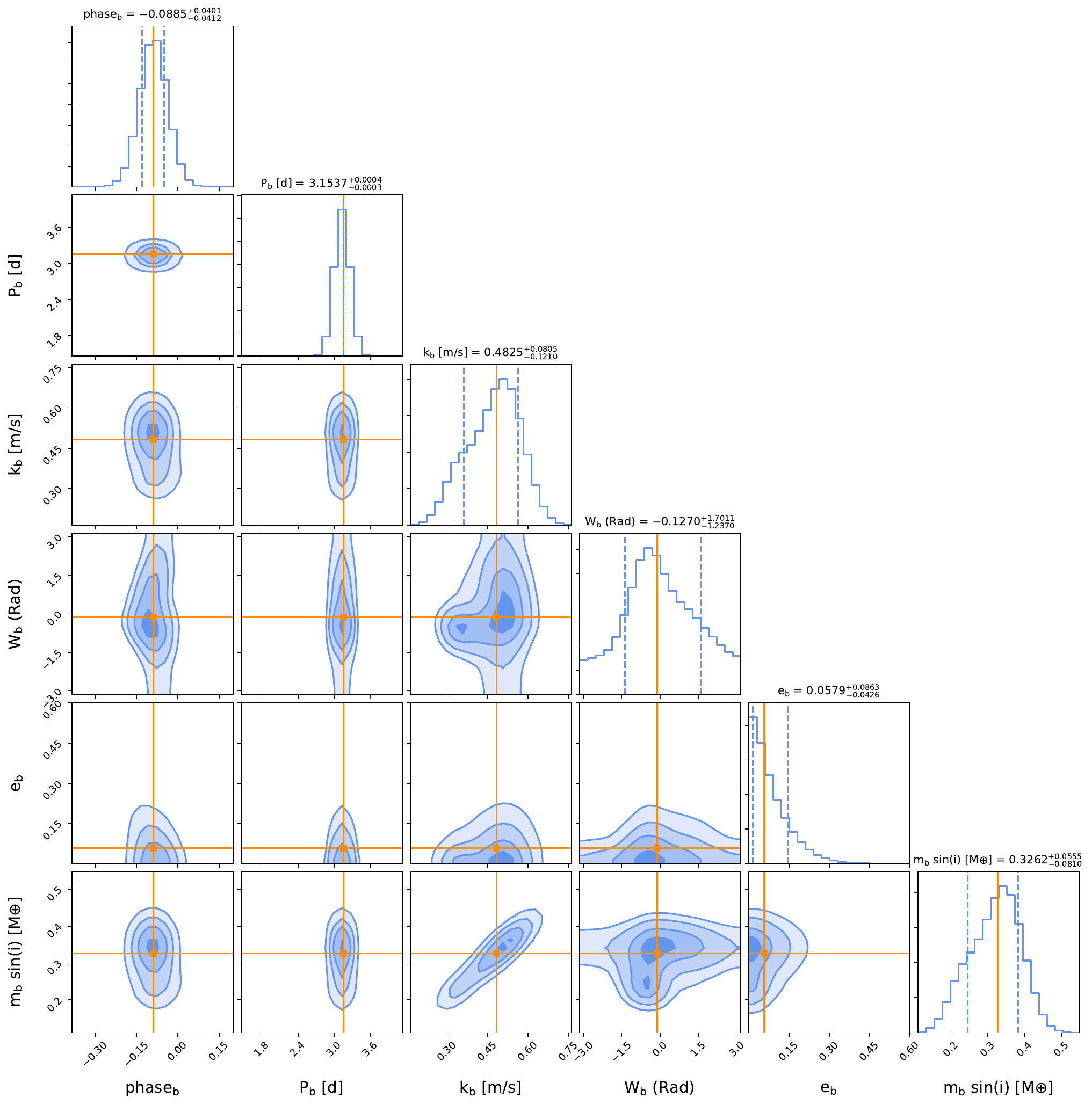}
\caption{Corner plot with the posterior distribution of the orbital parameters of the sub-Earth-mass planet of GJ~699 with a 3.15~d from the model using ESPRESSO, HARPS and HARPS-N data.
}
\label{gj699_esp_har_1pe_corner}
\end{figure}

\begin{figure*}[!h]
\includegraphics[width=18cm]{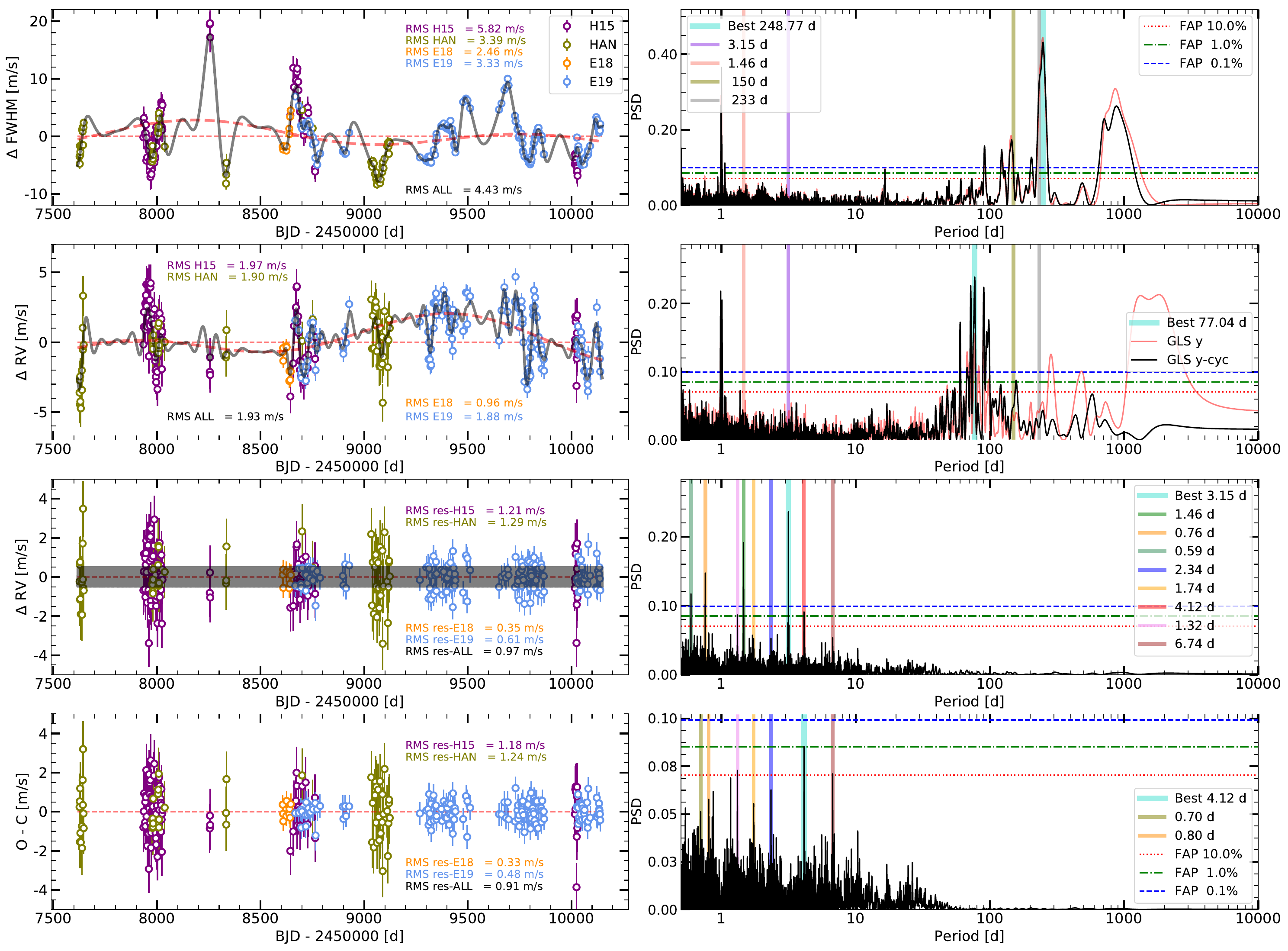}
\caption{FWHM measurements ({\it top}), RV measurements ({\it middle}), RV residuals from cycle and GP model ({\it next bottom}) and RV residuals from Keplerian model ({\it bottom}), of the ESPRESSO, HARPS and HARPS-N datasets using SHO ($P_{\rm ROT}$ and $P_{\rm ROT}/2$) GP model (grey solid line) to describe the activity caused by stellar rotation and including a double sinusoidal model to describe the long-term cycle (red thick dashed line), and a Keplerian model (grey shaded area), and GLS periodograms ({\it left}) of GJ~699 (model $L1$ in Table~\ref{tab:logz}). The uncertainties include the jitter term coming from the global model.
}
\label{gj699_esp_har}
\end{figure*}

In Fig.~\ref{gj699_esp_har} we show the FWHM and RV measurements, including ESPRESSO, HARPS and HARPS-N data, with the analysis of the GP+cycle model together with one Keplerian model using wide priors (model $J1$ in Table~\ref{tab:logz}). Figure~\ref{gj699_esp_har_1pe} depicts the RV curve in orbital phase of the sub-Earth mass planet at 3.15~d of model represented in Fig.~\ref{gj699_esp_har}. Figure~\ref{gj699_esp_har_1pe_corner} displays the corner plot of the orbit of the 3.15~d sub-Earth mass planet. In Fig.~\ref{gj699_esp_har_1pe_corner} we show the prior and posteriors distributions of all parameters of the model in Fig.~\ref{gj699_esp_har}. Finally, in Fig.~\ref{gj699_esp_har_car} we represent the FWHM and RV measurements, including ESPRESSO, HARPS, HARPS-N and CARMENES data, with the analysis of the GP+cycle model together with one Keplerian model using wide priors (model $L1$ in Table~\ref{tab:logz}). We also provide the prior, posterior and parameter values of this model $L1$ in Table~\ref{tab:modGPcyc1peCAR} and the model statistics in Table~\ref{tab:modGPcyc1peCARstat}. In Table~\ref{tab:alldataset} we provide RV and FWHM measurements of all datasets used in Fig.~\ref{gj699_esp_har_car}.

\begin{table}[!h]
\begin{center}
\caption{Parameters, priors and posteriors of model run $L1$ of Table~\ref{tab:logz} \label{tab:modGPcyc1peCAR}}
\begin{tabular}[centre]{lrr}
\hline \hline
Parameter  & Prior & Posterior   \\
\hline
\multicolumn{3}{c}{Offsets} \\
\hline
V0 FWHM$_{CAR}$         [\ms] & $\mathcal{N} (0,10)  $     & $ -1.16^{+1.84}_{-1.82} $ \\
V0 FWHM$_{H15}$         [\ms] & $\mathcal{N} (0,10)  $     & $ -1.05^{+1.30}_{-1.36} $ \\
V0 FWHM$_{HAN}$         [\ms] & $\mathcal{N} (0,10)  $     & $  2.48^{+1.31}_{-1.16} $ \\
V0 FWHM$_{E18}$         [\ms] & $\mathcal{N} (0,10)  $     & $ -0.20^{+1.66}_{-1.57} $ \\
V0 FWHM$_{E19}$         [\ms] & $\mathcal{N} (0,10)  $     & $ -0.78^{+1.33}_{-1.34} $ \\
V0 RV$_{CAR}$           [\ms] & $\mathcal{N} (0,3)   $     & $ -0.23^{+0.31}_{-0.21} $ \\
V0 RV$_{H15}$           [\ms] & $\mathcal{N} (0,3)   $     & $ -0.47^{+0.37}_{-0.27} $ \\
V0 RV$_{HAN}$           [\ms] & $\mathcal{N} (0,3)   $     & $ -1.10^{+0.40}_{-0.32} $ \\
V0 RV$_{E18}$           [\ms] & $\mathcal{N} (0,3)   $     & $  1.19^{+0.42}_{-0.30} $ \\
V0 RV$_{E19}$           [\ms] & $\mathcal{N} (0,3)   $     & $ -0.54^{+0.30}_{-0.23} $ \\
\hline
\multicolumn{3}{c}{Jitters} \\
\hline
$\ln$ Jit. FWHM$_{CAR}$ [\ms] & $\mathcal{N} (1.5,3) $     & $ -0.77^{+0.75}_{-1.47} $ \\
$\ln$ Jit. FWHM$_{H15}$ [\ms] & $\mathcal{N} (1.5,3) $     & $  0.28^{+0.16}_{-0.19} $ \\
$\ln$ Jit. FWHM$_{HAN}$ [\ms] & $\mathcal{N} (1.5,3) $     & $ -1.99^{+0.80}_{-0.92} $ \\
$\ln$ Jit. FWHM$_{E18}$ [\ms] & $\mathcal{N} (1.5,3) $     & $ -0.49^{+0.31}_{-0.37} $ \\
$\ln$ Jit. FWHM$_{E19}$ [\ms] & $\mathcal{N} (1.5,3) $     & $ -0.87^{+0.11}_{-0.13} $ \\
$\ln$ Jit. RV$_{CAR}$   [\ms] & $\mathcal{N} (0.5,1) $     & $ -0.07^{+0.07}_{-0.08} $ \\
$\ln$ Jit. RV$_{H15}$   [\ms] & $\mathcal{N} (0.5,1) $     & $  0.23^{+0.07}_{-0.07} $ \\
$\ln$ Jit. RV$_{HAN}$   [\ms] & $\mathcal{N} (0.5,1) $     & $  0.51^{+0.08}_{-0.08} $ \\
$\ln$ Jit. RV$_{E18}$   [\ms] & $\mathcal{N} (0.5,1) $     & $ -0.88^{+0.27}_{-0.32} $ \\
$\ln$ Jit. RV$_{E19}$   [\ms] & $\mathcal{N} (0.5,1) $     & $ -0.49^{+0.06}_{-0.08} $ \\
\hline
\multicolumn{3}{c}{Long-term cycle} \\
\hline
$\ln$ ACYC1 FWHM        [\ms] & $\mathcal{N} (1.5,1.5) $  & $  1.07^{+0.45}_{-0.56}   $ \\
$\ln$ ACYC1 RV          [\ms] & $\mathcal{N} (0.5,0.5) $  & $  0.29^{+0.17}_{-0.19}   $ \\
$\ln$ ACYC2 FWHM        [\ms] & $\mathcal{N} (1.5,1.5) $  & $  1.06^{+0.34}_{-0.45}   $ \\
$\ln$ ACYC2 RV          [\ms] & $\mathcal{N} (0.5,0.5) $  & $ -0.06^{+0.13}_{-0.17}   $ \\

$P_{\rm CYC}$             [d] & $\mathcal{U} (3250,300) $ & $ 3499^{+381}_{-483} $ \\

PH1 CYC FW                   & $\mathcal{U} (-0.5,0.5) $  & $ 0.23^{+0.05}_{-0.08}    $ \\
PH1 CYC RV                   & $\mathcal{U} (-0.1,0.6) $  & $ 0.33^{+0.05}_{-0.04}    $ \\
PH2 CYC FW                   & $\mathcal{U} ( 0.0,0.4) $  & $ 0.45^{+0.03}_{-0.03}    $ \\
PH2 CYC RV                   & $\mathcal{U} ( 0.0,0.3) $  & $ 0.33^{+0.02}_{-0.02}    $ \\
\hline
\multicolumn{3}{c}{SHO (P and P/2) GP} \\
\hline
$\ln$ A11 GP FWHM       [\ms] & $\mathcal{N} (1.5,3) $    & $  1.60^{+0.07}_{-0.08} $ \\
$\ln$ A12 GP FWHM       [\ms] & $\mathcal{N} (1.5,3) $    & $ -0.28^{+1.25}_{-1.49} $ \\
$\ln$ A21 GP FWHM       [\ms] & $\mathcal{N} (1.5,3) $    & $ -1.03^{+0.54}_{-1.26} $ \\
$\ln$ A22 GP FWHM       [\ms] & $\mathcal{N} (1.5,3) $    & $ -0.56^{+0.92}_{-1.43} $ \\
$\ln$ A11 GP RV         [\ms] & $\mathcal{N} (0.5,1) $    & $ -0.80^{+0.21}_{-0.38} $ \\
$\ln$ A12 GP RV         [\ms] & $\mathcal{N} (0.5,2) $    & $ -0.21^{+0.71}_{-1.02} $ \\
$\ln$ A21 GP RV         [\ms] & $\mathcal{N} (0.5,1) $    & $  0.39^{+0.09}_{-0.10} $ \\
$\ln$ A22 GP RV         [\ms] & $\mathcal{N} (0.5,2) $    & $ -0.22^{+1.05}_{-0.92} $ \\

$P_{\rm ROT}$             [d] & $\mathcal{U} (50,300) $    & $ 144.6^{+7.6}_{-9.2}   $ \\
$\ln T_{\rm ROT}$         [d] & $\mathcal{N} (3,2)    $    & $ 3.60^{+0.17}_{-0.16}  $ \\
\hline
\multicolumn{3}{c}{Keplerian orbit} \\
\hline
$\phi_b$                          & $\mathcal{U}  (-0.5,1) $ & $ -0.075 ^{+0.042 }_{-0.044}  $ \\
$\ln P_b$                     [d] & $\mathcal{LU} (0.5,50) $ & $ 1.1487 ^{+0.0001}_{-0.0001} $ \\
$k_b$                       [\ms] & $\mathcal{U}  (0,5)    $ & $ 0.373  ^{+0.065 }_{-0.059}  $ \\
$\sqrt{e_b} \cos(\omega_b)$       & $\mathcal{N}  (0,0.3)  $ & $ 0.08   ^{+0.13  }_{-0.19}   $ \\
$\sqrt{e_b} \sin(\omega_b)$       & $\mathcal{N}  (0,0.3)  $ & $ 0.08   ^{+0.14  }_{-0.15}   $ \\
\hline
\end{tabular}
\end{center}
\textbf{Notes:} Parameters with prior and posterior values of model $L1$ in Table~\ref{tab:logz} of the ESPRESSO, HARPS, HARPS-N and CARMENES RV and FWHM datasets, including a GP, cycle and Keplerian model, with model statistics given in Table~\ref{tab:modGPcyc1peCARstat}.
\end{table}

\begin{table}
\begin{center}
\caption{Statistics of model run $L1$ of Table~\ref{tab:logz} \label{tab:modGPcyc1peCARstat}}
\begin{tabular}[centre]{lr}
\hline \hline
Parameter  & Value    \\
\hline
$\ln \mathcal{Z}$              & $-4110.7$  \\
$\Delta \ln \mathcal{Z}$       & $+12.3$    \\
$N_{\rm par}$                  &   $44$     \\
$N_{\rm samples}$              &   $408244$ \\
\hline
RMS FWHM$_{CAR}$        [\ms]  &   10.96    \\
RMS FWHM$_{H15}$        [\ms]  &    5.70    \\
RMS FWHM$_{HAN}$        [\ms]  &    3.39    \\
RMS FWHM$_{E18}$        [\ms]  &    2.46    \\
RMS FWHM$_{E19}$        [\ms]  &    3.40    \\
RMS FWHM                [\ms]  &    8.94    \\
\hline
RMS  RV$_{CAR}$         [\ms]  &    2.18    \\
RMS  RV$_{H15}$         [\ms]  &    1.94    \\
RMS  RV$_{HAN}$         [\ms]  &    1.87    \\
RMS  RV$_{E18}$         [\ms]  &    0.96    \\
RMS  RV$_{E19}$         [\ms]  &    1.87    \\
RMS RV                  [\ms]  &    2.06    \\
\hline
RMS (O--C) FWHM$_{CAR}$ [\ms]  &    7.28    \\
RMS (O--C) FWHM$_{H15}$ [\ms]  &    2.13    \\
RMS (O--C) FWHM$_{HAN}$ [\ms]  &    0.93    \\
RMS (O--C) FWHM$_{E18}$ [\ms]  &    0.54    \\
RMS (O--C) FWHM$_{E19}$ [\ms]  &    0.45    \\
RMS (O--C) FWHM         [\ms]  &    5.73    \\
\hline
RMS (O--C)  RV$_{CAR}$  [\ms]  &    1.67    \\
RMS (O--C)  RV$_{H15}$  [\ms]  &    1.33    \\
RMS (O--C)  RV$_{HAN}$  [\ms]  &    1.64    \\
RMS (O--C)  RV$_{E18}$  [\ms]  &    0.29    \\
RMS (O--C)  RV$_{E19}$  [\ms]  &    0.51    \\
RMS (O--C) RV           [\ms]  &    1.47    \\
\hline
\end{tabular}
\end{center}
\end{table}

\begin{figure*}
\includegraphics[width=18cm]{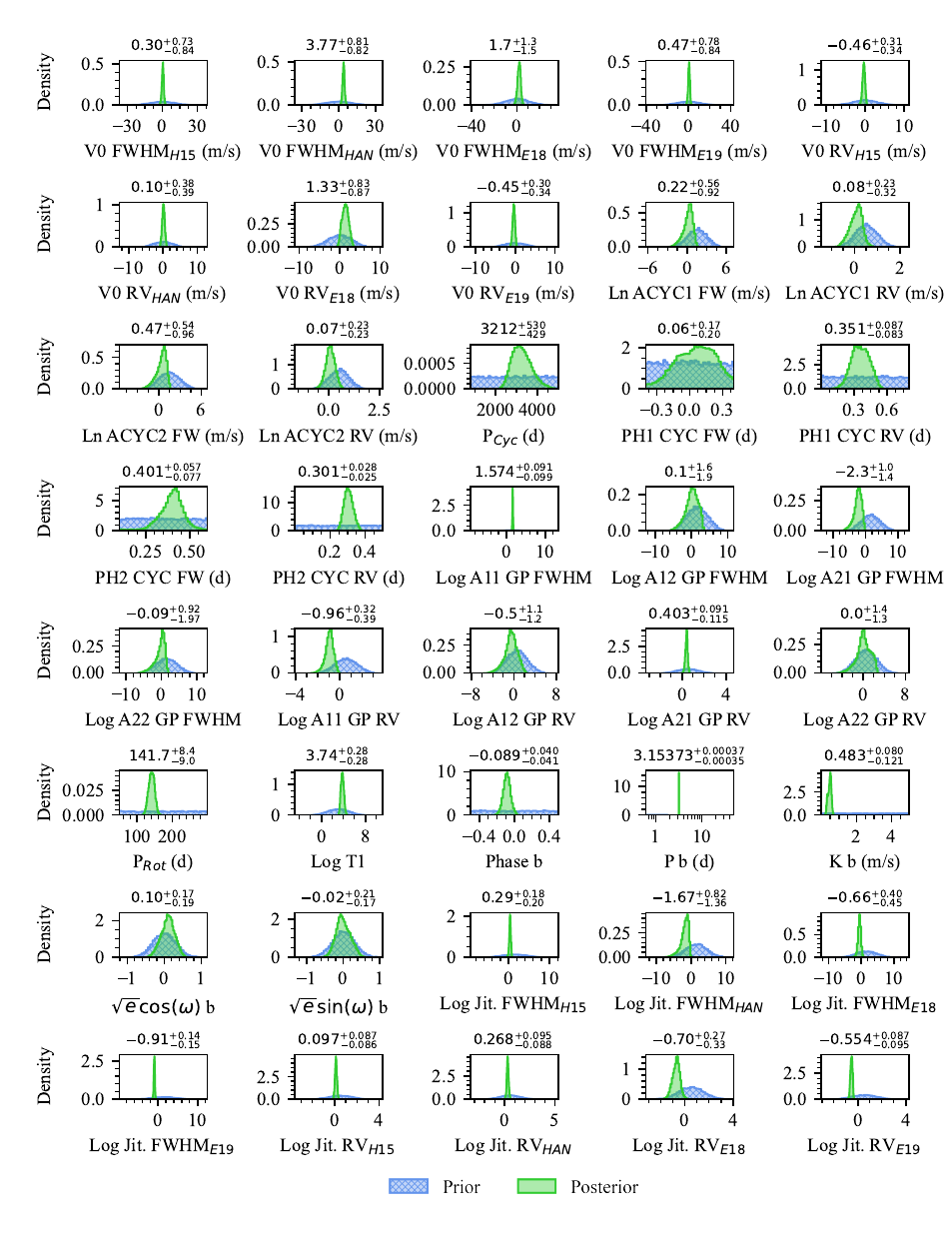}
\caption{
Prior and posterior distributions of all 40 parameters of model $J1$ in Table~\ref{tab:logz} of ESPRESSO, HARPS and HARPS-N data together, including offsets, jitter, the parameters of GP and the long-term cycle, and the parameters of the Keplerian model revealing the sub-Earth-mass planet of GJ~699 with a 3.15~d orbital period.
}
\label{gj699_esp_har_1pe_corner_all}
\end{figure*}

\FloatBarrier

\begin{figure*}
\includegraphics[width=18cm]{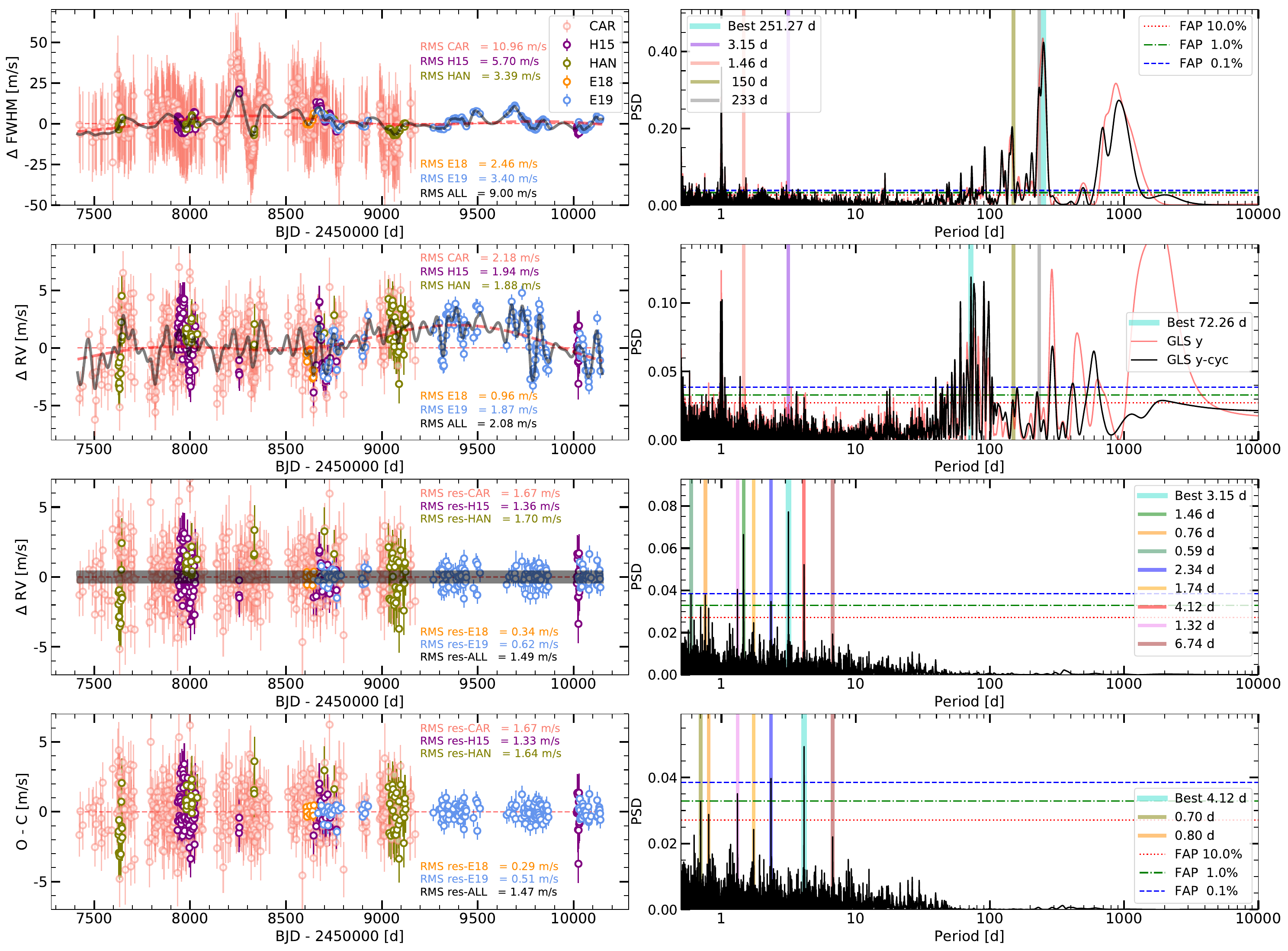}
\caption{
FWHM measurements ({\it top}), RV measurements ({\it middle}), RV residuals from cycle and GP model ({\it next bottom}) and RV residuals from Keplerian model ({\it bottom}), of the ESPRESSO, HARPS, HARPS-N and CARMENES datasets using SHO ($P_{\rm ROT}$ and $P_{\rm ROT}/2$) GP model (grey solid line) to describe the activity caused by stellar rotation and including a double sinusoidal model to describe the long-term cycle (red thick dashed line), and a Keplerian model (grey shaded area), and GLS periodograms ({\it left}) of GJ~699 (model $L1$ in Table~\ref{tab:logz} with prior and posterior parameters given in Table~\ref{tab:modGPcyc1peCAR}.). The uncertainties include the jitter term coming from the global model.
}
\label{gj699_esp_har_car}
\end{figure*}


\begin{table*}[!hb]
\begin{center}
\caption{TM RV and CCF FHWM measurements of all datasets \label{tab:alldataset}}
\begin{tabular}[centre]{lrrrrr}
\hline \hline
BJD-2450000  & RV (m/s)  & $\delta$RV (m/s)  & FWHM (m/s)  & $\delta$FWHM (m/s)  & SPEC  \\
\hline
7422.746621   &  -4.619540  &  1.215566  &  -1.813300  &  23.485450  &   CAR \\
7427.743544   &  -3.696041  &  1.056106  &  -7.264980  &  24.071170  &   CAR \\
7440.708032   &  -3.570952  &  1.457533  &  -5.332020  &  23.981810  &   CAR \\
...  &  ...  &  ...  &  ...  &  ...  &   ... \\
10135.524214  &  -0.598229  &  0.150586  &   2.058090  &   0.436766  &   E19 \\
10137.709347  &  -1.619432  &  0.132375  &   1.448422  &   0.403703  &   E19 \\
10139.512816  &  -2.547420  &  0.078461  &   2.590661  &   0.226382  &   E19 \\
\hline
\end{tabular}
\end{center}
\textbf{Notes:} Portion of the complete table at CDS of all datasets used in Fig.~\ref{gj699_esp_har_car} (model $L1$ in Table~\ref{tab:logz} with prior and posterior parameters given in Table~\ref{tab:modGPcyc1peCAR}) with the TM RVs computed with \texttt{S-BART} code for ESPRESSO, HARPS and HARPS-N data, and with \texttt{SERVAL} code for CARMENES data. CCF FWHMs are computed from DRS CCF profiles for ESPRESSO, HARPS and HARPS-N data, and with the \texttt{RACCOON} code for CARMENES data. CARMENES data is provided in the public DR1 in \citet{rib23}. Labels corresponding to the different spectra are given in column SPEC as CAR, E18, E19, H15 and HAN for CARMENES, ESPRESSO before and after the intervention in June 2019, HARPS after the intervention in 2015 and HARPS-N, respectively.
\end{table*}


\clearpage

\section{Candidate multi-planet system~\label{sec:4planet}}

The fact that the candidate signals are so weak makes it difficult to detect them in a blind search, so we decided to run guided search with normal priors around the period of the signals with a $\sigma = 0.3$~d (prior $\mathcal{N}(P_{\rm orb},0.3)$~d). We first search for the highest peak in the GLS periodogram after subtraction of the activity model, which is the 3.15~d signal. We adopt a prior on orbital period $P_{\rm orb}$ $\mathcal{N}(3.15,0.3)$~d and eccentricity $e$ with $\sqrt{e} ~\cos(\omega)$ and $\sqrt{e} ~\sin(\omega)$ $\mathcal{N}(0,0.05)$, basically an almost circular orbit, based on previous results. This is model $F1$ in Table~\ref{tab:logzN}. The Bayesian evidence, though, may be not that informative as compared with the blind search runs because the prior on period is very narrow. However, it indicates that the global fit model reduces significantly the RMS of the RV residuals from 0.60 to 0.45~\ms. The next peak in the RV residuals is 4.12~d, so we then run a two-planet model $F2$ with priors centered at period 3.15~d and 4.12d, with an increase in $\Delta \ln \mathcal{Z} = +3.9$ with respect to the single-planet model $F1$. In the RV residuals appears the peak at 2.34~d significantly above the 0.1\% FAP line.

\begin{table}[!h]
\begin{center}
\caption{Bayesian evidence of different models\label{tab:logzN}}
\begin{tabular}[centre]{llrrrr}
\hline \hline
Name & Model  & $N_{\rm pl}$ & $N_{\rm par}$ & $\ln \mathcal{Z}$ & $\Delta \ln \mathcal{Z}$ \\
\hline
\multicolumn{6}{c}{E18,E19 ($N_{\rm point}=149 \times 2$)} \\
\hline
{\bf D}      & {\bf A+cycN} & {\bf 0} & {\bf 26} & {\bf --447.2} & {\bf 0.0} \\
F1           & D+1pe0N0.3  & 1 & 31 &  $-437.3$ & $+9.9$  \\
F1c          & D+1pcN0.3   & 1 & 29 &  $-438.0$ & $+9.2$  \\
F2           & D+2pe0N0.3  & 2 & 36 &  $-433.4$ & $+13.8$ \\
F3           & D+3pe0N0.3  & 3 & 41 &  $-429.5$ & $+17.7$ \\
F4           & D+4pe0N0.3  & 4 & 46 &  $-429.0$ & $+18.2$ \\
F4c          & D+4pcN0.3   & 4 & 38 &  $-429.1$ & $+18.1$ \\
\hline
\end{tabular}
\end{center}
\textbf{Notes:} Model selection based on Bayesian evidence of the analysis of CCF FWHM and TM RV measurements. Same as in Table~\ref{tab:logz} but for normal priors around the candidate periods $N0.3$ for $\mathcal{N}(P_{\rm orb},\sigma_P)$ with $P_{\rm orb}$ the orbital period and $\sigma_P=0.3$~d. $npe$ and $npc$ indicate $n$ Keplerian and circular orbits, and $npe0$ indicates a Keplerian with narrow prior on eccentricity $e$ with $\sqrt{e} ~\cos(\omega)$ and $\sqrt{e} ~\sin(\omega)$ $\mathcal{N}(0,0.05)$. We highlight in bold fonts the reference activity-only model in this dataset, as in Table~\ref{tab:logz}.
\end{table}

We run a three-planet model for periods 3.15~d, 4.12d and 2.34~d, again with narrow priors on eccentricity. The resulting run $F3$ has a higher Bayesian evidence by $\Delta \ln \mathcal{Z} = +3.9$ over the two-planet model $F2$. We also check the possibility of a wider prior on eccentricity ($e$ with $\sqrt{e} ~\cos(\omega)$ and $\sqrt{e} ~\sin(\omega)$ $\mathcal{N}(0,0.3)$) of the three-planet model. This model tend to favor very low eccentricity (consistent with zero) for the signals 3.15~d and 4.12~d, but it goes to high eccentricity values for the signal at 2.34~d. In this case, the RV residuals do not leave any peak above the 0.1\% FAP line. However, such a system with short-period candidate planets with the inner one being a high eccentric planet would not be stable in the long term. In the case of low priors on eccentricity, model $F3$ reveals a peak at 6.74~d in the GLS periodogram of the RV residuals.

\begin{figure*}
\includegraphics[width=18cm]{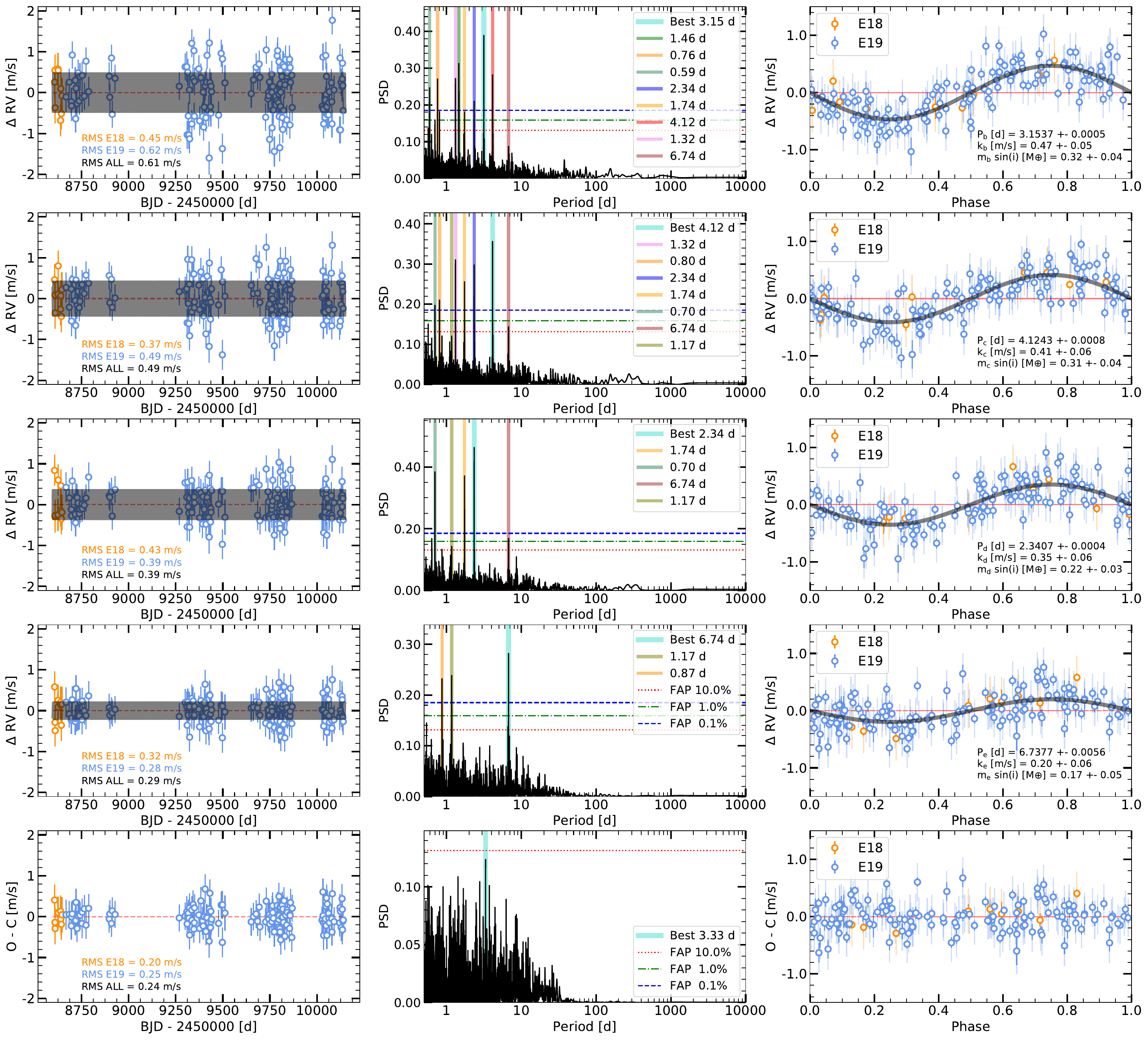}
\caption{
    ESPRESSO RV measurements versus BJD after removing the long-term cycle and the GP model ({\it Left}), GLS periodograms ({\it middle}), and RV curves of the sub-Earth-mass planet candidates ({\it right}) of GJ~699 with  3.15~d, 4.12~d,  2.34~d and 6.74~d orbital periods together with ESPRESSO RVs with uncertainties with (light colour) and without including the jitter term (dark colour) coming from the global model $F4$ in Table~\ref{tab:logzN}.
}
\label{gj699_4pe}
\end{figure*}

\begin{figure*}
\includegraphics[width=18cm]{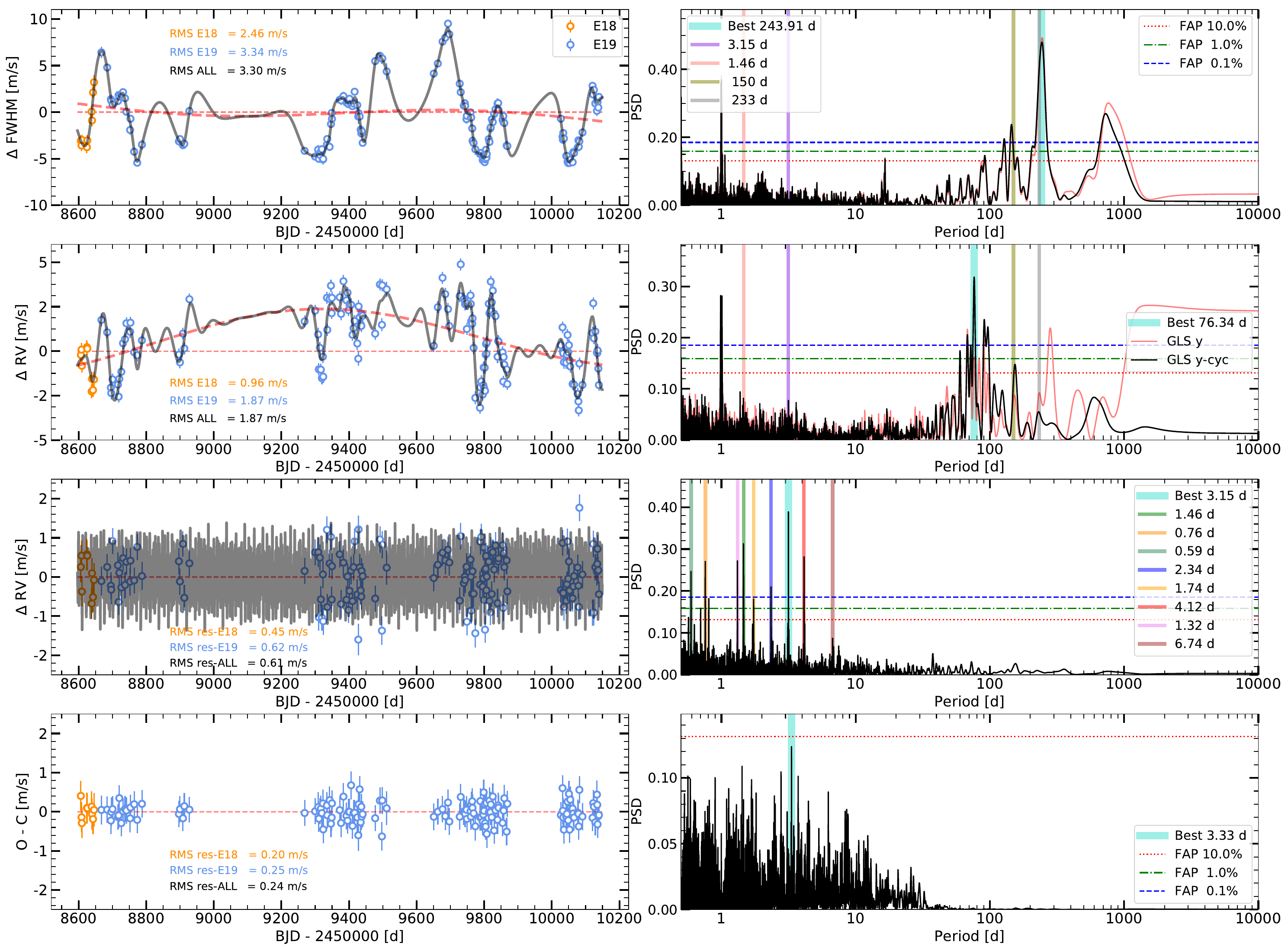}
\caption{
FWHM measurements ({\it top}), RV measurements ({\it middle}), RV residuals from cycle and GP model ({\it next bottom}) and RV residuals from four Keplerian model ({\it bottom}), of the ESPRESSO, HARPS, HARPS-N and CARMENES datasets using SHO ($P_{\rm ROT}$ and $P_{\rm ROT}/2$) GP model to describe the activity caused by stellar rotation and including a double sinusoidal model to describe the long-term cycle, and four Keplerian models, and GLS periodograms ({\it left}) of GJ~699 (model $F4$ in Table~\ref{tab:logzN}). The uncertainties include the jitter term coming from the global model.
}
\label{gj699_esp_4pe_gls}
\end{figure*}

We then run a four-planet model, with a narrow prior on eccentricity $e$ with $\sqrt{e} ~\cos(\omega)$ and $\sqrt{e} ~\sin(\omega)$ $\mathcal{N}(0,0.05)$, for the fourth signals with narrow priors centered at 3.15~d, 4.12~d, 2.34~d and 6.74~d. This model $F4$ has the same Bayesian evidence as $F3$, only increasing by $\Delta \ln \mathcal{Z} = +0.5$. We checked that using this narrow prior on eccentricity gives almost the same result as using sinusoids as circular orbits (models $F1c$ and $F4c$ in Table~\ref{tab:logzN}), for both models $F1$ and $F4$. Fig.~\ref{gj699_4pe} summarises the result of run $F4$, where we display just the RVs after subtracting the activity model displayed in Fig.~\ref{gj699_esp_4pe_gls}. Although $F4$ is a simultaneous fit of a four-planet Keplerian model, we represent in Fig.~\ref{gj699_4pe} the RVs after subtracting first the activity model, and then in each row, after removing the signal at the highest peak in the GLS periodograms given in middle panels, one by one. The left panels show the RVs in phase of the given planet period after subtracting the activity model and the rest of the fitted planetary signals. In the middle panels of Fig.~\ref{gj699_4pe}, the GLS periodograms of the RVs (displayed in left panels) show that after subtracting the last candidate signal at 6.74~d, there are no more signals above the 10\% FAP line. We also see, in the central panels of Fig.~\ref{gj699_esp_4pe_gls}, several peaks corresponding to the main signals and their 1 d alias, 3.15~d (with its 1 d alias 1.46~d and 0.76~d, and with 0.59~d as 1 d alias of 1.46~d), 4.12~d (with its 1 d alias 1.32~d and 0.80~d), 2.34~d (with its 1 d alias 1.74~d and 0.70~d), and 6.74~d (with its 1 d alias 1.17~d and 0.87~d). The RMS of the ESPRESSO RVs goes from 0.61~\ms in the RVs after subtracting the activity cycle+rotation model to 0.49, 0.39, 0.29, and 0.24~\ms\ after removing the candidate planet signals at 3.15, 4.12, 2.34, and 6.74~d (see Figs.~\ref{gj699_4pe} and~\ref{gj699_esp_4pe_gls}). The semi-amplitude velocities of run $F4$ are $k_p= 47$, 41, 35 and 20~\cms, thus providing the candidate planetary minimum masses of 0.32, 0.31, 0.22, 0.17~\mearth\ for candidate planets tentatively labelled as $b$, $c$, $d$ and $e$ (see Table~\ref{tab:4planet}).

\begin{figure*}
\includegraphics[width=18cm]{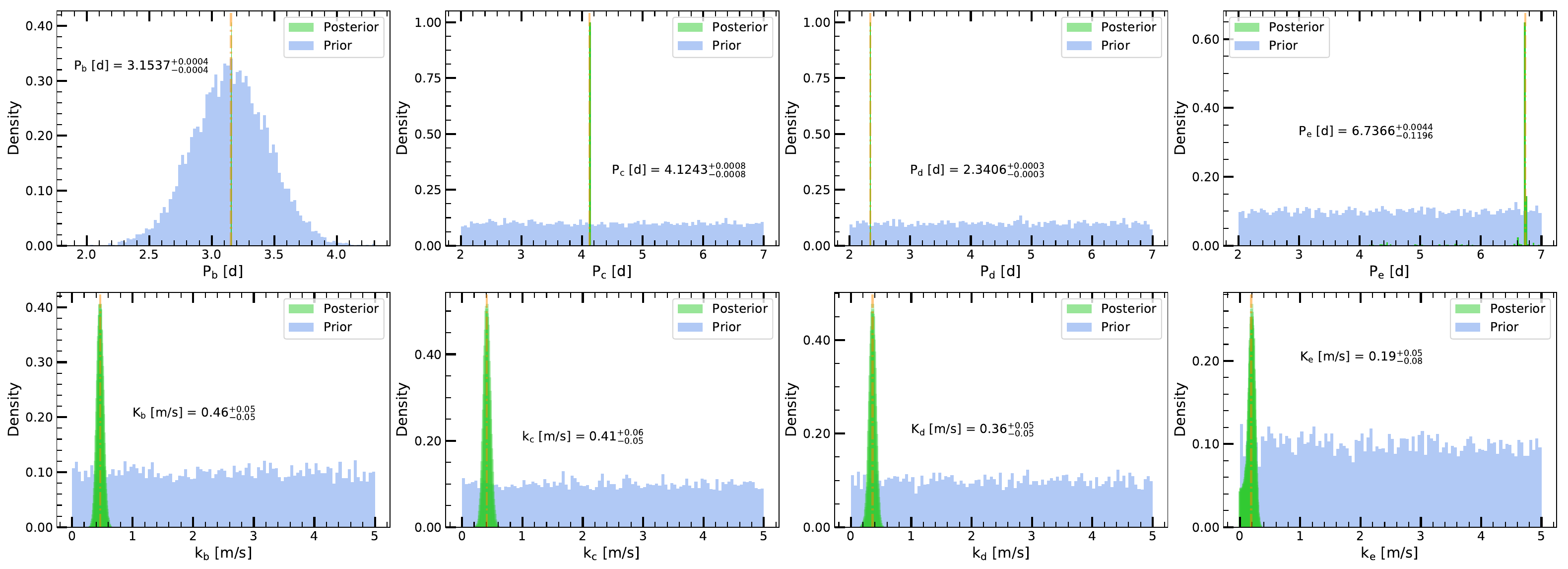}
\caption{
    Prior and posterior distributions of the period and semi-amplitude from the global analysis of FWHM and RV measurements of the blind search of the candidate planet system using ESPRESSO data. We use a normal prior on orbital period $\mathcal{N}(3.15,0.3)$~d for the confirmed planet b ({\it left}), and an uniform prior $\mathcal{U}(2,7)$~d for planet c ({\it center-left}), planet d ({\it center-right}), and planet e ({\it right}). For all planets we use an uniform prior $\mathcal{U}(0,5)$~\ms\ on semi-amplitude velocity.}
\label{gj699_4pc}
\end{figure*}

Finally, we also tested a run with the first confirmed planet as a circular orbit with a narrow prior $\mathcal{N}(3.15,0.3)$~d on orbital period and three additional planetary signals also as circular orbits but uniform priors $\mathcal{U}(2,7)$~d on orbital period. We use for all candidate planetary signals an uniform prior $\mathcal{U}(0,5)$~\ms\ on semi-amplitude velocity. We recover the same solution as in models $F4$ and $F4c$ in Table~\ref{tab:logzN}, but the Bayesian evidence of this model is just the same as for model $F1c$ in Table~\ref{tab:logzN}, i.e. $\ln \mathcal{Z} = -438.0$. Figure~\ref{gj699_4pc} shows the prior and posterior distributions of the orbital period and semi-amplitude velocities of the four candidate planets. The orbital periods are unequivocally retrieved for all planets, in particular for the first three planets, and for the fourth planet with just a few samples falling on both 1-yr alias (6.61~d and 6.86~d) of the main period 6.74~d. The semi-amplitudes are also unequivocally recovered for the first three planets at 9.2, 8.2 and 7.2~$\sigma$, whereas for the fourth planet it is only recovered at 2.4~$\sigma$.

\begin{table}
\begin{center}
\caption{Parameters of the candidate planetary system \label{tab:4planet}}
\begin{tabular}[centre]{l c}
\hline \hline
Parameter  & Candidate Planet $b$  \\
\hline
$T_{0,b}$ -- 2460139 [d]        & 0.290 $\pm$ 0.104   \\
$P_b$ [d]                       & 3.1537 $\pm$ 0.0005 \\
$k_b$ [\ms]                     & 0.47 $\pm$ 0.05     \\
$m_b \sin i$ [\mearth]          & 0.32 $\pm$ 0.04     \\
$a_b$ [AU]                      & 0.0229 $\pm$ 0.0003 \\
$e_b$                           & \textless~0.01      \\
$S_b$ [\searth]                 & 6.76 $\pm$ 0.05     \\
$T_{\rm eq,b,}$$_{A = 0.3}$ [K] & 400 $\pm$ 7         \\
\hline
Parameter  & Candidate Planet $c$  \\
\hline
$T_{0,c}$ -- 2460139 [d]        & 0.730 $\pm$ 0.129   \\
$P_c$ [d]                       & 4.1243 $\pm$ 0.0008 \\
$k_c$ [\ms]                     & 0.41 $\pm$ 0.06     \\
$m_c \sin i$ [\mearth]          & 0.31 $\pm$ 0.04     \\
$a_c$ [AU]                      & 0.0274 $\pm$ 0.0004 \\
$e_c$                           & \textless~0.01      \\
$S_c$ [\searth]                 & 4.73 $\pm$ 0.04     \\
$T_{\rm eq,c,}$$_{A = 0.3}$ [K] & 366 $\pm$ 6         \\
\hline
Parameter  & Candidate Planet $d$  \\
\hline
$T_{0,d}$ -- 2460138 [d]        & 0.458 $\pm$ 0.101   \\
$P_d$ [d]                       & 2.3407 $\pm$ 0.0004 \\
$k_d$ [\ms]                     & 0.35 $\pm$ 0.06     \\
$m_d \sin i$ [\mearth]          & 0.22 $\pm$ 0.03     \\
$a_d$ [AU]                      & 0.0188 $\pm$ 0.0003 \\
$e_d$                           & \textless~0.01      \\
$S_d$ [\searth]                 & 10.06 $\pm$ 0.08    \\
$T_{\rm eq,d,}$$_{A = 0.3}$ [K] & 441 $\pm$ 8         \\
\hline
Parameter  & Candidate Planet $e$  \\
\hline
$T_{0,e}$ -- 2460137 [d]        & 0.405 $\pm$ 0.551   \\
$P_e$ [d]                       & 6.7377 $\pm$ 0.0056 \\
$k_e$ [\ms]                     & 0.20 $\pm$ 0.06     \\
$m_e \sin i$ [\mearth]          & 0.17 $\pm$ 0.05     \\
$a_e$ [AU]                      & 0.0381 $\pm$ 0.0006 \\
$e_e$                           & \textless~0.01      \\
$S_e$ [\searth]                 & 2.45 $\pm$ 0.03     \\
$T_{\rm eq,e,}$$_{A = 0.3}$ [K] & 310 $\pm$ 5         \\
\hline
\end{tabular}
\end{center}
\end{table}

\subsection{Stability of the candidate multi-planet system}

We evaluate the stability of the 4-planet candidate system using the \texttt{SPOCK}\footnote{\url{https://github.com/dtamayo/spock}} tool~\citep{tam20}, together with the Nbody Regressor~\citep{hus20} and the \texttt{REBOUND}\footnote{\url{https://github.com/ hannorein/rebound}} code~\citep[e.g.][]{rei17}. We found that assuming the zero eccentricity\footnote{all planets in model $F4$ have posterior eccentricity values $e_p < 0.01$, by definition from narrow priors on eccentricity} for all planets, the system remains stable more than $10^9$ orbits of the inner candidate planet at 2.34~d, which is the maximum Nbody instability time explored by default by this particular Nbody regressor. We run this calculation using the planet masses assuming $m_p = m_p \sin i$ (thus $i=90$~degrees), but also we run it again assuming by $m_p = (m_p \sin i) \times 3$. This latter conservative case would mean an orbital inclination of $i=19.5$~degrees. In both cases the result is the same.
    
The candidate system architecture is displayed using the \texttt{REBOUND} plotting tool in Fig.~\ref{gj699_hz_4pe}. We see all these candidate planets in orbits inner to the habitable zone limits estimated in Section~\ref{sec:stepar}. The candidate four-planet system looks particularly compact, similar to other systems such as e.g. the TRAPPIST seven-planet system~\citep{gil17Natur}. \citet{dre24} have recently revisited the planetary system around the Teegarden's star with a detection and confirmation of one additional Earth-mass planet in a 26~d orbit, and additional candidate planets. These authors compare the Teegarden's system with other compact systems with their planets orbiting within 0.1~AU, similar to the candidate planets of Barnard's star. They highlight mutual Hill radius separation as a good indicator of the dynamic compactness of a planetary system, also called fractional orbital separation~\citep{gla93}. The mutual Hill radius separation is calculated from the stellar mass, $M_\star$, the planetary masses, $m_p$, and semi-major axes a of two neighbouring planets j and j+1 as in eq.~\ref{eq_rhill} :
    \begin{equation} \label{eq_rhill}
    \Delta (R_H) = 2 {{a_{j+1}-a_j} \over {a_{j+1}+a_j}} \left( {{3 M_\star} \over {m_{j+1}+m_j}} \right)^{1/3}
    \end{equation}
    \noindent \citet{dre24} compare the values of $\Delta (R_H)$ of planets in Teegarden's star with other six compact systems. Although the sample is small, they argue that planet systems with more massive planets have mutual Hill radius separations above $\Delta (R_H) \ge 30$ while the ones with low-mass, likely terrestrial, planets have $\Delta (R_H) \le 20$. \citet{wei18} indicated that systems with more planets tend to have smaller $\Delta (R_H)$ values, but only about 10\% of the systems have a $\Delta (R_H) < 10$. Simulations of the stability evolution of compact multi-planet systems suggest a log-linear dependency with the mutual Hill radius separation~\citep[e.g.][]{cha96}. \citet{cha96} found that systems with mutual Hill radius separation $\Delta (R_H)$ less than 10 are always unstable. These simulations suggest that at about 13 mutual Hill radius separations, the orbital crossing timescale seems to be longer than at least $10^9$ orbits for all systems~\citep{gra21,ric23}. Below this limit the orbital crossing timescale changes by more than one order of magnitude for small changes in mutual Hill separation. The seven-planet system TRAPPIST-1 with planet masses from 0.33 to 1.37~\mearth\ at orbital periods from 1.5~d to 18.8~d \citep{ago21} have all values $\Delta (R_H)$ below 13. Similarly, the three-planet system YZCeti with planets at $P_{\rm orb}=$~2.02, 3.06 and 4.66~d with minimum masses 0.70, 1.14, and 1.09 \mearth\, respectively \citep{ast17,sto20} have also all mutual Hill radius separations in the range $10 < \Delta (R_H) < 13$ \citep[see Fig.~6 in][]{dre24}. The three-planet system orbiting Teegarden's star at periods 4.9~d, 11.4~d, and 26.1~d with minimum masses 1.16, 1.05 and 0.82~\mearth\ have $\Delta (R_H)=19.3$ and 19.7, for planet pairs $bc$ and $cd$, respectively. They also suggest additional candidate planets at 7.7~d and 17~d, which would move the mutual Hill radius separations of all planet pairs of the candidate five-planet system at $\Delta (R_H) < 13$~\citep{dre24}. On the other hand, the two-planet systems GJ~1002 with planets at 10.3~d and 21.2~d with minimum masses 1.1~and 1.4~\mearth\ \citep{sua23} and Proxima Cen system with the planets Proxima~d and Proxima~b at periods 5.1~d and 11.2~d with minimum masses 0.26~\mearth\ and 1.1\mearth\ \citep{far22}, have mutual Hill radius separations larger than 13, specifically at $\Delta (R_H) = 17.2$ and 23.0 for planet pairs $bc$ in GJ 1002 and $db$ in Proxima Centauri. The sub-Earth mass candidate planets in Barnard's star system have $\Delta (R_H)$ of 13.3, 11.3 and 22.6 for planet pairs $db$, $bc$ and $ce$, respectively. The two inner and outer pairs of planets $db$ and $ce$ have a $\Delta (R_H) > 13$, whereas the middle pair of planets $bc$ would have $\Delta (R_H) < 13$ but larger than 10. Whether this could cause orbital instabilities to the candidate four-planet system would require further investigation.
    
\begin{figure}[!h]
\includegraphics[width=9cm]{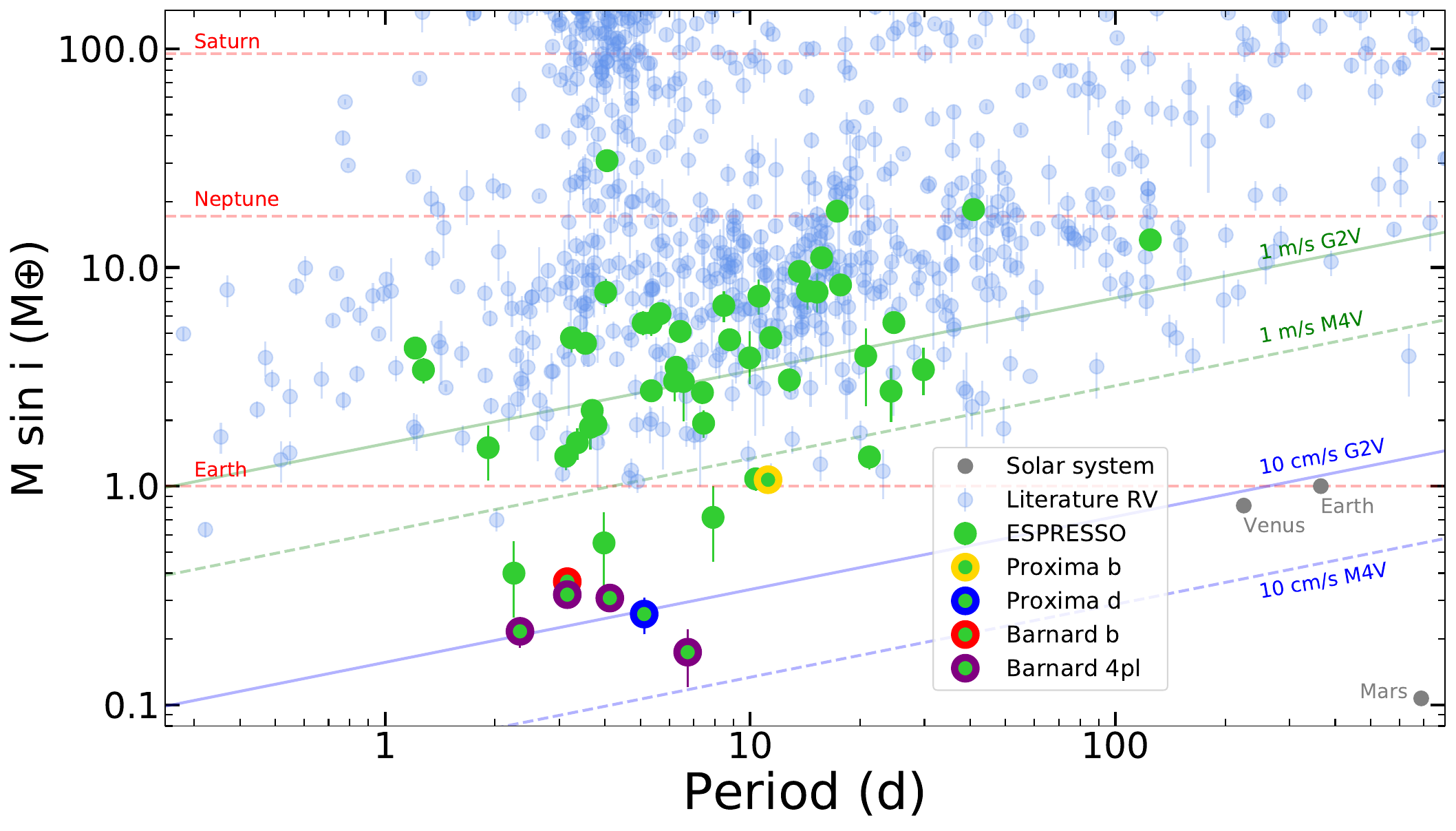}
\caption{Same as Fig.~\ref{gj699_planet_mass} with the confirmed planet Barnard b (red), and planet candidates (purple) of the 4-planet candidate system orbiting Barnard's star, together with the two planets orbiting Proxima Cen, Proxima b (yellow) and d (blue) are highlighted.
}
\label{gj699_planet_mass_4pl}
\end{figure}

\begin{figure}[!h]
\includegraphics[width=7.5cm]{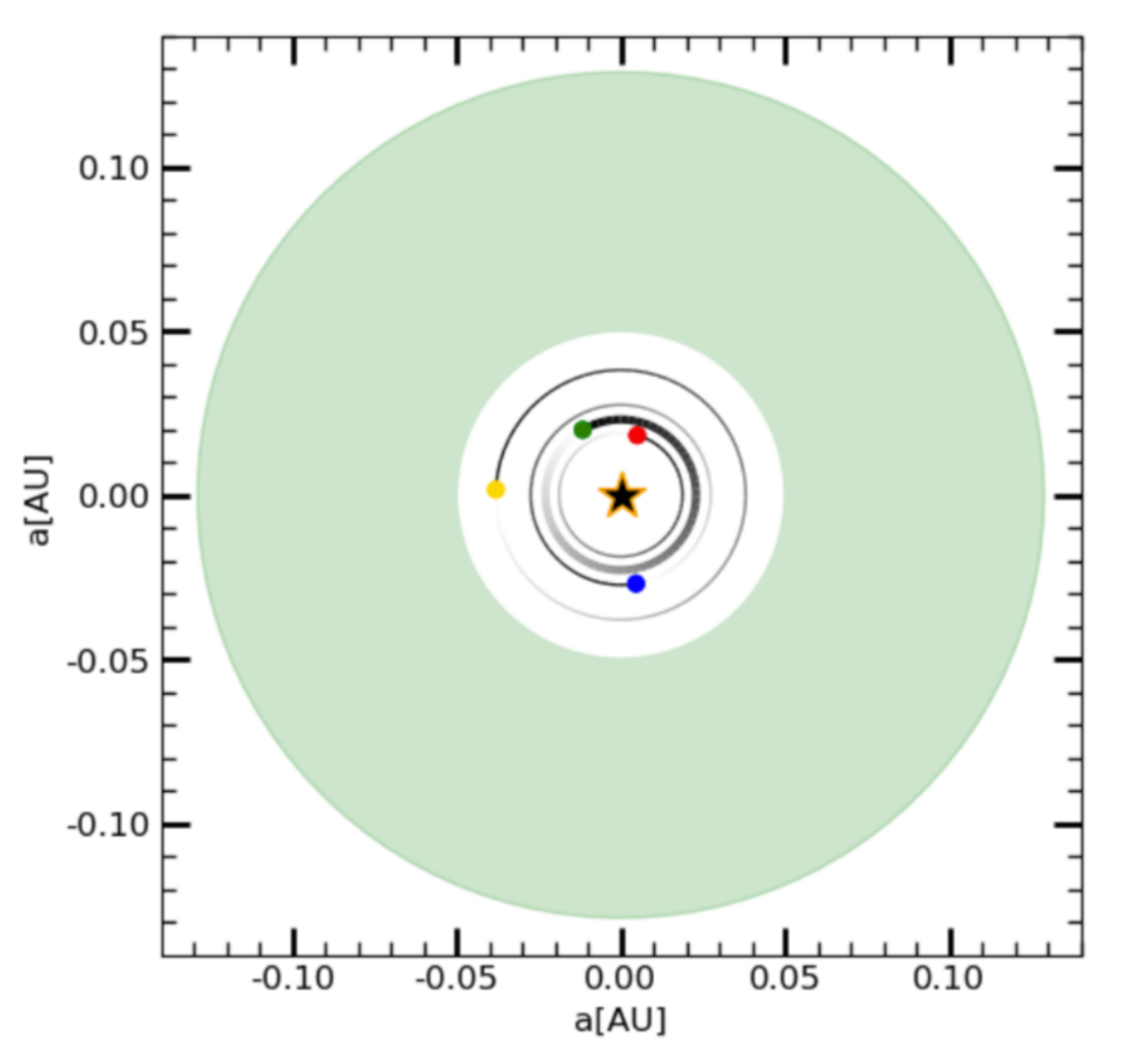}
\caption{Schematic view of the candidate planetary system of GJ 699 together with the habitable zone (light green region). The planets are depicted as green (planet b) circle, blue (planet c), red (planet d), and yellow (planet e).
}
\label{gj699_hz_4pe}
\end{figure}

The candidate planetary system orbiting our second closest neighbour Barnard's star consists of four sub-Earth mass planets at periods 2.34~d, 3.15~d, 4.12~d, and 6.74~d with minimum masses 0.22, 0.32, 0.31, 0.17~\mearth. Fig.~\ref{gj699_planet_mass_4pl} compares the confirmed planet Barnard b and the candidate planets with other planets in the exoplanet database.

\end{appendix}

\end{document}